\documentclass[twocolumn,twocolappendix]{aastex631}

\usepackage{amsmath}
\usepackage{xcolor}
\usepackage{comment}
\usepackage{graphicx}
\usepackage{float} 

\usepackage{array}
\newcolumntype{s}{@{\hskip 1pt}c@{\hskip 1pt}} \newsavebox\ltmcbox

\shorttitle{Dynamical Masses in Seven High-acceleration Star Systems}
\shortauthors{Giovinazzi et al.}
\DeclareOldFontCommand{\bf}{\normalfont\bfseries}{\mathbf} \providecommand{\DIFadd}[1]{{\bf #1}} \providecommand{\DIFdel}[1]{} \providecommand{\DIFaddbegin}{} \providecommand{\DIFaddend}{} \providecommand{\DIFdelbegin}{} \providecommand{\DIFdelend}{}   \providecommand{\DIFaddFL}[1]{\DIFadd{#1}} \providecommand{\DIFdelFL}[1]{\DIFdel{#1}} \providecommand{\DIFaddbeginFL}{} \providecommand{\DIFaddendFL}{} \providecommand{\DIFdelbeginFL}{} \providecommand{\DIFdelendFL}{} \newcommand{\DIFscaledelfig}{0.5}
\RequirePackage{settobox} \RequirePackage{letltxmacro} \newsavebox{\DIFdelgraphicsbox} \newlength{\DIFdelgraphicswidth} \newlength{\DIFdelgraphicsheight} \LetLtxMacro{\DIFOincludegraphics}{\includegraphics} \newcommand{\DIFaddincludegraphics}[2][]{{\color{blue}\fbox{\DIFOincludegraphics[#1]{#2}}}} \newcommand{\DIFdelincludegraphics}[2][]{\sbox{\DIFdelgraphicsbox}{\DIFOincludegraphics[#1]{#2}}\settoboxwidth{\DIFdelgraphicswidth}{\DIFdelgraphicsbox} \settoboxtotalheight{\DIFdelgraphicsheight}{\DIFdelgraphicsbox} \scalebox{\DIFscaledelfig}{\parbox[b]{\DIFdelgraphicswidth}{\usebox{\DIFdelgraphicsbox}\\[-\baselineskip] \rule{\DIFdelgraphicswidth}{0em}}\llap{\resizebox{\DIFdelgraphicswidth}{\DIFdelgraphicsheight}{\setlength{\unitlength}{\DIFdelgraphicswidth}\begin{picture}(1,1)\thicklines\linethickness{2pt} {\color[rgb]{1,0,0}\put(0,0){\framebox(1,1){}}}{\color[rgb]{1,0,0}\put(0,0){\line( 1,1){1}}}{\color[rgb]{1,0,0}\put(0,1){\line(1,-1){1}}}\end{picture}}\hspace*{3pt}}} } \LetLtxMacro{\DIFOaddbegin}{\DIFaddbegin} \LetLtxMacro{\DIFOaddend}{\DIFaddend} \LetLtxMacro{\DIFOdelbegin}{\DIFdelbegin} \LetLtxMacro{\DIFOdelend}{\DIFdelend} \DeclareRobustCommand{\DIFaddbegin}{\DIFOaddbegin \let\includegraphics\DIFaddincludegraphics} \DeclareRobustCommand{\DIFaddend}{\DIFOaddend \let\includegraphics\DIFOincludegraphics} \DeclareRobustCommand{\DIFdelbegin}{\DIFOdelbegin \let\includegraphics\DIFdelincludegraphics} \DeclareRobustCommand{\DIFdelend}{\DIFOaddend \let\includegraphics\DIFOincludegraphics} \LetLtxMacro{\DIFOaddbeginFL}{\DIFaddbeginFL} \LetLtxMacro{\DIFOaddendFL}{\DIFaddendFL} \LetLtxMacro{\DIFOdelbeginFL}{\DIFdelbeginFL} \LetLtxMacro{\DIFOdelendFL}{\DIFdelendFL} \DeclareRobustCommand{\DIFaddbeginFL}{\DIFOaddbeginFL \let\includegraphics\DIFaddincludegraphics} \DeclareRobustCommand{\DIFaddendFL}{\DIFOaddendFL \let\includegraphics\DIFOincludegraphics} \DeclareRobustCommand{\DIFdelbeginFL}{\DIFOdelbeginFL \let\includegraphics\DIFdelincludegraphics} \DeclareRobustCommand{\DIFdelendFL}{\DIFOaddendFL \let\includegraphics\DIFOincludegraphics} \RequirePackage{listings} \lstdefinelanguage{DIFcode}{ moredelim=[il][\color{white}\tiny]{\%DIF\ <\ }, moredelim=[il][\sffamily\bfseries]{\%DIF\ >\ } } \lstdefinestyle{DIFverbatimstyle}{ language=DIFcode, basicstyle=\ttfamily, columns=fullflexible, keepspaces=true } \lstnewenvironment{DIFverbatim}{\lstset{style=DIFverbatimstyle}}{} \lstnewenvironment{DIFverbatim*}{\lstset{style=DIFverbatimstyle,showspaces=true}}{} 

\renewcommand{\DIFadd}[1]{#1}
\renewcommand{\DIFaddFL}[1]{#1}
\renewcommand{\DIFdel}[1]{} 
\renewcommand{\DIFdelFL}[1]{}
\renewcommand{\DIFaddbegin}{} 
\renewcommand{\DIFaddend}{} 
\renewcommand{\DIFdelbegin}{} 
\renewcommand{\DIFdelend}{} 
\renewcommand{\DIFaddbeginFL}{} 
\renewcommand{\DIFaddendFL}{} 
\renewcommand{\DIFdelbeginFL}{} 
\renewcommand{\DIFdelendFL}{}

\begin{document}

\title{The NEID Earth Twin Survey. II. Dynamical Masses in Seven High-acceleration Star Systems}

\author[0000-0002-0078-5288]{Mark R. Giovinazzi}
\affiliation{Department of Physics and Astronomy, University of Pennsylvania, 209 South 33rd Street, Philadelphia, PA 19104 USA}
\affiliation{Department of Physics \& Astronomy, Amherst College, 25 East Drive, Amherst, MA 01002, USA}

\author[0000-0002-6096-1749]{Cullen H. Blake}
\affiliation{Department of Physics and Astronomy, University of Pennsylvania, 209 South 33rd Street, Philadelphia, PA 19104 USA}

\author[0000-0003-0149-9678]{Paul Robertson}
\affiliation{Department of Physics \& Astronomy, The University of California, Irvine, CA 92697, USA}

\author[0000-0002-9082-6337]{Andrea S.J. Lin}
\affiliation{Department of Astronomy \& Astrophysics, 525 Davey Laboratory, The Pennsylvania State University, University Park, PA 16802, USA}
\affiliation{Center for Exoplanets and Habitable Worlds, 525 Davey Laboratory, The Pennsylvania State University, University Park, PA 16802, USA}

\author[0000-0002-5463-9980]{Arvind F. Gupta}
\affiliation{U.S. National Science Foundation National Optical-Infrared Astronomy Research Laboratory, 950 N. Cherry Ave., Tucson, AZ 85719, USA}

\author[0000-0001-9596-7983]{Suvrath Mahadevan}
\affiliation{Department of Astronomy \& Astrophysics, 525 Davey Laboratory, The Pennsylvania State University, University Park, PA 16802, USA}
\affiliation{Center for Exoplanets and Habitable Worlds, 525 Davey Laboratory, The Pennsylvania State University, University Park, PA 16802, USA}

\author[0000-0001-6160-5888]{Jason T. Wright}
\affiliation{Department of Astronomy \& Astrophysics, 525 Davey Laboratory, The Pennsylvania State University, University Park, PA 16802, USA}
\affiliation{Center for Exoplanets and Habitable Worlds, 525 Davey Laboratory, The Pennsylvania State University, University Park, PA 16802, USA}

\author[0000-0001-8170-7072]{Daniella Bardalez Gagliuffi}
\affiliation{Department of Physics \& Astronomy, Amherst College, 25 East Drive, Amherst, MA 01002, USA}

\author[0000-0002-3610-6953]{Jiayin Dong}
\affiliation{Center for Computational Astrophysics, Flatiron Institute, 162 Fifth Avenue, New York, NY 10010, USA}
\affiliation{Department of Astronomy, University of Illinois at Urbana-Champaign, Urbana, IL 61801, USA}

\author[0000-0002-3853-7327]{Rachel B. Fernandes}
\affiliation{Department of Astronomy \& Astrophysics, 525 Davey Laboratory, The Pennsylvania State University, University Park, PA 16802, USA}
\affiliation{Center for Exoplanets and Habitable Worlds, 525 Davey Laboratory, The Pennsylvania State University, University Park, PA 16802, USA}

\author[0000-0003-0199-9699]{Evan Fitzmaurice}
\affiliation{Department of Astronomy \& Astrophysics, 525 Davey Laboratory, The Pennsylvania State University, University Park, PA 16802, USA}
\affiliation{Center for Exoplanets and Habitable Worlds, 525 Davey Laboratory, The Pennsylvania State University, University Park, PA 16802, USA}
\affiliation{Institute for Computational and Data Sciences, 224B Computer Building, Penn State University,  University Park, PA, 16802, USA}

\author[0000-0003-1312-9391]{Samuel Halverson}
\affiliation{Jet Propulsion Laboratory, California Institute of Technology, 4800 Oak Grove Drive, Pasadena, CA 91109, USA}

\author[0000-0001-8401-4300]{Shubham Kanodia}
\affiliation{Earth and Planets Laboratory, Carnegie Science, 5241 Broad Branch Road, NW, Washington, DC 20015, USA}

\author[0000-0002-9632-9382]{Sarah E. Logsdon}
\affiliation{U.S. National Science Foundation National Optical-Infrared Astronomy Research Laboratory, 950 N.\ Cherry Ave., Tucson, AZ 85719, USA}

\DIFaddbegin \author[0000-0002-4927-9925]{\DIFadd{Jacob K. Luhn}}
\affiliation{NASA JPL, 4800 Oak Grove Drive, Pasadena, CA 91109, USA}

\DIFaddend \author[0000-0003-0241-8956]{Michael W. McElwain}
\affiliation{Exoplanets and Stellar Astrophysics Laboratory, NASA Goddard Space Flight Center, Greenbelt, MD 20771, USA}

\author[0000-0002-0048-2586]{Andy Monson}
\affiliation{Steward Observatory, The University of Arizona, 933 N.\ Cherry Ave, Tucson, AZ 85721, USA}

\author[0000-0001-8720-5612]{Joe P. Ninan}
\affiliation{Department of Astronomy and Astrophysics, Tata Institute of Fundamental Research, Homi Bhabha Road, Colaba, Mumbai 400005, India}

\author[0000-0002-2488-7123]{Jayadev Rajagopal}
\affiliation{U.S. National Science Foundation National Optical-Infrared Astronomy Research Laboratory, 950 N.\ Cherry Ave., Tucson, AZ 85719, USA}

\author[0000-0001-8127-5775]{Arpita Roy}
\affiliation{Astrophysics \& Space Institute, Schmidt Sciences, New York, NY 10011, USA}

\author[0000-0002-4046-987X]{Christian Schwab}
\affiliation{School of Mathematical and Physical Sciences, Macquarie University, Balaclava Road, North Ryde, NSW 2109, Australia}

\author[0000-0001-7409-5688]{Gudmundur \DIFdelbegin \DIFdel{Stefansson}\DIFdelend \DIFaddbegin \DIFadd{Stef}{\DIFadd{\'a}}\DIFadd{nsson}\DIFaddend }
\affiliation{Anton Pannekoek Institute for Astronomy, 904 Science Park, University of Amsterdam, Amsterdam, 1098 XH, The Netherlands}

\author[0000-0002-4788-8858]{Ryan Terrien}
\affiliation{Carleton College, One North College Street, Northfield, MN 55057, USA}

\author[0000-0003-3773-5142]{Jason D. Eastman}
\affiliation{Center for Astrophysics \textbar \ Harvard \& Smithsonian, 60 Garden Street, Cambridge, MA 02138, USA}

\author[0000-0002-1160-7970]{Jonathan Horner}
\affiliation{Centre for Astrophysics, University of Southern Queensland, West Street, Toowoomba, QLD 4350, Australia}

\author[0000-0002-8864-1667]{Peter Plavchan}
\affiliation{George Mason University, 4400 University Drive, Fairfax, VA 22030, USA}

\author[0000-0002-6937-9034]{Sharon X. Wang}
\affiliation{Department of Astronomy, Tsinghua University, Beijing 100084, PR China}

\author[0000-0003-1928-0578]{Maurice L. Wilson}
\affiliation{High Altitude Observatory, National Center for Atmospheric Research, 3080 Center Green Drive, Boulder, CO 80301, USA}

\author[0000-0001-9957-9304]{Robert A. Wittenmyer}
\affiliation{Centre for Astrophysics, University of Southern Queensland, West Street, Toowoomba, QLD 4350, Australia}

\received{November 19, 2024}
\revised{March 17, 2025}
\accepted{April 28, 2025}
\submitjournal{The Astronomical Journal}

\begin{abstract}
We present a set of companion dynamical masses and orbital parameters of seven star systems from the NEID Earth Twin Survey with significant absolute astrometric accelerations between the epochs of Hipparcos and Gaia. These include four binary star systems (HD 68017 AB, 61 Cygni AB, HD 24496 AB, and HD 4614 AB) and three planetary systems (HD 217107, HD 190360, and HD 154345). Our analyses incorporate a long baseline of RVs that \DIFdelbegin \DIFdel{include over 900 }\DIFdelend \DIFaddbegin \DIFadd{includes over 1100 }\DIFaddend previously unpublished measurements from NEID and MINERVA, \DIFaddbegin \DIFadd{extending the overall RV baseline for each system by $\approx2.5$ years, }\DIFaddend as well as relative astrometry for the stellar binary systems where the positions of both stars are well-measured. In each case, the combination of astrometry and RVs constrains the three-dimensional acceleration of the host star and enables precise dynamical masses. We publish true masses for three planets whose measurements were previously entangled with their inclinations, four stellar masses with $\lesssim$1\% relative precision, and improved orbital solutions for all seven systems, including the first for HD 24496 AB. \DIFaddbegin \DIFadd{These solutions not only agree with previous estimates, but also improve their fidelity. }\DIFaddend We also explore each system for evidence of periodic signals in the residuals around our best-fit models, and discuss the potential that the three planetary systems have for being directly imaged. With dynamical mass estimates and reliable orbit ephemerides, these seven star systems represent promising benchmarks for future stellar and planetary characterization efforts, and are amenable for \DIFdelbegin \DIFdel{even }\DIFdelend further improvement with the upcoming release of Gaia epoch astrometry.

\end{abstract}

\keywords{Dynamical Masses, Binary Systems, Orbit Determination}

\section{Introduction} \label{sec:intro}

Long-term monitoring of a star's astrometric motion can refine the orbital solutions of known companions, disentangle tight binary star systems, or even reveal companions to which transit or radial velocity (RV) surveys are insensitive. For centuries, short-period visual binary systems where both stars could be resolved were the only laboratory for studying stellar orbits. The first observations of these orbits were among the earliest tests of Keplerian motion outside of the solar system \citep[e.g.,][]{1844MNRAS...6R.136B, 1877MmRAS..43....1K}. Today, measurements of the relative motion of binary stars as they orbit each other is among the most powerful of probes for model-independent stellar masses.

By building up a sample of precise stellar masses, stellar evolutionary tracks (e.g. MIST; \citealt{MIST0, MIST1}) can be improved, and more accurate mass-luminosity relationships can be derived \citep[e.g.,][]{2019ApJ...871...63M, 2022AJ....164..164G}, enhancing the ability to infer physical properties of single stars. Additionally, exoplanet characterization is largely dependent on the physical properties of the host stars. In fact, model-induced misinterpretations of stellar parameters have been shown to bias our physical characterization of planets \citep[e.g.,][]{2018ApJS..239...14J, 2020MNRAS.496..851W, 2021MNRAS.504.4968C, 2022MNRAS.510.2041C, KELT24_stellar_params}. More reliable methods of mass determination are therefore needed to simplify analyses that involve stellar modeling. Recent developments in precision astrometric and spectrographic instrumentation have refined our capabilities to infer masses from fundamental principles that do not rely on model-dependent assumptions.

In the last few decades, long-term ground-based astrometric programs like the Research Consortium on Nearby Stars (RECONS; \citealt{1994AJ....108.1437H, 2018AJ....155..265H}) and advancements in adaptive optics (AO; e.g., \citealt{Keck_AO, Robo_AO_instrument, gemini_AO, Hirsch_AO}) have revolutionized the field of precision ground-based astrometry. Space-based missions like Hipparcos \citep{Hipparcos_catalog} and Gaia \citep{Gaia_mission} have unveiled large samples of binary star systems by comparing the parallaxes and proper motions of nearby stars \citep[e.g.,][]{HIP_double1, HIP_double2, gaia_wide_binaries, GaiaDR3_astrometrc_binaries} with precisions at the level of milli- to micro-arcseconds, giving way to an emergence of population-level analyses on hierarchical stellar systems. In addition to using the relative positional measurements for resolved two-star systems over the 25-year time difference between Hipparcos and Gaia, observing a change in a single star's proper motion over that same baseline allows for a direct measurement of its projected acceleration due to its companion, which may be either spatially resolved or unresolved. This measure of a star's acceleration is a direct dynamical constraint on its companion's mass. \cite{hgca_dr2, hgca_edr3} published the Hipparcos-Gaia Catalog of Accelerations (HGCA) which serves as an inventory of stars observed to have significant accelerations via the detection of a proper motion anomaly between the epochs of the two missions. The HGCA has identified some of the best candidates for studying the dynamical properties of stellar systems, and has already been used for cases of binary star systems, substellar companions, and exoplanets previously only known via RV surveys \DIFdelbegin \DIFdel{\citep[e.g.,][]{dynamical_masses_HD68017, bd_masses_hgca_edr3, nine_RV_exoplanets}}\DIFdelend \DIFaddbegin \DIFadd{\citep[e.g.,][]{Brandt_68017, bd_masses_hgca_edr3, nine_RV_exoplanets}}\DIFaddend .

Doppler spectroscopy is capable of constraining the relative line-of-sight motion of a star due to its bound companions. Recent advancements in these RV instruments have not only extended the observational baseline of frequently-studied star systems to characterize long-term trends, but also enabled the detection of low-mass planets via their enhanced precision. While RV surveys have long been carried out to detect bound companions, they are limited in that true masses can only be solved for up to a factor of $\sin i$, where the inclination is defined to be $i=90^\circ$ for an edge-on, transiting orbit and $i=0^\circ$ for a face-on orbit. In some cases, bound companions known from RV detections can be mistakenly classified. For example, one of the first extrasolar planetary candidates identified through RVs (HD 114762~b; \citealt{HD114762b}) was long contested as a brown dwarf \citep[e.g.,][]{HD114762_BDconf1, HD114762_BDconf2} but was recently identified as a low-mass star \citep{HD114762_Gaiamass, 2022AJ....164..196W}. This misclassification arose because RV measurements alone cannot resolve the inclination ambiguity inherent in Doppler shifts. By combining systems where both astrometric and RV measurements are available, we can constrain the object's three-dimensional acceleration, thereby breaking the degeneracy that exists between a companion's mass and inclination \DIFdelbegin \DIFdel{\citep[e.g.,][]{2023MNRAS.522.5622L, 2024arXiv241005654S}
}\DIFdelend \DIFaddbegin \DIFadd{\citep[e.g.,][]{2023MNRAS.522.5622L, 2025AJ....169..107S}
}\DIFaddend 

NEID \citep{NEID_design} is a high-precision spectrometer that has already helped discover numerous exoplanet systems \citep[e.g.,][]{TOI3714_3629, TOI3757, TOI4127, Gupta_TIC, HD86728}. The NEID Earth-Twin Survey (NETS; \citealt{NETS}) is a guaranteed time observations program that is allotted a minimum of 240 hours per year with the ultimate goal of finding Earth-mass exoplanets in the habitable zones of the closest and brightest G, K, and M dwarfs. \DIFaddbegin \DIFadd{The NETS sample was curated by selecting bright ($V\lesssim8$), well-characterized stars from previous surveys, applying criteria to ensure precise RV measurements, and removing those with high activity or rapid rotation (see Section 3 and Section 4 of \citealt{NETS} for a complete description of how the NETS catalog was constructed). }\DIFaddend Planets with $m_\mathrm{p}=1~\mathrm{M_\oplus}$ in the habitable zones of GKM-type stars will induce RV semi-amplitudes between $\mathrm{5~cm~s^{-1}}$ and $\mathrm{3~m~s^{-1}}$, depending on the properties of the host star. While signals from low-amplitude planets such as these have historically been difficult to disentangle from instrumental and stellar noise sources, facilities like NEID have already demonstrated exciting new capabilities.

In this paper, we present seven star systems from NETS with significant astrometric accelerations ($>3\sigma$, or \DIFdelbegin \DIFdel{$\chi^2_\mathrm{HGCA}>11.8$ }\DIFdelend \DIFaddbegin \DIFadd{$\chi^2_\mathrm{HG}>11.8$ }\DIFaddend as reported in the HGCA; \citealt{hgca_edr3}), enabling dynamical masses and refined orbits for each. In Section \ref{sec:data}, we discuss the totality of the data used in our analyses, including the publication of new RV measurements from the precision Doppler spectrometers NEID and MINERVA. Section \ref{sec:systems} provides background on the systems considered in our work. In Section \ref{sec:analyses}, we outline the full fitting procedure used to solve for the physical parameters of our systems \DIFdelbegin \DIFdel{, and Section \ref{sec:results} details }\DIFdelend \DIFaddbegin \DIFadd{and present }\DIFaddend those results in \DIFdelbegin \DIFdel{the context of the current landscape of extrasolar systems}\DIFdelend \DIFaddbegin \DIFadd{Section \ref{sec:results}. Section \ref{sec:discussion} contextualizes our findings within the broader framework of stellar and planetary mass characterization, emphasizing their implications for evolutionary models and future direct imaging efforts}\DIFaddend . In Section \ref{sec:companions} we present new limits on hierarchical companions within our systems. A summary of our work is provided in Section \ref{sec:concl}.

\section{Data and Methodology} \label{sec:data}

We assembled our sample by searching the 41 targets in NETS for cases where the star was measured to have a statistically significant ($>3\sigma$) astrometric acceleration in the HGCA. We publish these RVs from NEID, as well as new RVs from MINERVA in cases where it observed our sample from NETS, and also include historical RV and astrometric data from the literature. For each system, we also searched the Transiting Exoplanet Survey Satellite (TESS; \citealt{TESS}) for any transits around all stars but found none.

\begin{figure*}\centering
    \includegraphics[width=\linewidth]{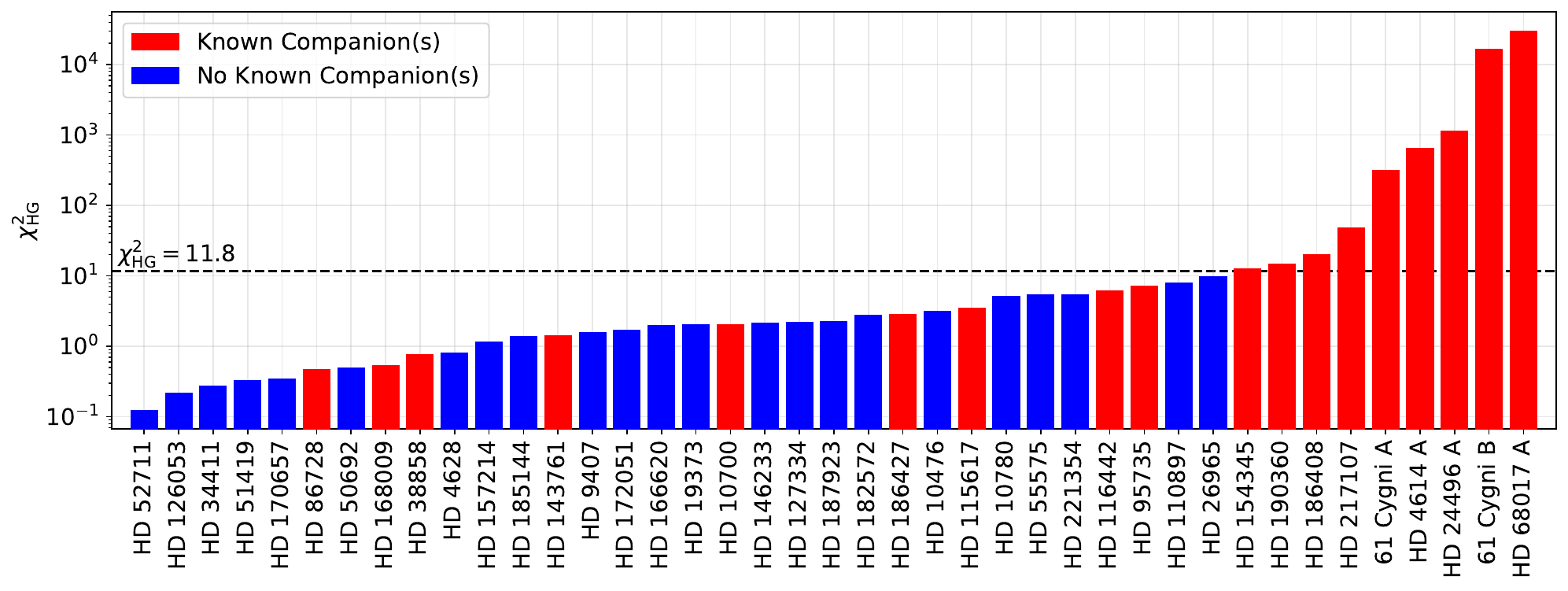}
    \caption{The full sample of NETS stars, ranked by their astrometric acceleration parameter, $\chi_\mathrm{HG}^2$. Stars plotted with red bars are known to have at least one planetary or stellar companion, while those plotted with blue \DIFdelbeginFL \DIFdelFL{bard }\DIFdelendFL \DIFaddbeginFL \DIFaddFL{bars }\DIFaddendFL are not known to have any bound companions. The \DIFaddbeginFL \DIFaddFL{horizontal }\DIFaddendFL dashed line indicates our cut for keeping only stars with $\chi^2_\mathrm{HG}>11.8$. \DIFaddbeginFL \DIFaddFL{The only NETS target not shown here is HD 179957, which is not cataloged by Hipparcos and therefore has no reported $\chi^2_\mathrm{HG}$ value. We note that its bound stellar companion, HD 179958, is cataloged by Hipparcos and does have a $>3\sigma$ astrometric acceleration. Since we do not have NEID RVs on the accelerating star, this system is not included in our analysis. }\DIFaddendFL 61 Cygni B\DIFaddbeginFL \DIFaddFL{, on the other hand, }\DIFaddendFL is a NEID target not in NETS; however, it is bound to 61 Cygni A, and so we include it in our analyses and highlight it here with a hatch.} \label{fig:nets_chi2s}
\end{figure*}

We identified seven NETS systems \DIFaddbegin \DIFadd{-- HD 68017 AB, 61 Cygni AB (HD 201091/2), HD 24496 AB, HD 4614 AB, HD 217107, HD 190360, and HD 154345 -- }\DIFaddend with significant accelerations (see Figure \ref{fig:nets_chi2s} for a depiction of all NETS targets and their respective measures of astrometric acceleration) and sufficient RV baselines to measure dynamical masses and orbital parameters. \DIFaddbegin \DIFadd{These systems are described further in Section \ref{sec:systems}. }\DIFaddend We note that one NETS target, 16 Cygni A (HD 186408\DIFaddbegin \DIFadd{, not to be confused with 61 Cygni}\DIFaddend ), is observed to have a significant acceleration ($\chi_\mathrm{HG}^2=20.35$), where the acceleration is driven largely by its nearby ($\approx3.25''$) bound companion 16 Cygni C. However, 16 Cygni is a triple star system where 16 Cygni B is close enough ($\rho_\mathrm{AB}\approx39.76''$) that attempting to solve for the full three-body fit introduces too many degeneracies given the available data. Excluding 16 Cygni B from our fits may bias the results of an otherwise two-body 16 Cygni AC solution. For these reasons, we do not publish a joint RV and astrometric solution for the 16 Cygni system. All data used as input to these analyses are provided in \DIFdelbegin \DIFdel{a }\DIFdelend corresponding machine-readable \DIFdelbegin \DIFdel{table}\DIFdelend \DIFaddbegin \DIFadd{tables}\DIFaddend .

\subsection{Absolute Astrometry} \label{subsec:absastr}

The total three-dimensional motion of a typical star through the galaxy is comprised of its space motion as it orbits the Milky Way and its orbital motion resulting from any bound bodies. Even for bright, nearby stars, changes in absolute astrometric motion of a single star due to orbiting companions is difficult to measure, particularly from the ground. However, with the release of data from space-based missions like Hipparcos and Gaia, which combine multiple years' worth of observations to make projected velocity measurements, precision astrometry can practically be used to detect bound companions. By comparing the proper motion of a star as measured by Hipparcos ($t=1991.25$) to that as measured by Gaia ($t=2016.0$), a direct observation of the absolute projected acceleration is made, assuming a known distance from Gaia. If the radial acceleration of the star can also be measured, then the measured three-dimensional acceleration can yield an independent constraint on the orbiting companion's mass.

The HGCA introduces a $\chi_\mathrm{HG}^2$ metric for quantifying the significance of the acceleration of a star between the epochs of Hipparcos and Gaia. This statistic is computed by considering three separate proper motion measurements of the target star: those made independently by Hipparcos and Gaia at their respective epochs, as well as the long-term proper motion estimated by comparing the two International Celestial Reference System (ICRS) positions in a stationary frame, time-stamped at the midpoint of the observations. The $\chi^2_\mathrm{HG}$ is then defined as the relative significance of the difference between the most discrepant of these three proper motion measurements, and serves as a snapshot of the star's acceleration between these two epochs. Systems with $\chi^2_\mathrm{HG}\gtrsim6$ have been shown to be reasonable candidates for making dynamical inferences based on the HGCA's detected measure of acceleration \citep{nine_RV_exoplanets}. Here, we select systems that have $\chi_\mathrm{HG}^2>11.8$, corresponding to a $3\sigma$ detection level, ensuring that the stars in our sample are significantly accelerating. 

Gaia reports a goodness-of-fit metric known as the renormalized unit weight error (RUWE) to assess a source's status as a single star. Stars with $\mathrm{RUWE > 1.4}$ are usually indicative of a poor astrometric fit to a single-star model, and therefore suggest a tightly-bound multiple (see, for example, \citealt{gaia_astrometric_params}). Stars that have $\mathrm{RUWE}\approx1$ are generally well-described as a single source with linear motion on the sky. Two sources in our sample, HD 68017 and HD 4614, have RUWE values inconsistent with our expectation for a single star. We provide rationales for this fact in Section \ref{sec:systems}. Table~\ref{tab:abs_ast_data} lists the \DIFdelbegin \DIFdel{relevant }\DIFdelend \DIFaddbegin \DIFadd{notable }\DIFaddend absolute astrometry considered in this work.

\DIFdelbegin \DIFdelend \DIFaddbegin \begin{deluxetable*}{lcccccc}
\tablecaption{Selected summary of absolute astrometry considered in our analyses. \label{tab:abs_ast_data}}
\DIFaddend \tablehead{
\colhead{System} & \colhead{$\chi^2_\mathrm{HG}$} & \colhead{$\mu_{\alpha,\mathrm{H}}~\left[\mathrm{mas~yr^{-1}}\right]$} & \colhead{$\mu_{\alpha,\mathrm{G}}~\left[\mathrm{mas~yr^{-1}}\right]$} & \colhead{$\mu_{\delta,\mathrm{H}}~\left[\mathrm{mas~yr^{-1}}\right]$} & \colhead{$\mu_{\delta,\mathrm{G}}~\left[\mathrm{mas~yr^{-1}}\right]$} & \colhead{RUWE}
}
\startdata
HD 68017 \DIFaddbegin \DIFadd{A }\DIFaddend & \DIFdelbegin \DIFdel{$3.02\times10^4$ }\DIFdelend \DIFaddbegin \DIFadd{$3.017\times10^4$ }\DIFaddend & $-461.481  \pm 0.923$ & \DIFdelbegin \DIFdel{$-484.99 \pm 0.065$ }\DIFdelend \DIFaddbegin \DIFadd{$-484.99  \pm 0.089$ }\DIFaddend & $-644.263 \pm 0.523$ & \DIFdelbegin \DIFdel{$-643.444 \pm 0.054$ }\DIFdelend \DIFaddbegin \DIFadd{$-643.444 \pm 0.075$ }\DIFaddend & 2.969 \\
61 Cygni B & $1.664\times10^4$ & $4111.874  \pm 0.401$ & \DIFdelbegin \DIFdel{$4105.976 \pm 0.026$ }\DIFdelend \DIFaddbegin \DIFadd{$4105.976 \pm 0.035$ }\DIFaddend & $3145.692 \pm 0.492$ & \DIFdelbegin \DIFdel{$3155.942 \pm 0.027$ }\DIFdelend \DIFaddbegin \DIFadd{$3155.942 \pm 0.037$ }\DIFaddend & 0.961 \\
HD 24496 \DIFaddbegin \DIFadd{A }\DIFaddend & $1.143\times10^3$ & $ 217.439  \pm 0.996$ & \DIFdelbegin \DIFdel{$214.119 \pm 0.023$ }\DIFdelend \DIFaddbegin \DIFadd{$214.119  \pm 0.032$ }\DIFaddend & $-165.024 \pm 0.706$ & \DIFdelbegin \DIFdel{$-167.427 \pm 0.014$ }\DIFdelend \DIFaddbegin \DIFadd{$-167.427 \pm 0.020$ }\DIFaddend & 0.977 \\
 HD 4614 \DIFaddbegin \DIFadd{A  }\DIFaddend & \DIFdelbegin \DIFdel{$6.59\times10^2$  }\DIFdelend \DIFaddbegin \DIFadd{$6.594\times10^2$ }\DIFaddend & $1086.607 \pm 0.451$ & \DIFdelbegin \DIFdel{$1078.609 \pm 0.127$ }\DIFdelend \DIFaddbegin \DIFadd{$1078.609 \pm 0.174$ }\DIFaddend & $-559.575 \pm 0.419$ & \DIFdelbegin \DIFdel{$-551.133 \pm 0.153$ }\DIFdelend \DIFaddbegin \DIFadd{$-551.133 \pm 0.210$ }\DIFaddend & 3.121 \\
HD 217107  & \DIFdelbegin \DIFdel{$4.92\times10^1$  }\DIFdelend \DIFaddbegin \DIFadd{$4.919\times10^1$ }\DIFaddend & $ -6.251   \pm 0.661$ & \DIFdelbegin \DIFdel{$ -6.819 \pm 0.025$ }\DIFdelend \DIFaddbegin \DIFadd{$-6.819   \pm 0.034$ }\DIFaddend & $ -15.839 \pm 0.493$ & \DIFdelbegin \DIFdel{$-15.04 \pm 0.023$  }\DIFdelend \DIFaddbegin \DIFadd{$-15.040   \pm 0.032$ }\DIFaddend & 0.876 \\
HD 190360  & \DIFdelbegin \DIFdel{$1.48\times10^1$  }\DIFdelend \DIFaddbegin \DIFadd{$1.483\times10^1$ }\DIFaddend & \DIFdelbegin \DIFdel{$ 683.48  \pm 0.373$ }\DIFdelend \DIFaddbegin \DIFadd{$ 683.480  \pm 0.373$ }\DIFaddend & \DIFdelbegin \DIFdel{$683.196 \pm 0.027$ }\DIFdelend \DIFaddbegin \DIFadd{$683.196  \pm 0.037$ }\DIFaddend & $-524.908 \pm 0.433$ & \DIFdelbegin \DIFdel{$-525.501 \pm 0.036$ }\DIFdelend \DIFaddbegin \DIFadd{$-525.501 \pm 0.050$ }\DIFaddend & 1.012 \\
HD 154345  & \DIFdelbegin \DIFdel{$1.28\times10^1$  }\DIFdelend \DIFaddbegin \DIFadd{$1.281\times10^1$ }\DIFaddend & $ 123.299  \pm 0.474$ & \DIFdelbegin \DIFdel{$123.274 \pm 0.019$ }\DIFdelend \DIFaddbegin \DIFadd{$123.274  \pm 0.026$ }\DIFaddend & \DIFdelbegin \DIFdel{$ 853.82  \pm 0.488$ }\DIFdelend \DIFaddbegin \DIFadd{$ 853.820  \pm 0.488$ }\DIFaddend & \DIFdelbegin \DIFdel{$853.639 \pm 0.023$ }\DIFdelend \DIFaddbegin \DIFadd{$853.639  \pm 0.032$ }\DIFaddend & 1.036 \\
\enddata
\DIFdelbegin \DIFdelend \DIFaddbegin \tablecomments{From left to right, the columns are primary accelerating star, $\chi^2$ detection of astrometric acceleration metric from the HGCA, proper motion in right ascension ($\mu_\alpha$) as reported by Hipparcos (H), proper motion in right ascension as reported by Gaia (G), proper motion in declination ($\mu_\delta$) as reported by Hipparcos, proper motion in declination as reported by Gaia, and the Gaia RUWE metric. The $\chi^2_\mathrm{HGCA}$ and proper motion values come from the HGCA eDR3 \citep{hgca_edr3}. These proper motions are the same as those reported by Hipparcos and Gaia, but use the calibrated uncertainties published in \citep{hgca_edr3}. The RUWE values come from the Gaia DR3 catalog \citep{gaia_dr3}.}
\DIFaddend \end{deluxetable*}

\subsection{Relative Astrometry} \label{subsec:relastr}

The relative astrometric position between two bodies is usually parameterized by their projected separation, $\rho$, and position angle, $\phi$, where the former is measured as the on-sky angular distance and the latter is the angle of the fainter companion with respect to the vector connecting the brighter companion and the north celestial pole. If the two bodies are bound, the observed changes in $\rho$ and $\phi$ will follow the equations of Keplerian motion and can therefore be used to infer their orbit.

The majority of relative astrometry measurements used in our analyses were collected from the Washington Double Star (WDS) catalog \citep{WDS}, which has carefully compiled over two million ground-based measurements of more than 150,000 multiple-star systems from the literature, with the earliest dating back to 1690. Additionally, we carefully investigated the literature to identify new observations since recent ground-based images have not all been incorporated into WDS. In total, we include 1721 relative astrometric measurements across our four binary systems, as none of the planets included in our analysis have been directly imaged. We assume that any systematic differences in plate scales or rotations between different ground-based surveys are small, and therefore do not attempt to correct the relative astrometry for such offsets for most of our systems. Two of the systems presented here, 61 Cygni and HD 4614, have a substantial number of relative astrometry measurements that motivate a more thorough treatment, which we outline below.

\subsubsection{Data Cleaning} \label{subsec:data_clean}

With more than \DIFdelbegin \DIFdel{2,000 and 1,000 }\DIFdelend \DIFaddbegin \DIFadd{2000 and 1000 }\DIFaddend WDS astrometry measurements between 61 Cygni and HD 4614, respectively, many of which are poorly documented and without reported errors, we elect to prune our data by selecting only observations from programs satisfying either of the following conditions\DIFdelbegin \DIFdel{:
}\DIFdelend \DIFaddbegin \DIFadd{.
}\DIFaddend 

\begin{enumerate}
    \item Programs that began operation post-1950 and have $>35$ observations
    \item Programs that ceased operation pre-1950 and have \DIFdelbegin \DIFdel{$>6$ }\DIFdelend \DIFaddbegin \DIFadd{$\geq6$ }\DIFaddend observations
\end{enumerate}

\DIFaddbegin \DIFadd{These conditions were chosen to reflect the natural distribution of observation counts, maximizing the number of measurements while maintaining a long baseline. }\DIFaddend Programs that do not meet the above criteria but have reported errors are individually inspected and kept if they are not an obvious outlier to the model. These are generally newer observations with detailed measurement uncertainties (e.g., Hipparcos and Gaia). \DIFdelbegin \DIFdel{We are careful to include programs that have contributed measurements }\DIFdelend \DIFaddbegin \DIFadd{Ultimately, these criteria allow us to include six programs that began }\DIFaddend prior to 1850 in order to incorporate as much coverage of the orbit as possible, as long-period systems with relatively short observational baselines can be misinterpreted (see, for example, \citealt{raghavan_2010}).

To internally calibrate programs with missing or no reported uncertainties, we perform an iterative routine to estimate the internal measurement precision, $\sigma$, of each observational program and thereafter remove outliers. For each program, we fit a 2nd-order polynomial to its dataset of relative astrometry, both for $\rho\left(t\right)$ and $\phi\left(t\right)$. For each parameter, we compute two quantities determined from the best-fit polynomial: $\Delta$, the difference between each measurement and polynomial model, and $\sigma$, the standard deviation of all the measurement residuals about the model. We then remove outliers in each program defined as $\lvert\Delta/\sigma\rvert>3$. As points are removed, this process is repeated iteratively until all observations in each program have $\lvert\Delta/\sigma\rvert<3$ for both $\rho$ and $\phi$.

Trimming the data in this way provides a self-consistent means of estimating uncertainties for larger programs, while also removing the need to calibrate poorly-reported single observations and preserving a long observational time baseline. A summary of these observations is given in Table \ref{tab:programs}. \DIFaddbegin \DIFadd{After cleaning, we are left with 1358 and 349 astrometric measurements for 61 Cygni and HD 4614, respectively.
}\DIFaddend 

\begin{longtable}{lcccc}
\caption{Relative Astrometry}
\label{tab:programs} \\ \hline
\textbf{Program} & \textbf{$N_\mathrm{obs}$} & \textbf{Time Range} & \textbf{$\bar{\sigma}_\mathrm{\rho}$ [$\arcsec$]} & \textbf{$\bar{\sigma}_\mathrm{\phi}$ [$^\circ$]} \\
\hline
\endfirsthead
\multicolumn{5}{l}{\textbf{Table \ref{tab:programs} (continued)}} \\
\hline
\endhead
\hline \multicolumn{5}{r}{\textit{Continued on next page}} \\
\endfoot
\hline
\endlastfoot

\multicolumn{5}{c}{\textbf{HD 68017}} \\
\hline
(1) & 2 & 2011.143 - 2012.018 & 0.00050 & 0.1000 \\ 
(2) & 1 & 2016.965 & 0.01600 & 1.8000 \\ 
\hline
\multicolumn{5}{c}{\textbf{61 Cygni}} \\
\hline
(3) & 7 & 1822.720 - 1837.710 & 0.08364 & 0.0688 \\ 
(4) & 7 & 1830.810 - 1839.690 & 0.18945 & 0.1819 \\ 
(5) & 10 & 1843.530 - 1874.740 & 0.06066 & 0.2250 \\ 
(6) & 9 & 1862.970 - 1878.580 & 0.03062 & 0.0652 \\ 
(7) & 6 & 1867.890 - 1875.950 & 0.05249 & 0.1195 \\ 
(8) & 6 & 1875.790 - 1876.620 & 0.10158 & 0.2410 \\ 
(9) & 13 & 1877.640 - 1925.720 & 0.13209 & 0.1525 \\ 
(10) & 12 & 1880.746 - 1891.631 & 0.05650 & 0.1781 \\ 
(11) & 10 & 1886.900 - 1903.720 & 0.09996 & 0.1085 \\ 
(12) & 10 & 1887.030 - 1897.840 & 0.06837 & 0.1633 \\ 
(13) & 6 & 1900.740 - 1904.670 & 0.03691 & 0.1079 \\ 
(14) & 9 & 1913.840 - 1916.790 & 0.03834 & 0.0542 \\ 
(15) & 12 & 1921.680 - 1928.820 & 0.03299 & 0.4927 \\ 
(16) & 6 & 1921.688 - 1932.720 & 0.01619 & 0.1415 \\ 
(17) & 12 & 1922.690 - 1928.670 & 0.02302 & 0.0589 \\ 
(18) & 6 & 1932.870 - 1938.790 & 0.01158 & 0.0510 \\ 
(19) & 6 & 1939.760 - 1946.730 & 0.01492 & 0.0746 \\ 
(20) & 12 & 1941.870 - 1945.550 & 0.00911 & 0.0274 \\ 
(21) & 174 & 1945.877 - 1955.570 & 0.01674 & 0.0344 \\ 
(22) & 136 & 1955.488 - 1962.797 & 0.02427 & 0.0390 \\ 
(23) & 103 & 1955.554 - 1967.733 & 0.01603 & 0.0524 \\ 
(24) & 48 & 1958.765 - 2006.699 & 0.00756 & 0.0202 \\ 
(25) & 141 & 1968.546 - 1972.846 & 0.00974 & 0.0280 \\ 
(26) & 96 & 1973.439 - 1976.664 & 0.01500 & 0.0300 \\ 
(27) & 171 & 1977.401 - 1981.868 & 0.00806 & 0.0142 \\ 
(28) & 1 & 1990.548 & 0.20000 & 0.2000 \\ 
(29) & 1 & 1991.250 & 0.00600 & 0.0090 \\ 
(30) & 1 & 2000.000 & 0.39000 & 0.6500 \\ 
(31) & 36 & 2003.582 - 2007.744 & 0.00923 & 0.0143 \\ 
(32) & \DIFdelbegin \DIFdel{8 }\DIFdelend \DIFaddbegin \DIFadd{7 }\DIFaddend & 2003.683 - 2013.705 & 0.00419 & 0.0090 \\ 
\DIFdelbegin \DIFdel{Izm2016 }\DIFdel{1 }\DIFdel{2003.686 }\DIFdel{0.00830 }\DIFdel{0.0100 }\DIFdel{Izm2017 }\DIFdel{1 }\DIFdel{2003.689 }\DIFdel{0.01580 }\DIFdel{0.0170 }\DIFdel{Izm2018 }\DIFdel{1 }\DIFdel{2003.711 }\DIFdel{0.00450 }\DIFdel{0.0130 }\DIFdel{(36}\DIFdelend \DIFaddbegin \DIFadd{(33}\DIFaddend ) & 1 & 2003.719 & 0.00400 & 0.0110 \\ 
(\DIFdelbegin \DIFdel{37}\DIFdelend \DIFaddbegin \DIFadd{34}\DIFaddend ) & 37 & 2006.364 - 2007.908 & 0.04865 & 0.0792 \\ 
(\DIFdelbegin \DIFdel{38}\DIFdelend \DIFaddbegin \DIFadd{35}\DIFaddend ) & 87 & 2007.689 - 2013.912 & 0.01004 & 0.0137 \\ 
(\DIFdelbegin \DIFdel{39}\DIFdelend \DIFaddbegin \DIFadd{36}\DIFaddend ) & 1 & 2010.767 & 0.80000 & 1.5000 \\ 
(\DIFdelbegin \DIFdel{40}\DIFdelend \DIFaddbegin \DIFadd{37}\DIFaddend ) & 1 & 2011.526 & 0.10000 & 0.3000 \\ 
(\DIFdelbegin \DIFdel{41}\DIFdelend \DIFaddbegin \DIFadd{38}\DIFaddend ) & 1 & 2011.618 & 0.09000 & 0.3100 \\ 
(\DIFdelbegin \DIFdel{42}\DIFdelend \DIFaddbegin \DIFadd{39}\DIFaddend ) & 130 & 2014.542 - 2019.725 & 0.00966 & 0.0131 \\ 
(\DIFdelbegin \DIFdel{43}\DIFdelend \DIFaddbegin \DIFadd{40}\DIFaddend ) & 6 & 2014.687 - 2019.648 & 0.00243 & 0.0043 \\ 
(\DIFdelbegin \DIFdel{44}\DIFdelend \DIFaddbegin \DIFadd{41}\DIFaddend ) & 1 & 2015.500 & 0.00016 & 0.0002 \\ 
(\DIFdelbegin \DIFdel{45}\DIFdelend \DIFaddbegin \DIFadd{42}\DIFaddend ) & 1 & 2016.000 & 0.00007 & 0.0001 \\ 
(\DIFdelbegin \DIFdel{46}\DIFdelend \DIFaddbegin \DIFadd{43}\DIFaddend ) & 2 & 2016.360 - 2016.819 & 0.09900 & 0.1785 \\ 
(\DIFdelbegin \DIFdel{47}\DIFdelend \DIFaddbegin \DIFadd{44}\DIFaddend ) & 1 & 2016.664 & 0.10800 & 0.2000 \\ 
(\DIFdelbegin \DIFdel{48}\DIFdelend \DIFaddbegin \DIFadd{45}\DIFaddend ) & 1 & 2017.578 & 0.20000 & 0.2000 \\ 
(\DIFdelbegin \DIFdel{49}\DIFdelend \DIFaddbegin \DIFadd{46}\DIFaddend ) & 1 & 2017.684 & 0.02600 & 0.0490 \\ 
(\DIFdelbegin \DIFdel{50}\DIFdelend \DIFaddbegin \DIFadd{47}\DIFaddend ) & 1 & 2017.724 & 0.15000 & 0.2360 \\ 
(\DIFdelbegin \DIFdel{51}\DIFdelend \DIFaddbegin \DIFadd{48}\DIFaddend ) & 1 & 2018.581 & 0.00700 & 0.0490 \\ 
(\DIFdelbegin \DIFdel{52}\DIFdelend \DIFaddbegin \DIFadd{49}\DIFaddend ) & 1 & 2018.773 & 0.15000 & 0.2360 \\ 
(\DIFdelbegin \DIFdel{53}\DIFdelend \DIFaddbegin \DIFadd{50}\DIFaddend ) & 1 & 2019.689 & 0.15000 & 0.2360 \\ 
(\DIFdelbegin \DIFdel{54}\DIFdelend \DIFaddbegin \DIFadd{51}\DIFaddend ) & 1 & 2019.722 & 0.20000 & 0.1000 \\ 
(\DIFdelbegin \DIFdel{55}\DIFdelend \DIFaddbegin \DIFadd{52}\DIFaddend ) & 1 & 2019.725 & 0.11314 & 0.2040 \\ 
(\DIFdelbegin \DIFdel{56}\DIFdelend \DIFaddbegin \DIFadd{53}\DIFaddend ) & 1 & 2021.775 & 0.15000 & 0.2360 \\ 
\hline
\multicolumn{5}{c}{\textbf{HD 24496}} \\
\hline
(\DIFdelbegin \DIFdel{57}\DIFdelend \DIFaddbegin \DIFadd{54}\DIFaddend ) & 1 & 1979.960 & 0.20000 & 3.3300 \\
(\DIFdelbegin \DIFdel{58}\DIFdelend \DIFaddbegin \DIFadd{55}\DIFaddend ) & 1 & 1988.970 & 0.30000 & 5.0000 \\
(\DIFdelbegin \DIFdel{59}\DIFdelend \DIFaddbegin \DIFadd{56}\DIFaddend ) & 1 & 2012.572 & 0.06000 & 1.0000 \\
(\DIFdelbegin \DIFdel{44}\DIFdelend \DIFaddbegin \DIFadd{41}\DIFaddend ) & 1 & 2015.500 & 0.00012 & 0.0014 \\
(2)  & 4 & 2015.842 - 2018.080 & 0.00800 & 0.1500 \\
(\DIFdelbegin \DIFdel{45}\DIFdelend \DIFaddbegin \DIFadd{42}\DIFaddend ) & 1 & 2016.000 & 0.00005 & 0.0007 \\
(\DIFdelbegin \DIFdel{60}\DIFdelend \DIFaddbegin \DIFadd{57}\DIFaddend ) & 1 & 2019.842 & 0.08000 & 1.2000 \\
(\DIFdelbegin \DIFdel{61}\DIFdelend \DIFaddbegin \DIFadd{58}\DIFaddend ) & 1 & 2020.964 & 0.17600 & 0.1300 \\
\hline
\multicolumn{5}{c}{\textbf{HD 4614}} \\
\hline
(5) & 12 & 1841.340 - 1875.150 & 0.08333 & 0.9537 \\ 
(\DIFdelbegin \DIFdel{62}\DIFdelend \DIFaddbegin \DIFadd{59}\DIFaddend ) & 11 & 1845.390 - 1862.710 & 0.15691 & 0.3469 \\ 
(\DIFdelbegin \DIFdel{63}\DIFdelend \DIFaddbegin \DIFadd{60}\DIFaddend ) & 9 & 1845.860 - 1856.070 & 0.06976 & 0.6335 \\ 
(6) & 17 & 1862.860 - 1878.580 & 0.02855 & 0.4819 \\ 
(9) & 23 & 1877.750 - 1926.700 & 0.16387 & 1.1752 \\ 
(10) & 7 & 1881.904 - 1890.789 & 0.04253 & 0.5429 \\ 
(\DIFdelbegin \DIFdel{58}\DIFdelend \DIFaddbegin \DIFadd{55}\DIFaddend ) & 8 & 1882.090 - 1883.160 & 0.14682 & 0.8351 \\ 
(12) & 7 & 1887.060 - 1896.090 & 0.01885 & 0.4955 \\ 
(\DIFdelbegin \DIFdel{64}\DIFdelend \DIFaddbegin \DIFadd{61}\DIFaddend ) & 7 & 1891.060 - 1900.790 & 0.14356 & 0.8203 \\ 
(\DIFdelbegin \DIFdel{65}\DIFdelend \DIFaddbegin \DIFadd{62}\DIFaddend ) & 7 & 1915.802 - 1923.738 & 0.06797 & 0.3551 \\ 
(19) & 7 & 1938.960 - 1946.900 & 0.02124 & 0.1702 \\ 
(22) & 55 & 1956.602 - 1962.948 & 0.01252 & 0.0799 \\ 
(23) & 42 & 1963.585 - 1966.890 & 0.01144 & 0.0567 \\ 
(25) & 72 & 1967.618 - 1973.000 & 0.00615 & 0.0464 \\ 
(26) & 54 & 1973.756 - 1976.825 & 0.00856 & 0.0557 \\ 
(30) & 1 & 1991.250 & 0.01580 & 0.0750 \\ 
(66) & 3 & 1993.635 - 1993.635 & 0.11333 & 0.1000 \\ 
(31) & 1 & 2000.030 & 0.28000 & 1.2200 \\ 
(\DIFdelbegin \DIFdel{45}\DIFdelend \DIFaddbegin \DIFadd{42}\DIFaddend ) & 1 & 2015.500 & 0.00111 & 0.0040 \\ 
(\DIFdelbegin \DIFdel{46}\DIFdelend \DIFaddbegin \DIFadd{43}\DIFaddend ) & 1 & 2016.000 & 0.00013 & 0.0010 \\ 
(\DIFdelbegin \DIFdel{67}\DIFdelend \DIFaddbegin \DIFadd{64}\DIFaddend ) & 1 & 2016.836 & 0.10100 & 0.3100 \\ 
(\DIFdelbegin \DIFdel{68}\DIFdelend \DIFaddbegin \DIFadd{65}\DIFaddend ) & 1 & 2016.903 & 0.10000 & 0.1000 \\ 
(\DIFdelbegin \DIFdel{56}\DIFdelend \DIFaddbegin \DIFadd{53}\DIFaddend ) & 1 & 2019.731 & 0.10630 & 0.4490 \\ 
(\DIFdelbegin \DIFdel{69}\DIFdelend \DIFaddbegin \DIFadd{66}\DIFaddend ) & 1 & 2019.732 & 0.14000 & 0.5100 \\ 
\end{longtable}
\begin{minipage}{0.46\textwidth}
\footnotesize
\textbf{Note:} This table lists the relative astrometry data for selected programs. The columns represent the program name, number of observations ($N_\mathrm{obs}$), time range, average angular separation uncertainty ($\bar{\sigma}_\mathrm{\rho}$), and average position angle uncertainty ($\bar{\sigma}_\mathrm{\phi}$). References are as follows: (1) \citealt{2012ApJ...761...39C} (2) \citealt{Hirsch_AO} (3) \citealt{StF1837} (4) \citealt{Smy1844} (5) \citealt{Stt1878} (6) \citealt{D__1884} (7) \citealt{Du_1876} (8) \citealt{WS_1877} (9) \citealt{Dob1927} (10) \citealt{Hl_1892c} (11) \citealt{Cel1923} (12) \citealt{Sp_1909} (13) \citealt{StH1911a} (14) \citealt{Rab1923} (15) \citealt{Prz1926} (16) \citealt{StG1962a} (17) \citealt{Lbz1929} (18) \citealt{Rab1939} (19) \citealt{Rab1953} (20) \citealt{Jef1951} (21) \citealt{DeO1957} (22) \citealt{USN1963} (23) \citealt{USN1969} (24) \citealt{PkO2017b} (25)  \citealt{USN1974} (26) \citealt{USN1978} (27) \citealt{USN1984} (28) \citealt{Vie1992} (29) \citealt{Hipparcos_catalog} (30) \citealt{2mass} (31) \citealt{Izm2010} (32) \citealt{IZM2015} (33) \DIFdelbegin \DIFdel{Izm2016 }\DIFdelend \DIFaddbegin \DIFadd{\citealt{IZM2019} }\DIFaddend (34) \DIFdelbegin \DIFdel{Izm2017 }\DIFdelend \DIFaddbegin \DIFadd{\citealt{VLM2008} }\DIFaddend (35) \DIFdelbegin \DIFdel{Izm2018 }\DIFdelend \DIFaddbegin \DIFadd{\citealt{Pulkovo2009} }\DIFaddend (36) \DIFdelbegin \DIFdel{\citealt{IZM2019} }\DIFdelend \DIFaddbegin \DIFadd{\citealt{Nug2012} }\DIFaddend (37) \DIFdelbegin \DIFdel{\citealt{VLM2008} }\DIFdelend \DIFaddbegin \DIFadd{\citealt{BMA2012b} }\DIFaddend (38) \DIFdelbegin \DIFdel{\citealt{Pulkovo2009} }\DIFdelend \DIFaddbegin \DIFadd{\citealt{Cao2011} }\DIFaddend (39) \DIFdelbegin \DIFdel{\citealt{Nug2012} }\DIFdelend \DIFaddbegin \DIFadd{\citealt{2021Ap.....64..160I} }\DIFaddend (40) \DIFdelbegin \DIFdel{\citealt{BMA2012b} }\DIFdelend \DIFaddbegin \DIFadd{\citealt{Izm2020} }\DIFaddend (41) \DIFdelbegin \DIFdel{\citealt{Cao2011} }\DIFdelend \DIFaddbegin \DIFadd{\citealt{gaia_dr2} }\DIFaddend (42) \DIFdelbegin \DIFdel{\citealt{2021Ap.....64..160I} }\DIFdelend \DIFaddbegin \DIFadd{\citealt{gaia_dr3} }\DIFaddend (43) \DIFdelbegin \DIFdel{\citealt{Izm2020} }\DIFdelend \DIFaddbegin \DIFadd{\citealt{Kpp2017j} }\DIFaddend (44) \DIFdelbegin \DIFdel{\citealt{gaia_dr2} }\DIFdelend \DIFaddbegin \DIFadd{\citealt{WSI2017b} }\DIFaddend (45) \DIFdelbegin \DIFdel{\citealt{gaia_dr3} }\DIFdelend \DIFaddbegin \DIFadd{\citealt{Wbt2018} }\DIFaddend (46) \DIFdelbegin \DIFdel{\citealt{Kpp2017j} }\DIFdelend \DIFaddbegin \DIFadd{\citealt{WSI2018a} }\DIFaddend (47) \DIFdelbegin \DIFdel{\citealt{WSI2017b} }\DIFdelend \DIFaddbegin \DIFadd{\citealt{Smr2018c} }\DIFaddend (48) \DIFdelbegin \DIFdel{\citealt{Wbt2018} }\DIFdelend \DIFaddbegin \DIFadd{\citealt{WSI2021} }\DIFaddend (49) \DIFdelbegin \DIFdel{\citealt{WSI2018a} }\DIFdelend \DIFaddbegin \DIFadd{\citealt{Smr2019} }\DIFaddend (50) \DIFdelbegin \DIFdel{\citealt{Smr2018c} }\DIFdelend \DIFaddbegin \DIFadd{\citealt{Smr2021b} }\DIFaddend (51) \DIFdelbegin \DIFdel{\citealt{WSI2021} }\DIFdelend \DIFaddbegin \DIFadd{\citealt{Wbt2020} }\DIFaddend (52) \DIFdelbegin \DIFdel{\citealt{Smr2019} }\DIFdelend \DIFaddbegin \DIFadd{\citealt{Kpp2020f} }\DIFaddend (53) \DIFdelbegin \DIFdel{\citealt{Smr2021b} }\DIFdelend \DIFaddbegin \DIFadd{\citealt{JDSO_link} }\DIFaddend (54) \DIFdelbegin \DIFdel{\citealt{Wbt2020} }\DIFdelend \DIFaddbegin \DIFadd{\citealt{1980ApJS...44..111H} }\DIFaddend (55) \DIFdelbegin \DIFdel{\citealt{Kpp2020f} }\DIFdelend \DIFaddbegin \DIFadd{\citealt{Hei1990b} }\DIFaddend (56) \DIFdelbegin \DIFdel{\citealt{JDSO_link} }\DIFdelend \DIFaddbegin \DIFadd{\citealt{2022AJ....163..200S} }\DIFaddend (57) \DIFdelbegin \DIFdel{\citealt{1980ApJS...44..111H} (58) \citealt{Hei1990b} (59) \citealt{2022AJ....163..200S} (60) }\DIFdelend this work (\DIFdelbegin \DIFdel{61}\DIFdelend \DIFaddbegin \DIFadd{58}\DIFaddend ) \citealt{SHS2022b} (\DIFdelbegin \DIFdel{62}\DIFdelend \DIFaddbegin \DIFadd{59}\DIFaddend ) \citealt{1937AnLei..18B...1S} (M$\ddot{\mathrm{a}}$der) (\DIFdelbegin \DIFdel{63}\DIFdelend \DIFaddbegin \DIFadd{60}\DIFaddend ) \citealt{1937AnLei..18B...1S} (Jacob) (\DIFdelbegin \DIFdel{64}\DIFdelend \DIFaddbegin \DIFadd{61}\DIFaddend ) \citealt{Gld1901} (\DIFdelbegin \DIFdel{65}\DIFdelend \DIFaddbegin \DIFadd{62}\DIFaddend ) \citealt{VBs1927a} (\DIFdelbegin \DIFdel{66}\DIFdelend \DIFaddbegin \DIFadd{63}\DIFaddend ) \citealt{Tok1995} (\DIFdelbegin \DIFdel{67}\DIFdelend \DIFaddbegin \DIFadd{64}\DIFaddend ) \citealt{Pnt2017b} (\DIFdelbegin \DIFdel{68}\DIFdelend \DIFaddbegin \DIFadd{65}\DIFaddend ) \citealt{Wbt2017} (\DIFdelbegin \DIFdel{69}\DIFdelend \DIFaddbegin \DIFadd{66}\DIFaddend ) \citealt{MuR2020a}. \DIFaddbegin \DIFadd{All relative astrometry is provided in a machine-readable table.
}\DIFaddend \end{minipage}

\subsubsection{Correcting for Physical Effects} \label{subsec:data_refactor}

Beyond selecting a subset of observations deemed to be reliable, we must consider several physical effects that could be impacting the data. Stars with large absolute radial velocities with respect to Earth will have their apparent angular separations change over time due to the change in their distance from Earth. The 61 Cygni system, for example, has a significant radial motion towards Earth ($\sim65$~km~s$^{-1}$; \citealt{2022arXiv220800211G}), and since it is so close to Earth, its measured separation will be artificially inflated across the astrometric baseline of $\sim$200 years. The correction of this type to a separation measurement of $\rho_0$ made at time $t_0$ can be made according to Equation \ref{eq:perspective_accel}.

\begin{equation}
    \label{eq:perspective_accel}
    \rho = \frac{d_{2016}}{d_{2016} + v_\mathrm{r}\left(2016 - t_0\right)}~\rho_0
\end{equation}

Here, we are choosing to anchor the relative astrometry of 61 Cygni to the Gaia DR3 epoch of 2016.0, since we have the distance, $d_{2016}$, and radial velocity, $v_\mathrm{r}$ at this time. This effect, while generally small ($\sim1~\mathrm{mas}$), is straightforward to implement and is an important consideration for nearby, high line-of-sight motion systems with long observational baselines.

As the Earth precesses, the reference direction of the north celestial pole will appear to rotate relative to a binary system, altering the position angle between the two stars. This effect is less trivial to incorporate and is generally small ($\lesssim0.1^\circ$). For these reasons, we do not consider precession-related effects in the relative astrometry used in our analyses. However, the Pulkovo Observatory in Saint Petersburg, Russia, which has steadily observed 61 Cygni across much of the last century, adopted a method for rotating their position angles to the J2000 epoch to account for precession (see Equation 1 from \DIFdelbegin \DIFdel{\citealt{2019AstL...45...30I}}\DIFdelend \DIFaddbegin \DIFadd{\citealt{IZM2019}}\DIFaddend ). For these Pulkovo measurements, we undo their correction to position angle to maintain continuity across all 59 programs included in the 61 Cygni system. Similarly, we note that reference frames like J1950, FK5, and J2000 are rotated from the new standard of the ICRS; however, this effect is also small ($\sim0.1^\circ$), and thus we effectively assume all observations were made with respect to the ICRS equinox. The adoption of the ICRS reference frame will make future astrometric surveys easier to calibrate.

\subsection{Radial Velocities} \label{subsec:rvs}

In order to constrain the line-of-sight acceleration of our stars due to their bound companions, we performed a thorough search through the literature for published RV measurements\DIFaddbegin \DIFadd{, primarily via known instrument data archives, bulk data release papers like \cite{2021ApJS..255....8R}, and public data repositories like DACE}\footnote{\DIFadd{The Data \& Analysis Center for Exoplanets (DACE) platform is available at }\href{https://dace.unige.ch.}{\DIFadd{https://dace.unige.ch.}}}\DIFaddend . Additionally, we publish new RVs from MINERVA and NEID. In total, we have collected 4963 RVs that span more than 35 years. A summary of all RV observations used in our analyses are presented in Table \ref{tab:rv_programs}, and displayed in Figure \ref{fig:rv_coverage}.

\subsubsection{ELODIE}

ELODIE is a fiber-fed echelle spectrograph installed on the 1.93~m reflector telescope at the Haute-Provence Observatory sensitive to a wavelength range of $\lambda=390.6-681.1$~nm with a resolution of $R\approx42,000$ \citep{ELODIE}. \cite{HD154345_sophie_elodie} published 49 RVs from ELODIE to recover HD 154345~b. \cite{HD190360_elodie_afoe} used the spectrograph to report the discovery of HD 190360~b using 56 RVs reduced from ELODIE spectra.

\subsubsection{SOPHIE}

The Spectrographe pour l’Observation des Phénomènes des Intérieurs stellaires et des Exoplanètes (SOPHIE) is a fiber-fed echelle spectrograph that replaced ELODIE at the Haute-Provence Observatory in France \citep{SOPHIE}. SOPHIE has two modes: high-resolution ($R=75,000$) and high-efficiency ($R=40,000$), both spanning the wavelength range $\lambda=387.2-694.3$~nm. \cite{HD154345_sophie_elodie} collected 10 RVs in the high-resolution mode for HD 154345 to refine the orbital solution for HD 154345~b.

\subsubsection{AFOE}

The Advanced Fiber Optic Echelle (AFOE) \citep{AFOE} is a fiber-fed echelle spectrograph with a maximum resolution of $R\approx56,000$ and a sensitivity to wavelengths between $\lambda\approx400-700$~nm installed on the Tillinghast 1.5~m telescope at the Fred Lawrence Whipple Observatory. \cite{HD190360_elodie_afoe} also published 13 RVs taken using AFOE to aid in the discovery of HD 190360~b.

\subsubsection{CORALIE}

CORALIE is an echelle spectrograph ($R\approx60,000$) installed on the Euler Swiss telescope at La Silla Observatory that is an improved version of ELODIE and a precursor to SOPHIE. \cite{HD2171707_coralie} published 63 RVs on HD 217107 to confirm the presence of the hot Jupiter HD 217107~b. These observations also revealed a long-term residual trend that was ultimately determined to be due to the outer planet HD 217107~c. 

\subsubsection{Tull}

The Tull spectrograph ($R\approx60,000$) is sensitive to a wavelength range of $\lambda=340-1000$~nm and is installed at the McDonald Observatory's Harlan J. Smith telescope \citep{Tull}. \cite{hd217107_tull} published 23 RVs using Tull on the HD 217107 planetary system to refine the system parameters.

\subsubsection{Hamilton}

The Hamilton spectrograph \citep{Hamilton} was a pioneer in high-resolution Doppler spectroscopy, acquiring its first light in 1986. It was first installed at the Lick Observatory's \citep{Lick_Observatory} Shane $3~\mathrm{m}$ telescope, and was capable of making broadband ($\lambda=0.34-1.10~\mu\mathrm{m}$) spectroscopic measurements up to a resolution of $R=60,000$, and is still available today. We include 626 RVs published by \cite{2021ApJS..255....8R} that were taken using Hamilton across 61 Cygni A/B, HD 4614, HD 217107, and HD 190360.

\subsubsection{HIRES}

The high-resolution echelle spectrometer (HIRES; \citealt{1994SPIE.2198..362V}) has $R=67,000$ with a wavelength range spanning $\lambda=0.31-1.10~\mu\mathrm{m}$ located at Keck Observatory on the $10~\mathrm{m}$ Keck I telescope and was commissioned in 1993. In 2004, the instrument's CCD was upgraded, leading to a $\approx3\times$ enhancement in RV precision \citep{HIRES_upgrade}. The pre- and post-upgrade RVs from HIRES are treated here as two independent datasets, allowing for unique RV offsets to be fit for each as a free parameter. Collectively, we use 228 RV measurements from HIRES taken prior to the CCD upgrade (between HD 68017, HD 154345, HD 190360, HD 24496, and HD 217107), and 1514 taken after (spanning all seven systems). By treating these two datasets as separate instruments in our fits, all of our RVs account for the intra-night drift corrections introduced by \cite{HIRES_RV_corrections}. The HIRES RVs we use were all published by \cite{2021ApJS..255....8R}.

\subsubsection{APF}

The Automated Planet Finder (APF; \citealt{2014SPIE.9145E..2BR}) is an optical spectrograph with a wavelength range of $\lambda=0.37-0.97~\mu\mathrm{m}$ configured with the 2.4~m telescope on Mount Hamilton at Lick Observatory that can achieve a resolution up to $R=150,000$. It was built to be among the first of robotic projects with the goal of detecting and characterizing exoplanets. We use 2195 RVs from APF that were published by \cite{2021ApJS..255....8R} for 61 Cygni A/B, HD 68017, HD 190360, HD 4614, and HD 217107.

\subsubsection{MINERVA}

The MINiature Exoplanet Radial Velocity Array (MINERVA; \citealt{swift2015}) is a set of four PlaneWave CDK700, 0.7~m telescopes at the Fred Lawrence Whipple Observatory on Mt. Hopkins, each feeding the array's $R=75,000$ echelle spectrograph via optical fiber. As a result of MINERVA's lone exposure meter which reports only an average flux from all four telescopes, each telescope has a slightly different barycentric correction and is therefore treated as a separate instrument for fitting purposes. We publish 543 MINERVA RVs across all four telescopes with typical exposure times of 1800~s, reduced according to \cite{wilson2019}. These include 114 measurements of HD 217107, 429 of HD 4614, and 158 of 61 Cygni A. Some of those from HD 217107 were originally published in \cite{2020AN....341..870G} but were binned nightly and not treated as separate instruments. Here, we publish all unbinned MINERVA RVs and account for individual telescope offsets.

\subsubsection{NEID}

As detailed in Section \ref{sec:intro}, NEID is a high-resolution ($R\approx110,000$) spectrometer with a broad spectral grasp ($\lambda=380-930~\mathrm{nm}$) located at the WIYN 3.5~m Telescope\footnote{The WIYN Observatory is a joint facility of the NSF's National Optical-Infrared Astronomy Research Laboratory, Indiana University, the University of Wisconsin-Madison, Pennsylvania State University, Purdue University and Princeton University.} on Kitt Peak. It is capable of measuring RVs with an instrumental error budget of 27~$\mathrm{cm~s^{-1}}$ \citep{NEID_error_budget, 2024arXiv240813318F}. In this work, the NEID 2D echelle spectra were obtained with signal-to-noise (SNR) ratios ranging between SNR=150-500 and reduced via version 1.3.0 of the NEID Data Reduction Pipeline\footnote{\href{https://neid.ipac.caltech.edu/docs/NEID-DRP/}{https://neid.ipac.caltech.edu/docs/NEID-DRP/}} (DRP). We removed subsequent RV data points that were flagged by the DRP Level 2 products as outliers and are left with 417 new NEID RVs from 31 May 2021 to 12 June 2022 across all seven of our systems. On 2022 June 13, all activity on Kitt Peak was shut down due to the Contreras fire. NEID resumed operation in November 2022, but a noticeable instrumental RV offset is observed between data taken pre- and post-fire (see \citealt{HD86728} for a more detailed discussion on this offset). Unlike the pre-fire data, the post-fire RVs do not significantly enhance the RV baselines and are still in proprietary period. As a result, we choose to exclude any post-fire data from our analyses. Since two of our systems (HD 217107 and HD 190360) have short-period planets with relatively large RV semi-amplitudes, we do not apply any binning schemes to NEID, nor any of our RV data, as intra-night RV variation may therefore be physical.

\begin{longtable}{lccc}
\caption{Relative Radial Velocity Measurements}
\label{tab:rv_programs} \\ \hline
\textbf{Instrument} & \textbf{$N_\mathrm{obs}$} & \textbf{Time Range} & \textbf{$\sigma_\mathrm{RV}$ [m/s]} \\
\hline
\endfirsthead
\multicolumn{4}{l}{\textbf{Table \ref{tab:rv_programs} (continued)}} \\
\hline
\textbf{System} & \textbf{$N_\mathrm{obs}$} & \textbf{Time Range} & \textbf{$\sigma_\mathrm{RV}$ [m/s]} \\
\hline
\endhead
\hline \multicolumn{4}{r}{\textit{Continued on next page}} \\
\endfoot
\hline
\endlastfoot
\multicolumn{4}{c}{\textbf{HD 68017 A}} \\
\hline
HIRES-pre & 34 & 1997.034 - 2003.952 & 1.48 \\
HIRES-post & 166 & 2004.813 - 2020.097 & \DIFdelbegin \DIFdel{1.0 }\DIFdelend \DIFaddbegin \DIFadd{1.00 }\DIFaddend \\
APF & 12 & 2013.971 - 2014.297 & 2.85 \\
NEID & 31 & 2021.714 - 2022.389 & 0.43 \\
\multicolumn{4}{c}{\textbf{61 Cygni B}} \\
\hline
Hamilton & 70 & 1987.445 - 2007.891 & 8.15 \\
HIRES-post & 232 & 2004.638 - 2019.987 & 1.07 \\
APF & 259 & 2013.833 - 2019.201 & 2.63 \\
NEID & 40 & 2021.762 - 2022.439 & 0.26 \\
\multicolumn{4}{c}{\textbf{HD 24496 A}} \\
\hline
HIRES-pre & 15 & 1998.067 - 2003.705 & 1.33 \\
HIRES-post & 217 & 2004.638 - 2019.697 & 1.14 \\
NEID & 31 & 2021.724 - 2022.211 & 0.38 \\
\multicolumn{4}{c}{\textbf{HD 4614 A}} \\
\hline
Hamilton & 75 & 1987.814 - 2005.058 & 10.07 \\
HIRES-post & 75 & 2004.81 - 2019.719 & 1.33 \\
APF & 495 & 2013.793 - 2019.566 & 3.04 \\
MINERVA T1 & 116 & 2017.713 - 2019.064 & 2.95 \\
MINERVA T2 & 110 & 2017.713 - 2019.074 & 6.64 \\
MINERVA T4 & 140 & 2017.713 - 2019.031 & 2.56 \\
MINERVA T3 & 63 & 2017.727 - 2019.074 & 4.16 \\
NEID & 157 & 2021.571 - 2022.445 & 0.33 \\
\multicolumn{4}{c}{\textbf{HD 217107}} \\
\hline
Hamilton & 226 & 1998.525 - 2008.872 & 6.09 \\
CORALIE & 63 & 1998.694 - 1999.984 & 9.00 \\
HIRES-pre & 63 & 1998.697 - 2004.523 & 1.41 \\
Tull & 23 & 1999.741 - 2005.872 & 5.30 \\
HIRES-post & 112 & 2004.638 - 2020.009 & 1.02 \\
APF & 330 & 2013.738 - 2019.762 & 2.30 \\
MINERVA T1 & 32 & 2016.392 - 2018.877 & 4.44 \\
MINERVA T2 & 23 & 2016.392 - 2018.866 & 6.48 \\
MINERVA T3 & 14 & 2016.392 - 2017.902 & 11.55 \\
MINERVA T4 & 45 & 2016.392 - 2018.877 & 4.11 \\
NEID & 21 & 2021.71 - 2022.434 & 0.28 \\
\multicolumn{4}{c}{\textbf{HD 190360}} \\
\hline
ELODIE & 56 & 1994.707 - 2002.884 & 7.00 \\
AFOE & 13 & 1996.411 - 2001.683 & 9.00 \\
HIRES-pre & 91 & 1996.771 - 2004.523 & 1.24 \\
Hamilton & 146 & 2002.647 - 2006.694 & 3.99 \\
HIRES-post & 252 & 2004.637 - 2020.185 & 1.17 \\
APF & 754 & 2013.579 - 2020.291 & 2.27 \\
NEID & 29 & 2021.412 - 2022.33 & 0.32 \\
\multicolumn{4}{c}{\textbf{HD 154345}} \\
\hline
ELODIE & 49 & 1994.304 - 2006.545 & 8.00 \\
HIRES-pre & 25 & 1997.267 - 2004.52 & 1.57 \\
HIRES-post & 224 & 2005.155 - 2020.185 & 1.16 \\
SOPHIE & 10 & 2008.134 - 2011.299 & 4.20 \\
NEID & 23 & 2021.721 - 2022.445 & 0.39 \\
\end{longtable}

\begin{minipage}{0.45\textwidth}
\footnotesize
\textbf{Note:} Columns are, from left to right, the instrument used to collect data, the number of observations or individual exposures used by it in our analysis, the range between the first and last of those observations, and the median of each instrument's reported error per star system. All RVs used in this analysis are made available in a corresponding machine-readable table.
\end{minipage}

\begin{figure*}
    \centering
    \includegraphics[width=\linewidth]{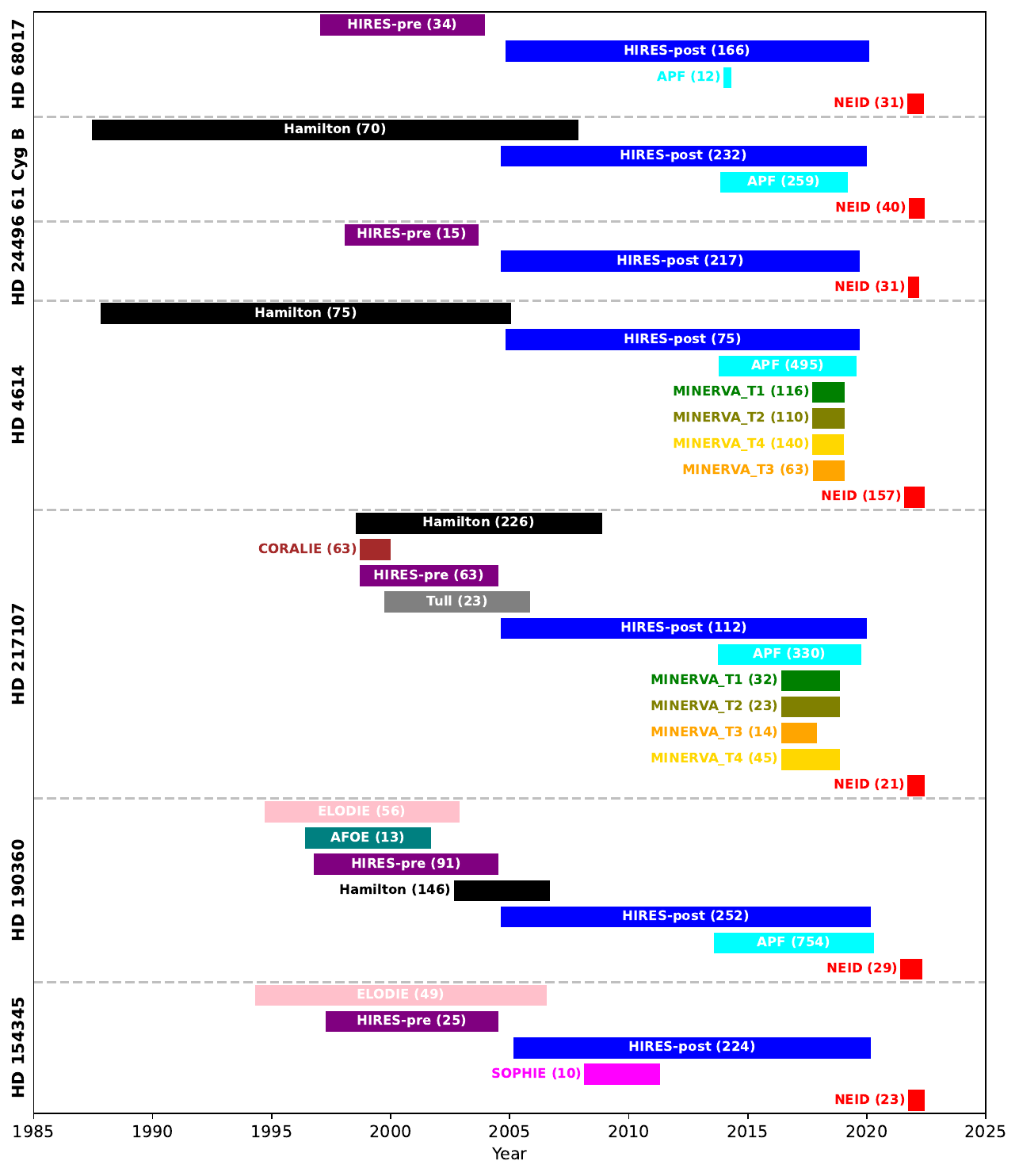}
    \caption{Bar chart of all RVs used for the seven accelerating stars considered in our analysis. Temporal overlap of RV datasets from different instruments helps to constrain instrumental offsets and improves the accuracy of orbital fits by providing a more continuous coverage across the total observation baseline.}
    \label{fig:rv_coverage}
\end{figure*}

\section{Systems} \label{sec:systems}

\begin{figure*}
    \DIFaddendFL \centering
    \includegraphics[width=\linewidth]{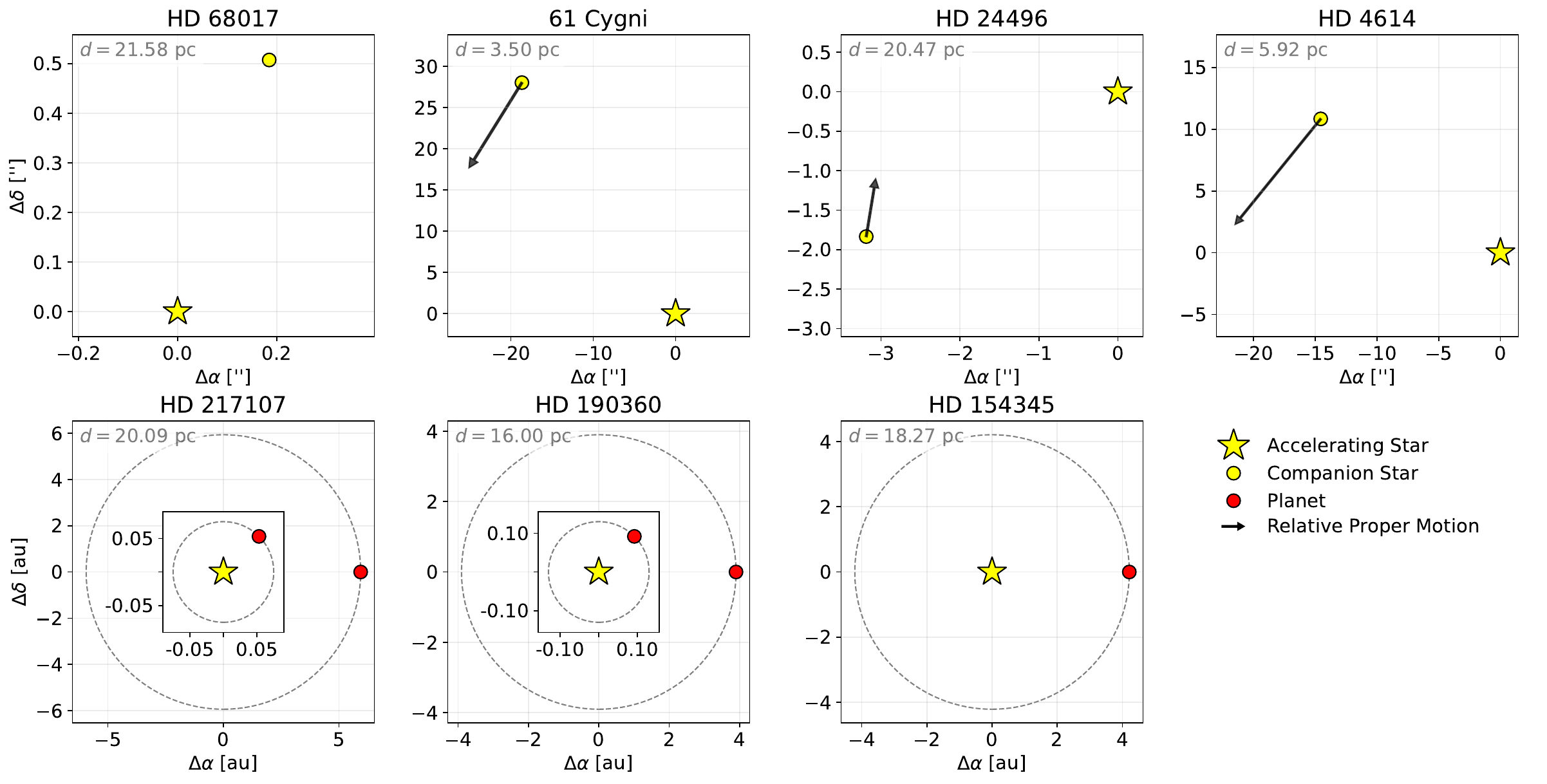}
    \caption{Grid of the seven systems included in our analyses. Each system is centered relative to the position of the primary star at $t=2016.0$ and is viewed with an equal aspect ratio. Primary stars are denoted by yellow stars, bound stellar companions by yellow circles, and their relative proper motions by arrows. The three planetary systems are shown in the bottom row, with the planets marked by red circles. \DIFdelbeginFL \DIFdelFL{The planets' }\DIFdelendFL \DIFaddbeginFL \DIFaddFL{For simplicity, their }\DIFaddendFL orbits are \DIFdelbeginFL \DIFdelFL{drawn }\DIFdelendFL \DIFaddbeginFL \DIFaddFL{sketched }\DIFaddendFL as circles, with \DIFdelbeginFL \DIFdelFL{their }\DIFdelendFL radii corresponding to their semimajor axes\DIFaddbeginFL \DIFaddFL{, and placed at arbitrary position angles}\DIFaddendFL . Systems with two planets feature zoomed-in inset axes to highlight the inner planet. Since HD 68017 AB is not resolved by Gaia, we do not include its relative proper motion information.} \label{fig:systems}
\end{figure*}

Our targets consist of four binary star systems with no known hierarchical companions (61 Cygni, HD 68017, HD 24496, and HD 4614) and three single stars with known planets (HD 217107, HD 190360, and HD 154345). A complete summary of the properties of these systems are given in Table \ref{tab:systems}. We offer a depiction of our seven systems in Figure \ref{fig:systems}.

\subsection{HD 68017} \label{subsec:HD68017}

HD 68017 is a tightly-separated ($\rho\sim0.5\arcsec$) binary star system in a $P\approx70$~yr orbit. The system's primary, HD 68017 A, is a G3 star while the secondary, HD 68017 B, is a mid-M dwarf. The pair was unresolved until \cite{2012ApJ...761...39C} captured two AO images of the system using NIRC2 at Keck and discovered the fainter HD 68017 B ($\Delta K=4.92\pm0.10$). Though HD 68017 A is too close to its bound counterpart to be resolved by most spectrographs, \cite{2012ApJ...761...39C} showed that their contrast in the visible to near-infrared wavelength range is $>100$, limiting HD 68017 B's spectral contributions to Doppler observations. \cite{Brandt_68017} used the Gaia DR2 version of the HGCA to report a moderately eccentric ($e=0.353^{+0.025}_{-0.030}$), $P=70^{+6}_{-8}~\mathrm{yr}$ orbit with $M_\mathrm{A}=0.93\pm0.05~\mathrm{M_\odot}$ and $M_\mathrm{B}=0.1495^{+0.0022}_{-0.0021}~\mathrm{M_\odot}$. Since then, \cite{hgca_edr3} has unveiled a new version of the HGCA that leverages the most recent ephemerides from Gaia eDR3. We incorporate this new version of the HGCA, an additional relative astrometry measurement from \cite{Hirsch_AO}, and 31 new sub-1~$\mathrm{m~s^{-1}}$ RVs data points from NEID.

The pair remains unresolved in Gaia DR3, which has a nominal resolving threshold of $\approx0.23-0.70''$ \citep{2015A&A...576A..74D}. Gaia therefore registers it as a single source, but reports that $\mathrm{RUWE=2.969}$. In this case, since the source is known to be a stellar binary, HD 68017 AB, a comprehensive two-star fit that includes all available astrometry and RVs is sufficient. While it is possible that the reported proper motion measurements are biased in both Hipparcos and Gaia due to the blending of the two stars, the astrometric discrepancy between the two epochs is substantial in comparison, and so effects at this level are ignored. We note that Gaia DR5 projects to resolve this pair, as it will be able to achieve a resolution of $\sim0.1''$\footnote{\url{https://www.cosmos.esa.int/web/gaia/science-performance}}. Here, we include an additional resolved image of HD 68017 AB from \cite{Hirsch_AO}, as well as 67 new RVs from NEID to refine the orbit of the system and produce one of the most precise mass estimates of a main sequence star to-date.

\subsection{61 Cygni} \label{subsec:61Cyg}

As one of the oldest known binary star systems, 61 Cygni has nearly three hundred years worth of observations, starting in 1753 \citep{1833MmRAS...5...13H}. The system comprises 61 Cygni A, a $V=5.2$, K5V star, and 61 Cygni B, a $V=6.0$, K7V star. \cite{kervella_61cyg} measured interferometric radii of the two stars to determine $R_\mathrm{A}=0.665\pm0.005~\mathrm{R_\odot}$ and $R_\mathrm{B}=0.595\pm0.008~\mathrm{R_\odot}$, and used those to fit for an estimated system age of $6.1\pm1.1$~Gyr.

61 Cygni is situated just $\sim$3.5~pc from the Sun, and was the first extrasolar system to have its distance measured \citep{1838MNRAS...4..152B}. \cite{1838MNRAS...4..152B} used their parallax estimate of the bound pair and several early relative astrometry measurements to estimate $P>540$~yr and $M_\mathrm{tot}<0.5~\mathrm{M_\odot}$. \cite{1886AN....113..321P} later published an orbit for the system with $P=782.6$~yr and $i=63.55^\circ$. \cite{1931MNRAS..92..121F} then estimated the pair to be eccentric with $e\sim0.4$ and have an orbital period of order $\sim1000$~yr. Most recently, \cite{2021Ap.....64..160I} used 200 years' worth of relative astrometric observations to determine an orbit with $P=704.858\pm40.221$~yr, $e=0.435\pm0.044$, and $M_\mathrm{tot}=1.286\pm0.107~\mathrm{M_\odot}$. 61 Cygni has also been the subject of many previous analyses on the basis of reconciling an apparent periodic residual in the relative astrometry (see, for example \citealt{61cyg_residual_search1, 61cyg_residual_search2, 61cyg_residual_search3, 61cyg_residual_search4, 61cyg_residual_search5}). 

\cite{2021AJ....162..230B} found that 61 Cygni A is one of 38 sources in the HGCA known to have a corrupted astrometric solution from Hipparcos. For this reason, a comprehensive analysis centered on 61 Cygni A that incorporates a prior on the astrometric acceleration is not feasible. Instead, the primary subject of our analysis is its stellar counterpart 61 Cygni B, which is targeted by the NEID program Searching for Nearby Exoplanets Around K-dwarfs (SNEAK, designed as a complementary survey to NETS; \citealt{SNEAK}) and has a similarly large observed absolute astrometric acceleration. Therefore, we consider the 61 Cygni B's observed acceleration in order to further constrain both the system's orbital parameters and the dynamical mass of 61 Cygni A. We still publish the NETS RVs for 61 Cygni A and use them in Section \ref{sec:companions} to place limits on substellar companions.

\subsection{HD 24496} \label{subsec:HD24496}

HD 24496 is a late G-type star with a K-type stellar companion (HD 24496 B). The pair was first observed to be comoving across nine years by \cite{1980ApJS...44..111H, Hei1990b}. \cite{2021ApJS..255....8R} later identified a linear trend across $\approx20$ years of HIRES RVs of $\approx-7.3~\mathrm{m~s^{-1}~yr^{-1}}$. \cite{hd24496_activity_I} observed HD 24496 to be chromospherically active, and \cite{2004ApJS..152..261W} later used activity indicators from 15 measurements spanning five years to determine a rotational period of 29~d and estimate an age of $\approx3.4~\mathrm{Gyr}$. This estimate is corroborated by \cite{2007ApJS..168..297T}, which interpolated a grid of stellar evolutionary tracks to find an age of $3.16^{+3.88}_{-3.16}$~Gyr. This system previously has had no published orbit.

\subsection{HD 4614} \label{subsec:HD4614}

HD 4614 ($\mathrm{eta}$ Cassiopeiae) is a visual binary system with astrometric measurements spanning nearly 200 years. The primary, HD 4614 A, is a late F-type star, while the secondary is a mid K-type star. \cite{2012ApJ...746..101B} fit temperature and luminosity measurements for HD 4614 A to stellar isochrones to estimate its age to be $5.4\pm0.9$~Gyr.

\begin{figure}
    \DIFaddendFL \centering
    \includegraphics[width=\linewidth]{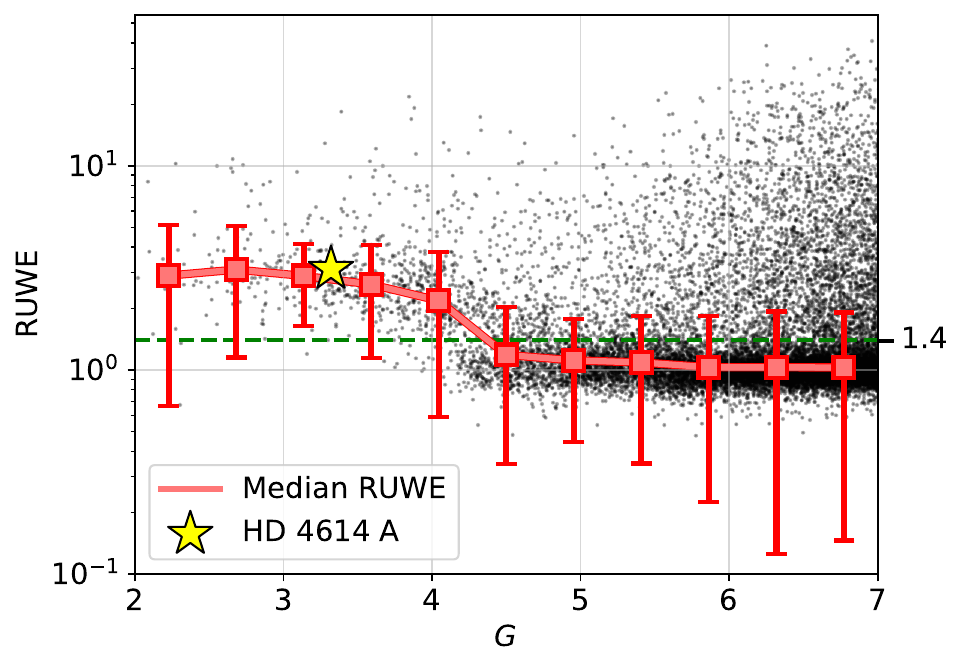}
    \caption{Scatter of all sources in Gaia with apparent magnitude $G<7$. Here, we show that sources with $G~\lesssim~4$ are far less frequently observed to have $\mathrm{RUWE}~<~1.4$, a strong indicator that the RUWE values of these bright sources may be artificially inflated.} \label{fig:brightruwes}
\end{figure}

HD 4614 A is observed by Gaia to have $\mathrm{RUWE}=3.121$, with its stellar companion, HD 4614 B, having a RUWE near unity ($\mathrm{RUWE}=1.039$). The photocenters of bright sources, however, are known to be more difficult to monitor as a result of CCD saturation, which can result in artificially-inflated goodness-of-fit statistics like the RUWE \citep[e.g.,][]{gaia_astrometric_params}. This effect is demonstrated in Figure \ref{fig:brightruwes}. \cite{2012ApJ...746..101B} used the CHARA array \citep{chara} to infer interferometric radii for 44 AFG stars; we use the limb-darkened angular diameter they determined in conjunction with the Gaia DR3 parallax to estimate a radius for HD 4614 A of $1.0336\pm0.0027~\mathrm{R_\odot}$. By comparing HD 4614 A to other stars of similar magnitude and color, as well as to MIST evolutionary tracks for a 5.4~Gyr, $\mathrm{\left[Fe/H\right]}=-0.23$~dex \citep{Hirsch_AO} star, we do not find any considerable evidence that the source is an unresolved stellar multiple and therefore treat it as a single star. We do, however, explore the system for gravitationally-bound substellar companions in Section \ref{sec:companions}.

\DIFdelbegin \DIFdelendFL \DIFaddbeginFL 

\subsection{HD 217107} \label{subsec:HD217107}

HD 217107 is a late G-type star with no stellar companions, but two planets known from RVs, HD 217107~b \citep{1999PASP..111...50F} and HD 217107~c \citep{2005ApJ...632..638V}. The inner planet, HD 217107~b, has $m\sin i \approx 1.4~\mathrm{M_J}$ with $P=7.13$~d, while HD 217107~c is a super Jupiter-mass ($m\sin i \gtrsim 4~\mathrm{M_J}$) planet with $P\approx14$~yr \cite[e.g.,][]{2020AN....341..870G, 2021ApJS..255....8R}. Using an interferometrically-determined angular diameter that considered limb-darkening via \cite{2013ApJ...771...40B}, as well as the Gaia DR3 parallax, we estimate HD 217107 to have a stellar radius of $1.2245\pm0.0173~\mathrm{R_\odot}$.
\cite{2013ApJ...771...40B} also estimated a system age for HD 217107 of $11.9\pm2.0$~Gyr.

\subsection{HD 190360} \label{subsec:HD190360}

HD 190360 is a late G-type star with two planets. \DIFaddbegin \DIFadd{From RVs, }\DIFaddend HD 190360~b \DIFdelbegin \DIFdel{is a $1.5$}\DIFdelend \DIFaddbegin \DIFadd{has a minimum mass of $m\sin i \approx 1.5$}\DIFaddend ~M$_\mathrm{J}$ \DIFdelbegin \DIFdel{planet on }\DIFdelend \DIFaddbegin \DIFadd{with }\DIFaddend a $7.92~\mathrm{yr}$ orbit \citep{HD190360_elodie_afoe, 2021ApJS..255....8R}, \DIFdelbegin \DIFdel{and }\DIFdelend \DIFaddbegin \DIFadd{while }\DIFaddend HD 190360~c is a super Neptune (\DIFdelbegin \DIFdel{$m\sim21.5~\mathrm{M_\oplus}$}\DIFdelend \DIFaddbegin \DIFadd{$m\sin i \approx21.5~\mathrm{M_\oplus}$}\DIFaddend ) on a $17.12~\mathrm{d}$ orbit \citep{2005ApJ...632..638V, 2021ApJS..255....8R}. Following the same method as in the cases of HD 4614 and HD 217107, but using an angular diameter determined via \cite{2008ApJ...680..728B}, we estimate HD 190360's stellar radius to be $1.2010\pm0.0327~\mathrm{R_\odot}$.

HD 190630 is in a wide ($178''$) orbit with a mid M-type stellar companion, HD 190360 B. \cite{mamajek_activity_age_estimation} estimates a system age of 8.5-8.6~Gyr using activity indicators. We note, however, that \cite{2013ApJ...771...40B} estimated a system age of $11.3\pm3.5$~Gyr, highlighting a need for more reliable age estimators for field stars. We estimate HD 190360 B's mass (via the relation derived in \citealt{2022AJ....164..164G}) to be $M=0.20\pm0.02~\mathrm{M_\odot}$. To estimate the expected orbital period and induced relative RV trend, we use \texttt{binary\_mc} \citep{zenodo15352084}\footnote{\url{https://github.com/markgiovinazzi/binary_mc}}, an open-source software that simulates 10 million randomly drawn orbits between two Gaia sources using their source IDs and assumed instantaneous position measurements from the Gaia catalog. This returns an orbital period of the pair to be $\sim100,000$ years. Given its low mass and long orbit, HD 190360 B is only expected to induce an RV slope of $10^{-3}~\mathrm{m~s^{-1}~yr^{-1}}$ on HD 190360 A. Not only is it too distant to be the source of astrometric acceleration, but also its RV trend is too weak to detect given even our best RV precision and the relatively short baseline when compared to the estimated length of the orbit. Therefore, we exclude HD 190360 B from the remainder of our analysis, and hereafter treat HD 190360 as an isolated star. \cite{2017AJ....153..136S} notes the possibility of HD 190360 having a very nearby, unresolved stellar companion, as evidenced by a poor fit to the SED, though we note that Gaia reports $\mathrm{RUWE=1.012}$ for the source. We do not explore this apparent discrepancy in the stellar fit to HD 190360, and simply assume it to be a single star.

\subsection{HD 154345} \label{subsec:HD154345}

HD 154345 is a nearby G8 dwarf with no stellar companions and one known planet, HD 154345~b, which was first reported by \cite{2008ApJ...683L..63W}. HD 154345~b, known only through RVs, is a Jupiter-analog, most recently found to have a minimum mass of \DIFdelbegin \DIFdel{$m\sin i=0.905^{+0.071}_{-0.089}~\mathrm{M_\odot}$ }\DIFdelend \DIFaddbegin \DIFadd{$m\sin i=0.905^{+0.071}_{-0.089}~\mathrm{M_\mathrm{J}}$ }\DIFaddend and an orbital period of $P=3420^{+61}_{-46}~\mathrm{d}$ \citep{2021ApJS..255....8R}. \cite{2008ApJ...683L..63W} found the magnetic activity cycle of the star to be nearly coincidental with that of the induced RV signal from the planet. While they ultimately determined the nature of the periodic signal to be a bona fide planet, both \cite{2008ApJ...683L..63W} and \cite{2009A&A...494..769T} noted that astrometric measurements of the host star could not only confirm the existence of HD 154345~b via dynamical inference, but also determine its true mass.

\DIFdelbegin \DIFdelend \DIFaddbegin \begin{deluxetable*}{cccc|cc|cc}
\DIFaddend \caption{Observed Properties of Stars} \label{tab:systems}
\tablehead{
    \multicolumn{4}{c}{\textbf{Identifiers}} & 
    \multicolumn{2}{c}{\textbf{Astrometry}} & 
    \multicolumn{2}{c}{\textbf{Magnitudes}} \\
    \colhead{\textbf{HD}} & 
    \colhead{\textbf{HIP}} & 
    \colhead{\textbf{Alt ID}} & 
    \colhead{\textbf{Gaia DR3}} & 
\colhead{\textbf{$\pi$}} & 
\colhead{\textbf{RV}} & 
\colhead{\textbf{$G$}} & 
    \colhead{\textbf{$K_\mathrm{s}$}}
}
\startdata
\textbf{68017 \DIFaddbegin \DIFadd{A}\DIFaddend } & 40118 & GJ 9256 & 902007870203594880 & $46.3415\pm0.0609$ & $29.5850\pm0.0037$ & 6.618883 & 5.090 \\
\textbf{201092} & 104217 & 61 Cygni B & 1872046574983497216 & $286.0054\pm0.0289$ & $-64.2480\pm0.0009$ & 5.450645 & 2.320 \\
\textbf{24496 \DIFaddbegin \DIFadd{A}\DIFaddend } & 18267 & GJ 3255 & 43499295632961152 & $48.8490\pm0.0207$ & $18.8710\pm0.0008$ & 6.654032 & 4.995 \\
\textbf{4614 \DIFaddbegin \DIFadd{A}\DIFaddend } & 3821 & $\eta$ Cas A & 425040000962559616 & $168.8322\pm0.1663$ & $8.4040\pm0.0004$& 3.320067& 1.990 \\
\textbf{217107} & 113421 & HR 8734 & 2648914040357320576 & $49.7846\pm0.0263$ & $-13.1480\pm0.0002$ & 5.996743 & 4.536 \\
\textbf{190360} & 98767 & GJ 777 A & 2029433521248546304 & $62.4865\pm0.0354$ & $-45.2540\pm0.0005$ & 5.552787 & 4.076 \\
\textbf{154345} & 83389 & GJ 651 & 1359938520253565952 & $54.7359\pm0.0176$ & $-46.8830\pm0.0007$ & 6.583667 & 5.003 \\
\enddata
\DIFdelbegin \DIFdelend \DIFaddbegin \tablecomments{Observed properties of the seven accelerating stars at the focus of our analyses. The parallaxes, absolute RVs, and $G$ magnitudes come directly from the Gaia DR3 catalog \citep{gaia_dr3}, while the $K_\mathrm{s}$ magnitudes come from 2MASS \citep{2mass}.}
\DIFaddend \end{deluxetable*}

\DIFdelbegin \DIFdelendFL \DIFaddbeginFL 

\section{MCMC Analyses and Orbit Fitting} \label{sec:analyses}

To estimate the orbital parameters of our targets, we employ the Python package \texttt{orvara} (Orbits from Radial Velocity, Absolute, and/or Relative Astrometry; \citealt{orvara}), a robust and efficient orbit-fitting software that incorporates the absolute acceleration derived from the proper motion anomaly between Hipparcos and Gaia measurements, RVs, and relative astrometry between the central body and one or more gravitationally bound companions, all within a Bayesian framework. \texttt{orvara} is a parallel-tempering Markov Chain Monte Carlo (MCMC) sampler using \texttt{ptemcee} \citep{emcee, ptemcee}.

We adopt Gaussian parallax priors based on the reported values in Gaia DR3, but follow \cite{gaia_wide_binaries} in inflating the error by a factor of 1.28 to account for underestimated uncertainties. \texttt{orvara} also uses Gaia proper motions for bound stellar companions in order to estimate the relative motion of the two stars. We then follow \cite{hgca_edr3} in inflating the proper motion uncertainties of resolved companions by a factor of 1.37. \DIFaddbegin \DIFadd{This factor was determined by fitting a Gaussian mixture model to a set of 374 reference stars, treating error inflation as a free parameter. }\DIFaddend When enough relative astrometry is available, no prior information about the primary mass is needed, allowing us to determine true dynamical masses with minimal assumptions \citep{Brandt_68017}. However, for systems where the companion is unresolved or the observational coverage is short compared to the orbital period, and the derived parameters rely only on RVs and absolute astrometry, the masses cannot be reliably determined from our data alone. For these stars, we adopt \DIFdelbegin \DIFdel{mass priors }\DIFdelend \DIFaddbegin \DIFadd{Gaussian priors on mass derived }\DIFaddend from the \texttt{SpecMatch-syn} \citep{specmatch_syn} analyses reported in \cite{Hirsch_AO}. \DIFdelbegin \DIFdel{Our mass priors are provided in Table \ref{tab:priors}. For each spectrograph, we also input a hard-bound region for our RV jitter terms of $\left[10^{-5},~50\right]~\mathrm{m~s^{-1}}$, which is more than sufficient given the relatively good precision of each of our RV instruments. A complete summary of all }\DIFdelend \DIFaddbegin \DIFadd{The remaining priors used in our analysis are outlined in Table 4 of \cite{orvara}, including log-flat priors on semimajor axis (between $10^{-5}-2\times10^5~\mathrm{au}$), secondary mass (between $10^{-4}-10^3~\mathrm{M_\odot}$), and RV jitter (between $10^{-5}-50~\mathrm{m~s^{-1}}$). The prior on primary mass is also log-flat (between $10^{-4}-10^3~\mathrm{M_\odot}$) unless a Gaussian prior is applied. Inclination follows a $\sin i$ prior, where $0^\circ<i<180^\circ$, to account for randomly oriented orbits. A detailed summary of target-specific }\DIFaddend priors assumed in this analysis is provided in Table \ref{tab:priors}.
\DIFdelbegin \DIFdel{Using these priors, we fit orbits }\DIFdelend \DIFaddbegin 

\DIFadd{Our MCMC fits }\DIFaddend for each target star \DIFdelbegin \DIFdel{using }\DIFdelend \DIFaddbegin \DIFadd{use }\DIFaddend 30 temperatures, 100 walkers, and 500,000 steps, keeping every 50th step with a \DIFdelbegin \DIFdel{subsequent }\DIFdelend burn-in of the first 5,000 overall steps from each walker\DIFdelbegin \DIFdel{to ensure that all chains are well-mixed}\DIFdelend \DIFaddbegin \DIFadd{. To assess convergence, we analyze the remaining chains using trace plots, which depict the evolution of sampled parameter values over iterations. Well-mixed, stationary traces without discernible trends or drifts indicate thorough exploration of the parameter space and suggest that the chains have likely converged}\DIFaddend .

\begin{deluxetable*}{lcccc}
\caption{System priors used in this analysis}
\label{tab:priors}
\tablehead{\colhead{System} & \colhead{Primary Mass [$M_\odot$]} & \colhead{System Parallax [mas]} & \colhead{Companion pmRA~$\mathrm{\left[mas~yr^{-1}\right]}$} & \colhead{Companion pmDec~$\mathrm{\left[mas~yr^{-1}\right]}$}}
\startdata
HD 68017  & $\mathcal{U}\left[10^{-4},~10^3\right]$ & $\mathcal{N}\left(46.3415,0.0780\right)$ & -- & -- \\
61 Cygni & $\mathcal{U}\left[10^{-4},~10^3\right]$ & $\mathcal{N}\left(286.0054,0.0370\right)$ & $\mathcal{N}\left(4164.209,0.075\right)$ & $\mathcal{N}\left(3249.614,0.075\right)$ \\
HD 24496  & $\mathcal{N}\left(0.93,0.05\right)$ & $\mathcal{N}\left(48.8490,0.0265\right)$ & $\mathcal{N}\left(209.739,0.062\right)$ & $\mathcal{N}\left(-194.285,0.041\right)$ \\
HD 4614   & $\mathcal{U}\left[10^{-4},~10^3\right]$ & $\mathcal{N}\left(168.8322,0.2129\right)$ & $\mathcal{N}\left(1144.693,0.025\right)$ & $\mathcal{N}\left(-469.668,0.026\right)$ \\
HD 217107 & $\mathcal{N}\left(1.06,0.04\right)$ & $\mathcal{N}\left(49.7846,0.0337\right)$ & -- & -- \\
HD 190360 & $\mathcal{N}\left(0.99,0.04\right)$ & $\mathcal{N}\left(62.4865,0.0453\right)$ & -- & -- \\
HD 154345 & $\mathcal{N}\left(0.89,0.04\right)$ & $\mathcal{N}\left(54.7359,0.0225\right)$ & -- & -- \\
\enddata
\DIFdelbegin \DIFdelend \DIFaddbegin \tablecomments{From left to right, the prior columns are for the system's primary mass, the primary star's parallax, followed by the bound stellar companion's in proper motion (when it is resolved by Gaia) in right ascension and declination. The Gaussian mass priors used for HD 24496 A, HD 217107, HD 190360, and HD 154345 come from \cite{Hirsch_AO}, while the hardbound mass priors (when our analysis has enough information to dynamically determine the mass) are the default setting in \texttt{orvara} left in place to prevent walkers from finding solutions that are not physically motivated. The parallax and proper motion priors both come from Gaia, and include the inflation factors of 1.28 and 1.37, respectively, as outlined in Section \ref{sec:analyses}.}
\DIFaddend \end{deluxetable*}

\section{Results\DIFdelbegin \DIFdel{and Discussion}\DIFdelend } \label{sec:results}

\begin{figure*}
    \centering
\begin{minipage}{0.42\linewidth}
        \centering
        \includegraphics[width=\linewidth]{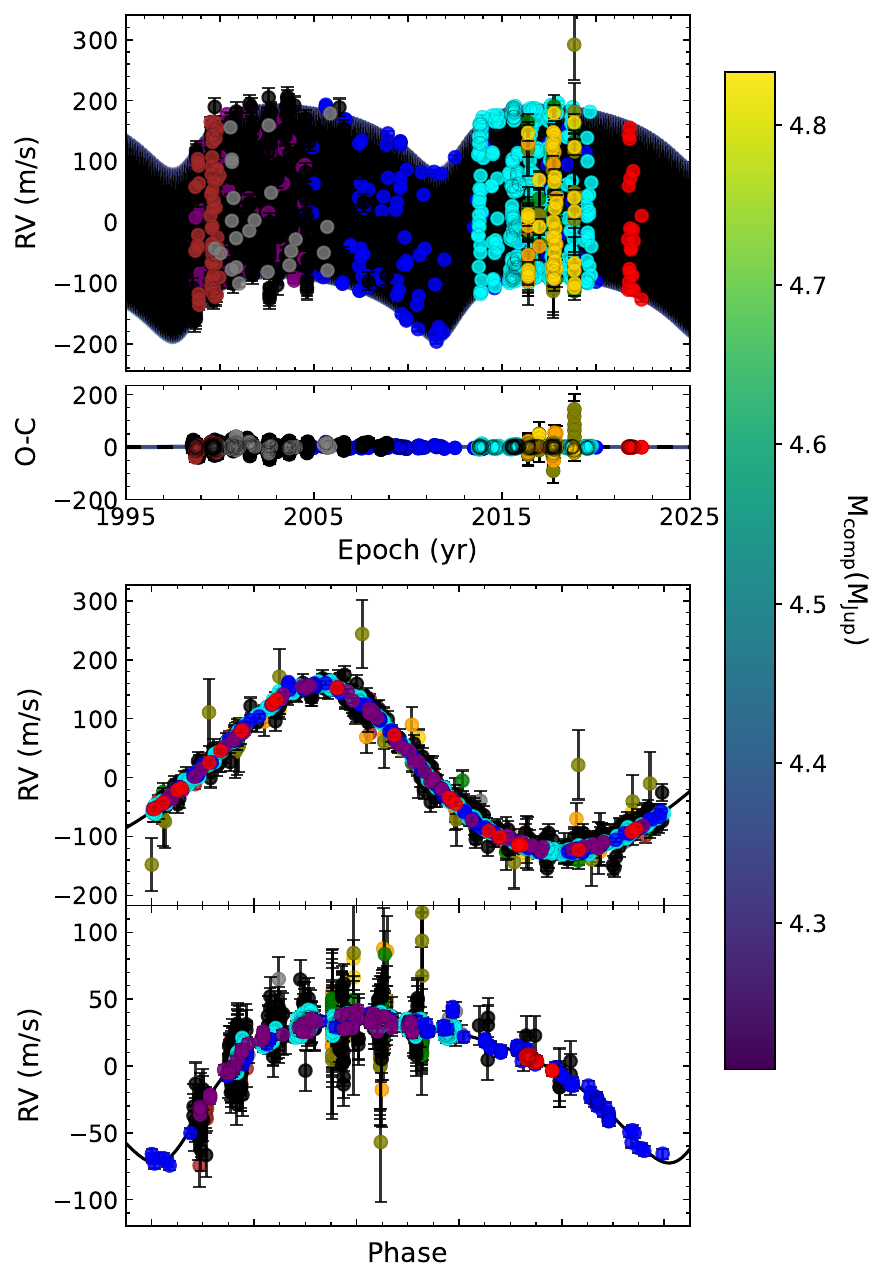}
        {\DIFaddFL{\textbf{(a) HD 217107:} Hamilton, CORALIE, HIRES-pre, Tull, HIRES-post, APF, MINERVA T1-T4, NEID}}
    \end{minipage} \hfill
    \begin{minipage}{0.42\linewidth}
        \centering
        \includegraphics[width=\linewidth]{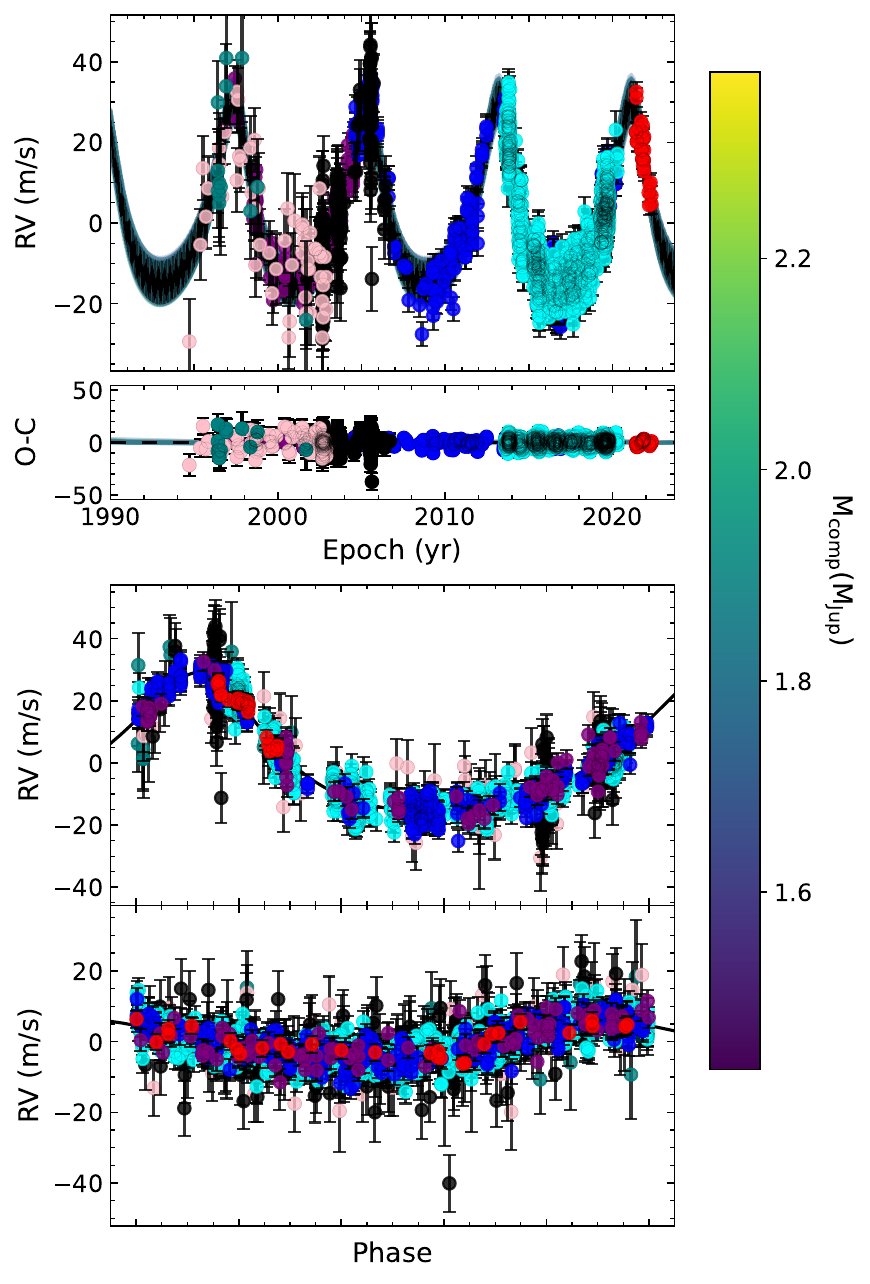}
        {\DIFaddFL{\textbf{(b) HD 190360:} ELODIE, AFOE, HIRES-pre, Hamilton, HIRES-post, APF, NEID}}
    \end{minipage}

    \medskip

\begin{minipage}{0.42\linewidth}
        \centering
        \vspace{-0.3cm}
        \includegraphics[width=\linewidth]{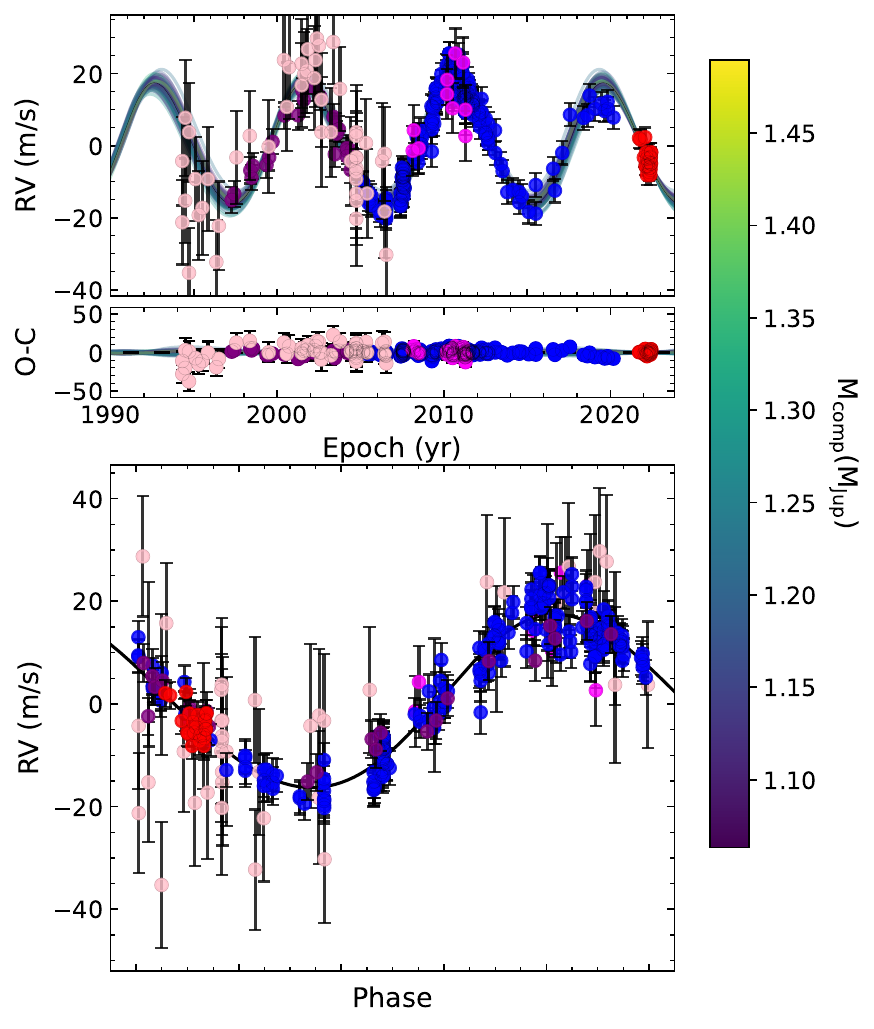}
        {\DIFaddFL{\textbf{(c) HD 154345:} ELODIE, HIRES-pre, HIRES-post, SOPHIE, NEID}}
    \end{minipage}

    \caption{\DIFaddFL{Best-fit RV models for the three planetary systems presented in this analysis (top left: HD 217107; top right: HD 190360; bottom middle: HD 154345). Each panel also includes phase-folded RV curves for planets. Both two-planet systems (HD 217107 and HD 190360) show planet b on top of planet c. The error bars in each panel include instrumental RV jitter terms which have been independently fit for, and the colors of the data points corresponding to those provided in Figure \ref{fig:rv_coverage}.}}
    \label{fig:combined_RVs_planets}
\end{figure*}

For three of our four binary star systems (HD 68017, 61 Cygni, and HD 4614), we rely solely on astrometric and RV data to determine both masses without any prior information on the primary, allowing for a model-independent solution; the exception is HD 24496, with a 600-year orbit and only 40 years of data. In contrast, for our three planetary system, we impose mass priors on the host stars to enable a complete orbital solution despite the lack of relative astrometry. The RV curves for our binary star systems and planetary systems are given in Figure \ref{fig:combined_RVs_binaries} and Figure \ref{fig:combined_RVs_planets}, respectively. This approach uniquely resolves the $m\sin i$ degeneracy typical of RV-only planet detections, allowing us to confirm the planetary nature of these substellar companions. Proper motion anomalies between \DIFaddbegin \DIFadd{the }\DIFaddend Hipparcos and Gaia epochs are provided in \DIFdelbegin \DIFdel{Figures \ref{fig:combined_pms} and }\DIFdelend \DIFaddbegin \DIFadd{Figure \ref{fig:combined_pms} and Figure }\DIFaddend \ref{fig:combined_additional_pms} for binary \DIFaddbegin \DIFadd{star }\DIFaddend and planetary systems, respectively. \DIFaddbegin \DIFadd{These plots depict the absolute astrometry as measured by the two space missions, directly showing the astrometric accelerations that they have constrained. }\DIFaddend For cases with available relative astrometry, we show the relative orbital motions in Figure \ref{fig:combined_orbits}\DIFdelbegin \DIFdel{, with main system orbital posteriors summarized in Table \ref{tab:posteriors}}\DIFdelend \DIFaddbegin \DIFadd{. A complete summary of all free parameters fitted in our models, along with all derived parameters and posterior solutions, is provided in Table \ref{tab:stellar_posteriors} and \ref{tab:planet_posteriors} for the binary star and planetary systems, respectively}\DIFaddend .

\subsection{HD 68017}

Our analysis updates the HD 68017 system orbital period to $P={71.5}_{-4.1}^{+5.0}$~yr, semimajor axis to $a=17.31^{+0.67}_{-0.57}$~au, eccentricity to $e=0.348^{+0.019}_{-0.016}$, and finds masses of $M_\mathrm{A}={0.861}_{-0.028}^{+0.027}~\mathrm{M_\odot}$ and $M_\mathrm{B}=0.1548\pm0.0014~\mathrm{M_\odot}$. This results in relative mass precisions for HD 68017 A and HD 68017 B of $\sigma_{M_\mathrm{A}}/M_\mathrm{A}=3.19\%$ and $\sigma_{M_\mathrm{B}}/M_\mathrm{B}=0.90\%$, respectively. The system's orbital parameters are in excellent agreement with those from \cite{Brandt_68017}, which found $P={70}_{-8}^{+6}$~yr, $a=17.4^{+0.9}_{-1.2}$~au, and $e=0.353^{+0.025}_{-0.030}$. The masses they found, however ($M_\mathrm{A}={0.93}\pm0.05~\mathrm{M_\odot}$ and $M_\mathrm{B}=0.1495^{+0.0022}_{-0.0021}~\mathrm{M_\odot}$) are $\gtrsim1~\sigma$ discrepant. This is likely a direct result of the updated HGCA astrometry between the DR2 and eDR3 releases. We also note that since Gaia does not resolve the two components of the HD 68017 binary system, the photocenter (while dominated by HD 68017 A) is a blend of both sources. Therefore, the proper motion estimates for HD 68017 A may be slightly compromised \DIFaddbegin \DIFadd{(see top panel of Figure \ref{fig:combined_pms})}\DIFaddend . A corner plot highlighting some of the key physical parameters of the system's orbit is provided in Figure \ref{fig:HD68017_corner}.

\subsection{61 Cygni}

\begin{figure*}
    \centering
\begin{minipage}{0.49\textwidth}
        \centering
        \includegraphics[width=\linewidth]{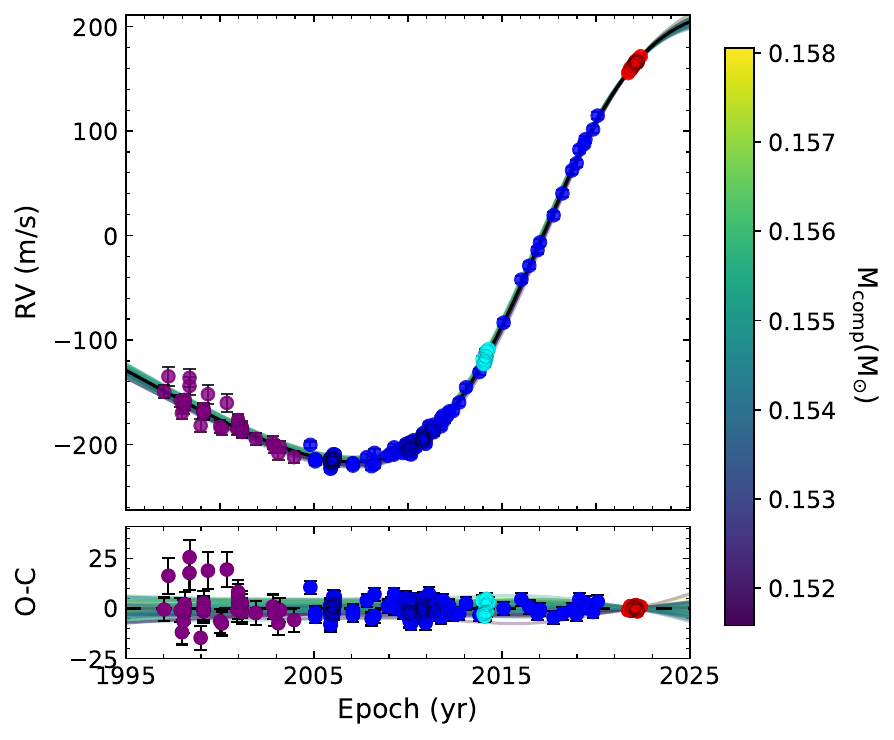}
        {\DIFaddFL{\textbf{(a) HD 68017:} HIRES-pre, HIRES-post, APF, NEID}}
    \end{minipage} \hfill
    \begin{minipage}{0.49\textwidth}
        \centering
        \label{fig:61CygB_RVs}
        \includegraphics[width=\linewidth]{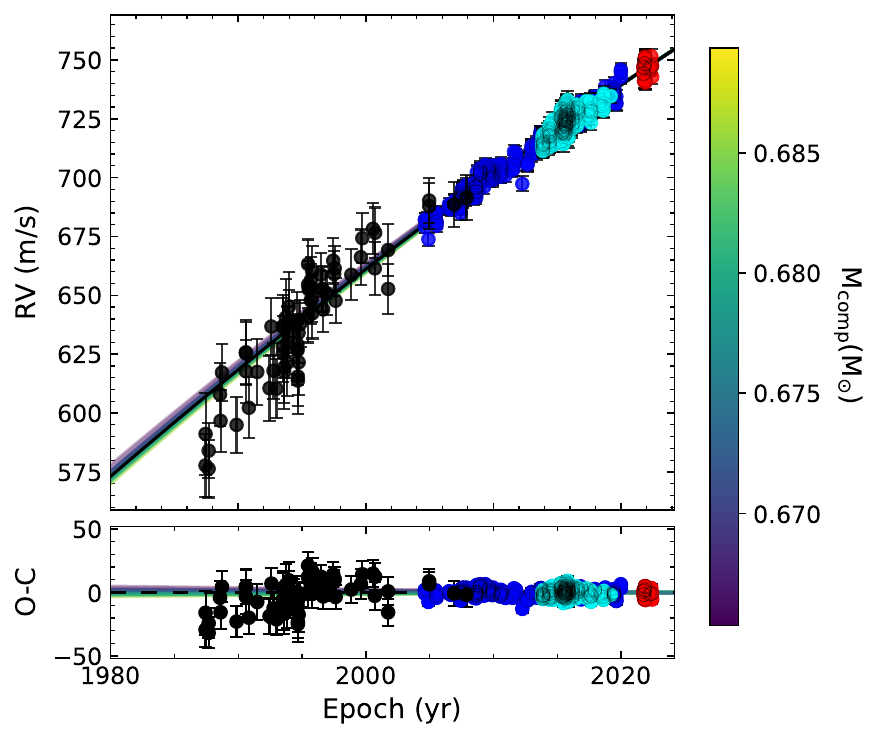}
        {\DIFaddFL{\textbf{(b) 61 Cygni:} Hamilton, HIRES-post, APF, NEID}}
    \end{minipage}
\DIFaddendFL 

    \DIFaddbeginFL \medskip

\begin{minipage}{0.49\textwidth}
        \centering
        \includegraphics[width=\linewidth]{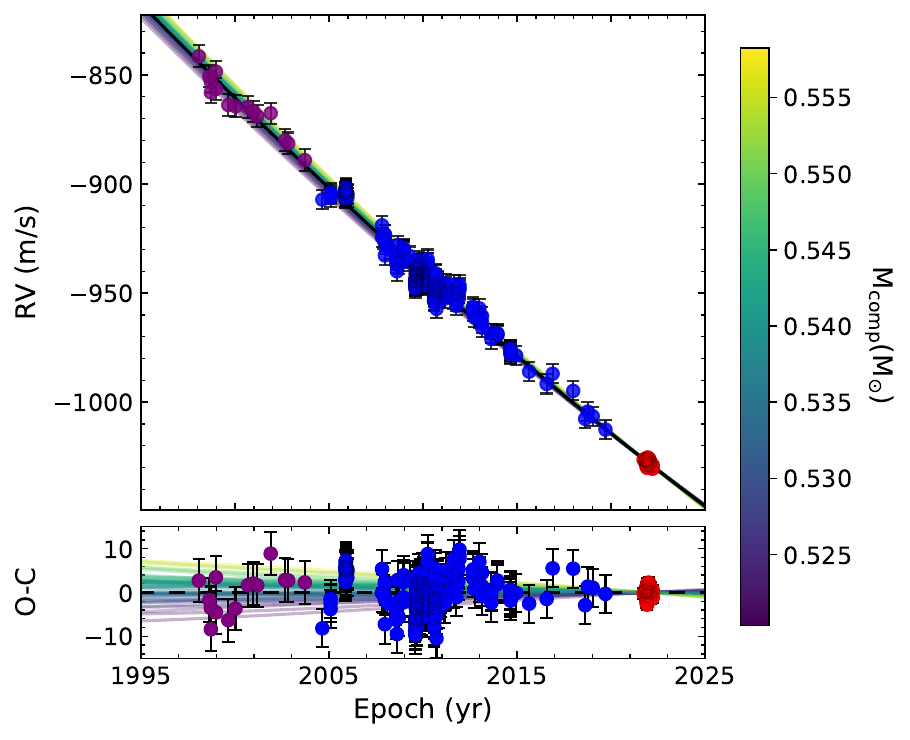}
        {\DIFaddFL{\textbf{(c) HD 24496:} HIRES-pre, HIRES-post, NEID}}
    \end{minipage} \hfill
    \begin{minipage}{0.49\textwidth}
        \centering
        \includegraphics[width=\linewidth]{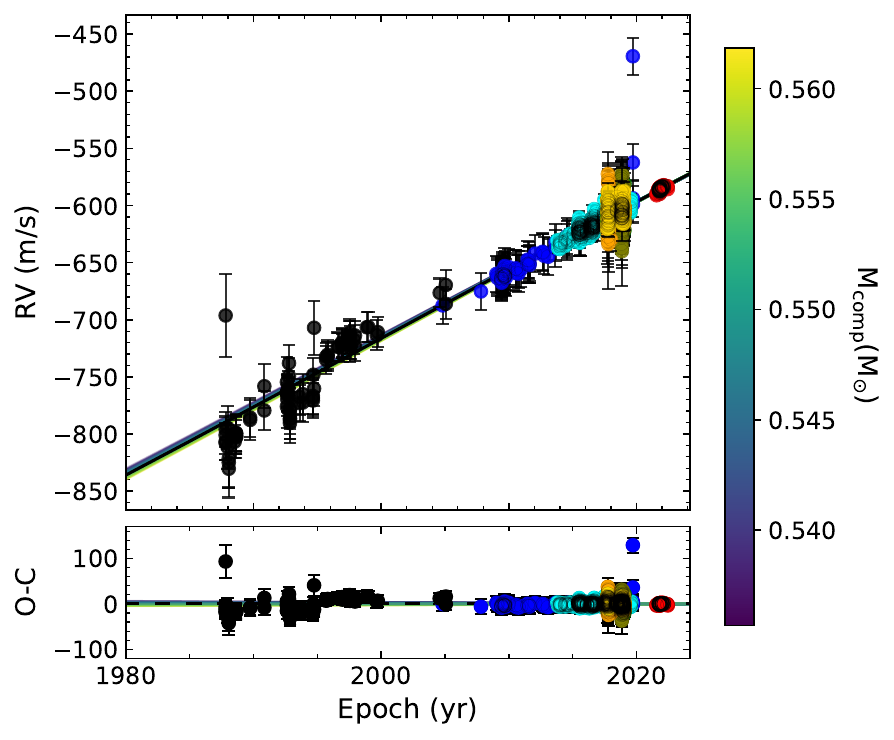}
        {\DIFaddFL{\textbf{(d) HD 4614:} Hamilton, HIRES-post, APF, ~~~~~~~~MINERVA T1-T4, NEID}}
    \end{minipage}

    \caption{\DIFaddFL{Best-fit RV models for the four binary star systems presented in this analysis. The RVs shown here are those collected for the accelerating star in each system (top left: HD 68017 A; top right: 61 Cygni B; bottom left: HD 24496 A; bottom right: HD 4614 A). The error bars in each panel include instrumental RV jitter terms which have been independently fit for, and the colors of the data points corresponding to those provided in Figure \ref{fig:rv_coverage}. In each case, the best-fit model is then subtracted off and the subsequent residuals are consistent with random scatter about zero.}}
    \label{fig:combined_RVs_binaries}
\end{figure*}

We find $P={707.3}_{-4.4}^{+4.5}$~yr with $e=0.4422\pm{0.0056}$ and masses of $M_\mathrm{B}={0.6289}_{-0.0092}^{+0.0094}~\mathrm{M_\odot}$ and $M_\mathrm{A}={0.6772}\pm{+0.0051}~\mathrm{M_\odot}$ for 61 Cygni B and 61 Cygni A, respectively. This translates into 1.48\% and 0.75\% relative precisions in the system's masses, as well as a refined system mass of $M_\mathrm{tot}=1.3060_{-0.0114}^{+0.0118}~\mathrm{M_\odot}$. The long period of the two stars is highlighted by the locally linear line-of-sight trend captured by our RVs, shown in Figure \ref{fig:61CygB_RVs}. Explicit curvature in observed RVs is not necessary to constrain a star's radial acceleration. \cite{gaia_fgk_standards} identified both 61 Cygni A and 61 Cygni B as members of its Gaia FGK standard stars catalog and published two mass estimates for each using two different evolutionary track models (BaSTI: \citealt{BaSTI_I, BaSTI_II}; STAREVOL: \citealt{starevol_I, starevol_II}). The individual masses that we determine are in broad agreement with those derived by these models; using BaSTI, they found $M_\mathrm{B}={0.613}\pm0.011~\mathrm{M_\odot}$ and $M_\mathrm{A}={0.695}\pm0.007~\mathrm{M_\odot}$, while using STAREVOL, they found $M_\mathrm{B}={0.617}\pm0.009~\mathrm{M_\odot}$ and $M_\mathrm{A}={0.679}\pm0.009~\mathrm{M_\odot}$. Figure \ref{fig:61Cyg_corner} shows a select set of system posteriors that we publish for 61 Cygni.

Based on the interferometric radii reported by \cite{kervella_61cyg} and our updated, extremely precise stellar mass estimates, we infer the stellar densities for the two stars in 61 Cygni are $\rho_\mathrm{A}=0.553\pm0.013~\rho_\odot$ and $\rho_\mathrm{B}=0.644\pm0.028~\rho_\odot$.

\subsection{HD 24496}

Here, we publish for the first time a complete orbital solution for the HD 24496 system. \cite{2014AJ....147...86T, 2014AJ....147...87T} used $V-$band mass-magnitude relations to estimate individual masses of $M_\mathrm{A}=0.96~\mathrm{M_\odot}$ and $M_\mathrm{B}=0.53~\mathrm{M_\odot}$, as well a known angular separation to estimate a period of $P=335.541$~yr. HD 24496 represents an excellent example of how a small number of high-precision observations from multiple modes of data can place tight constraints on even a long-period system's orbit. For our best-fit model we find that the two stars have an orbital period of $P={589}_{-84}^{+57}$~yr with masses of $M_\mathrm{A}={0.941}\pm0.053~\mathrm{M_\odot}$ and $M_\mathrm{B}={0.5389}_{-0.0081}^{+0.0082}~\mathrm{M_\odot}$. As described, our analysis includes a prior on HD 24496 A's primary mass. This is because the HD 24496 system has too few relative astrometric points to independently determine a complete orbital solution with high fidelity. However, the joint RV and astrometric fitting provided here enables a robust estimation of an orbit that was previously unsolvable due to a lack of data. Of the four binary star systems presented in this work, HD 24496 has the least eccentric orbit ($e={0.099}_{-0.054}^{+0.130}$), though more data will be needed to enhance the statistical significance of this measurement (see Figure \ref{fig:HD24496_corner}).

\subsection{HD 4614}

\begin{figure*}
    \centering
\begin{minipage}{0.49\textwidth}
        \centering
        \includegraphics[width=\linewidth]{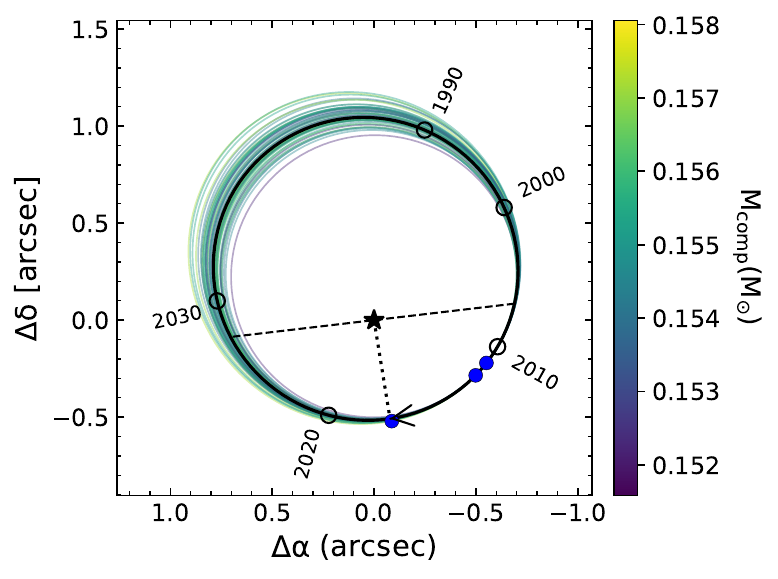}
        \DIFaddFL{\textbf{(a) HD 68017}
        }\label{fig:68017_relast}
    \end{minipage} \hfill
    \begin{minipage}{0.49\textwidth}
        \centering
        \includegraphics[width=\linewidth]{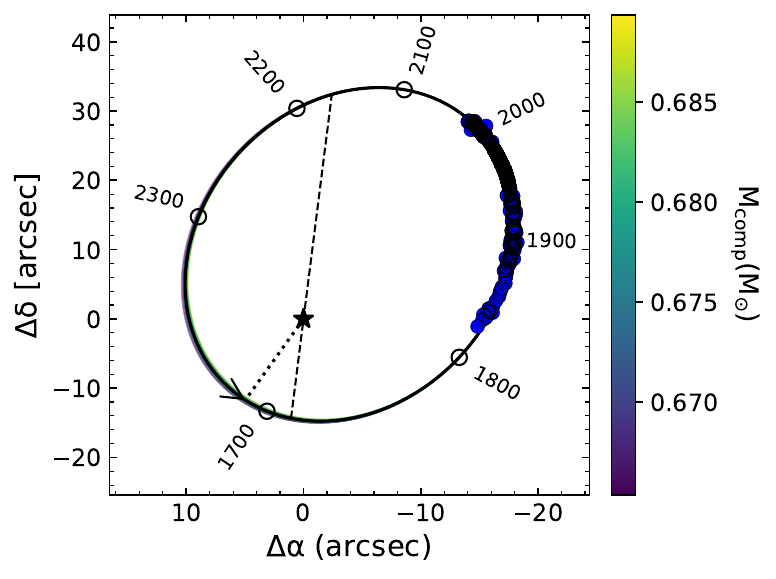}
        \DIFaddFL{\textbf{(b) 61 Cygni}
        }\label{fig:61Cyg_orbit}
    \end{minipage}

    \medskip

\begin{minipage}{0.49\textwidth}
        \centering
        \includegraphics[width=\linewidth]{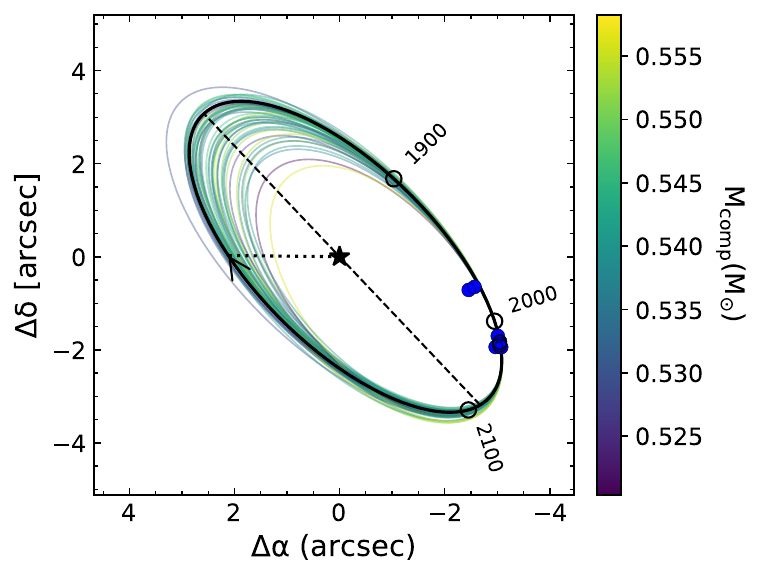}
        \DIFaddFL{\textbf{(c) HD 24496}
        }\label{fig:HD24496_orbit}
    \end{minipage} \hfill
    \begin{minipage}{0.49\textwidth}
        \centering
        \includegraphics[width=\linewidth]{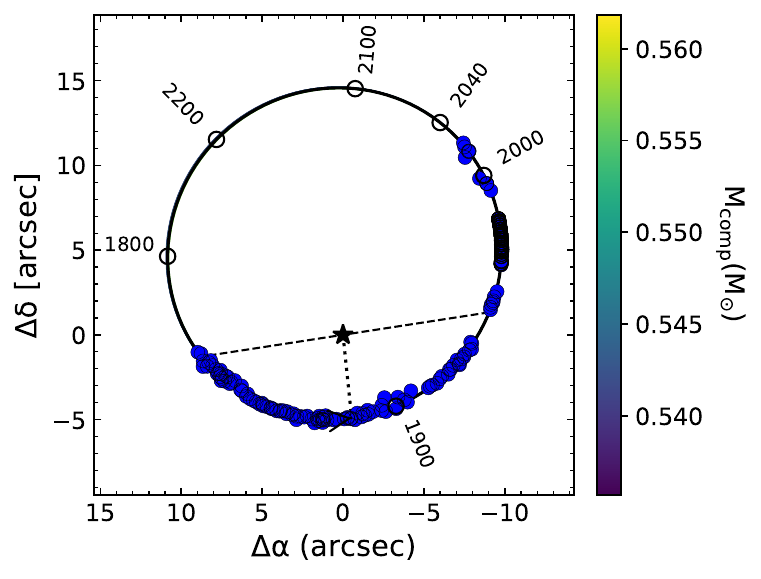}
        \DIFaddFL{\textbf{(d) HD 4614}
        }\label{fig:HD4614_orbit}
    \end{minipage}

    \caption{\DIFaddFL{Projected relative orbits for the four binary star systems considered here (top left: HD 68017 B around HD 68017 A; top right: 61 Cygni A around 61 Cygni B; bottom left: HD 24496 B around HD 24496 A; bottom right: HD 4614 B around HD 4614 A). The filled blue points represent the relative astrometric measurements included in our analysis, while the open circles are the predicted positions of the maximum likelihood orbital models at past and future specified epochs.}}
    \label{fig:combined_orbits}
\end{figure*}

With nearly 200 years' worth of relative astrometry measurements included in our analysis, the observed arc in the HD 4614 system spans close to half the orbit (see Figure \ref{fig:HD4614_orbit}). \cite{1876AN.....88...45D} first estimated the bound pair to have a most likely orbit with $P=222.435$~yr and $e=0.5763$. \cite{1969AJ.....74..760S} later used additional relative astrometry measures and a more robustly-determined parallax estimate to publish an orbit with $P=480$~yr, $e=0.497$, $M_\mathrm{A}=0.86~\mathrm{M_\odot}$, and $M_\mathrm{B}=0.56~\mathrm{M_\odot}$, and this remains one of the literature's most recent publications of the orbit. We find strong agreement with their orbit. Specifically, our analysis estimates $P=472.2\pm1.1$~yr and $e={0.49416}\pm{0.00070}$, $M_\mathrm{A}={1.0258}_{-0.0069}^{+0.0070}~\mathrm{M_\odot}$, and $M_\mathrm{B}={0.5487}\pm{0.0056}~\mathrm{M_\odot}$, yielding relative mass precisions of 0.68\% and 1.02\%, respectively. These individual masses are $\gtrsim2\sigma$ discrepant from those reported by \cite{Hirsch_AO}, which found $M_\mathrm{A}=0.91\pm0.04~\mathrm{M_\odot}$, and $M_\mathrm{B}={0.59}\pm{0.02}~\mathrm{M_\odot}$. On this difference, we note two things. First, stellar mass derivation from spectral characterization efforts like those presented by \cite{Hirsch_AO} are model-dependent. Conversely, while Figure \ref{fig:brightruwes} suggests bright sources like HD 4614 A may have artificially-high RUWE values, it does not preclude the presence of a bound unresolved ultracool dwarf, for example. In this case, it is possible that the dynamical mass of HD 4614 A is actually that of HD 4614 A plus some unknown companion, which would work to bridge the mass discrepancies observed between our analysis (see Figure~\ref{fig:HD4614_corner}) and that of \cite{Hirsch_AO}. \DIFaddbegin \DIFadd{They surveyed 205 binary star companions within 25 pc and found 24 (11.7\%) to have separations $<1$~au across all mass ratios, which for HD 4614 A is likely too close to expect to have resolved any nearby stellar sources given its brightness. \cite{Hirsch_AO} also estimated a planet occurrence rate of $0.12\pm0.04$ planets per star in a binary star system. Together, these findings leave open the possibility that HD 4614 A hosts an unresolved companion. }\DIFaddend Figure \ref{fig:HD4614_corner} shows a corner plot of several physical properties of the HD 4614 system's orbit.

\DIFdelbegin \DIFdel{We have achieved precise constraints on the relative mass of these stars, providing valuable benchmarks for stellar evolutionary models, providing calibration points across a wide range of stellar parameters. Since stellar mass is a fundamental parameter that underpins our understanding of stellar evolution, atmospheric composition, and even the characterization of orbiting exoplanets, the precision of these dynamical masses, particularly at the $1\%$ level, ensures that these stars will serve as critical reference points in future mass-calibration schemes, further constraining the physical properties of stars and their companions. For most main-sequence stars, mass and luminosity are closely correlated (e.g., \citealt{old_mass_lum}). Our refined dynamical mass estimates offer a direct comparison to photometric masses from recent works, such as \cite{2019ApJ...871...63M} and \cite{2022AJ....164..164G}. Figure \ref{fig:phot_mass_comp} illustrates this comparison.
}

{\DIFdelFL{Comparison of our dynamical stellar mass estimates with recent photometric mass relationships. \cite{2019ApJ...871...63M} (MK) use the 2MASS $K_\mathrm{s}$-band, while \cite{2022AJ....164..164G} (GORP) use the Gaia $G_\mathrm{RP}$-band. The HD 68017 AB pair is unresolved in both 2MASS and Gaia and therefore has no reported $K_\mathrm{s}$ or $G_\mathrm{RP}$ magnitudes. However, \cite{2012ApJ...761...39C} used NIRC2 to resolve the two stars in its $K'$ filter, which is nearly identical to that of 2MASS $K_\mathrm{s}$ \citep[e.g.,][]{2010ApJ...718..810S, KELT24_stellar_params}, allowing us to deconvolve the system's composite $K_\mathrm{s}$ magnitude to estimate photometric masses for HD 68017 A and HD 68017 B using the MK method. Similarly, Gaia (but not 2MASS) resolves the HD 24496 AB pair, and so the GORP method is shown for HD 24496 B. HD 24496 A is not provided because a mass prior is used, while HD 4614 A is not shown because it is too massive for either of these two relations. In all cases, the photometric relationships are found to be in good agreement with the dynamical mass estimates.}} 

\DIFdelend \subsection{HD 217107}

We find a true dynamical mass for HD 217107~c of $m_\mathrm{c}=4.37^{+0.13}_{-0.10}~\mathrm{M_J}$, confirming its planetary nature with an orbit close to edge-on ($i_\mathrm{c}=88^{+14\circ}_{-12}$), and a period of $P_c={14.069}\pm0.031$~yr (a $1\sigma$ uncertainty of 11.3~d). Figure~\ref{fig:HD217107c_corner} shows a sample of HD 217107~c's posteriors. For the inner planet, we find a minimum mass of $m_\mathrm{b}\sin i_\mathrm{b}={1.370}_{-0.020}^{+0.016}~\mathrm{M_J}$ and a period of $P_\mathrm{b}=7.126870132\pm0.000004748$~d (a $1\sigma$ uncertainty of 0.41~s). Our updated analysis is consistent with the most recent publication for the HD 217107 planetary system by \cite{2021ApJS..255....8R}, which for the two planets found $m_\mathrm{b}\sin i_\mathrm{b}=1.385\pm0.039~\mathrm{M_J}$, and $m_\mathrm{c}\sin i_\mathrm{c}=4.31\pm0.13~\mathrm{M_J}$.

\begin{figure}
    \DIFaddendFL \centering
    \includegraphics[width=\linewidth]{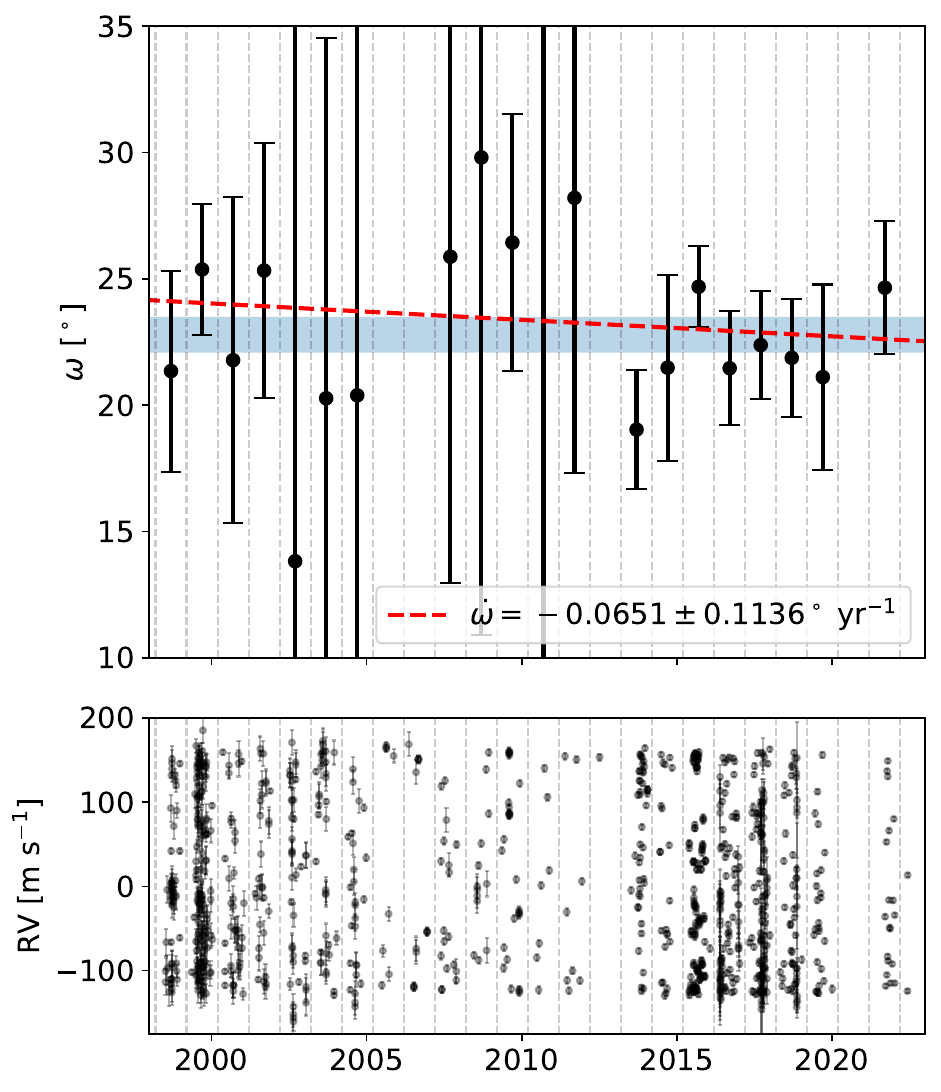}
    \caption{Top: resulting arguments of periastra of HD 217107~b from MCMC fits using subsets of the RVs from our analysis. The rightmost bin of observations consist solely of those taken from NEID. \DIFaddbeginFL \DIFaddFL{Both high-cadence observations at lesser precision and fewer RVs at higher precision are shown to be capable of achieving the same level of constraint. }\DIFaddendFL The best-fit slope representing precession is shown as the solid red line. \DIFaddbeginFL \DIFaddFL{The blue stripe corresponds to the $1\sigma$ region for HD 217107~c's argument of periastron from Table \ref{tab:planet_posteriors}. }\DIFaddendFL Bottom: RVs of the HD 217017 system with the best-fit model for HD 217107~c subtracted off. The vertical dashed lines show the corresponding bins of RVs used to search for $\dot{\omega}$.} \label{fig:HD217107_precess}
\end{figure}

\cite{2020AN....341..870G} observed that the hot Jupiter HD 217107~b is expected to undergo long-term orbital precession at the level of $\dot{\omega}=0.008^\circ~\mathrm{yr}^{-1}$ as a result of general relativistic effects. Although they were unable to measure the effect, they noted that a longer baseline of more precise RVs was needed. Our analysis for HD 217107 uses 2.5 more years of precision RVs, and so we attempt to recover the effect. We subtracted off the best-fit RV model for HD 217107~c, including instrumental RV zero-point offsets and jitter terms, and reran a one-planet fit for HD 217107~b using \texttt{orvara} for yearly subsets of RV measurements to attempt to measure the time-evolution of the inner planet's argument of periastron (see Figure \ref{fig:HD217107_precess}). Ultimately, this analysis can only determine $\dot{\omega}=-0.0651\pm0.1136^\circ~\mathrm{yr}^{-1}$ \DIFaddbegin \DIFadd{and therefore }\DIFaddend cannot confidently measure the predicted GR-induced orbital precession of HD 217107~b. The NEID-only epoch of RVs considered here constrains $\omega_\mathrm{b}=24.65\pm2.63^\circ$, suggesting that subsequent \DIFaddbegin \DIFadd{high-cadence }\DIFaddend RV monitoring of the system over the next decade from \DIFdelbegin \DIFdel{high-precision RV instruments }\DIFdelend \DIFaddbegin \DIFadd{precision spectrographs }\DIFaddend like NEID may be \DIFdelbegin \DIFdel{capable of observing }\DIFdelend \DIFaddbegin \DIFadd{able to observe }\DIFaddend this effect.

\DIFdelbegin \DIFdelendFL \DIFaddbeginFL 

\subsection{HD 190360}

For HD 190360, our joint fitting determines a true dynamical mass of $m_\mathrm{b}={1.68}_{-0.16}^{+0.26}~\mathrm{M_J}$, $P_\mathrm{b}={7.917}_{-0.010}^{+0.011}$~yr (a $1\sigma$ uncertainty of 3.84~d), and $e_\mathrm{b}={0.3184}_{-0.0063}^{+0.0065}$. In this case, the distribution of accepted steps for the planet's inclination is bimodal around the edge-on, $i=90^\circ$ threshold, resulting from an inability to determine \DIFdelbegin \DIFdel{whether the orbit is prograde }\DIFdelend \DIFaddbegin \DIFadd{the relative grade of the orbit }\DIFaddend ($i<90^\circ$ \DIFdelbegin \DIFdel{) or retrograde (}\DIFdelend \DIFaddbegin \DIFadd{or }\DIFaddend $i>90^\circ$) without relative astrometry. Our \DIFdelbegin \DIFdel{solution }\DIFdelend \DIFaddbegin \DIFadd{dynamical mass }\DIFaddend for HD 190360~b is in \DIFdelbegin \DIFdel{strong }\DIFdelend \DIFaddbegin \DIFadd{broad }\DIFaddend agreement with \cite{feng_190360}, who also considered Hipparcos and Gaia astrometry and found $m_\mathrm{b}=1.8\pm0.2~\mathrm{M_J}$\DIFdelbegin \DIFdel{, $P_\mathrm{b}=7.815\pm0.035$~yr, $e_\mathrm{b}=0.386\pm0.025$, and $i_\mathrm{b}=80.2\pm23.2^\circ$}\DIFdelend . As for the inner planet, HD 190360~c, the astrometric signal it induces on its host star is negligible when compared to the precision and observational baseline between Hipparcos and Gaia. However, our comprehensive analysis presented here offers enhanced clarity on the planet's ephemerides by including new RVs from NEID. We find HD 190360~c to have $m_\mathrm{c}\sin i_\mathrm{c}=1.371\pm0.018~\mathrm{M_J}$, $P_\mathrm{c}=17.1169664\pm0.0005113$~d (a $1\sigma$ uncertainty of 44.18~s), and $e_\mathrm{c}={0.165}_{-0.024}^{+0.023}$. As is the case for planets known through RVs, the correlation between mass and inclination is still apparent for HD 190360~b, though the proper motion anomaly is such that a disentanglement of these two parameters is beginning to emerge (see Figure \ref{fig:HD190360 B_corner}). Additional astrometric data in the form of imaging or extended proper motion monitoring from future releases of Gaia, for example, will be needed to more clearly unveil the architecture of HD 190360 planetary system.

\subsection{HD 154345}

As the only known planet in its system, HD 154345~b is the sole contributor to its star's observed acceleration. Ultimately, our analysis not only finds a true dynamical mass for HD 154345~b of $m_\mathrm{b}={1.186}_{-0.059}^{+0.095}~\mathrm{M_J}$, but also refines its orbit to determine an orbital period of $P_\mathrm{b}={8.981}_{-0.076}^{+0.079}$~yr (a $1\sigma$ uncertainty of 28.31~d) and an eccentricity of $e_\mathrm{b}={0.058}\pm0.019$. \cite{2021ApJS..255....8R} was the most recent RV-only analysis for HD 154345, and found $m_\mathrm{b}\sin i_\mathrm{b}=0.905^{+0.071}_{-0.089}~\mathrm{M_J}$\DIFaddbegin \DIFadd{, $P_\mathrm{b}=9.36^{+0.17}_{-0.13}~\mathrm{yr}$ }\DIFaddend and $e_\mathrm{b}=0.038^{+0.036}_{-0.027}$. \cite{xiao_154345} since used the Gaia eDR3 version of the HGCA \DIFaddbegin \DIFadd{and a reduction of HIRES RVs published by \cite{2019MNRAS.484L...8T} }\DIFaddend to estimate a true dynamical mass for HD 154345~b of $m_\mathrm{b}={1.19}_{-0.11}^{+0.14}~\mathrm{M_J}$, as well as an orbital period of $P_\mathrm{b}=9.15\pm0.11$~yr and an eccentricity of $e_\mathrm{b}=0.157^{+0.030}_{-0.029}$. Our mass estimate for HD 154345~b agrees well with that from \cite{xiao_154345}, but the eccentricity of our orbit is more in line with that from \cite{2021ApJS..255....8R}. We \DIFdelbegin \DIFdel{note, though, that the instrumental upgrade for HIRES in 2004 treated here as separate instruments was ignored in the analysis published by \cite{xiao_154345}, which could explain the discrepancy in eccentricity}\DIFdelend \DIFaddbegin \DIFadd{conclude that this discrepancy is the result of the choice in reduction of HIRES RVs between those from \cite{2019MNRAS.484L...8T} and those from \cite{2021ApJS..255....8R}. Our choice to use the HIRES RVs published in \cite{2021ApJS..255....8R} comes from their recommendation in Section 3.1 of their work}\DIFaddend . The astrometric signal detected for HD 154345 is consistent with the idea that the observed RV signal about this star is driven by a physical companion, and works to not only confirm the planetary nature of HD 154345~b, but also to refine its orbit. A select set of orbital parameters describing the HD 154345 system is provided in the corner plot in Figure \ref{fig:HD154345b_corner}.

\DIFdelbegin \DIFdel{Each of the three cold Jupiters presented here, HD 217107~c, HD 190360~b, and HD 154345~b, are both massive and widely-separated enough from their stars that they not only induce a measurable astrometric reflex motion on their host stars, but also may be amenable for direct imaging with missions like JWST \citep{jwst}. Figure \ref{fig:combined_relsep_planets} shows the best-fit relative separation models for each of these three planets between 2020-2030. Each planet has at least one window between 2025-2030 where the angular separation from its star projects to be $\rho\gtrsim0\farcs20$. Using our reported true dynamical masses for these planets, and assuming ages of 5~Gyr for each, we use the SpeX Prism Spectral Analysis Toolkit (SPLAT; \citealt{splat}) to estimate their effective temperatures. For HD 217107~c, HD 190360~b, and HD 154345~b, we find $T_\mathrm{eff}=197.89\pm3.19$~K, $T_\mathrm{eff}\approx125$~K, and $T_\mathrm{eff}\approx125$~K, respectively.
}

\DIFdelend \DIFaddbegin \begin{deluxetable*}{lcccc}
\DIFaddend \caption{Posterior \DIFdelbegin \DIFdel{Distributions }\DIFdelend \DIFaddbegin \DIFadd{distributions }\DIFaddend for \DIFdelbegin \DIFdel{Binary and Planetary Systems}\DIFdelend \DIFaddbegin \DIFadd{the four binary star systems}\DIFaddend }
\DIFdelbegin \DIFdelend \DIFaddbegin \label{tab:stellar_posteriors}
\tablehead{\colhead{Parameter} & \colhead{HD 68017} & \colhead{61 Cygni} & \colhead{HD 24496} & \colhead{HD 4614}}
\DIFaddend \startdata
\DIFdelbegin \DIFdelend $M_\mathrm{pri}$ \DIFdelbegin \DIFdel{($M_\odot$) }\DIFdelend \DIFaddbegin [\DIFadd{$\mathrm{M}_\odot$}] \DIFaddend & ${0.861}_{-0.028}^{+0.027}$ & ${0.6289}_{-0.0092}^{+0.0094}$ & \DIFdelbegin \DIFdel{${0.941}_{-0.053}^{+0.053}$ }\DIFdelend \DIFaddbegin \DIFadd{${0.941}\pm{0.053}$ }\DIFaddend & ${1.0258}_{-0.0069}^{+0.0070}$ \\
$M_\mathrm{sec}$ \DIFdelbegin \DIFdel{($M_\odot$) }\DIFdelend \DIFaddbegin [\DIFadd{$\mathrm{M}_\odot$}] \DIFaddend & \DIFdelbegin \DIFdel{${0.1548}_{-0.0014}^{+0.0014}$ }\DIFdelend \DIFaddbegin \DIFadd{${0.1548}\pm{0.0014}$ }\DIFaddend & ${0.6771}_{-0.0051}^{+0.0052}$ & ${0.5389}_{-0.0081}^{+0.0082}$ & \DIFdelbegin \DIFdel{${0.5487}_{-0.0056}^{+0.0056}$ }\DIFdelend \DIFaddbegin \DIFadd{${0.5487}\pm{0.0056}$ }\DIFaddend \\
$a$ \DIFdelbegin \DIFdel{(au) }\DIFdelend \DIFaddbegin [\DIFadd{au}] \DIFaddend & ${17.31}_{-0.57}^{+0.67}$ & ${86.76}_{-0.15}^{+0.16}$ & ${80.2}_{-7.8}^{+4.6}$ & \DIFdelbegin \DIFdel{${70.55}_{-0.15}^{+0.15}$ }\DIFdelend \DIFaddbegin \DIFadd{${70.55}\pm{0.15}$ }\DIFaddend \\
$\sqrt{e}\cos\omega$ & ${-0.022}_{-0.036}^{+0.035}$ & \DIFdelbegin \DIFdel{${-0.5794}_{-0.0018}^{+0.0018}$ }\DIFdelend \DIFaddbegin \DIFadd{${-0.5794}\pm{0.0018}$ }\DIFaddend & ${0.23}_{-0.17}^{+0.18}$ & \DIFdelbegin \DIFdel{${0.0203}_{-0.0030}^{+0.0030}$ }\DIFdelend \DIFaddbegin \DIFadd{${0.0203}\pm{0.0030}$ }\DIFaddend \\
$\sqrt{e}\sin\omega$ & \DIFdelbegin \DIFdel{${-0.589}_{-0.014}^{+0.014}$ }\DIFdelend \DIFaddbegin \DIFadd{${-0.589}\pm{0.014}$ }\DIFaddend & \DIFdelbegin \DIFdel{${0.3265}_{-0.0087}^{+0.0087}$ }\DIFdelend \DIFaddbegin \DIFadd{${0.3265}\pm{0.0087}$ }\DIFaddend & ${-0.220}_{-0.038}^{+0.044}$ & ${0.70267}_{-0.00059}^{+0.00058}$ \\
$i$ \DIFdelbegin \DIFdel{(}\DIFdelend \DIFaddbegin [\DIFaddend $^\circ$\DIFdelbegin \DIFdel{) }\DIFdelend \DIFaddbegin ] \DIFaddend & \DIFdelbegin \DIFdel{${169.46}_{-0.14}^{+0.14}$ }\DIFdelend \DIFaddbegin \DIFadd{${169.46}\pm{0.14}$ }\DIFaddend & \DIFdelbegin \DIFdel{${52.99}_{-0.12}^{+0.12}$ }\DIFdelend \DIFaddbegin \DIFadd{${52.99}\pm{0.12}$ }\DIFaddend & \DIFdelbegin \DIFdel{${117.11}_{-0.91}^{+2.2}$ }\DIFdelend \DIFaddbegin \DIFadd{${117.11}_{-0.91}^{+2.20}$ }\DIFaddend & \DIFdelbegin \DIFdel{${34.938}_{-0.078}^{+0.078}$ }\DIFdelend \DIFaddbegin \DIFadd{${34.938}\pm{0.078}$ }\DIFaddend \\
$\Omega$ \DIFdelbegin \DIFdel{(}\DIFdelend \DIFaddbegin [\DIFaddend $^\circ$\DIFdelbegin \DIFdel{) }\DIFdelend \DIFaddbegin ] \DIFaddend & \DIFdelbegin \DIFdel{${97.06}_{-0.35}^{+0.35}$ }\DIFdelend \DIFaddbegin \DIFadd{${97.06}\pm{0.35}$ }\DIFaddend & \DIFdelbegin \DIFdel{${355.73}_{-0.30}^{+0.30}$ }\DIFdelend \DIFaddbegin \DIFadd{${355.73}\pm{0.30}$ }\DIFaddend & ${38.7}_{-2.4}^{+1.2}$ & \DIFdelbegin \DIFdel{${98.31}_{-0.15}^{+0.15}$ }\DIFdelend \DIFaddbegin \DIFadd{${98.31}\pm{0.15}$ }\DIFaddend \\
$\lambda$ \DIFdelbegin \DIFdel{(}\DIFdelend \DIFaddbegin [\DIFaddend $^\circ$\DIFdelbegin \DIFdel{) }\DIFdelend \DIFaddbegin ] \DIFaddend & ${233.02}_{-0.64}^{+0.65}$ & \DIFdelbegin \DIFdel{${313.66}_{-0.33}^{+0.33}$ }\DIFdelend \DIFaddbegin \DIFadd{${313.66}\pm{0.33}$ }\DIFaddend & ${138.1}_{-8.5}^{+3.4}$ & \DIFdelbegin \DIFdel{${180.232}_{-0.073}^{+0.073}$ }\DIFdelend \DIFaddbegin \DIFadd{${180.232}\pm{0.073}$ }\DIFaddend \\
$\pi$ \DIFdelbegin \DIFdel{(mas) }\DIFdelend \DIFaddbegin [\DIFadd{mas}] \DIFaddend & ${46.342}\pm0.055$ & ${286.006}\pm0.0290$ & ${48.848}_{-0.021}^{+0.020}$ & ${168.81}_{-0.17}^{+0.16}$ \\
$d$ \DIFdelbegin \DIFdel{(pc) }\DIFdelend \DIFaddbegin [\DIFadd{pc}] \DIFaddend & ${21.579}\pm0.011$ & ${3.496}\pm0.0004$ & ${20.472}\pm0.004$ & ${5.92}\pm0.01$ \\
$P$ \DIFdelbegin \DIFdel{(yr) }\DIFdelend \DIFaddbegin [\DIFadd{yr}] \DIFaddend & ${71.5}_{-4.1}^{+5.0}$ & ${707.1}_{-4.4}^{+4.5}$ & ${589}_{-84}^{+57}$ & \DIFdelbegin \DIFdel{${472.2}_{-1.1}^{+1.1}$ }\DIFdelend \DIFaddbegin \DIFadd{${472.2}\pm{1.1}$ }\DIFaddend \\
$\omega$ \DIFdelbegin \DIFdel{(}\DIFdelend \DIFaddbegin [\DIFaddend $^\circ$\DIFdelbegin \DIFdel{) }\DIFdelend \DIFaddbegin ] \DIFaddend & \DIFdelbegin \DIFdel{${267.8}_{-3.4}^{+3.4}$ }\DIFdelend \DIFaddbegin \DIFadd{${267.8}\pm{3.4}$ }\DIFaddend & ${150.61}_{-0.67}^{+0.68}$ & ${318}_{-32}^{+13}$ & \DIFdelbegin \DIFdel{${88.34}_{-0.25}^{+0.25}$ }\DIFdelend \DIFaddbegin \DIFadd{${88.34}\pm{0.25}$ }\DIFaddend \\
$e$ & ${0.348}_{-0.016}^{+0.019}$ & ${0.4424}_{-0.0057}^{+0.0056}$ & ${0.099}_{-0.054}^{+0.13}$ & \DIFdelbegin \DIFdel{${0.49416}_{-0.00070}^{+0.00070}$ }\DIFdelend \DIFaddbegin \DIFadd{${0.49416}\pm{0.00070}$ }\DIFaddend \\
$a$ \DIFdelbegin \DIFdel{(mas) }\DIFdelend \DIFaddbegin [\DIFadd{mas}] \DIFaddend & ${802}_{-26}^{+31}$ & ${24814}_{-43}^{+45}$ & ${3919}_{-383}^{+224}$ & ${11909}_{-21}^{+22}$ \\
\DIFdelbegin \DIFdel{$T_0$ (JD) }\DIFdelend \DIFaddbegin \DIFadd{$T_\mathrm{p}$ }[\DIFadd{JD}] \DIFaddend & \DIFdelbegin \DIFdel{${2457722}_{-99}^{+99}$ }\DIFdelend \DIFaddbegin \DIFadd{${2457722}\pm{99}$ }\DIFaddend & ${2596496}_{-1233}^{+1255}$ & ${2557107}_{-10043}^{+8299}$ & ${2583650}_{-419}^{+421}$ \\
$q$ & ${0.1798}_{-0.0055}^{+0.0061}$ & \DIFdelbegin \DIFdel{${1.077}_{-0.016}^{+0.016}$ }\DIFdelend \DIFaddbegin \DIFadd{${1.077}\pm{0.016}$ }\DIFaddend & ${0.573}_{-0.031}^{+0.035}$ & \DIFdelbegin \DIFdel{${0.5350}_{-0.0084}^{+0.0084}$ }\DIFdelend \DIFaddbegin \DIFadd{${0.5350}\pm{0.0084}$ }\DIFaddend \\
\DIFdelbegin \DIFdelend \DIFaddbegin \DIFadd{$\sigma_\mathrm{NEID}$ }[\DIFadd{$\mathrm{m~s^{-1}}$}] & \DIFadd{${0.78}_{-0.12}^{+0.15}$ }& \DIFadd{${3.54}_{-0.37}^{+0.44}$ }&  \DIFadd{${1.19}_{-0.15}^{+0.19}$ }& \DIFadd{${1.142}_{-0.067}^{+0.073}$ }\DIFaddend \\
\DIFdelbegin \DIFdelend \DIFaddbegin \DIFadd{$\sigma_\mathrm{MINERVA_{T1}}$ }[\DIFadd{$\mathrm{m~s^{-1}}$}] \DIFaddend & \DIFdelbegin \DIFdelend \DIFaddbegin \DIFadd{-- }\DIFaddend & \DIFdelbegin \DIFdelend \DIFaddbegin \DIFadd{-- }\DIFaddend & \DIFdelbegin \DIFdelend \DIFaddbegin \DIFadd{-- }\DIFaddend & \DIFdelbegin \DIFdelend \DIFaddbegin \DIFadd{${8.11}_{-0.62}^{+0.68}$ }\\
\DIFadd{$\sigma_\mathrm{MINERVA_{T2}}$ }[\DIFadd{$\mathrm{m~s^{-1}}$}] \DIFaddend & \DIFdelbegin \DIFdelend \DIFaddbegin \DIFadd{-- }& \DIFadd{-- }& \DIFadd{-- }& \DIFadd{${12.3}_{-1.1}^{+1.2}$ }\DIFaddend \\
\DIFdelbegin \DIFdelend \DIFaddbegin \DIFadd{$\sigma_\mathrm{MINERVA_{T3}}$ }[\DIFadd{$\mathrm{m~s^{-1}}$}] & \DIFadd{-- }& \DIFadd{-- }& \DIFadd{-- }& \DIFadd{${5.5}_{-1.0}^{+1.1}$ }\\
\DIFadd{$\sigma_\mathrm{MINERVA_{T4}}$ }[\DIFadd{$\mathrm{m~s^{-1}}$}] & \DIFadd{-- }& \DIFadd{-- }& \DIFadd{-- }& \DIFadd{${5.68}_{-0.42}^{+0.46}$ }\\
\DIFadd{$\sigma_\mathrm{HIRES_\mathrm{pre}}$ }[\DIFadd{$\mathrm{m~s^{-1}}$}] & \DIFadd{${6.2}_{-1.0}^{+1.2}$ }& \DIFadd{-- }& \DIFadd{${4.55}_{-0.83}^{+1.10}$ }& \DIFadd{-- }\\
\DIFadd{$\sigma_\mathrm{HIRES_\mathrm{post}}$ }[\DIFadd{$\mathrm{m~s^{-1}}$}] & \DIFadd{${2.87}_{-0.19}^{+0.20}$ }& \DIFadd{${2.62}_{-0.14}^{+0.15}$ }& \DIFadd{${4.19}_{-0.22}^{+0.23}$ }& \DIFadd{${15.6}_{-1.2}^{+1.4}$ }\\
\DIFadd{$\sigma_\mathrm{APF}$ }[\DIFadd{$\mathrm{m~s^{-1}}$}] & \DIFadd{${0.0045}_{-0.0045}^{+0.3100}$ }& \DIFadd{$2.05\pm0.23$ }& \DIFadd{-- }& \DIFadd{${1.82}_{-0.23}^{+0.22}$ }\\
\DIFadd{$\sigma_\mathrm{Hamilton}$ }[\DIFadd{$\mathrm{m~s^{-1}}$}] & \DIFadd{-- }& \DIFadd{${9.1}_{-1.3}^{+1.4}$ }& \DIFadd{-- }& \DIFadd{${11.8}_{-1.5}^{+1.7}$ }\\
\DIFadd{$\mathrm{RVZP_\mathrm{NEID}}$ }[\DIFadd{$\mathrm{m~s^{-1}}$}] & \DIFadd{164.11 }& \DIFadd{752.93 }& \DIFadd{-1027.33 }& \DIFadd{-585.44 }\\
\DIFadd{$\mathrm{RVZP_\mathrm{MINERVA_{T1}}}$ }[\DIFadd{$\mathrm{m~s^{-1}}$}] & \DIFadd{-- }& \DIFadd{-- }& \DIFadd{-- }& \DIFadd{-592.27 }\\
\DIFadd{$\mathrm{RVZP_\mathrm{MINERVA_{T2}}}$ }[\DIFadd{$\mathrm{m~s^{-1}}$}] & \DIFadd{-- }& \DIFadd{-- }& \DIFadd{-- }& \DIFadd{-596.94 }\\
\DIFadd{$\mathrm{RVZP_\mathrm{MINERVA_{T3}}}$ }[\DIFadd{$\mathrm{m~s^{-1}}$}] & \DIFadd{-- }& \DIFadd{-- }& \DIFadd{-- }& \DIFadd{-609.50 }\\
\DIFadd{$\mathrm{RVZP_\mathrm{MINERVA_{T4}}}$ }[\DIFadd{$\mathrm{m~s^{-1}}$}] & \DIFadd{-- }& \DIFadd{-- }& \DIFadd{-- }& \DIFadd{-604.57 }\\
\DIFadd{$\mathrm{RVZP_\mathrm{HIRES_\mathrm{pre}}}$ }[\DIFadd{$\mathrm{m~s^{-1}}$}] & \DIFadd{-180.26 }& \DIFadd{-- }& \DIFadd{-936.30 }& \DIFadd{-- }\\
\DIFadd{$\mathrm{RVZP_\mathrm{HIRES_\mathrm{post}}}$ }[\DIFadd{$\mathrm{m~s^{-1}}$}] & \DIFadd{-181.03 }& \DIFadd{712.32 }& \DIFadd{-940.22 }& \DIFadd{-644.31 }\\
\DIFadd{$\mathrm{RVZP_\mathrm{APF}}$ }[\DIFadd{$\mathrm{m~s^{-1}}$}] & \DIFadd{-117.49 }& \DIFadd{729.62 }& \DIFadd{-- }& \DIFadd{-615.50 }\\
\DIFadd{$\mathrm{RVZP_\mathrm{Hamilton}}$ }[\DIFadd{$\mathrm{m~s^{-1}}$}] & \DIFadd{-- }& \DIFadd{666.86 }& \DIFadd{-- }& \DIFadd{-715.63 }\\
\enddata
\tablecomments{In the case of 61 Cygni, the primary and secondary masses refer to 61 Cygni B and 61 Cygni A, respectively. From top to bottom, the parameters are primary mass, secondary mass, semimajor axis, $\sqrt{e}\cos\omega$ and $\sqrt{e}\sin\omega$ fitting terms, inclination, longitude of the ascending node, mean longitude at a reference epoch of 2455197.5 JD, parallax, distance, orbital period, argument of periastron, eccentricity, on-sky projected separation, time of periastron passage, mass ratio ($M_\mathrm{sec}/M_\mathrm{pri}$), followed by instrumental jitter terms and RV zero-point offsets (RVZP). The RVZP values are reported as maximum-likelihood estimate from our analyses. We refer to \cite{orvara} for further details on the selection of the fitted parameters used by \texttt{orvara}.}
\end{deluxetable*}

\begin{deluxetable*}{lccccc}
\caption{\DIFadd{Posterior distributions for the three planetary systems}}
\label{tab:planet_posteriors}
\tablehead{\colhead{Parameter} & \colhead{HD 217107 (b)} & \colhead{HD 217107 (c)} & \colhead{HD 190360 (b)} & \colhead{HD 190360 (c)} & \colhead{HD 154345 (b)}}
\startdata
\DIFaddend $M_\mathrm{pri}$ \DIFdelbegin \DIFdel{($M_\odot$) }\DIFdelend \DIFaddbegin [\DIFadd{$\mathrm{M}_\odot$}] \DIFaddend & ${1.045}_{-0.023}^{+0.018}$ & ${1.045}_{-0.023}^{+0.018}$ & ${0.991}_{-0.040}^{+0.039}$ & ${0.991}_{-0.040}^{+0.039}$ & \DIFdelbegin \DIFdel{${0.890}_{-0.040}^{+0.040}$ }\DIFdelend \DIFaddbegin \DIFadd{${0.890}\pm{0.040}$ }\DIFaddend \\
$M_\mathrm{sec}$ \DIFdelbegin \DIFdel{(}\DIFdelend \DIFaddbegin [\DIFaddend $\mathrm{M_J}$\DIFdelbegin \DIFdel{) }\DIFdelend \DIFaddbegin ] \DIFaddend & \DIFdelbegin \DIFdel{${1.446}_{-0.067}^{+0.18}$ }\DIFdelend \DIFaddbegin \DIFadd{${1.446}_{-0.067}^{+0.180}$ }\DIFaddend & ${4.37}_{-0.10}^{+0.13}$ & ${1.68}_{-0.16}^{+0.26}$ & ${0.080}_{-0.011}^{+0.049}$ & ${1.186}_{-0.059}^{+0.095}$ \\
$a$ \DIFdelbegin \DIFdel{(au) }\DIFdelend \DIFaddbegin [\DIFadd{au}] \DIFaddend & ${0.07359}_{-0.00053}^{+0.00042}$ & ${5.922}_{-0.044}^{+0.035}$ & ${3.963}_{-0.054}^{+0.051}$ & ${0.1296}_{-0.0018}^{+0.0017}$ & ${4.158}_{-0.067}^{+0.066}$ \\
$\sqrt{e}\sin\omega$ & \DIFdelbegin \DIFdel{${0.1388}_{-0.0042}^{+0.0042}$ }\DIFdelend \DIFaddbegin \DIFadd{${0.1388}\pm{0.0042}$ }\DIFaddend & \DIFdelbegin \DIFdel{${-0.264}_{-0.011}^{+0.011}$ }\DIFdelend \DIFaddbegin \DIFadd{${-0.264}\pm{0.011}$ }\DIFaddend & \DIFdelbegin \DIFdel{${0.144}_{-0.013}^{+0.013}$ }\DIFdelend \DIFaddbegin \DIFadd{${0.144}\pm{0.013}$ }\DIFaddend & ${-0.241}_{-0.053}^{+0.058}$ & ${-0.180}_{-0.046}^{+0.058}$ \\
$\sqrt{e}\cos\omega$ & ${0.3305}_{-0.0025}^{+0.0023}$ & ${-0.5674}_{-0.0074}^{+0.0078}$ & \DIFdelbegin \DIFdel{${0.5456}_{-0.0069}^{+0.0069}$ }\DIFdelend \DIFaddbegin \DIFadd{${0.5456}\pm{0.0069}$ }\DIFaddend & ${0.324}_{-0.046}^{+0.041}$ & ${0.150}_{-0.078}^{+0.061}$ \\
$i$ \DIFdelbegin \DIFdel{(}\DIFdelend \DIFaddbegin [\DIFaddend $^\circ$\DIFdelbegin \DIFdel{) }\DIFdelend \DIFaddbegin ] \DIFaddend & ${93}_{-26}^{+25}$ & ${88}_{-12}^{+14}$ & ${69}_{-17}^{+42}$ & ${90}_{-44}^{+45}$ & \DIFdelbegin \DIFdel{${88}_{-20}^{+20}$ }\DIFdelend \DIFaddbegin \DIFadd{${88}\pm{20}$ }\DIFaddend \\
$\Omega$ \DIFdelbegin \DIFdel{(}\DIFdelend \DIFaddbegin [\DIFaddend $^\circ$\DIFdelbegin \DIFdel{) }\DIFdelend \DIFaddbegin ] \DIFaddend & ${180}_{-121}^{+122}$ & ${178}_{-18}^{+22}$ & ${107}_{-20}^{+57}$ & \DIFdelbegin \DIFdel{${180}_{-122}^{+122}$ }\DIFdelend \DIFaddbegin \DIFadd{${180}\pm{122}$ }\DIFaddend & ${68}_{-31}^{+26}$ \\
$\lambda$ \DIFdelbegin \DIFdel{(}\DIFdelend \DIFaddbegin [\DIFaddend $^\circ$\DIFdelbegin \DIFdel{) }\DIFdelend \DIFaddbegin ] \DIFaddend & ${200.465}_{-0.094}^{+0.093}$ & ${155.00}_{-0.49}^{+0.50}$ & ${221.59}_{-0.44}^{+0.46}$ & \DIFdelbegin \DIFdel{${54.9}_{-1.8}^{+1.8}$ }\DIFdelend \DIFaddbegin \DIFadd{${54.9}\pm{1.8}$ }\DIFaddend & \DIFdelbegin \DIFdel{${332.4}_{-1.2}^{+1.2}$ }\DIFdelend \DIFaddbegin \DIFadd{${332.4}\pm{1.2}$ }\DIFaddend \\
$\pi$ \DIFdelbegin \DIFdel{(mas) }\DIFdelend \DIFaddbegin [\DIFadd{mas}] \DIFaddend & ${49.784586}_{-0.000082}^{+0.000042}$ & ${49.784586}_{-0.000082}^{+0.000042}$ & ${62.486533}_{-0.000071}^{+0.000034}$ & ${62.486533}_{-0.000071}^{+0.000034}$ & \DIFdelbegin \DIFdel{${54.7358942}_{-0.000017}^{+0.0000086}$ }\DIFdelend \DIFaddbegin \DIFadd{${54.7358942}_{-0.0000170}^{+0.0000086}$ }\DIFaddend \\
$d$ \DIFdelbegin \DIFdel{(pc) }\DIFdelend \DIFaddbegin [\DIFadd{pc}] \DIFaddend & ${20.086543}_{-0.000010}^{+{0.000012}}$ & ${20.086543}_{-0.000010}^{+{0.000012}}$ & ${16.003451}_{-0.000005}^{+{0.000007}}$ & ${16.003451}_{-0.000005}^{+{0.000007}}$ & ${18.2695479}_{-0.0000016}^{+{0.0000021}}$ \\
$P$ \DIFdelbegin \DIFdel{(yr) }\DIFdelend \DIFaddbegin [\DIFadd{yr}] \DIFaddend & \DIFdelbegin \DIFdel{${0.019512307}_{-0.000000013}^{+0.000000013}$ }\DIFdelend \DIFaddbegin \DIFadd{${0.019512708}\pm{0.000000013}$ }\DIFaddend & \DIFdelbegin \DIFdel{${14.069}_{-0.031}^{+0.031}$ }\DIFdelend \DIFaddbegin \DIFadd{${14.069}\pm{0.031}$ }\DIFaddend & ${7.917}_{-0.010}^{+0.011}$ & \DIFdelbegin \DIFdel{${0.0468637}_{-0.0000014}^{+0.0000014}$ }\DIFdelend \DIFaddbegin \DIFadd{${0.0468647}\pm{+0.0000014}$ }\DIFaddend & ${8.981}_{-0.076}^{+0.079}$ \\
$\omega$ \DIFdelbegin \DIFdel{(}\DIFdelend \DIFaddbegin [\DIFaddend $^\circ$\DIFdelbegin \DIFdel{) }\DIFdelend \DIFaddbegin ] \DIFaddend & ${22.78}_{-0.70}^{+0.71}$ & \DIFdelbegin \DIFdel{${205.0}_{-1.1}^{+1.1}$ }\DIFdelend \DIFaddbegin \DIFadd{${205.0}\pm1.1$ }\DIFaddend & \DIFdelbegin \DIFdel{${14.8}_{-1.4}^{+1.4}$ }\DIFdelend \DIFaddbegin \DIFadd{${14.8}\pm{1.4}$ }\DIFaddend & ${323.4}_{-9.1}^{+9.2}$ & ${309}_{-21}^{+18}$ \\
$e$ & ${0.1284}_{-0.0014}^{+0.0015}$ & ${0.3918}_{-0.0067}^{+0.0064}$ & ${0.3184}_{-0.0063}^{+0.0065}$ & ${0.165}_{-0.024}^{+0.023}$ & \DIFdelbegin \DIFdel{${0.058}_{-0.019}^{+0.019}$ }\DIFdelend \DIFaddbegin \DIFadd{${0.058}\pm{0.019}$ }\DIFaddend \\
$a$ \DIFdelbegin \DIFdel{(mas)  }\DIFdelend \DIFaddbegin [\DIFadd{mas}]  \DIFaddend & ${3.663}_{-0.027}^{+0.021}$ & ${294.8}_{-2.2}^{+1.7}$ & ${247.6}_{-3.4}^{+3.2}$ & ${8.10}_{-0.11}^{+0.10}$ & ${227.6}_{-3.7}^{+3.6}$ \\
\DIFdelbegin \DIFdel{$T_0$ (JD) }\DIFdelend \DIFaddbegin \DIFadd{$T_\mathrm{p}$ }[\DIFadd{JD}] \DIFaddend & \DIFdelbegin \DIFdel{${2455201.109}_{-0.014}^{+0.014}$ }\DIFdelend \DIFaddbegin \DIFadd{${2455201.109}\pm{0.014}$ }\DIFaddend & \DIFdelbegin \DIFdel{${2455911}_{-15}^{+15}$ }\DIFdelend \DIFaddbegin \DIFadd{${2455911}\pm{15}$ }\DIFaddend & ${2456428}_{-10}^{+10}$ & \DIFdelbegin \DIFdel{${2455210.27}_{-0.42}^{+0.42}$ }\DIFdelend \DIFaddbegin \DIFadd{${2455210.27}\pm{0.42}$ }\DIFaddend & ${2458230}_{-263}^{+147}$ \\
$q$ & \DIFdelbegin \DIFdel{${0.001322}_{-0.000061}^{+0.00017}$ }\DIFdelend \DIFaddbegin \DIFadd{${0.001322}_{-0.000061}^{+0.000170}$ }\DIFaddend & \DIFdelbegin \DIFdel{${0.003991}_{-0.000078}^{+0.00013}$ }\DIFdelend \DIFaddbegin \DIFadd{${0.003991}_{-0.000078}^{+0.000130}$ }\DIFaddend & ${0.00162}_{-0.00016}^{+0.00024}$ & ${0.000077}_{-0.000010}^{+0.000048}$ & \DIFdelbegin \DIFdel{${0.001268}_{-0.000050}^{+0.00010}$
}\DIFdelend \DIFaddbegin \DIFadd{${0.001268}_{-0.000050}^{+0.000100}$ }\\
\DIFadd{$\sigma_\mathrm{NEID}$ }[\DIFadd{$\mathrm{m~s^{-1}}$}] & \DIFadd{${2.74}_{-0.28}^{+0.31}$ }& \DIFadd{${2.74}_{-0.28}^{+0.31}$ }& \DIFadd{${2.21}_{-0.28}^{+0.35}$ }& \DIFadd{${2.21}_{-0.28}^{+0.35}$ }& \DIFadd{${2.52}_{-0.35}^{+0.45}$ }\\
\DIFadd{$\sigma_\mathrm{MINERVA_{T1}}$ }[\DIFadd{$\mathrm{m~s^{-1}}$}] & \DIFadd{${12.0}_{-1.7}^{+2.1}$ }& \DIFadd{${12.0}_{-1.7}^{+2.1}$ }& \DIFadd{-- }& \DIFadd{-- }& \DIFadd{-- }\\
\DIFadd{$\sigma_\mathrm{MINERVA_{T2}}$ }[\DIFadd{$\mathrm{m~s^{-1}}$}] & \DIFadd{${39.4}_{-7.3}^{+10}$ }& \DIFadd{${39.4}_{-7.3}^{+10}$ }& \DIFadd{-- }& \DIFadd{-- }& \DIFadd{-- }\\
\DIFadd{$\sigma_\mathrm{MINERVA_{T3}}$ }[\DIFadd{$\mathrm{m~s^{-1}}$}] & \DIFadd{${20.8}_{-5.8}^{+7.4}$ }& \DIFadd{${20.8}_{-5.8}^{+7.4}$ }& \DIFadd{-- }& \DIFadd{-- }& \DIFadd{-- }\\
\DIFadd{$\sigma_\mathrm{MINERVA_{T4}}$ }[\DIFadd{$\mathrm{m~s^{-1}}$}] & \DIFadd{${12.2}_{-1.4}^{+1.6}$ }& \DIFadd{${12.2}_{-1.4}^{+1.6}$ }& \DIFadd{-- }& \DIFadd{-- }& \DIFadd{-- }\\
\DIFadd{$\sigma_\mathrm{HIRES_\mathrm{pre}}$ }[\DIFadd{$\mathrm{m~s^{-1}}$}] & \DIFadd{${2.74}_{-0.28}^{+0.31}$ }& \DIFadd{${2.74}_{-0.28}^{+0.31}$ }& \DIFadd{${2.92}_{-0.26}^{+0.28}$ }& \DIFadd{${2.92}_{-0.26}^{+0.28}$ }& \DIFadd{${3.14}_{-0.57}^{+0.69}$ }\\
\DIFadd{$\sigma_\mathrm{HIRES_\mathrm{post}}$ }[\DIFadd{$\mathrm{m~s^{-1}}$}] & \DIFadd{${4.16}_{-0.30}^{+0.34}$ }& \DIFadd{${4.16}_{-0.30}^{+0.34}$ }& \DIFadd{${2.80}_{-0.15}^{+0.16}$ }& \DIFadd{${2.80}_{-0.15}^{+0.16}$ }& \DIFadd{${3.01}_{-0.16}^{+0.18}$ }\\
\DIFadd{$\sigma_\mathrm{APF}$ }[\DIFadd{$\mathrm{m~s^{-1}}$}] & \DIFadd{${2.38}_{-0.18}^{+0.20}$ }& \DIFadd{${2.38}_{-0.18}^{+0.20}$ }& \DIFadd{${2.32}\pm0.12$ }& \DIFadd{${2.32}\pm{0.12}$ }& \DIFadd{-- }\\
\DIFadd{$\sigma_\mathrm{Hamilton}$ }[\DIFadd{$\mathrm{m~s^{-1}}$}] & \DIFadd{${13.20}_{-0.75}^{+0.82}$ }& \DIFadd{${13.20}_{-0.75}^{+0.82}$ }& \DIFadd{${6.33}_{-0.52}^{+0.58}$ }& \DIFadd{${6.33}_{-0.52}^{+0.58}$ }& \DIFadd{-- }\\
\DIFadd{$\sigma_\mathrm{CORALIE}$ }[\DIFadd{$\mathrm{m~s^{-1}}$}] & \DIFadd{${0.0084}_{-0.0083}^{+0.5500}$ }& \DIFadd{${0.0084}_{-0.0083}^{+0.5500}$ }& \DIFadd{-- }& \DIFadd{-- }& \DIFadd{-- }\\
\DIFadd{$\sigma_\mathrm{Tull}$ }[\DIFadd{$\mathrm{m~s^{-1}}$}] & \DIFadd{${0.030}_{-0.030}^{+3.300}$ }& \DIFadd{${0.030}_{-0.030}^{+3.300}$ }& \DIFadd{-- }& \DIFadd{-- }& \DIFadd{-- }\\
\DIFadd{$\sigma_\mathrm{ELODIE}$ }[\DIFadd{$\mathrm{m~s^{-1}}$}] & \DIFadd{-- }& \DIFadd{-- }& \DIFadd{${0.059}_{-0.059}^{+3.500}$ }& \DIFadd{${0.059}_{-0.059}^{+3.500}$ }& \DIFadd{${9.5}_{-1.6}^{+1.8}$ }\\
\DIFadd{$\sigma_\mathrm{AFOE}$ }[\DIFadd{$\mathrm{m~s^{-1}}$}] & \DIFadd{-- }& \DIFadd{-- }& \DIFadd{${6.0}_{-6.0}^{+4.7}$ }& \DIFadd{${6.0}_{-6.0}^{+4.7}$ }& \DIFadd{-- }\\
\DIFadd{$\sigma_\mathrm{SOPHIE}$ }[\DIFadd{$\mathrm{m~s^{-1}}$}] & \DIFadd{-- }& \DIFadd{-- }& \DIFadd{-- }& \DIFadd{-- }& \DIFadd{${3.9}_{-3.9}^{+3.0}$ }\\
\DIFadd{$\mathrm{RVZP}_\mathrm{NEID}$ }[\DIFadd{$\mathrm{m~s^{-1}}$}] & \DIFadd{-31.14 }& \DIFadd{-31.14 }& \DIFadd{18.06 }& \DIFadd{18.06 }& \DIFadd{-4.56 }\\
\DIFadd{$\mathrm{RVZP}_\mathrm{MINERVA_{T1}}$ }[\DIFadd{$\mathrm{m~s^{-1}}$}] & \DIFadd{16.57 }& \DIFadd{16.57 }& \DIFadd{-- }& \DIFadd{-- }& \DIFadd{-- }\\
\DIFadd{$\mathrm{RVZP}_\mathrm{MINERVA_{T2}}$ }[\DIFadd{$\mathrm{m~s^{-1}}$}] & \DIFadd{14.32 }& \DIFadd{14.32 }& \DIFadd{-- }& \DIFadd{-- }& \DIFadd{-- }\\
\DIFadd{$\mathrm{RVZP}_\mathrm{MINERVA_{T3}}$ }[\DIFadd{$\mathrm{m~s^{-1}}$}] & \DIFadd{2.27 }& \DIFadd{2.27 }& \DIFadd{-- }& \DIFadd{-- }& \DIFadd{-- }\\
\DIFadd{$\mathrm{RVZP}_\mathrm{MINERVA_{T4}}$ }[\DIFadd{$\mathrm{m~s^{-1}}$}] & \DIFadd{6.02 }& \DIFadd{6.02 }& \DIFadd{-- }& \DIFadd{-- }& \DIFadd{-- }\\
\DIFadd{$\mathrm{RVZP}_\mathrm{HIRES_\mathrm{pre}}$ }[\DIFadd{$\mathrm{m~s^{-1}}$}] & \DIFadd{9.59 }& \DIFadd{9.59 }& \DIFadd{1.88 }& \DIFadd{1.88 }& \DIFadd{5.21 }\\
\DIFadd{$\mathrm{RVZP}_\mathrm{HIRES_\mathrm{post}}$ }[\DIFadd{$\mathrm{m~s^{-1}}$}] & \DIFadd{5.55 }& \DIFadd{5.55 }& \DIFadd{2.42 }& \DIFadd{2.42 }& \DIFadd{4.95 }\\
\DIFadd{$\mathrm{RVZP}_\mathrm{APF}$ }[\DIFadd{$\mathrm{m~s^{-1}}$}] & \DIFadd{26.99 }& \DIFadd{26.99 }& \DIFadd{-1.07 }& \DIFadd{-1.07 }& \DIFadd{-- }\\
\DIFadd{$\mathrm{RVZP}_\mathrm{Hamilton}$ }[\DIFadd{$\mathrm{m~s^{-1}}$}] & \DIFadd{16.94 }& \DIFadd{16.94 }& \DIFadd{3.81 }& \DIFadd{3.81 }& \DIFadd{-- }\\
\DIFadd{$\mathrm{RVZP}_\mathrm{CORALIE}$ }[\DIFadd{$\mathrm{m~s^{-1}}$}] & \DIFadd{417.04 }& \DIFadd{417.04 }& \DIFadd{-- }& \DIFadd{-- }& \DIFadd{-- }\\
\DIFadd{$\mathrm{RVZP}_\mathrm{Tull}$ }[\DIFadd{$\mathrm{m~s^{-1}}$}] & \DIFadd{1.62 }& \DIFadd{1.62 }& \DIFadd{-- }& \DIFadd{-- }& \DIFadd{-- }\\
\DIFadd{$\mathrm{RVZP}_\mathrm{ELODIE}$ }[\DIFadd{$\mathrm{m~s^{-1}}$}] & \DIFadd{-- }& \DIFadd{-- }& \DIFadd{45347.58 }& \DIFadd{45347.58 }& \DIFadd{46950.74 }\\
\DIFadd{$\mathrm{RVZP}_\mathrm{AFOE}$ }[\DIFadd{$\mathrm{m~s^{-1}}$}] & \DIFadd{-- }& \DIFadd{-- }& \DIFadd{45349.94 }& \DIFadd{45349.94 }& \DIFadd{-- }\\
\DIFadd{$\mathrm{RVZP}_\mathrm{SOPHIE}$ }[\DIFadd{$\mathrm{m~s^{-1}}$}] & \DIFadd{-- }& \DIFadd{-- }& \DIFadd{-- }& \DIFadd{-- }& \DIFadd{46841.77 }\\
\DIFaddend \enddata
\DIFdelbegin \DIFdelend \DIFaddbegin \tablecomments{All fitted and derived parameters here are the same as those described in Table \ref{tab:stellar_posteriors}.}
\DIFaddend \end{deluxetable*}

\DIFdelbegin \DIFdelend \DIFaddbegin 

\begin{figure*}
    \centering

\begin{minipage}{0.2\linewidth}
        \centering
        \DIFaddFL{\textbf{(a) HD 68017 A}
    }\end{minipage} \hfill
    \begin{minipage}{0.65\linewidth}
        \centering
        \includegraphics[width=\linewidth]{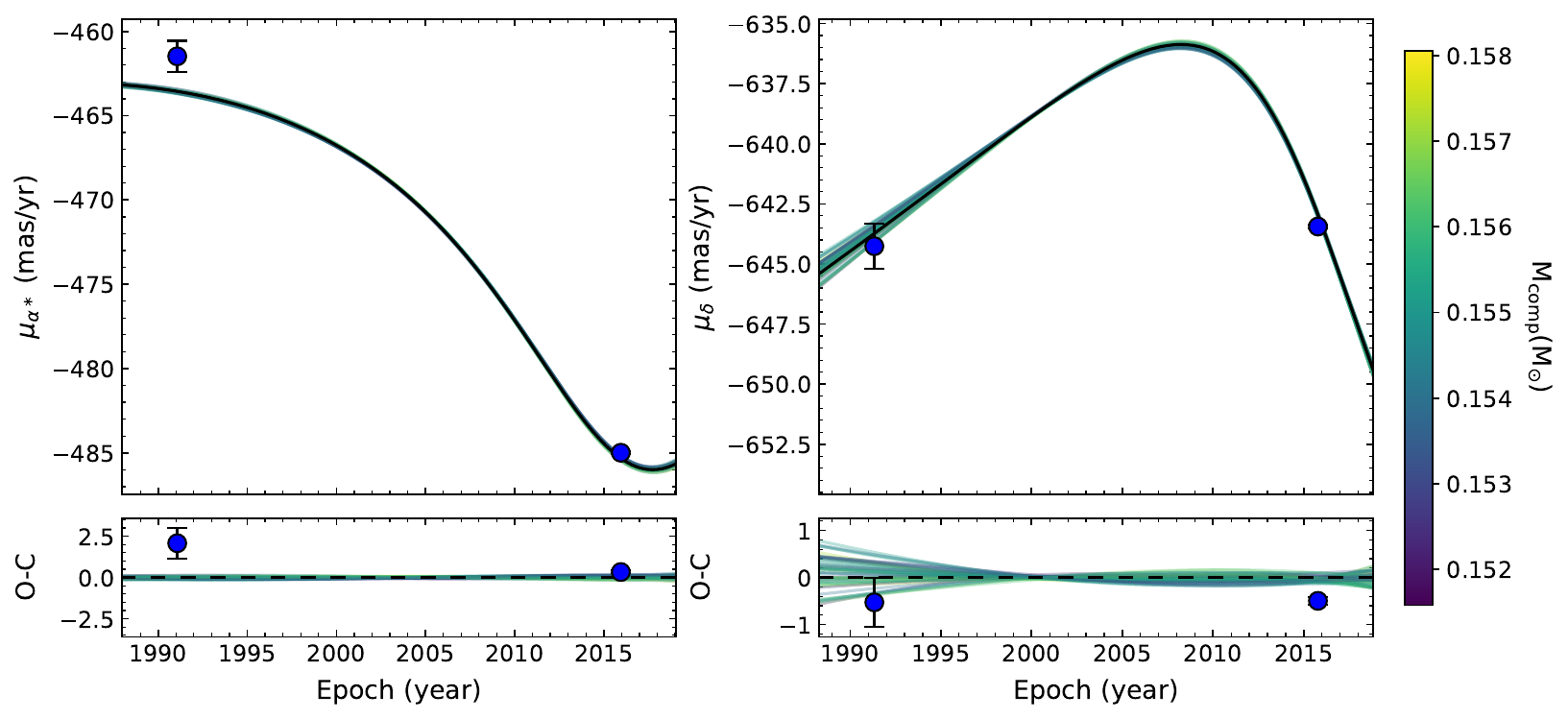}
        \label{fig:68017_pms}
    \end{minipage}

    \vspace{-0.5cm}

\begin{minipage}{0.2\linewidth}
        \centering
        \DIFaddFL{\textbf{(b) 61 Cygni B}
    }\end{minipage} \hfill
    \begin{minipage}{0.65\linewidth}
        \centering
        \includegraphics[width=\linewidth]{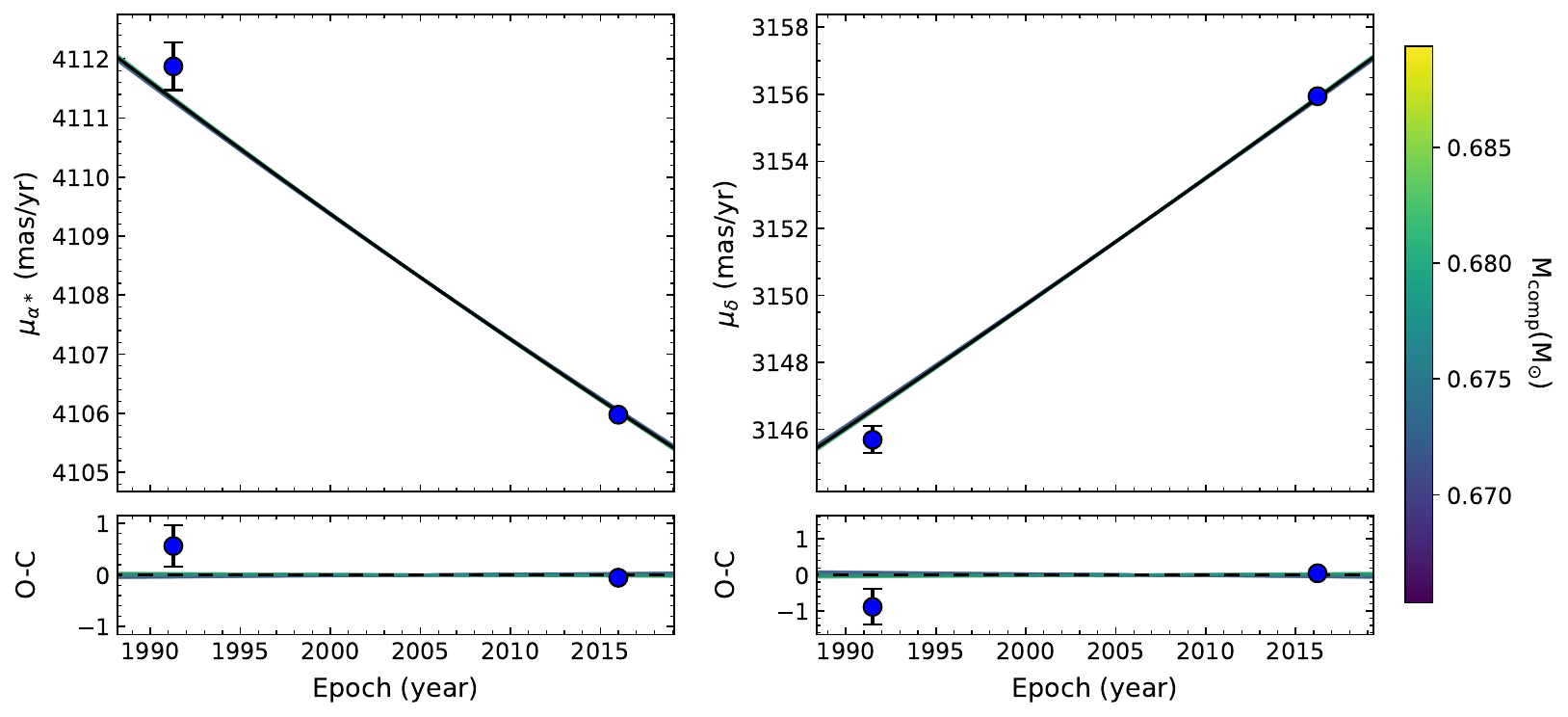}
        \label{fig:61CygB_pms}
    \end{minipage}

        \vspace{-0.5cm}

\begin{minipage}{0.2\linewidth}
        \centering
        \DIFaddFL{\textbf{(c) HD 24496 A}
    }\end{minipage} \hfill
    \begin{minipage}{0.65\linewidth}
        \centering
        \includegraphics[width=\linewidth]{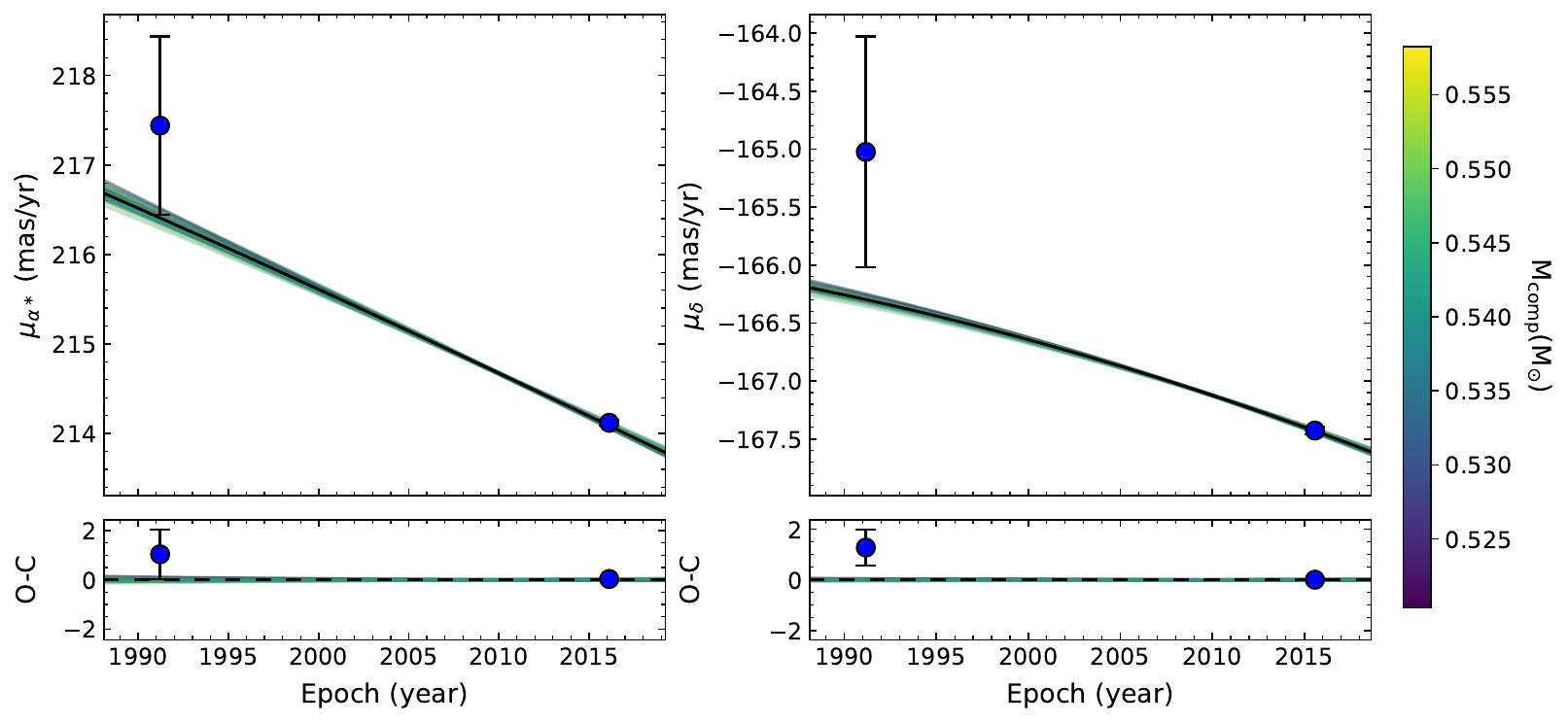}
        \label{fig:HD24496_pms}
    \end{minipage}

    \vspace{-0.5cm}

\begin{minipage}{0.2\linewidth}
        \centering
        \DIFaddFL{\textbf{(d) HD 4614 A}
    }\end{minipage} \hfill
    \begin{minipage}{0.65\linewidth}
        \centering
        \includegraphics[width=\linewidth]{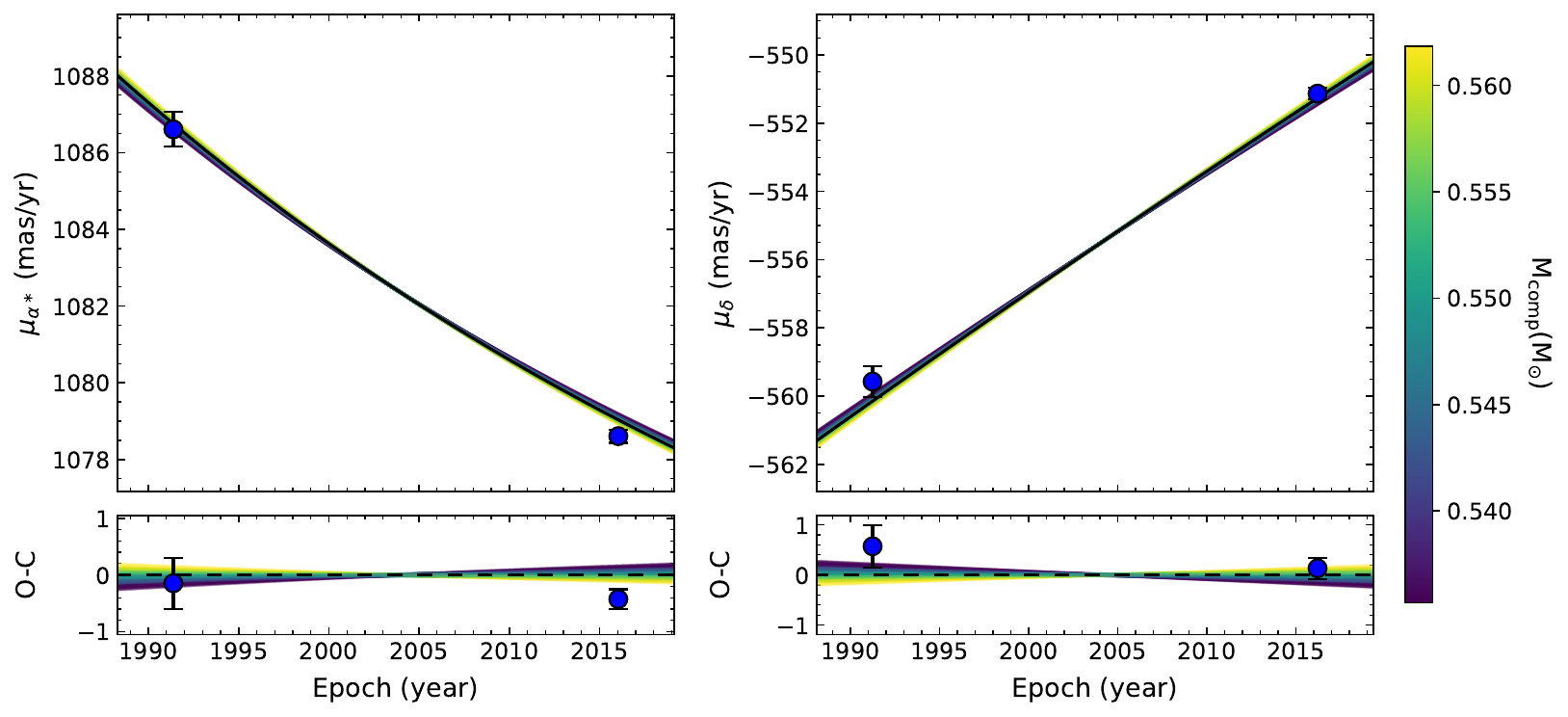}
        \label{fig:HD4614_pms}
    \end{minipage}

    \vspace{-0.5cm}

    \caption{\DIFaddFL{Observed proper motion values for the accelerating stars in our four binary systems: HD 68017 A, 61 Cygni B, HD 24496 A, and HD 4614 A. In each panel, the left plot shows the proper motion in right ascension as measured by Hipparcos (left point at $t\approx1991.25$) and Gaia (right point at $t\approx2016.0$), and the right plot shows the proper motion in declination for the same epochs. The best-fit models and colored trial orbits that correspond to companion mass are in broad agreement with the astrometric measurements for all systems. Though in the case of HD 68017, the offsets observed primarily in Hipparcos' proper motion in right ascension and Gaia's proper motion in declination are likely due to the mission's inability to resolve the stellar pair.}}
    \label{fig:combined_pms}
\end{figure*}

\begin{figure*}
    \centering

\begin{minipage}{0.15\linewidth}
        \centering
        \DIFaddFL{\textbf{(a) HD 217107}
    }\end{minipage} \hfill
    \begin{minipage}{0.75\linewidth}
        \centering
        \includegraphics[width=\linewidth]{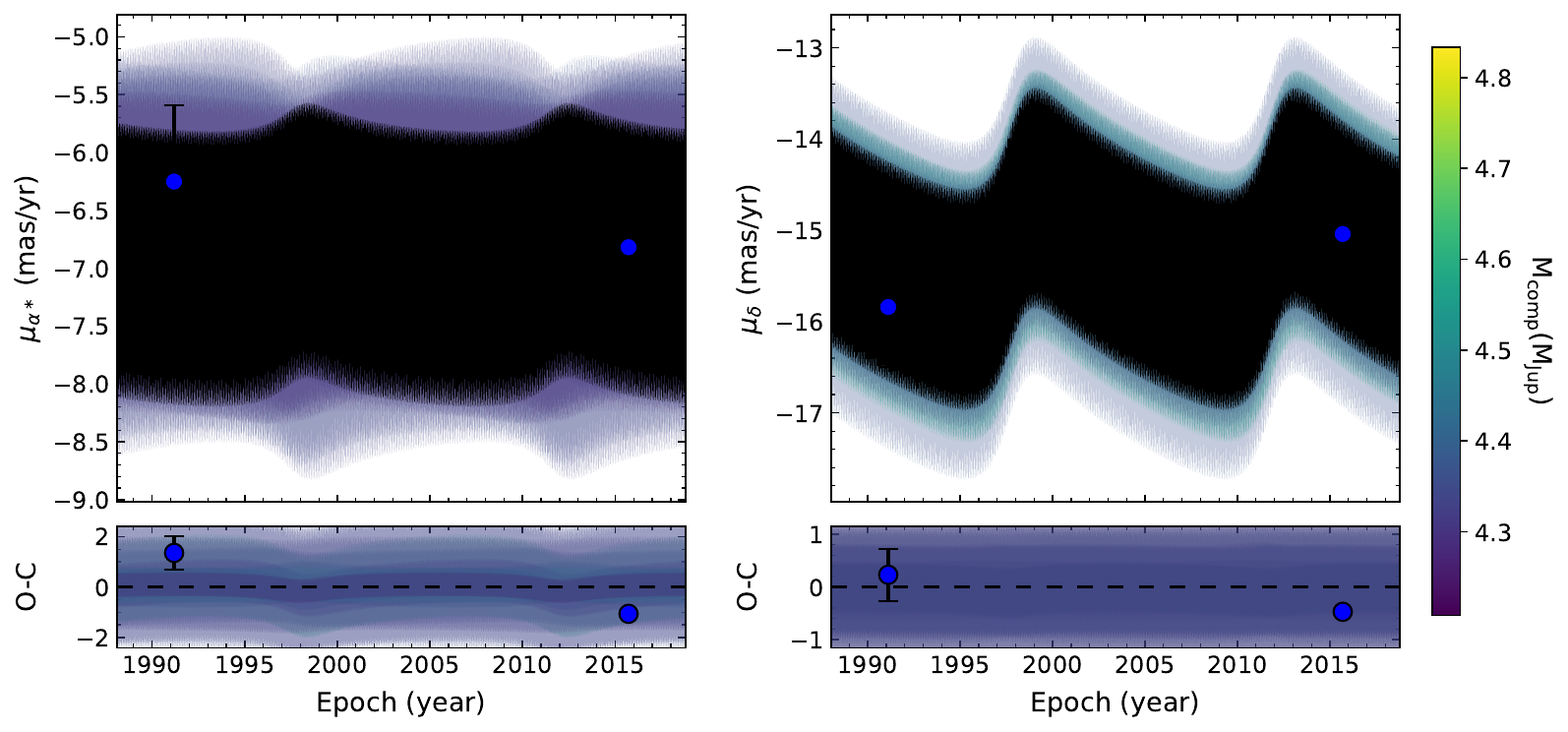}
        \label{fig:HD217107_pms}
    \end{minipage}

    \medskip

\begin{minipage}{0.15\linewidth}
        \centering
        \DIFaddFL{\textbf{(b) HD 190360}
    }\end{minipage} \hfill
    \begin{minipage}{0.75\linewidth}
        \centering
        \includegraphics[width=\linewidth]{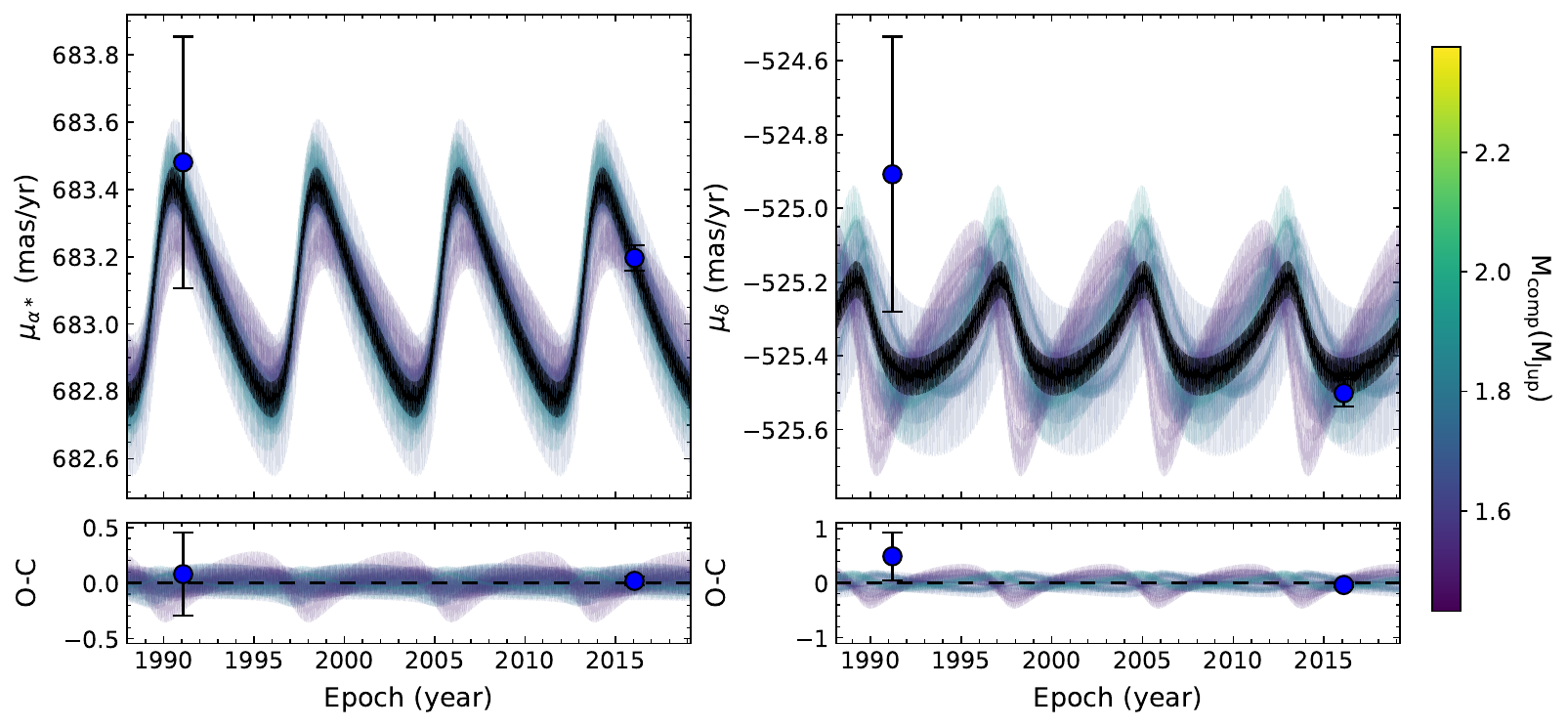}
        \label{fig:HD190360_pms}
    \end{minipage}

    \medskip

\begin{minipage}{0.15\linewidth}
        \centering
        \DIFaddFL{\textbf{(c) HD 154345}
    }\end{minipage} \hfill
    \begin{minipage}{0.75\linewidth}
        \centering
        \includegraphics[width=\linewidth]{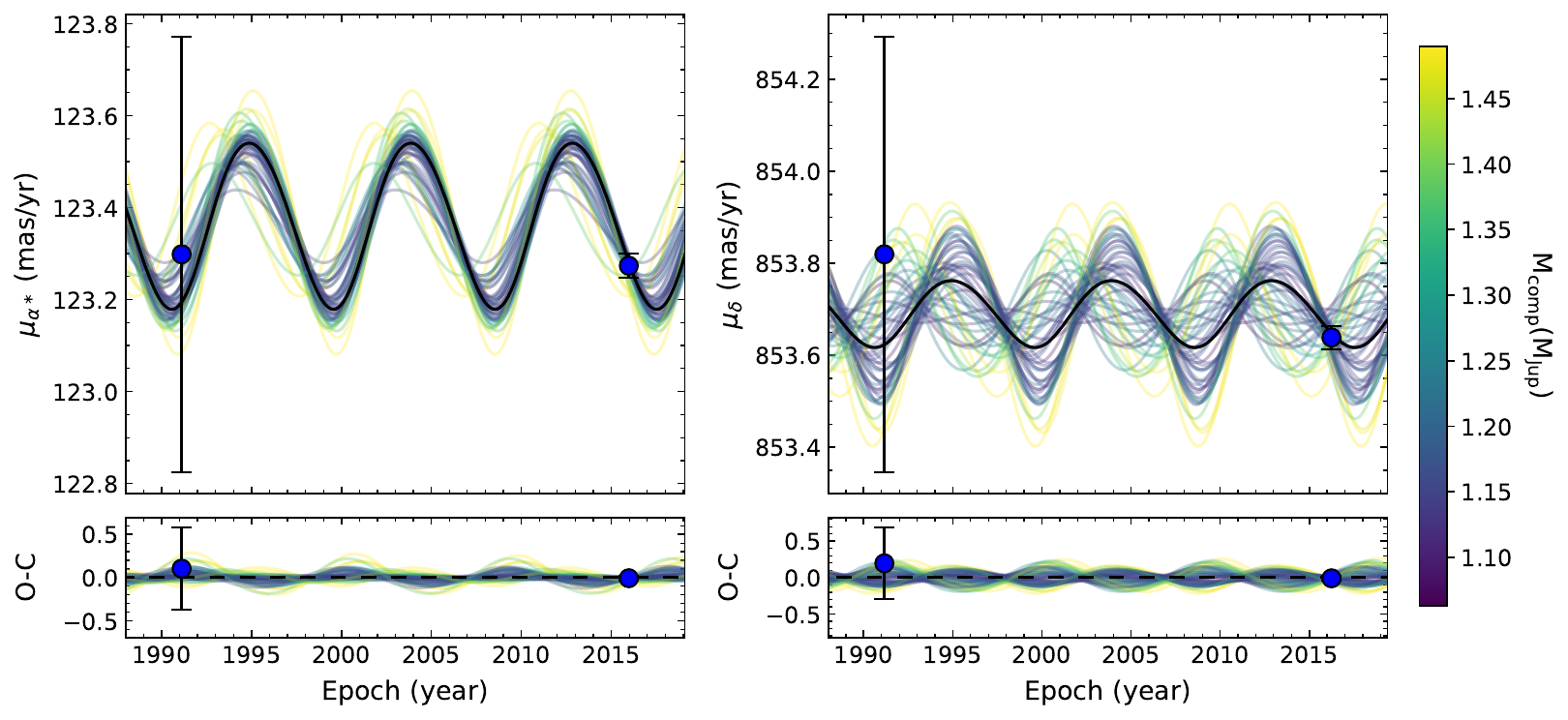}
        \label{fig:HD154345_pms}
    \end{minipage}

    \caption{\DIFaddFL{Observed proper motion values for the accelerating stars in our three planetary systems: HD 217107, HD 190360, and HD 154345. Each panel shows the proper motion values as measured by Hipparcos and Gaia, as well as colored trial orbits that correspond to companion mass, in the same way as Figure \ref{fig:combined_pms}.}}
    \label{fig:combined_additional_pms}
\end{figure*}

\section{\DIFadd{Discussion}} \label{sec:discussion}

\begin{figure}
    \centering
    \includegraphics[width=\linewidth]{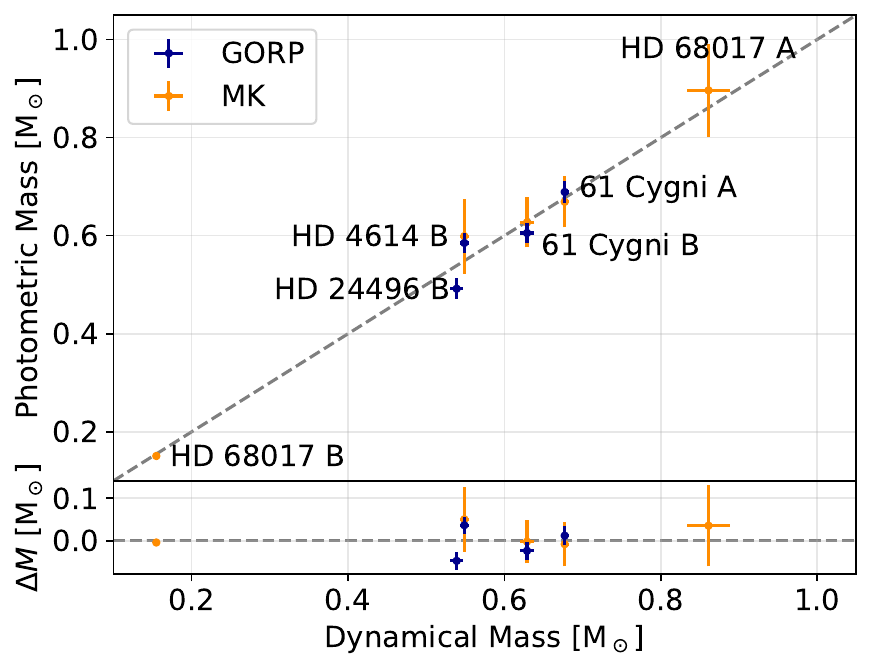}
    \caption{\DIFaddFL{Comparison of our dynamical stellar mass estimates with recent photometric mass relationships. The HD 68017 AB pair is unresolved in both 2MASS and Gaia and therefore has no reported $K_\mathrm{s}$ or $G_\mathrm{RP}$ magnitudes. However, \cite{2012ApJ...761...39C} used NIRC2 to resolve the two stars in its $K'$ filter, which is nearly identical to that of 2MASS $K_\mathrm{s}$ \citep[e.g.,][]{2010ApJ...718..810S, KELT24_stellar_params}, allowing us to deconvolve the system's composite $K_\mathrm{s}$ magnitude to estimate photometric masses for HD 68017 A and HD 68017 B using the MK method. Similarly, Gaia (but not 2MASS) resolves the HD 24496 AB pair, and so the GORP method is shown for HD 24496 B. HD 24496 A is not provided because a mass prior is used, while HD 4614 A is not shown because it is too massive for either of these two relations.}} \label{fig:phot_mass_comp}
\end{figure}

\DIFadd{We have achieved precise constraints on the individual masses within HD 68017 AB, 61 Cygni AB, HD 24496 AB, and HD 4614 AB, providing valuable benchmarks for stellar evolutionary models and calibration points across a wide range of stellar parameters. Since stellar mass is a fundamental parameter that underpins our understanding of stellar evolution, atmospheric composition, and even the characterization of orbiting exoplanets, the precision of these dynamical masses, particularly at the $1\%$ level, ensures that these stars will serve as critical reference points in future mass-calibration schemes, further constraining the physical properties of stars and their companions. For most main sequence stars, mass and luminosity are closely correlated (e.g., \citealt{old_mass_lum}). Our refined dynamical mass estimates offer a direct comparison with photometric masses from recent works, such as \cite{2019ApJ...871...63M} and \cite{2022AJ....164..164G}. \cite{2019ApJ...871...63M} (MK) use the 2MASS $K_\mathrm{s}$-band, while \cite{2022AJ....164..164G} (GORP) use the Gaia $G_\mathrm{RP}$-band. Figure \ref{fig:phot_mass_comp} illustrates this comparison, and in all cases, the photometric relationships are found to be in good agreement with the dynamical mass estimates.
}

\begin{figure*}
    \centering

    \includegraphics[width=\textwidth]{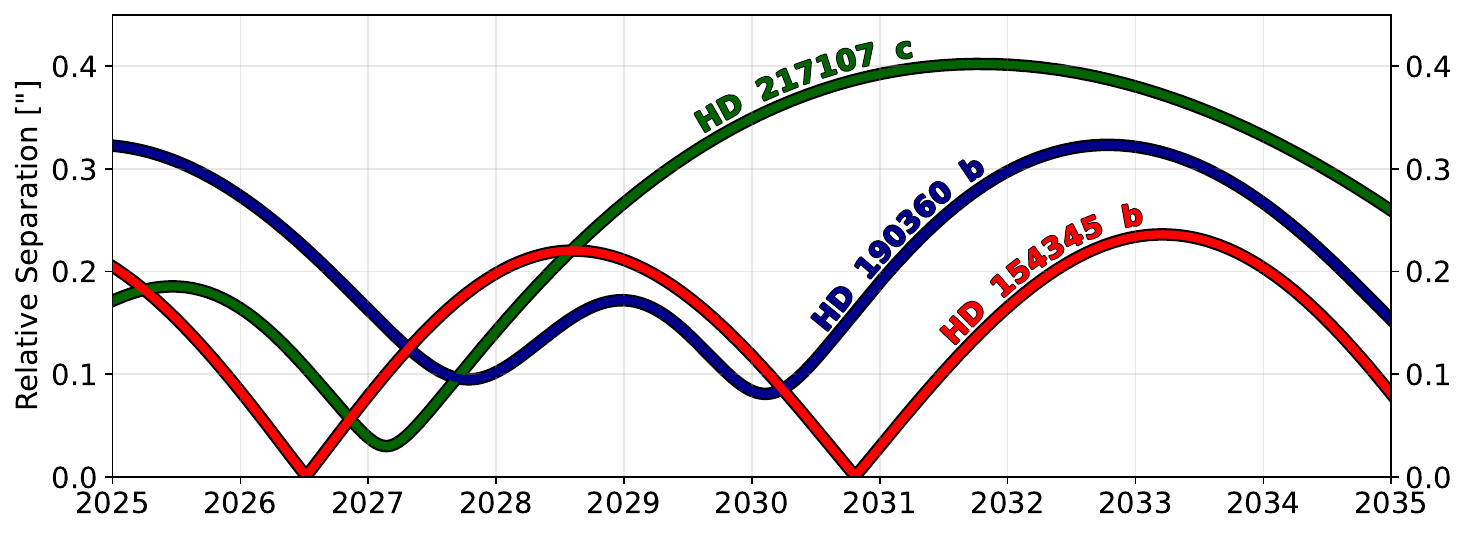}
    \caption{\DIFaddFL{Best-fit relative separation models for the three giant outer planets presented in this analysis.}}
    \label{fig:combined_relsep_planets}
\end{figure*}

\DIFadd{The true masses and refined orbits of the three cold Jupiters discussed here, HD 217107~c, HD 190360~b, and HD 154345~b, provide new opportunities to evaluate their viability for direct imaging. To do so, we consider their orbital separations, estimated temperatures, and expected contrast ratios in thermal emission and reflected light. Figure \ref{fig:combined_relsep_planets} shows their best-fit relative separation models from 2025-2035. Using our reported true dynamical masses and assuming ages of 5~Gyr for each planet, we estimate their effective temperatures using the SpeX Prism Spectral Analysis Toolkit (SPLAT; \citealt{splat}) and models published in \cite{baraffe2003}. For HD 217107~c, HD 190360~b, and HD 154345~b, we find $T_\mathrm{eff}=197.87\pm3.06$~K, $T_\mathrm{eff}=121.38\pm4.27$~K, and $T_\mathrm{eff}=110.57\pm1.91$~K, respectively. The models from \cite{baraffe2003} only go down to $\approx2~\mathrm{M_J}$, so for the two colder planets we extrapolate their temperatures using a linear fit to the model predictions at the lowest available masses. We then find that all three planets are widely separated from their stars and large enough to be viable targets for direct imaging with missions such as JWST \citep{jwst} or the Nancy Grace Roman Space Telescope \citep{2015arXiv150303757S}. JWST's Mid-Infrared Instrument (MIRI; \citealt{2023PASP..135d8003W}) observes at wavelengths between $5-28.5~\mathrm{\mu m}$, making it well-suited for detecting planetary thermal emission. MIRI's coronagraph, particularly its F1065C filter with a four-quadrant phase mask \citep{4QPM}, has an inner working angle of $0.33\arcsec$ and has demonstrated contrast capabilities $\gtrsim10^{-5}$ \citep{2022A&A...667A.165B}. Given this, HD 217107~c (which has a contrast of $\approx10^{-4}$ at this wavelength) is detectable using MIRI's coronagraph, but HD 190360~b and HD 154345~b (which have corresponding contrasts $\approx10^{-6}$) are too faint and orbit too closely. Roman's Coronagraph Instrument (CGI; \citealt{2023SPIE12680E..0TB}) operates in the optical regime ($\sim525-840$~nm) and has an inner working angle of $0.15\arcsec$. Since planets this cold are thermally invisible at these wavelengths, their detectability relies on reflected light. Assuming Jupiter-radii, an albedo similar to Jupiter (0.5), and an average orbital distance of 4~au, each planet is expected to have a contrast ratio $>10^{-9}$ in reflected light, making them promising targets for the CGI under optimal circumstances. As shown in Figure \ref{fig:combined_relsep_planets}, all three planets are expected to maintain angular separations of $\rho>0.15\arcsec$ for several years within the next decade. Adding relative astrometry to these three planets will significantly enhance their dynamical mass measurements and directly enable temperature estimations, establishing a pathway for these planets to calibrate future evolutionary models.
}

\DIFaddend \section{Limits on Additional Companions} \label{sec:companions}

\begin{figure}
    \DIFaddendFL \centering
    \includegraphics[width=\linewidth]{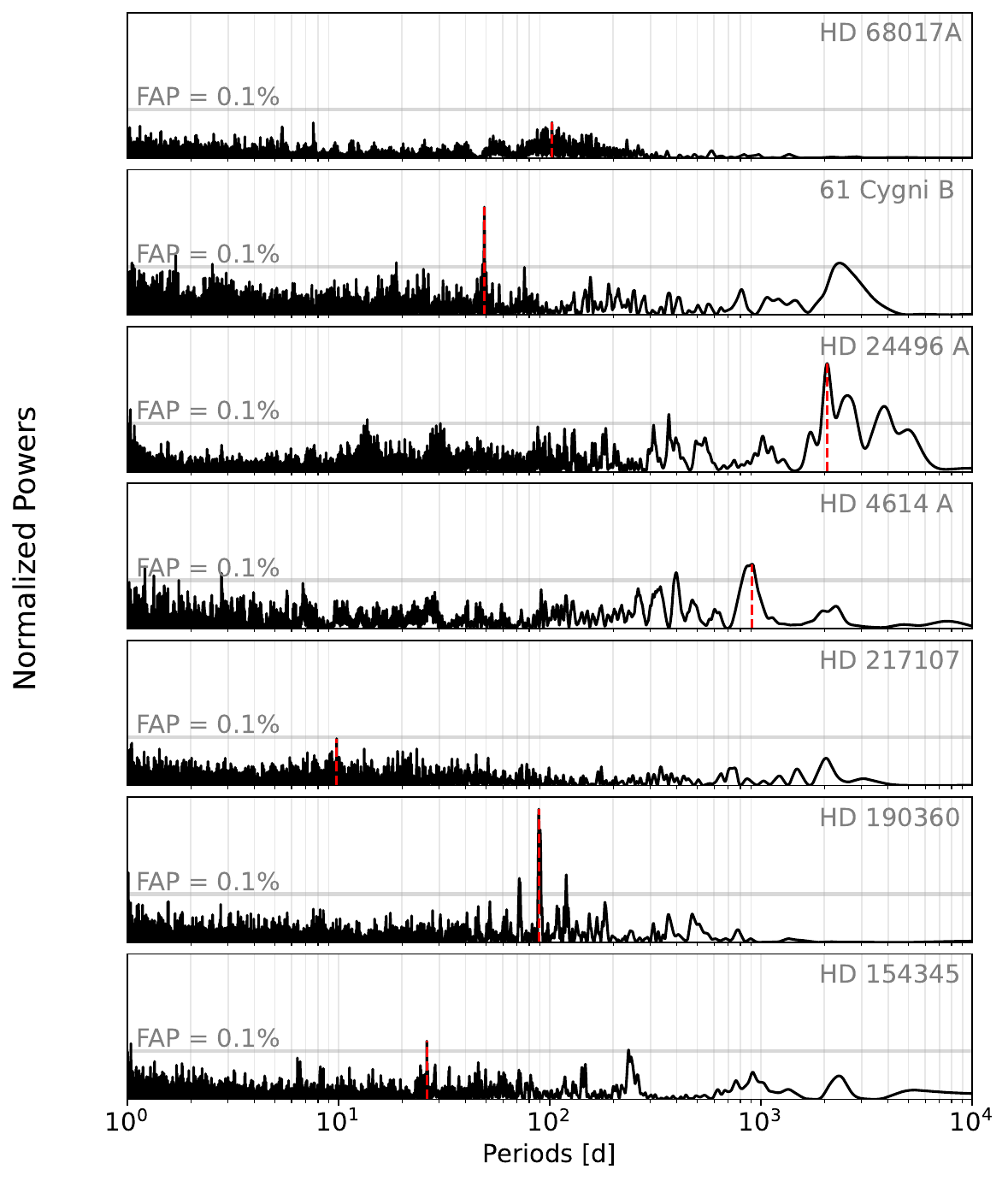}
    \caption{Panel of \DIFdelbeginFL \DIFdelFL{Lomb-scargle }\DIFdelendFL periodograms \DIFdelbeginFL \DIFdelFL{\citep{lomb, scargle} }\DIFdelendFL of RV residuals for all accelerating stars from this analysis. Dashed, vertical red lines are denoted for every system's highest-signal periodicity, and false alarm probabilities at a level of 0.1\% are given for each case. \DIFaddbeginFL \DIFaddFL{The period range searched spans from $1$~d to $27.4$~yr.}\DIFaddendFL } \label{fig:rv_periodograms}
\end{figure}

After fitting each system using \texttt{orvara}, we analyze the residuals of both relative astrometry and RVs. The only systems with a long enough baseline of astrometric measurements to consider a periodic search are 61 Cygni and HD \DIFdelbegin \DIFdel{4614. We fit periodograms to the RV residuals , and }\DIFdelend \DIFaddbegin \DIFadd{4614, though we do not identify any significant signals in either case. To analyze each system's RV residuals and search for periodic signals, we use Lomb-scargle periodograms \citep{lomb, scargle} computed between periods from $1-10^4~$~days with a log normalization and fine sampling of 2000 points per peak. For each system, we }\DIFaddend assess the nature of any peaks that \DIFdelbegin \DIFdel{surpass a }\DIFdelend \DIFaddbegin \DIFadd{exceed a 0.1\% }\DIFaddend false alarm probability\DIFdelbegin \DIFdel{threshold of 0.1\% }\DIFdelend \DIFaddbegin \DIFadd{, calculated via bootstrap resampling, and identify the period with the maximum power as the most significant signal }\DIFaddend (see Figure \ref{fig:rv_periodograms}). \DIFaddbegin \DIFadd{In cases where this signal corresponds to a known stellar activity cycle, we examine H$\alpha$ and Ca II H \& K activity indicators from our NEID spectra to confirm this. For stars where the dominant residual signal has no known origin, we also check Mount Wilson chromospheric activity surveys \citep{Mt_Wilson_I, Mt_Wilson_II}, as well as our NEID activity indicators, for possible correlations.
}\DIFaddend 

\DIFdelbegin \DIFdelendFL \DIFaddbeginFL 

\subsection{61 Cygni B}

\begin{figure}
    \DIFaddendFL \centering
    \includegraphics[width=\linewidth]{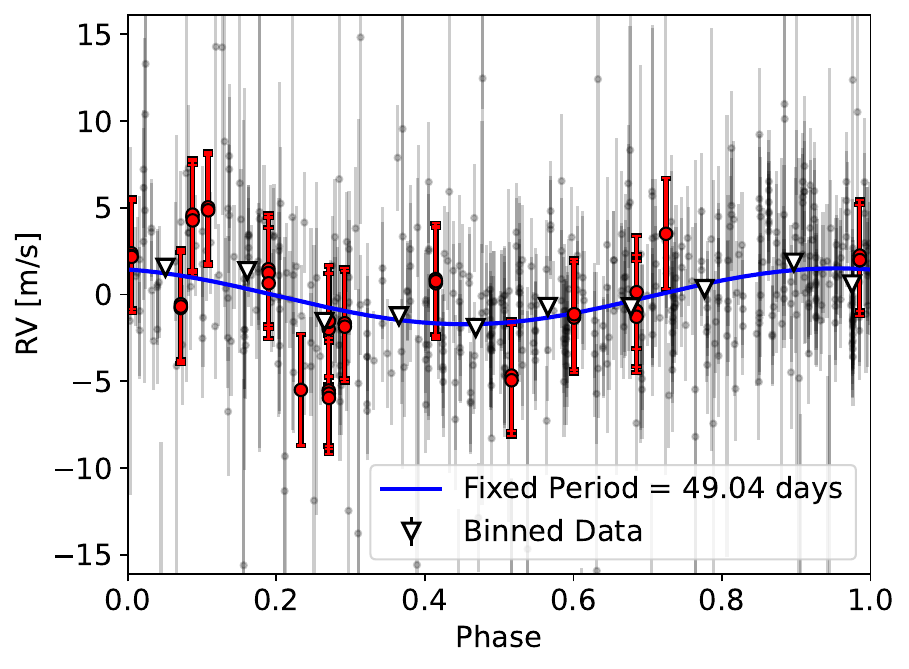}
    \caption{Phase-folded RV curve for 61 Cygni B. The best-fit period from our periodogram is fit for and shown in blue. NEID RVs, which are our analysis' most precise, are highlighted in red, while the remaining data are shown in light black. We bin all data by phase in increments of 0.1, and show those as white triangles.} \label{fig:61CygB_residual_periodogram}
\end{figure}

For 61 Cygni B, \cite{1981ApJ...250..276V} noted a rotational modulation in Ca II H \& K emission of $48.0\pm0.7$~d that was later corroborated by \cite{2021ApJS..255....8R}, which reported a $49.038^{+0.036}_{-0.032}$~d rotation period. A peak at this period is strongly evident in the RV residuals from panel two of Figure \ref{fig:rv_periodograms}. Figure \ref{fig:61CygB_residual_periodogram} shows the phase-folded RVs at that strongest peak in our periodogram, which is 49.043~d. \DIFaddbegin \DIFadd{This signal is also evident in the NEID H$\alpha$ and Ca II H \& K indicators.
}\DIFaddend 

\DIFdelbegin \DIFdelendFL \DIFaddbeginFL 

\subsection{61 Cygni A}

\begin{figure}
    \DIFaddendFL \centering
    \includegraphics[width=\linewidth]{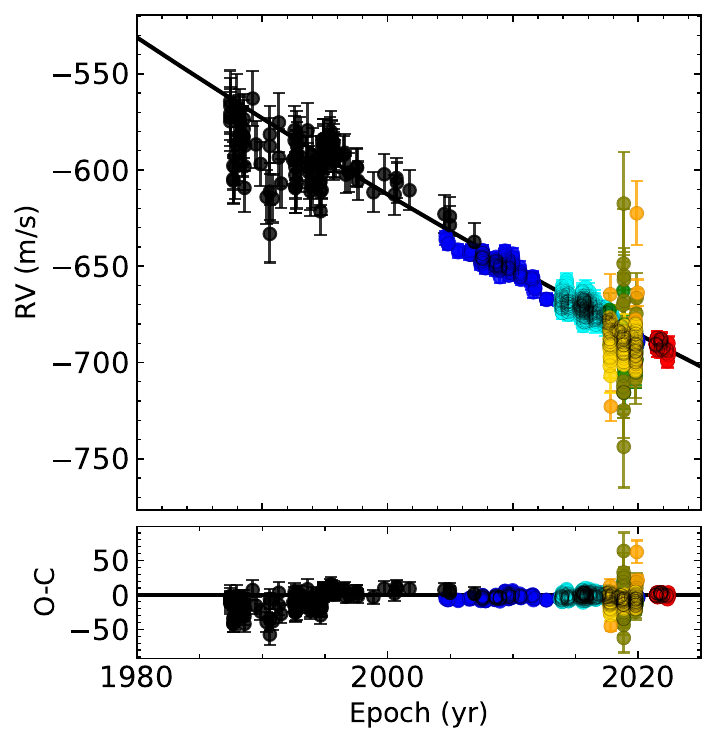}
    \caption{Same as Figure \ref{fig:combined_RVs_binaries} but for 61 Cygni A. Here, we only show the best-fit model, given as a black line. \DIFaddbeginFL \DIFaddFL{The colored points correspond to RV instruments Hamilton, HIRES-post, APF, MINERVA T1-T4, and NEID, as shown in Figure \ref{fig:rv_coverage}}\DIFaddendFL } \label{fig:61CygA_RVs}
\end{figure}

61 Cygni A is a guaranteed time observations target in the NETS program that was identified to be accelerating between the epochs of Hipparcos and Gaia by \cite{hgca_dr2, hgca_edr3}, though its acceleration solution was corrupted. \DIFdelbegin \DIFdel{Since we were able to leverage 61 Cygni B's observed proper motion anomaly and new RVs, we can use the maximum-likelihood orbit from the }\DIFdelend \DIFaddbegin \DIFadd{However, using the }\DIFaddend result described in Section \ref{subsec:61Cyg}\DIFdelbegin \DIFdel{to }\DIFdelend \DIFaddbegin \DIFadd{, we }\DIFaddend subtract off the \DIFaddbegin \DIFadd{system's maximum-likelihood orbit from the }\DIFaddend RVs available for 61 Cygni A \DIFaddbegin \DIFadd{to search for residual signals}\DIFaddend . In total, we use 933 RVs; 690 from \cite{2021ApJS..255....8R}, and the remaining are published here for the first time, 158 from MINERVA and 85 from NEID. Figure \ref{fig:61CygA_RVs} shows the RVs for 61 Cygni A and their corresponding residuals after subtracting the best-fit model from 61 Cygni B. RV zero-points and jitter terms are adopted from those determined from the 61 Cygni B analysis.

\DIFdelbegin \DIFdelendFL \DIFaddbeginFL 

\DIFdelbegin \DIFdelendFL \DIFaddbeginFL \begin{figure}
    \DIFaddendFL \centering
    \includegraphics[width=\linewidth]{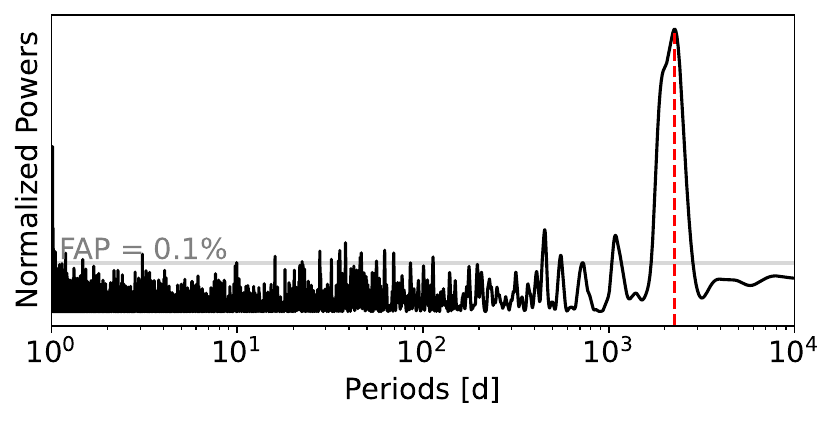}
    \caption{Same as Figure \ref{fig:rv_periodograms} but for 61 Cygni A}. \label{fig:61CygA_residual_periodogram}
\end{figure}

\begin{figure}
    \DIFaddendFL \centering
    \includegraphics[width=\linewidth]{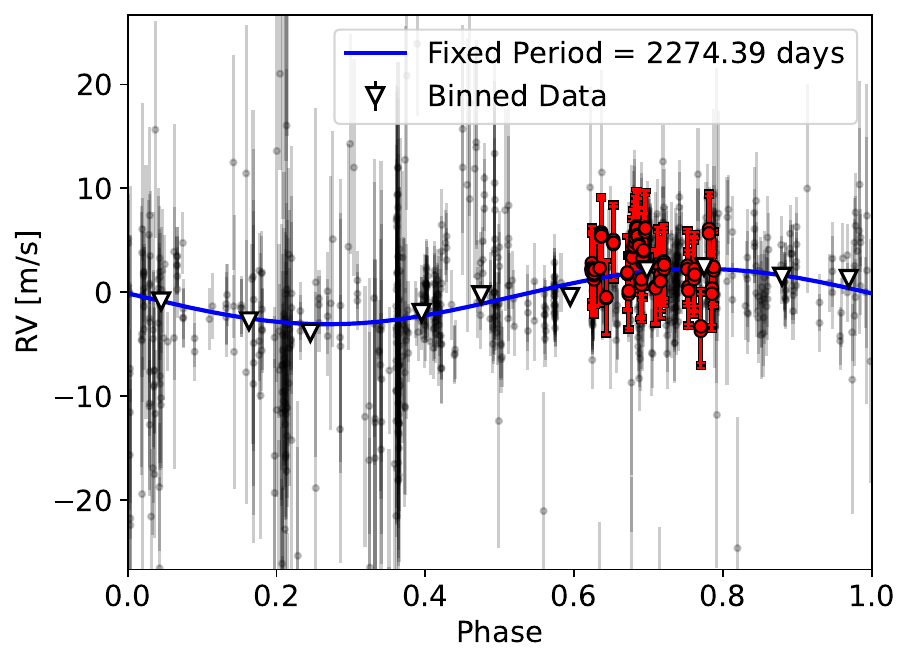}
    \caption{Same as Figure \ref{fig:61CygB_residual_periodogram} but for 61 Cygni A.} \label{fig:61CygA_pfp}
\end{figure}

We search the 61 Cygni A RV residuals after the physical effects from 61 Cygni B had been subtracted off for hierarchical companions with our periodogram (Figure \ref{fig:61CygA_residual_periodogram}). The most significant period we find is at $\approx6.2$~yr, which we identify to be near to known variability in chromospheric activity and X-ray luminosity \DIFdelbegin \DIFdel{\citep[e.g.,][]{1995ApJ...438..269B, 2006A&A...460..261H, 2016A&A...594A..29B}}\DIFdelend \DIFaddbegin \DIFadd{\citep[e.g.,][]{Mt_Wilson_II, 2006A&A...460..261H, 2016A&A...594A..29B}}\DIFaddend . \cite{2016A&A...594A..29B} also used chromospheric activity indicators, as well as Zeeman-Doppler imaging, to identify a rotation period for 61 Cygni A of $\approx35$~d which is also observed to eclipse the 0.1\% false alarm probability threshold in our periodogram. \DIFaddbegin \DIFadd{The H$\alpha$ and Ca II H \& K activity indicators from our NEID spectra corroborate this rotation signal, also identifying a periodicity near 35~d. }\DIFaddend We show the agreement between the strongest peak in our periodogram with RVs phase-folded at that \DIFaddbegin \DIFadd{$6.2~\mathrm{yr}$ }\DIFaddend period in Figure \ref{fig:61CygA_pfp}.

\DIFdelbegin \DIFdelendFL \DIFaddbeginFL 

\subsection{Activity Cycle of HD 24496} \DIFaddbegin \label{subsec:HD24496_activity}
\DIFaddend 

\begin{figure}
    \DIFaddendFL \centering
    \includegraphics[width=\linewidth]{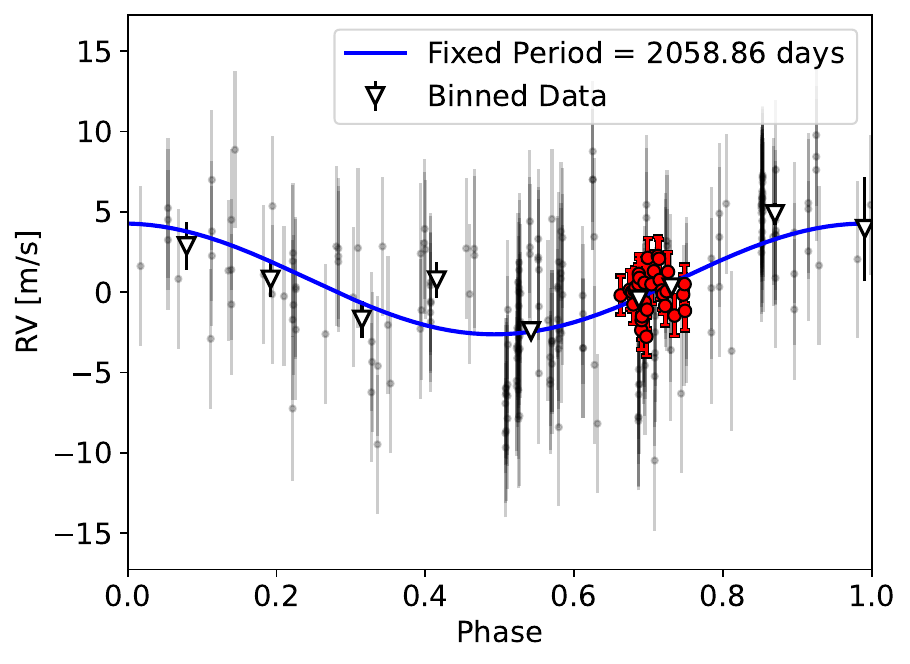}
    \caption{Same as Figure \ref{fig:61CygB_residual_periodogram} but for HD 24496.} \label{fig:HD24496_residual_periodogram}
\end{figure}

We find strong evidence for a periodic signal ($P=5.64$~yr, $K=2.5~\mathrm{m~s^{-1}}$) in the RV residuals of HD 24496, but \cite{2024arXiv240617332I} observe that it is highly correlated to periodicity in $S$- and $\mathrm{log_{10}~R'_{HK}}$-values, and therefore believed to be associated with an activity cycle. Figure \ref{fig:HD24496_residual_periodogram} shows the phase-folded RV diagram for HD 24496 for all data used here. \DIFaddbegin \DIFadd{This signal is too short to be explored in our NEID data, and this star is not studied in the Mount Wilson surveys.
}\DIFaddend 

\DIFdelbegin \DIFdelendFL \DIFaddbeginFL 

\subsection{2.4~yr Periodicity Around HD 4614}

\begin{figure}
    \DIFaddendFL \centering
    \includegraphics[width=\linewidth]{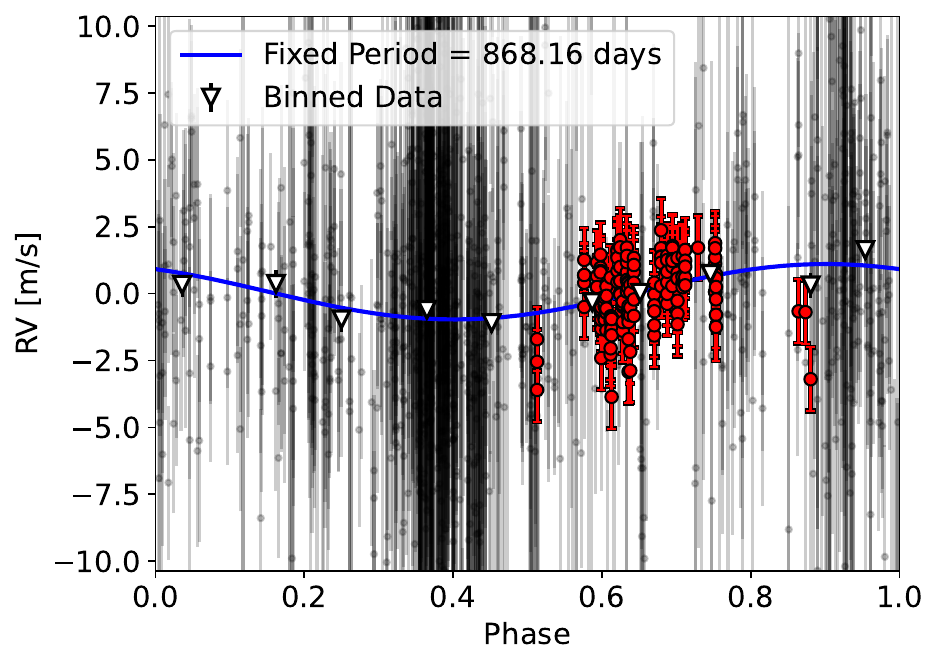}
    \caption{Same as Figure \ref{fig:61CygB_residual_periodogram} but for HD 4614.} \label{fig:HD4614_residual_periodogram}
\end{figure}

Figure \ref{fig:rv_periodograms} shows significant periodicity at 2.4~yr (868\,d) that is not correlated to any known activity cycles (Figure \ref{fig:HD4614_residual_periodogram}). If this signal were a bound companion, its signature in our RVs would imply a 22~$\mathrm{M_\oplus}$ planet. \DIFaddbegin \DIFadd{As is the case in Section \ref{subsec:HD24496_activity}, the 868~d residual period observed here is too short to be explored in our NEID data, and is similarly not published in the Mount Wilson surveys. Future study will be needed to determine whether the origin of this signal is a long-term activity cycle or a physical companion.
}\DIFaddend 

\DIFdelbegin \DIFdelendFL \DIFaddbeginFL 

\subsection{90~d Periodicity in HD 190360}

\begin{figure}
    \DIFaddendFL \centering
    \includegraphics[width=\linewidth]{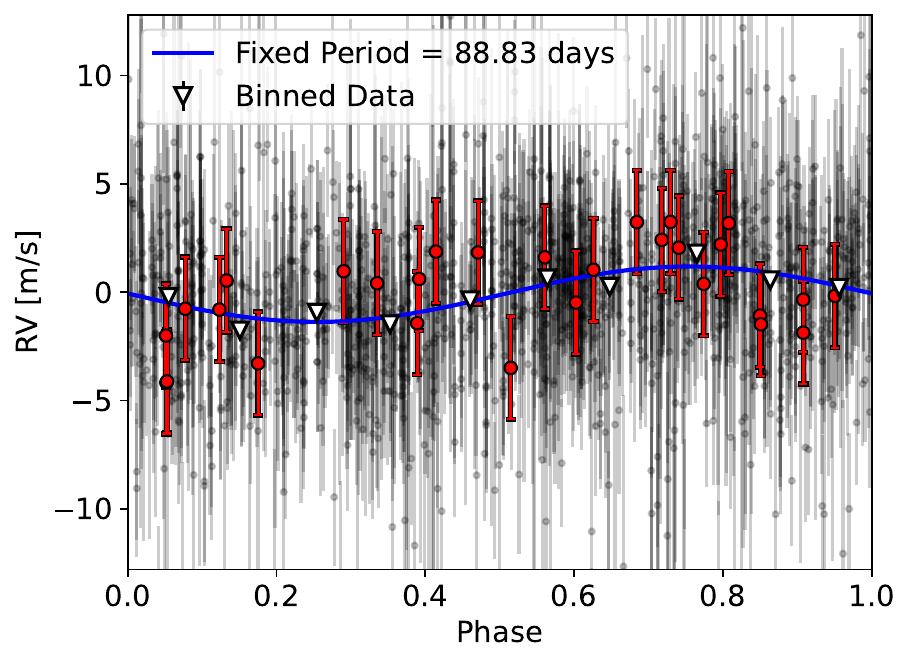}
    \caption{Same as Figure \ref{fig:61CygB_residual_periodogram} but for HD 190360.} \label{fig:HD190360_residual_periodogram}
\end{figure}

As observed in panel six of Figure \ref{fig:rv_periodograms}, HD 190360 has a strong signature of periodicity at $\sim90$~d. A period in this range is not consistent with any known \DIFdelbegin \DIFdel{cycles (e.g. rotation, activity ) of the star}\DIFdelend \DIFaddbegin \DIFadd{stellar cycles (\citealt{Mt_Wilson_II} studied HD 190360 and observed no significant periodicities in chromospheric activity indicators)}\DIFaddend , and \cite{Hirsch_AO} identified the period as a possible planet candidate. \cite{2021ApJS..255....8R}, however, later ruled the source of periodicity as a systematic instrumental artifact, given that it is near the 1/4 annual harmonic. We find \DIFaddbegin \DIFadd{that }\DIFaddend this $\approx90$~d period \DIFdelbegin \DIFdel{to also be }\DIFdelend \DIFaddbegin \DIFadd{is also }\DIFaddend evident in the residuals of our NEID RVs (see Figure \ref{fig:HD190360_residual_periodogram})\DIFaddbegin \DIFadd{, but not in any of our NEID activity indicators, }\DIFaddend and so in an effort to determine the physical nature of this residual signal, we fit a three-planet model to all available data. In this case, not only are the star's two known planets recovered, but we find good agreement with a Keplerian signal consistent with a \DIFdelbegin \DIFdel{$14.3~\mathrm{M_\oplus}$ }\DIFdelend \DIFaddbegin \DIFadd{$\approx10~\mathrm{M_\oplus}$ }\DIFaddend planet at $P=88.818\pm0.069$~d. \DIFdelbegin \DIFdel{We cannot confidently rule on the true nature of this signal, and continued high-cadence RV monitoring will be needed to do so. }\DIFdelend Assuming the nature of this signal is indeed planetary, HD 190360~d would become one of the longest-period super-Earths \DIFdelbegin \DIFdel{around a Sun-like starin the solar neighborhood. }

\DIFdelend \DIFaddbegin \DIFadd{($m<20~\mathrm{M_\oplus}$) around a solar-mass ($0.9<M~\left[\mathrm{M_\odot}\right]<1.1$) star. The true nature of this signal remains uncertain, but continued high-cadence RV monitoring and careful analysis of existing data across different instruments may provide clarity on its origin.
}\DIFaddend \DIFdelbegin \DIFdelendFL \DIFaddbeginFL 

\subsection{HD 154345 Rotation Period}

\begin{figure}
    \DIFaddendFL \centering
    \includegraphics[width=\linewidth]{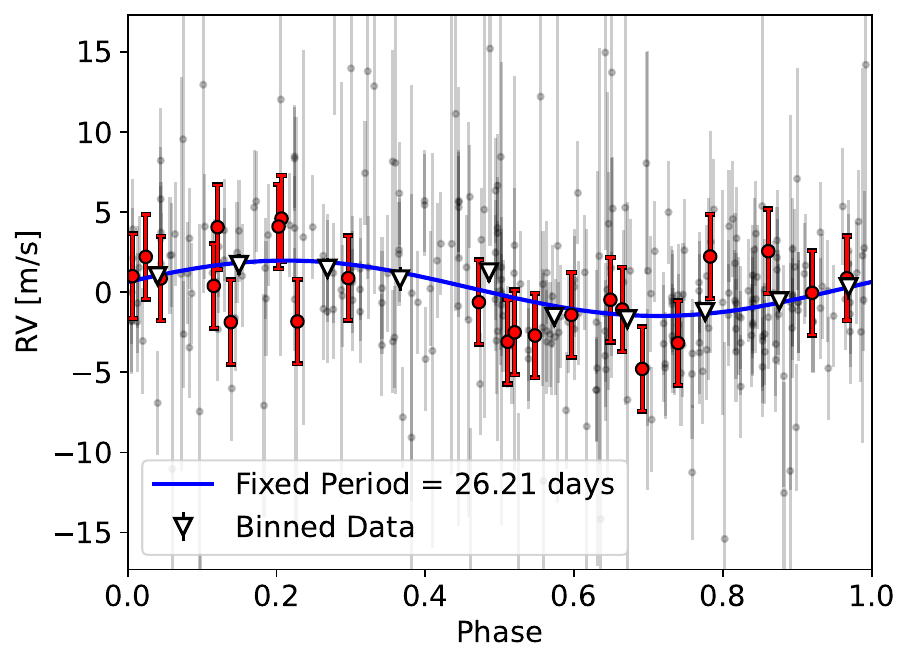}
    \caption{Same as Figure \ref{fig:61CygB_residual_periodogram} but for HD 154345.} \label{fig:HD154345_residual_periodogram}
\end{figure}

We explored the RV residuals of HD 154345 and find periodicity at 26.2~d. \cite{2010MNRAS.408.1666S} found that HD 154345 has a rotation period of $P_\mathrm{rot}=27.8\pm1.7$~d, which is consistent with the peak above the \DIFdelbegin \DIFdel{0.01}\DIFdelend \DIFaddbegin \DIFadd{0.1}\DIFaddend \% false alarm probability in Figure \ref{fig:rv_periodograms}. \DIFdelbegin \DIFdel{We also note that there are longstanding activity cycles consistent with 10~yr , which in conjunction with the }\DIFdelend \DIFaddbegin \DIFadd{This cycle is also in broad agreement with periodicity observed in the H$\alpha$ and Ca II H \& K NEID activity indicators. Beyond that signal, we also recover a residual period of 6.4~yr consistent with the longstanding magnetic cycle confirmed by \cite{2021ApJS..255....8R}. This activity cycle is statistically different from the }\DIFaddend period of HD 154345~b \DIFdelbegin \DIFdel{could influence the signal we observe }\DIFdelend \DIFaddbegin \DIFadd{($P={8.981}_{-0.076}^{+0.079}$~yr) found }\DIFaddend here. Figure \ref{fig:HD154345_residual_periodogram} shows the phase-folded RV residuals \DIFdelbegin \DIFdel{from our analysis for HD 154345.
}\DIFdelend \DIFaddbegin \DIFadd{about the recovered rotation period.
}\DIFaddend 

\DIFaddbegin 

\DIFaddend \DIFdelbegin \DIFdelendFL \DIFaddbeginFL 

\section{Conclusion} \label{sec:concl}

We have demonstrated the power of combining long baselines of RVs with astrometric data to characterize nearby, accelerating star systems. This coalescence of datasets constrains a star system's three-dimensional motion, allowing for reliable determinations of orbital parameters and model-independent mass estimations. Each of these stellar mass estimates \DIFdelbegin \DIFdel{represent }\DIFdelend \DIFaddbegin \DIFadd{represents a }\DIFaddend significant improvement from previous publications in the literature. In particular, three of our star's individual relative mass precisions (HD 68017 B, 61 Cygni A, HD 4614 A) are $<1\%$ (HD 4614 B has $\sigma_{M_B}/\mathrm{M_B}=1.02\%$), paving the way for enhanced parameter calibration schemes that rely on some assumption about the mass \citep[e.g.,][]{2015ApJ...804...64M, 2019ApJ...871...63M}.

We have also published the most comprehensive analyses of three of the longest-baseline RV exoplanet systems in HD 217107, HD 190360, and HD 154345. In each case, we have placed stringent constraints on possible orbital orientations for the outer planets. We find that HD 217107~c ($i=88^{+14\circ}_{-12}$) and HD 154345~b ($i=88\pm20^\circ$) are likely near edge-on planets, while HD 190360~b is notably inclined, with possible maximum-likelihood inclinations of $60^\circ$ or $120^\circ$, depending on the direction of the orbit. This allows us to make direct estimates of each planet's true dynamical mass, something not previously possible without the consideration of Hipparcos and Gaia astrometry.

Additionally, the NEID and MINERVA RVs that we publish here contribute to an ever-growing compilation of publicly available Doppler spectroscopy measurements for nearby stars that make future efforts aimed at surveying these star systems more feasible. Our NEID data in particular extend the RV baseline of our systems to, in some cases, more than 35 years. RV baselines this long push the bounds on what is possible regarding the search for ultra long-period planets. These NEID data are also among the most precise data ever released for exoplanet systems like these, enabling a robust exploration for shorter-period, low-amplitude Keplerian signals that may be physical. \DIFdelbegin \DIFdel{While we }\DIFdelend \DIFaddbegin \DIFadd{We }\DIFaddend find no direct evidence \DIFdelbegin \DIFdel{for additional bodies in our systems, the data published }\DIFdelend \DIFaddbegin \DIFadd{of hierarchical companions in HD 68017, 61 Cygni, HD 24496, HD 217107, or HD 154345. However, our fits for HD 4614 and HD 190360 reveal periodic residual RV signals, neither of which can be definitively attributed to stellar activity or a physical companion. The data presented }\DIFaddend here will be crucial for future \DIFdelbegin \DIFdel{studies of }\DIFdelend \DIFaddbegin \DIFadd{searches for low-amplitude companions in }\DIFaddend these accelerating star systems.

We highlight that NETS has several other interesting star systems known to have relatively significant $\chi_\mathrm{HG}^2$ values that will be \DIFaddbegin \DIFadd{the subject of similar analyses in the future}\DIFaddend . These include a triple \DIFdelbegin \DIFdel{stars }\DIFdelend \DIFaddbegin \DIFadd{star }\DIFaddend system with $\chi_\mathrm{HG}^2>11.8$ (16 Cygni), as well as HD 26965, a hierarchical multiple-star system previously thought to host a planet which was in fact stellar activity \citep{2024AJ....167..243B}. NETS also includes a sample of mildly-accelerating stars with $\chi_\mathrm{HG}^2<11.8$, and low-signal detections of astrometric accelerations like these have been shown to be useful in constraining the physical properties of extrasolar systems, and in some cases even breaking the $m\sin i$ degeneracy in RV-only exoplanet systems \citep{nine_RV_exoplanets}. Future studies of these low-accelerators may reveal the true dynamical masses of undiscovered companions that would otherwise remain hidden without the combination of RVs and astrometry.

The \DIFaddbegin \DIFadd{Gaia mission has already discovered new substellar objects via astrometric monitoring \cite[e.g.,][]{2025AJ....169..107S}, and is expected to publish its }\DIFaddend fourth data release \DIFdelbegin \DIFdel{of the Gaia mission }\DIFdelend (DR4) \DIFdelbegin \DIFdel{is expected by early }\DIFdelend \DIFaddbegin \DIFadd{by the end of }\DIFaddend 2026. An anticipated data product from Gaia DR4 is a catalog of \DIFdelbegin \DIFdel{the mission's }\DIFdelend epoch astrometry, which will \DIFdelbegin \DIFdel{comprise all }\DIFdelend \DIFaddbegin \DIFadd{provide multiple }\DIFaddend positional and projected motion measurements \DIFdelbegin \DIFdel{of targets across multiple epochs from Gaia's observing baseline}\DIFdelend \DIFaddbegin \DIFadd{for all targets across 66 months of observations}\DIFaddend . This will not only enhance the mass precisions of known companions to accelerating stars \DIFdelbegin \DIFdel{, but }\DIFdelend \DIFaddbegin \DIFadd{but will }\DIFaddend also enable the detection of new objects that are likely drivers of the measured astrometric accelerations. These discoveries will explore previously inaccessible regions of substellar parameter space, offering unparalleled opportunities for direct imaging with current and next-generation space-based telescopes to infer effective temperatures and bulk compositions. These new objects will also contribute to the completeness of multiple-systems in the solar neighborhood.

\section*{Acknowledgements}

\DIFaddbegin \DIFadd{The authors would like to thank the anonymous referee for the exceptional care and effort dedicated to reviewing this manuscript. Their thoughtful feedback has greatly improved both its clarity and overall quality.
}

\DIFaddend NEID data presented herein were obtained at the WIYN Observatory from telescope time allocated to NN-EXPLORE through the scientific partnership of the National Aeronautics and Space Administration, the National Science Foundation, and the National Optical Astronomy Observatory. Based in part on observations at Kitt Peak National Observatory, NSF’s NOIRLab (Prop. ID 2021A-2015, 2021B-2015, 2022A-2015, PI: S. Mahadevan; 2021B-0225, 2022A-923895, PI: A. Lin), managed by the Association of Universities for Research in Astronomy (AURA) under a cooperative agreement with the National Science Foundation. The authors are honored to be permitted to conduct astronomical research on Iolkam Du\'ag (Kitt Peak), a mountain with particular significance to the Tohono O\'odham. We thank the NEID Queue Observers and WIYN Observing Associates for their skillful execution of the NEID observations.

This research has made use of the SIMBAD database and the VizieR catalogue access tool, both operated at CDS, Strasbourg, France. Additionally, our work has made use of the NASA Exoplanet Archive \citep{https://doi.org/10.26133/nea1}, which is operated by the California Institute of Technology, under contract with the National Aeronautics and Space Administration under the Exoplanet Exploration Program. This work has made use of data from the European Space Agency (ESA) mission Gaia (\href{https://www.cosmos.esa.int/gaia}{https://www.cosmos.esa.int/gaia}), processed by the Gaia Data Processing and Analysis Consortium (DPAC, \href{https://www.cosmos.esa.int/web/gaia/dpac/consortium}{https://www.cosmos.esa.int/web/gaia/dpac/consortium}). Funding for the DPAC has been provided by national institutions, in particular the institutions participating in the Gaia Multilateral Agreement.

MINERVA is a collaboration among the Harvard-Smithsonian Center for Astrophysics, The Pennsylvania State University, the University of Montana, the University of Southern Queensland, University of Pennsylvania, George Mason University, and the University of New South Wales. It is made possible by generous contributions from its collaborating institutions and Mt. Cuba Astronomical Foundation, The David \& Lucile Packard Foundation, National Aeronautics and Space Administration (EPSCOR grant NNX13AM97A, XRP 80NSSC22K0233), the Australian Research Council (LIEF grant LE140100050), and the National Science Foundation (grants 1516242, 1608203, and 2007811). MRG would like to thank C. M. Dedrick, M. Cornachione, J. A. Johnson, S. A. Johnson, N. McCrady, and D. H. Sliski for their contributions to MINERVA that helped to make this work possible.

\DIFaddbegin \DIFadd{This publication makes use of The Data \& Analysis Center for Exoplanets (DACE), which is a facility based at the University of Geneva (CH) dedicated to extrasolar planets data visualisation, exchange and analysis. DACE is a platform of the Swiss National Centre of Competence in Research (NCCR) PlanetS, federating the Swiss expertise in Exoplanet research. The DACE platform is available at }\href{https://dace.unige.ch}{\DIFadd{https://dace.unige.ch}}\DIFadd{.
}

\DIFaddend The Center for Exoplanets and Habitable Worlds 
is
supported by Penn State and its Eberly College of Science.

This work was performed for the Jet Propulsion Laboratory, California Institute of Technology, sponsored by the United States Government under the Prime Contract 80NM0018D0004 between Caltech and NASA.

This research has made use of the Washington Double Star Catalog maintained at the U.S. Naval Observatory. Additionally, this work is not possible without the great efforts provided by the many observers who have studied these star systems in the centuries leading up to this publication. We are also grateful to the many lands on which these observations were taken.


\appendix
\DIFaddbegin \setcounter{figure}{0} \renewcommand{\thefigure}{A\arabic{figure}} \DIFaddend 

\DIFdelbegin {\DIFdelFL{HD 217107~c}}
\DIFdelFL{\hspace{-2.5cm}
    }{\DIFdelFL{HD 190360~b}}
\DIFdelFL{\hspace{-2.5cm}
    }{\DIFdelFL{HD 154345~b}}

{\DIFdelFL{Best-fit relative separation models for the three giant outer planets presented in this analysis. The panels show the relative separation observed for each system (left: HD 217107; center: HD 190360; right: HD 154345).}}

{\DIFdelFL{HD 68017}}
{\DIFdelFL{61 Cygni}}
{\DIFdelFL{HD 24496}}
{\DIFdelFL{HD 4614}}

{\DIFdelFL{Best-fit RV models for the four binary star systems presented in this analysis. The RVs shown here are those collected for the accelerating star in each system (top left: HD 68017 A; top right: 61 Cygni B; bottom left: HD 24496 A; bottom right: HD 4614 A). The error bars in each panel include instrumental RV jitter terms which have been independently fit for, and the colors of the data points corresponding to those provided in Figure \ref{fig:rv_coverage}. In each case, the best-fit model is then subtracted off and the subsequent residuals are consistent with random scatter about zero.}}

{\DIFdelFL{HD 217107}}
{\DIFdelFL{HD 190360}}

{\DIFdelFL{HD 154345}}

{\DIFdelFL{Best-fit RV models for the three planetary systems presented in this analysis (top left: HD 217107; top right: HD 190360; bottom middle: HD 154345). Each panel also includes phase-folded RV curves for planets. Both two-planet systems (HD 217107 and HD 190360) show planet b on top of planet c. The error bars in each panel include instrumental RV jitter terms which have been independently fit for, and the colors of the data points corresponding to those provided in Figure \ref{fig:rv_coverage}.}}

\DIFdelFL{0.1}

\DIFdelFL{0.1}

\DIFdelFL{0.1}

{\DIFdelFL{Observed proper motion values for the accelerating stars in our four binary systems: HD 68017 A, 61 Cygni B, HD 24496 A, and HD 4614 A. In each panel, the left plot shows the proper motion in right ascension as measured by Hipparcos (left point at $t\approx1991.25$) and Gaia (right point at $t\approx2016.0$), and the right plot shows the proper motion in declination for the same epochs.}}

\DIFdelFL{0.1}

\DIFdelFL{0.1}

{\DIFdelFL{Observed proper motion values for the accelerating stars in our three planetary systems: HD 217107, HD 190360, and HD 154345. Each panel shows the proper motion values as measured by Hipparcos and Gaia in the same way as Figure \ref{fig:combined_pms}.}}

{\DIFdelFL{HD 68017}}
{\DIFdelFL{61 Cygni}}
{\DIFdelFL{HD 24496}}
{\DIFdelFL{HD 4614}}

{\DIFdelFL{Projected relative orbits for the four binary star systems considered here (top left: HD 68017 B around HD 68017 A; top right: 61 Cygni A around 61 Cygni B; bottom left: HD 24496 B around HD 24496 A; bottom right: HD 4614 B around HD 4614 A). The filled blue points represent the relative astrometric measurements included in our analysis, while the open circles are the predicted positions of the maximum likelihood orbital models at past and future specified epochs.}}

\DIFdelend \begin{figure*}
    \centering
    \includegraphics[width=\linewidth]{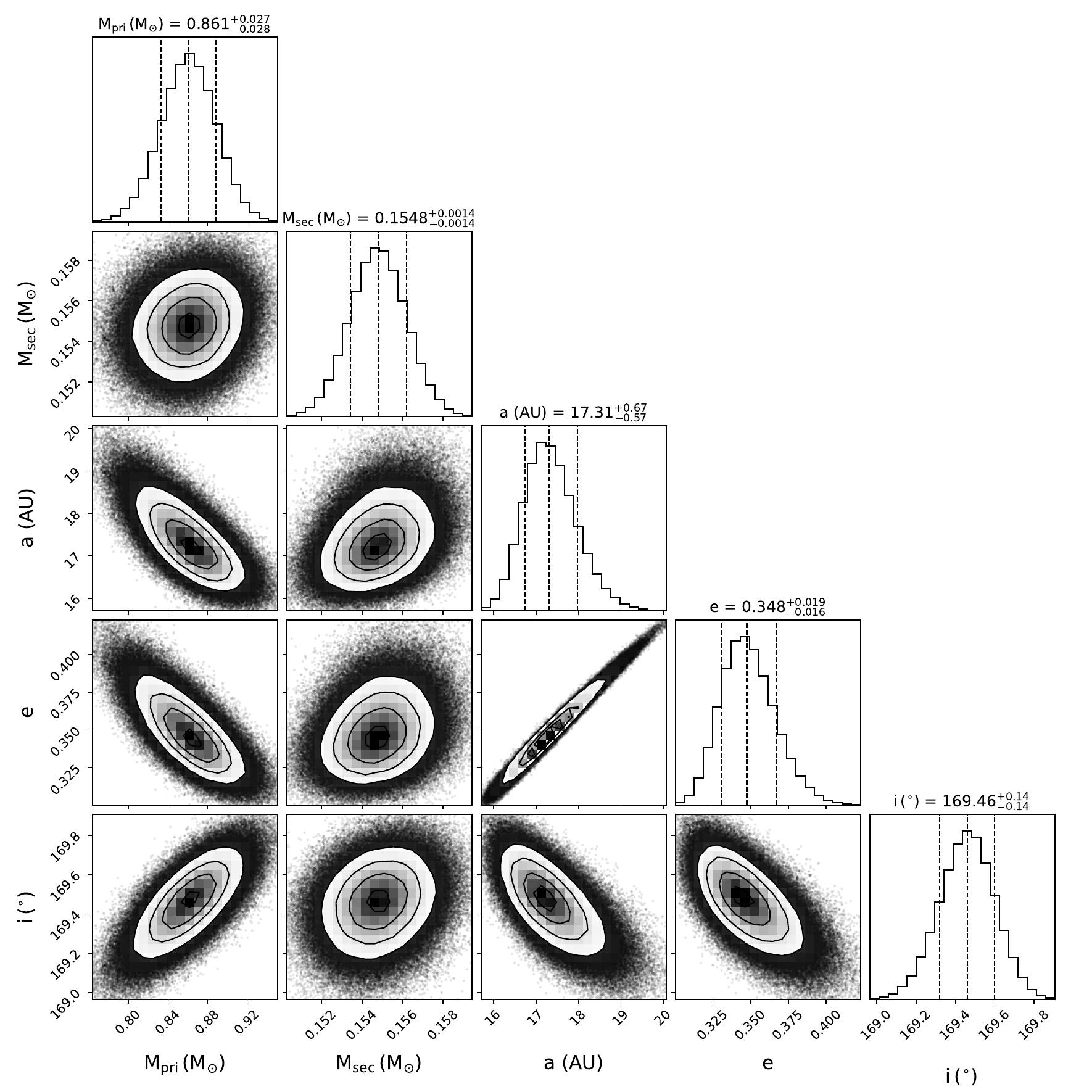}
    \caption{Posterior solution set for the orbital parameters of the HD 68017 binary star system. From left to right in the corner plot, the physical properties are primary mass, secondary mass, semimajor axis, eccentricity, and inclination.} \label{fig:HD68017_corner}
\end{figure*}

\begin{figure*}[p!]
    \centering
    \includegraphics[width=\linewidth]{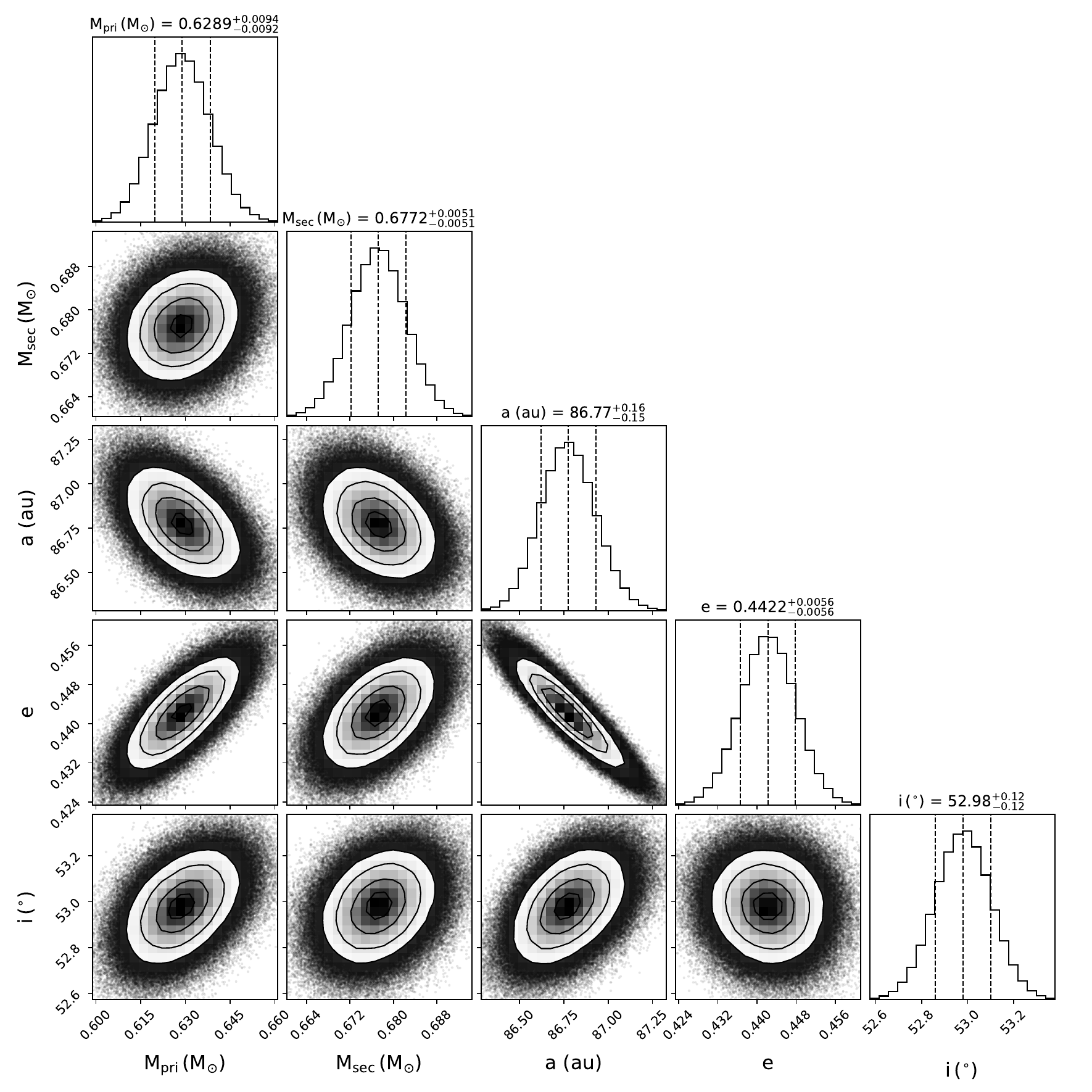}
    \caption{Same as Figure \ref{fig:HD68017_corner}, but for 61 Cygni AB.} \label{fig:61Cyg_corner}
\end{figure*}

\begin{figure*}[p!]
    \centering
    \includegraphics[width=\linewidth]{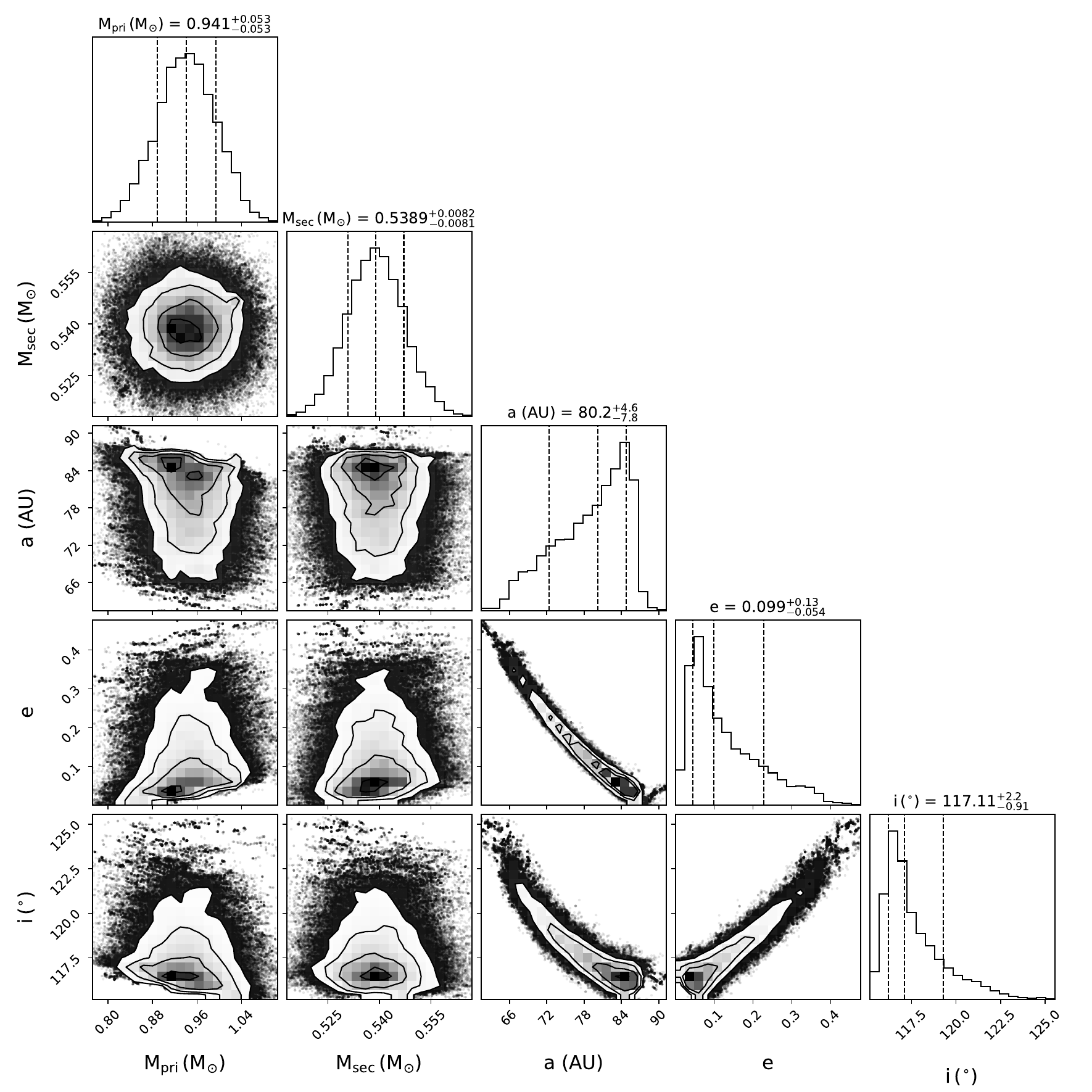}
    \caption{Same as Figure \ref{fig:HD68017_corner}, but for HD 24496 AB.} \label{fig:HD24496_corner}
\end{figure*}

\begin{figure*}[p!]
    \centering
    \includegraphics[width=\linewidth]{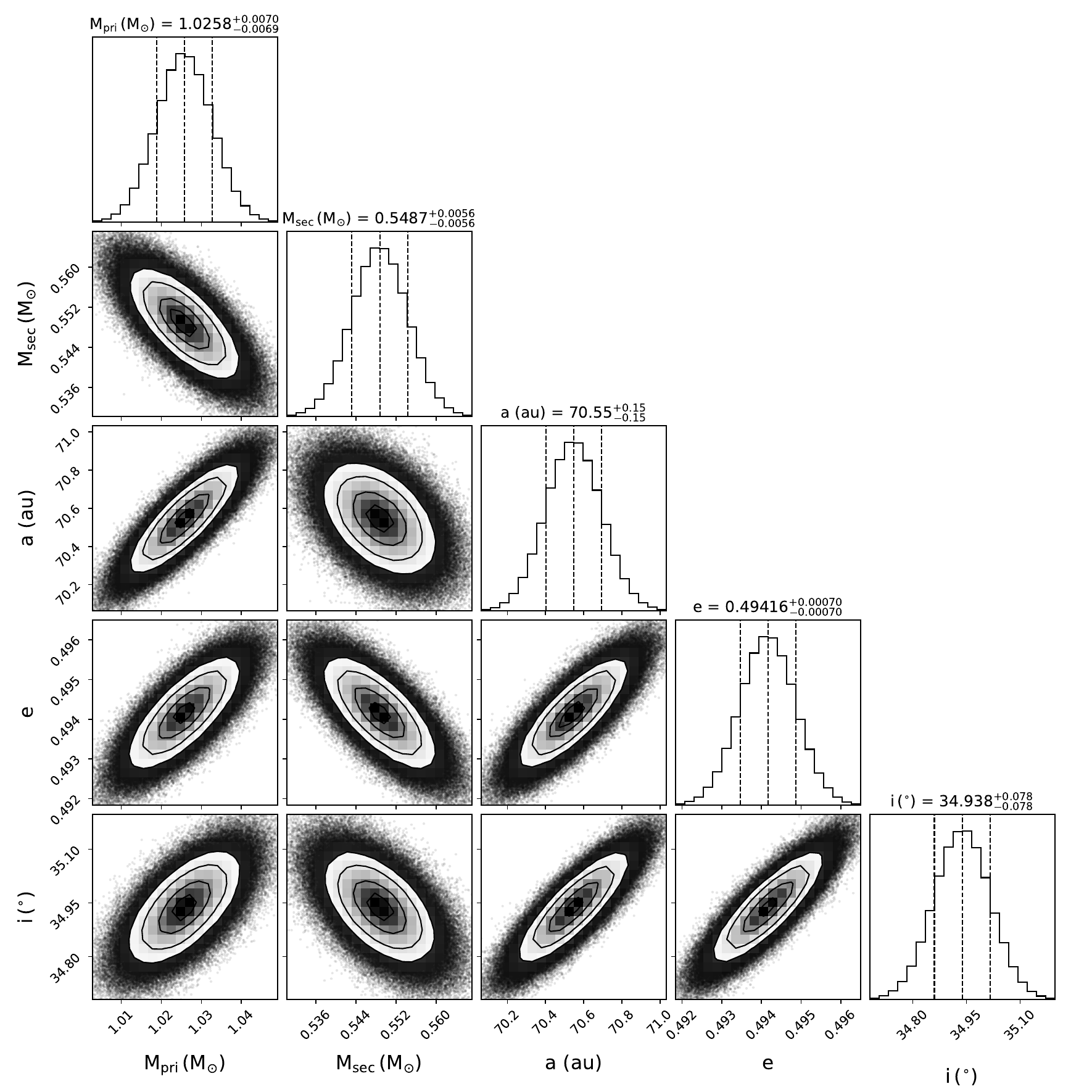}
    \caption{Same as Figure \ref{fig:HD68017_corner}, but for HD 4614 AB.} \label{fig:HD4614_corner}
\end{figure*}

\begin{figure*}[p!]
    \centering
    \includegraphics[width=\linewidth]{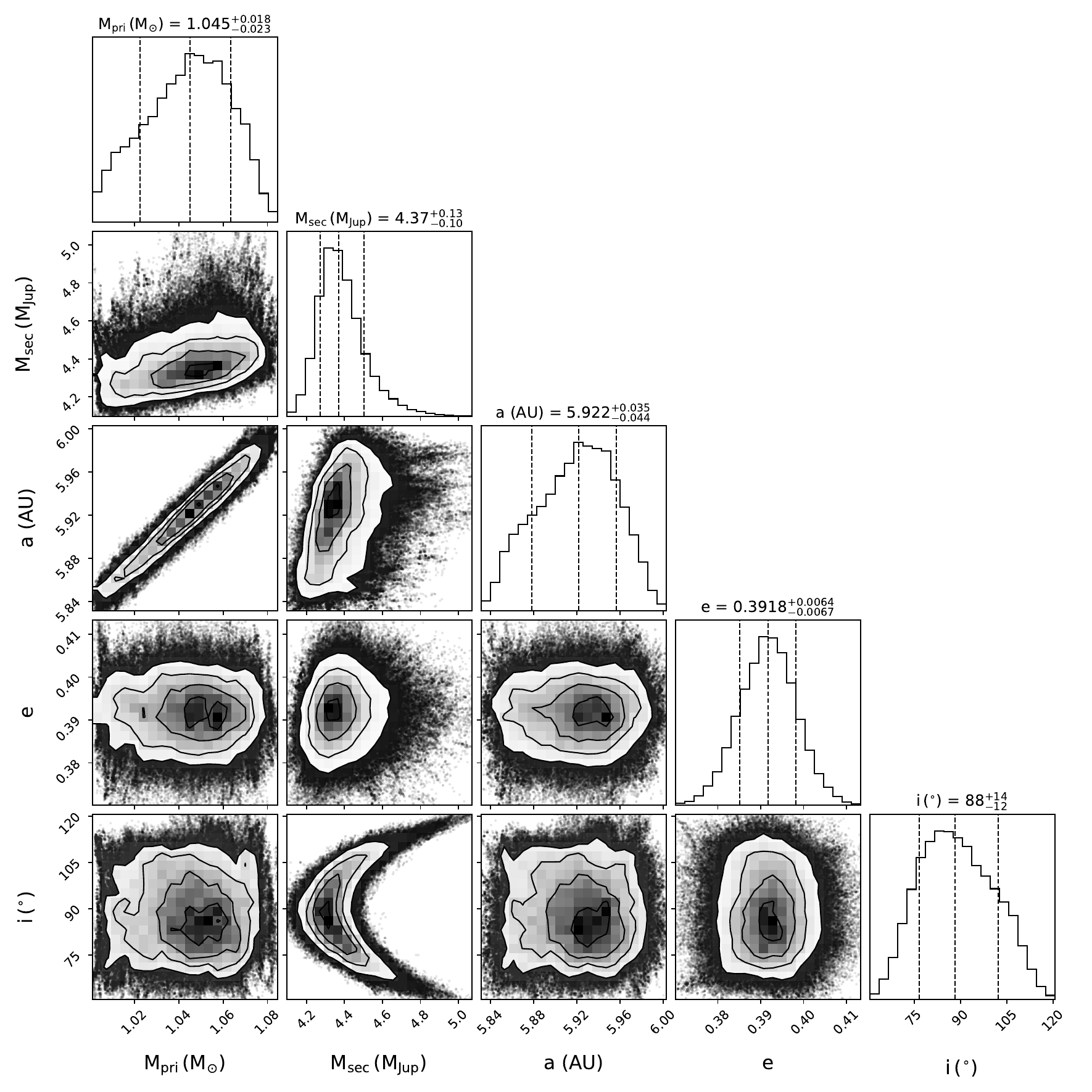}
    \caption{Same as Figure \ref{fig:HD68017_corner}, but for HD 217107~c.} \label{fig:HD217107c_corner}
\end{figure*}

\begin{figure*}[p!]
    \centering
    \includegraphics[width=\linewidth]{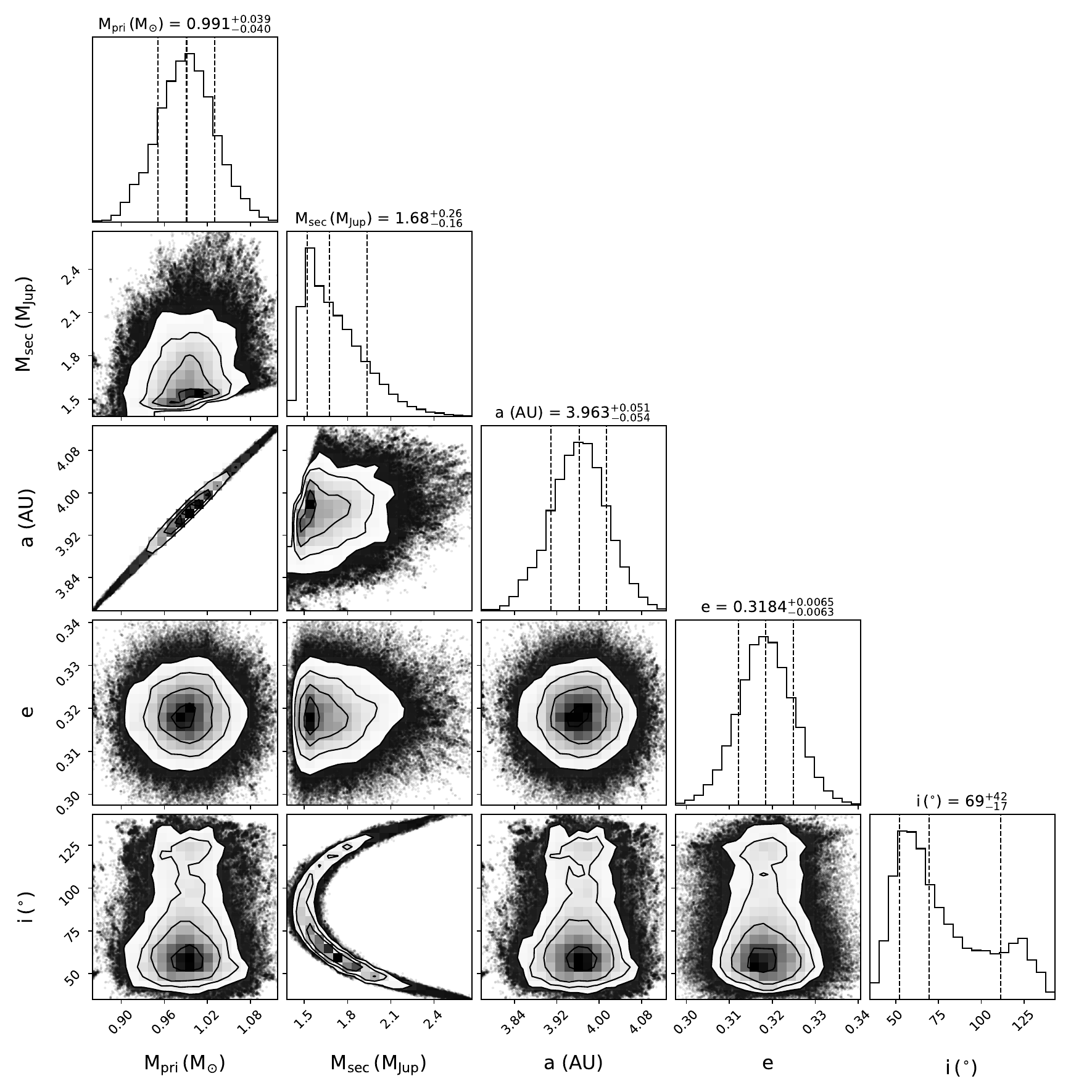}
    \caption{Same as Figure \ref{fig:HD68017_corner}, but for HD 190360~b.} \label{fig:HD190360 B_corner}
\end{figure*}

\begin{figure*}[p!]
    \centering
    \includegraphics[width=\linewidth]{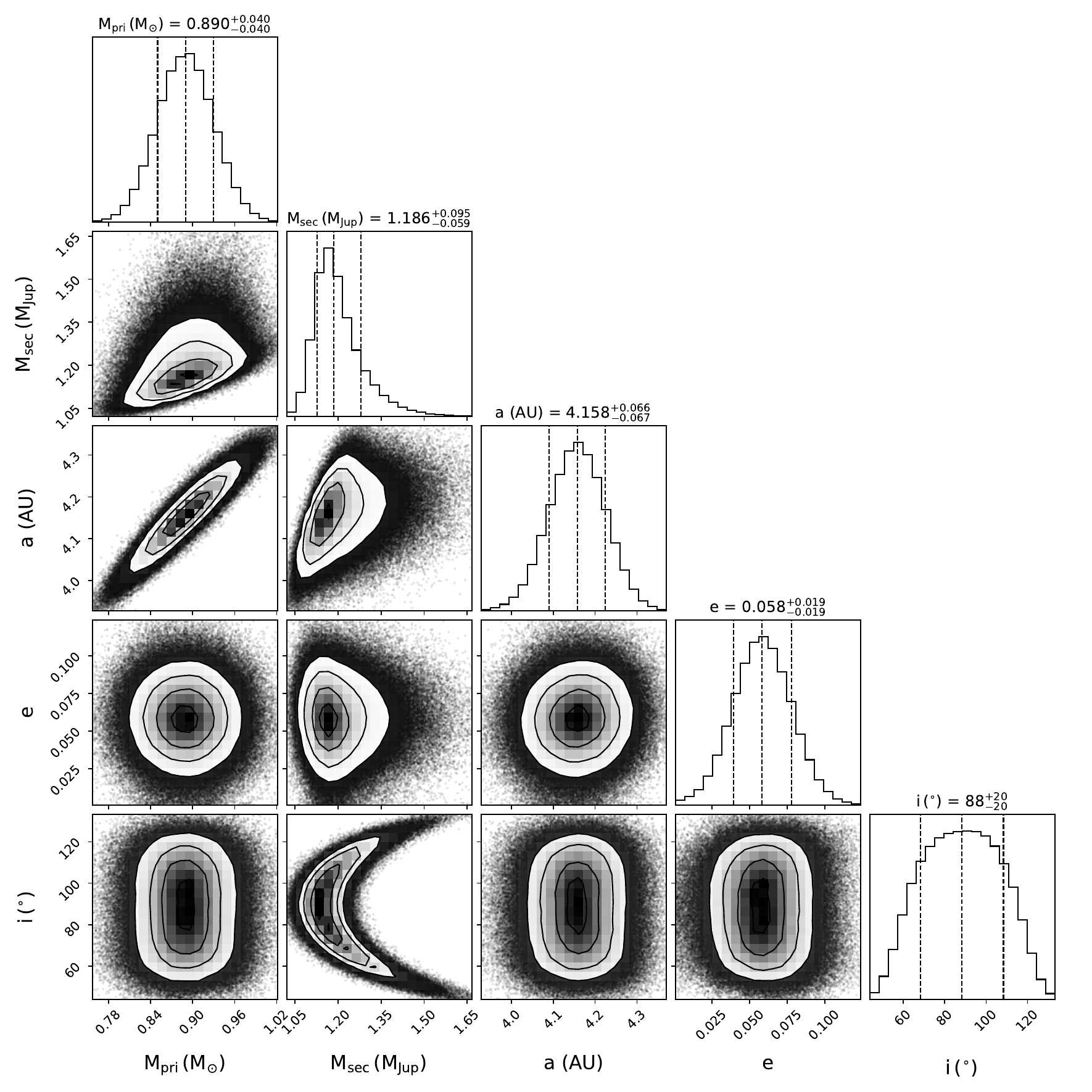}
    \caption{Same as Figure \ref{fig:HD68017_corner}, but for HD 154345~b.} \label{fig:HD154345b_corner}
\end{figure*}

\DIFdelbegin \DIFdelend \DIFaddbegin \facilities{Gaia, HIPPARCOS, WIYN (NEID), MINERVA}
\DIFaddend 

\DIFdelbegin \DIFdelend \DIFaddbegin \software{\texttt{astropy} \citep{astropy:2013, astropy:2018, astropy:2022}, \texttt{orvara} \citep{orvara}, NumPy (\href{https://numpy.org}{https://numpy.org}), matplotlib (\href{https://matplotlib.org}{https://matplotlib.org})}
\DIFaddend 

\clearpage

\bibliography{main.bib}{}

\begin{thebibliography}{}
\expandafter\ifx\csname natexlab\endcsname\relax\def\natexlab#1{#1}\fi
\providecommand{\url}[1]{\href{#1}{#1}}
\providecommand{\dodoi}[1]{doi:~\href{http://doi.org/#1}{\nolinkurl{#1}}}
\providecommand{\doeprint}[1]{\href{http://ascl.net/#1}{\nolinkurl{http://ascl.net/#1}}}
\providecommand{\doarXiv}[1]{\href{https://arxiv.org/abs/#1}{\nolinkurl{https://arxiv.org/abs/#1}}}

\bibitem[{{Astropy Collaboration} {et~al.}(2013){Astropy Collaboration},
  {Robitaille}, {Tollerud}, {Greenfield}, {Droettboom}, {Bray}, {Aldcroft},
  {Davis}, {Ginsburg}, {Price-Whelan}, {Kerzendorf}, {Conley}, {Crighton},
  {Barbary}, {Muna}, {Ferguson}, {Grollier}, {Parikh}, {Nair}, {Unther},
  {Deil}, {Woillez}, {Conseil}, {Kramer}, {Turner}, {Singer}, {Fox}, {Weaver},
  {Zabalza}, {Edwards}, {Azalee Bostroem}, {Burke}, {Casey}, {Crawford},
  {Dencheva}, {Ely}, {Jenness}, {Labrie}, {Lim}, {Pierfederici}, {Pontzen},
  {Ptak}, {Refsdal}, {Servillat}, \& {Streicher}}]{astropy:2013}
{Astropy Collaboration}, {Robitaille}, T.~P., {Tollerud}, E.~J., {et~al.} 2013,
  \aap, 558, A33, \dodoi{10.1051/0004-6361/201322068}

\bibitem[{{Astropy Collaboration} {et~al.}(2018){Astropy Collaboration},
  {Price-Whelan}, {Sip{\H{o}}cz}, {G{\"u}nther}, {Lim}, {Crawford}, {Conseil},
  {Shupe}, {Craig}, {Dencheva}, {Ginsburg}, {Vand erPlas}, {Bradley},
  {P{\'e}rez-Su{\'a}rez}, {de Val-Borro}, {Aldcroft}, {Cruz}, {Robitaille},
  {Tollerud}, {Ardelean}, {Babej}, {Bach}, {Bachetti}, {Bakanov}, {Bamford},
  {Barentsen}, {Barmby}, {Baumbach}, {Berry}, {Biscani}, {Boquien}, {Bostroem},
  {Bouma}, {Brammer}, {Bray}, {Breytenbach}, {Buddelmeijer}, {Burke},
  {Calderone}, {Cano Rodr{\'\i}guez}, {Cara}, {Cardoso}, {Cheedella}, {Copin},
  {Corrales}, {Crichton}, {D'Avella}, {Deil}, {Depagne}, {Dietrich}, {Donath},
  {Droettboom}, {Earl}, {Erben}, {Fabbro}, {Ferreira}, {Finethy}, {Fox},
  {Garrison}, {Gibbons}, {Goldstein}, {Gommers}, {Greco}, {Greenfield},
  {Groener}, {Grollier}, {Hagen}, {Hirst}, {Homeier}, {Horton}, {Hosseinzadeh},
  {Hu}, {Hunkeler}, {Ivezi{\'c}}, {Jain}, {Jenness}, {Kanarek}, {Kendrew},
  {Kern}, {Kerzendorf}, {Khvalko}, {King}, {Kirkby}, {Kulkarni}, {Kumar},
  {Lee}, {Lenz}, {Littlefair}, {Ma}, {Macleod}, {Mastropietro}, {McCully},
  {Montagnac}, {Morris}, {Mueller}, {Mumford}, {Muna}, {Murphy}, {Nelson},
  {Nguyen}, {Ninan}, {N{\"o}the}, {Ogaz}, {Oh}, {Parejko}, {Parley}, {Pascual},
  {Patil}, {Patil}, {Plunkett}, {Prochaska}, {Rastogi}, {Reddy Janga},
  {Sabater}, {Sakurikar}, {Seifert}, {Sherbert}, {Sherwood-Taylor}, {Shih},
  {Sick}, {Silbiger}, {Singanamalla}, {Singer}, {Sladen}, {Sooley},
  {Sornarajah}, {Streicher}, {Teuben}, {Thomas}, {Tremblay}, {Turner},
  {Terr{\'o}n}, {van Kerkwijk}, {de la Vega}, {Watkins}, {Weaver}, {Whitmore},
  {Woillez}, {Zabalza}, \& {Astropy Contributors}}]{astropy:2018}
{Astropy Collaboration}, {Price-Whelan}, A.~M., {Sip{\H{o}}cz}, B.~M., {et~al.}
  2018, \aj, 156, 123, \dodoi{10.3847/1538-3881/aabc4f}

\bibitem[{{Astropy Collaboration} {et~al.}(2022){Astropy Collaboration},
  {Price-Whelan}, {Lim}, {Earl}, {Starkman}, {Bradley}, {Shupe}, {Patil},
  {Corrales}, {Brasseur}, {N{"o}the}, {Donath}, {Tollerud}, {Morris},
  {Ginsburg}, {Vaher}, {Weaver}, {Tocknell}, {Jamieson}, {van Kerkwijk},
  {Robitaille}, {Merry}, {Bachetti}, {G{"u}nther}, {Aldcroft},
  {Alvarado-Montes}, {Archibald}, {B{'o}di}, {Bapat}, {Barentsen}, {Baz{'a}n},
  {Biswas}, {Boquien}, {Burke}, {Cara}, {Cara}, {Conroy}, {Conseil}, {Craig},
  {Cross}, {Cruz}, {D'Eugenio}, {Dencheva}, {Devillepoix}, {Dietrich},
  {Eigenbrot}, {Erben}, {Ferreira}, {Foreman-Mackey}, {Fox}, {Freij}, {Garg},
  {Geda}, {Glattly}, {Gondhalekar}, {Gordon}, {Grant}, {Greenfield}, {Groener},
  {Guest}, {Gurovich}, {Handberg}, {Hart}, {Hatfield-Dodds}, {Homeier},
  {Hosseinzadeh}, {Jenness}, {Jones}, {Joseph}, {Kalmbach}, {Karamehmetoglu},
  {Ka{l}uszy{'n}ski}, {Kelley}, {Kern}, {Kerzendorf}, {Koch}, {Kulumani},
  {Lee}, {Ly}, {Ma}, {MacBride}, {Maljaars}, {Muna}, {Murphy}, {Norman},
  {O'Steen}, {Oman}, {Pacifici}, {Pascual}, {Pascual-Granado}, {Patil},
  {Perren}, {Pickering}, {Rastogi}, {Roulston}, {Ryan}, {Rykoff}, {Sabater},
  {Sakurikar}, {Salgado}, {Sanghi}, {Saunders}, {Savchenko}, {Schwardt},
  {Seifert-Eckert}, {Shih}, {Jain}, {Shukla}, {Sick}, {Simpson},
  {Singanamalla}, {Singer}, {Singhal}, {Sinha}, {Sip{H{o}}cz}, {Spitler},
  {Stansby}, {Streicher}, {{{S}}umak}, {Swinbank}, {Taranu}, {Tewary},
  {Tremblay}, {Val-Borro}, {Van Kooten}, {Vasovi{'c}}, {Verma}, {de Miranda
  Cardoso}, {Williams}, {Wilson}, {Winkel}, {Wood-Vasey}, {Xue}, {Yoachim},
  {Zhang}, {Zonca}, \& {Astropy Project Contributors}}]{astropy:2022}
{Astropy Collaboration}, {Price-Whelan}, A.~M., {Lim}, P.~L., {et~al.} 2022,
  \apj, 935, 167, \dodoi{10.3847/1538-4357/ac7c74}

\bibitem[{{Bailey} {et~al.}(2023){Bailey}, {Bendek}, {Monacelli}, {Baker},
  {Bedrosian}, {Cady}, {Douglas}, {Groff}, {Hildebrandt}, {Kasdin}, {Krist},
  {Macintosh}, {Mennesson}, {Morrissey}, {Poberezhskiy}, {Subedi}, {Rhodes},
  {Roberge}, {Ygouf}, {Zellem}, {Zhao}, \& {Zimmerman}}]{2023SPIE12680E..0TB}
{Bailey}, V.~P., {Bendek}, E., {Monacelli}, B., {et~al.} 2023, in Society of
  Photo-Optical Instrumentation Engineers (SPIE) Conference Series, Vol. 12680,
  Society of Photo-Optical Instrumentation Engineers (SPIE) Conference Series,
  126800T, \dodoi{10.1117/12.2679036}

\bibitem[{{Baines} {et~al.}(2008){Baines}, {McAlister}, {ten Brummelaar},
  {Turner}, {Sturmann}, {Sturmann}, {Goldfinger}, \&
  {Ridgway}}]{2008ApJ...680..728B}
{Baines}, E.~K., {McAlister}, H.~A., {ten Brummelaar}, T.~A., {et~al.} 2008,
  \apj, 680, 728, \dodoi{10.1086/588009}

\bibitem[{{Baliunas} {et~al.}(1995){Baliunas}, {Donahue}, {Soon}, {Horne},
  {Frazer}, {Woodard-Eklund}, {Bradford}, {Rao}, {Wilson}, {Zhang}, {Bennett},
  {Briggs}, {Carroll}, {Duncan}, {Figueroa}, {Lanning}, {Misch}, {Mueller},
  {Noyes}, {Poppe}, {Porter}, {Robinson}, {Russell}, {Shelton}, {Soyumer},
  {Vaughan}, \& {Whitney}}]{Mt_Wilson_II}
{Baliunas}, S.~L., {Donahue}, R.~A., {Soon}, W.~H., {et~al.} 1995, \apj, 438,
  269, \dodoi{10.1086/175072}

\bibitem[{{Baraffe} {et~al.}(2003){Baraffe}, {Chabrier}, {Allard}, \&
  {Hauschildt}}]{baraffe2003}
{Baraffe}, I., {Chabrier}, G., {Allard}, F., \& {Hauschildt}, P. 2003, in IAU
  Symposium, Vol. 211, Brown Dwarfs, ed. E.~{Mart{\'\i}n}, 41

\bibitem[{{Baranec} {et~al.}(2012){Baranec}, {Riddle}, {Ramaprakash}, {Law},
  {Tendulkar}, {Kulkarni}, {Dekany}, {Bui}, {Davis}, {Burse}, {Das},
  {Hildebrandt}, {Punnadi}, \& {Smith}}]{Robo_AO_instrument}
{Baranec}, C., {Riddle}, R., {Ramaprakash}, A.~N., {et~al.} 2012, in Society of
  Photo-Optical Instrumentation Engineers (SPIE) Conference Series, Vol. 8447,
  Adaptive Optics Systems III, ed. B.~L. {Ellerbroek}, E.~{Marchetti}, \& J.-P.
  {V{\'e}ran}, 844704, \dodoi{10.1117/12.924867}

\bibitem[{{Baranne} {et~al.}(1996){Baranne}, {Queloz}, {Mayor}, {Adrianzyk},
  {Knispel}, {Kohler}, {Lacroix}, {Meunier}, {Rimbaud}, \& {Vin}}]{ELODIE}
{Baranne}, A., {Queloz}, D., {Mayor}, M., {et~al.} 1996, \aaps, 119, 373

\bibitem[{{Bessel}(1838)}]{1838MNRAS...4..152B}
{Bessel}, F.~W. 1838, \mnras, 4, 152, \dodoi{10.1093/mnras/4.17.152}

\bibitem[{{Bessel}(1844)}]{1844MNRAS...6R.136B}
---. 1844, \mnras, 6, 136, \dodoi{10.1093/mnras/6.11.136}

\bibitem[{{Boccaletti} {et~al.}(2022){Boccaletti}, {Cossou}, {Baudoz},
  {Lagage}, {Dicken}, {Glasse}, {Hines}, {Aguilar}, {Detre}, {Nickson},
  {Noriega-Crespo}, {G{\'a}sp{\'a}r}, {Labiano}, {Stark}, {Rouan}, {Reess},
  {Wright}, {Rieke}, {Garcia Marin}, {Lajoie}, {Girard}, {Perrin}, {Soummer},
  \& {Pueyo}}]{2022A&A...667A.165B}
{Boccaletti}, A., {Cossou}, C., {Baudoz}, P., {et~al.} 2022, \aap, 667, A165,
  \dodoi{10.1051/0004-6361/202244578}

\bibitem[{{Boisse} {et~al.}(2012){Boisse}, {Pepe}, {Perrier}, {Queloz},
  {Bonfils}, {Bouchy}, {Santos}, {Arnold}, {Beuzit}, {D{\'\i}az}, {Delfosse},
  {Eggenberger}, {Ehrenreich}, {Forveille}, {H{\'e}brard}, {Lagrange}, {Lovis},
  {Mayor}, {Moutou}, {Naef}, {Santerne}, {S{\'e}gransan}, {Sivan}, \&
  {Udry}}]{HD154345_sophie_elodie}
{Boisse}, I., {Pepe}, F., {Perrier}, C., {et~al.} 2012, \aap, 545, A55,
  \dodoi{10.1051/0004-6361/201118419}

\bibitem[{{Boro Saikia} {et~al.}(2016){Boro Saikia}, {Jeffers}, {Morin},
  {Petit}, {Folsom}, {Marsden}, {Donati}, {Cameron}, {Hall}, {Perdelwitz},
  {Reiners}, \& {Vidotto}}]{2016A&A...594A..29B}
{Boro Saikia}, S., {Jeffers}, S.~V., {Morin}, J., {et~al.} 2016, \aap, 594,
  A29, \dodoi{10.1051/0004-6361/201628262}

\bibitem[{{Boyajian} {et~al.}(2012){Boyajian}, {McAlister}, {van Belle},
  {Gies}, {ten Brummelaar}, {von Braun}, {Farrington}, {Goldfinger}, {O'Brien},
  {Parks}, {Richardson}, {Ridgway}, {Schaefer}, {Sturmann}, {Sturmann},
  {Touhami}, {Turner}, \& {White}}]{2012ApJ...746..101B}
{Boyajian}, T.~S., {McAlister}, H.~A., {van Belle}, G., {et~al.} 2012, \apj,
  746, 101, \dodoi{10.1088/0004-637X/746/1/101}

\bibitem[{{Boyajian} {et~al.}(2013){Boyajian}, {von Braun}, {van Belle},
  {Farrington}, {Schaefer}, {Jones}, {White}, {McAlister}, {ten Brummelaar},
  {Ridgway}, {Gies}, {Sturmann}, {Sturmann}, {Turner}, {Goldfinger}, \&
  {Vargas}}]{2013ApJ...771...40B}
{Boyajian}, T.~S., {von Braun}, K., {van Belle}, G., {et~al.} 2013, \apj, 771,
  40, \dodoi{10.1088/0004-637X/771/1/40}

\bibitem[{{Brandt} {et~al.}(2021{\natexlab{a}}){Brandt}, {Michalik}, {Brandt},
  {Li}, {Dupuy}, \& {Zeng}}]{2021AJ....162..230B}
{Brandt}, G.~M., {Michalik}, D., {Brandt}, T.~D., {et~al.} 2021{\natexlab{a}},
  \aj, 162, 230, \dodoi{10.3847/1538-3881/ac12d0}

\bibitem[{{Brandt} {et~al.}(2021{\natexlab{b}}){Brandt}, {Dupuy}, {Li}, {Chen},
  {Brandt}, {Wong}, {Currie}, {Bowler}, {Liu}, {Best}, \&
  {Phillips}}]{bd_masses_hgca_edr3}
{Brandt}, G.~M., {Dupuy}, T.~J., {Li}, Y., {et~al.} 2021{\natexlab{b}}, \aj,
  162, 301, \dodoi{10.3847/1538-3881/ac273e}

\bibitem[{{Brandt}(2018)}]{hgca_dr2}
{Brandt}, T.~D. 2018, \apjs, 239, 31, \dodoi{10.3847/1538-4365/aaec06}

\bibitem[{{Brandt}(2021)}]{hgca_edr3}
---. 2021, \apjs, 254, 42, \dodoi{10.3847/1538-4365/abf93c}

\bibitem[{{Brandt} {et~al.}(2019){Brandt}, {Dupuy}, \& {Bowler}}]{Brandt_68017}
{Brandt}, T.~D., {Dupuy}, T.~J., \& {Bowler}, B.~P. 2019, \aj, 158, 140,
  \dodoi{10.3847/1538-3881/ab04a8}

\bibitem[{{Brandt} {et~al.}(2021{\natexlab{c}}){Brandt}, {Dupuy}, {Li},
  {Brandt}, {Zeng}, {Michalik}, {Bardalez Gagliuffi}, \&
  {Raposo-Pulido}}]{orvara}
{Brandt}, T.~D., {Dupuy}, T.~J., {Li}, Y., {et~al.} 2021{\natexlab{c}}, \aj,
  162, 186, \dodoi{10.3847/1538-3881/ac042e10.48550/arXiv.2105.11671}

\bibitem[{{Brewer} {et~al.}(2012{\natexlab{a}}){Brewer}, {Rogers}, {Harder},
  {Lazak}, {Gillette}, {Gillette}, {Sweatt}, {Keele}, {Keele}, {Smith},
  {Mercado}, {Cheske}, {Zoltan}, \& {Stewart}}]{BMA2012b}
{Brewer}, M., {Rogers}, A., {Harder}, H., {et~al.} 2012{\natexlab{a}}, Journal
  of Double Star Observations, 8, 210

\bibitem[{{Brewer} {et~al.}(2012{\natexlab{b}}){Brewer}, {Rogers}, {Harder},
  {Lazak}, {Gillette}, {Gillette}, {Sweatt}, {Keele}, {Keele}, {Smith},
  {Mercado}, {Cheske}, {Zoltan}, \& {Stewart}}]{Cao2011}
---. 2012{\natexlab{b}}, Journal of Double Star Observations, 8, 210

\bibitem[{{Brown} {et~al.}(1994){Brown}, {Noyes}, {Nisenson}, {Korzennik}, \&
  {Horner}}]{AFOE}
{Brown}, T.~M., {Noyes}, R.~W., {Nisenson}, P., {Korzennik}, S.~G., \&
  {Horner}, S. 1994, \pasp, 106, 1285, \dodoi{10.1086/133506}

\bibitem[{{Burgasser} \& {Splat Development Team}(2017)}]{splat}
{Burgasser}, A.~J., \& {Splat Development Team}. 2017, in Astronomical Society
  of India Conference Series, Vol.~14, Astronomical Society of India Conference
  Series, 7--12, \dodoi{10.48550/arXiv.1707.00062}

\bibitem[{{Burrows} {et~al.}(2024){Burrows}, {Halverson}, {Siegel},
  {Gilbertson}, {Luhn}, {Burt}, {Bender}, {Roy}, {Terrien}, {Vangstein},
  {Mahadevan}, {Wright}, {Robertson}, {Ford}, {Stef{\'a}nsson}, {Ninan},
  {Blake}, {McElwain}, {Schwab}, \& {Zhao}}]{2024AJ....167..243B}
{Burrows}, A., {Halverson}, S., {Siegel}, J.~C., {et~al.} 2024, \aj, 167, 243,
  \dodoi{10.3847/1538-3881/ad34d5}

\bibitem[{{Butler} {et~al.}(2006){Butler}, {Wright}, {Marcy}, {Fischer},
  {Vogt}, {Tinney}, {Jones}, {Carter}, {Johnson}, {McCarthy}, \&
  {Penny}}]{HIRES_upgrade}
{Butler}, R.~P., {Wright}, J.~T., {Marcy}, G.~W., {et~al.} 2006, \apj, 646,
  505, \dodoi{10.1086/504701}

\bibitem[{{Ca{\~n}as} {et~al.}(2022){Ca{\~n}as}, {Kanodia}, {Bender},
  {Mahadevan}, {Stef{\'a}nsson}, {Cochran}, {Lin}, {Hwang}, {Powers}, {Monson},
  {Green}, {Parker}, {Swaby}, {Kobulnicky}, {Wisniewski}, {Gupta}, {Everett},
  {Jones}, {Anjakos}, {Beard}, {Blake}, {Diddams}, {Dong}, {Fredrick},
  {Hakemiamjad}, {Hebb}, {Libby-Roberts}, {Logsdon}, {McElwain}, {Metcalf},
  {Ninan}, {Rajagopal}, {Ramsey}, {Robertson}, {Roy}, {Ruhle}, {Schwab},
  {Terrien}, \& {Wright}}]{TOI3714_3629}
{Ca{\~n}as}, C.~I., {Kanodia}, S., {Bender}, C.~F., {et~al.} 2022, \aj, 164,
  50, \dodoi{10.3847/1538-3881/ac7804}

\bibitem[{{Celoria} \& {Gabba}(1923)}]{Cel1923}
{Celoria}, G., \& {Gabba}, L. 1923, {Misure DI stelle doppie eseguite col
  refrattore ``Merz'' DI 8 pollici negli anni 1886-1900 E col refrattore
  ``Merz-Repsold'' DI 18 pollici negli anni 1902-1905 DA G. Celoria, calcolate
  DA L. Gabba.} (Italy: Hoepli)

\bibitem[{{Choi} {et~al.}(2016){Choi}, {Dotter}, {Conroy}, {Cantiello},
  {Paxton}, \& {Johnson}}]{MIST1}
{Choi}, J., {Dotter}, A., {Conroy}, C., {et~al.} 2016, \apj, 823, 102,
  \dodoi{10.3847/0004-637X/823/2/102}

\bibitem[{{Clark} {et~al.}(2021){Clark}, {Clert{\'e}}, {Hinkel}, {Unterborn},
  {Wittenmyer}, {Horner}, {Wright}, {Carter}, {Morton}, {Spina}, {Asplund},
  {Buder}, {Bland-Hawthorn}, {Casey}, {De Silva}, {D'Orazi}, {Duong}, {Hayden},
  {Freeman}, {Kos}, {Lewis}, {Lin}, {Lind}, {Martell}, {Sharma}, {Simpson},
  {Zucker}, {Zwitter}, {Tinney}, {Ting}, {Nordlander}, \&
  {Amarsi}}]{2021MNRAS.504.4968C}
{Clark}, J.~T., {Clert{\'e}}, M., {Hinkel}, N.~R., {et~al.} 2021, \mnras, 504,
  4968, \dodoi{10.1093/mnras/stab1052}

\bibitem[{{Clark} {et~al.}(2022){Clark}, {Wright}, {Wittenmyer}, {Horner},
  {Hinkel}, {Clert{\'e}}, {Carter}, {Buder}, {Hayden}, {Bland-Hawthorn},
  {Casey}, {De Silva}, {D'Orazi}, {Freeman}, {Kos}, {Lewis}, {Lin}, {Lind},
  {Martell}, {Schlesinger}, {Sharma}, {Simpson}, {Stello}, {Zucker}, {Zwitter},
  {Munari}, \& {Nordlander}}]{2022MNRAS.510.2041C}
{Clark}, J.~T., {Wright}, D.~J., {Wittenmyer}, R.~A., {et~al.} 2022, \mnras,
  510, 2041, \dodoi{10.1093/mnras/stab3498}

\bibitem[{{Cochran} {et~al.}(1991){Cochran}, {Hatzes}, \&
  {Hancock}}]{HD114762_BDconf1}
{Cochran}, W.~D., {Hatzes}, A.~P., \& {Hancock}, T.~J. 1991, \apjl, 380, L35,
  \dodoi{10.1086/186167}

\bibitem[{{Crepp} {et~al.}(2012){Crepp}, {Johnson}, {Howard}, {Marcy},
  {Fischer}, {Hillenbrand}, {Yantek}, {Delaney}, {Wright}, {Isaacson}, \&
  {Montet}}]{2012ApJ...761...39C}
{Crepp}, J.~R., {Johnson}, J.~A., {Howard}, A.~W., {et~al.} 2012, \apj, 761,
  39, \dodoi{10.1088/0004-637X/761/1/39}

\bibitem[{{de Bruijne} {et~al.}(2015){de Bruijne}, {Allen}, {Azaz},
  {Krone-Martins}, {Prod'homme}, \& {Hestroffer}}]{2015A&A...576A..74D}
{de Bruijne}, J.~H.~J., {Allen}, M., {Azaz}, S., {et~al.} 2015, \aap, 576, A74,
  \dodoi{10.1051/0004-6361/201424018}

\bibitem[{{de Pont}(2017)}]{Pnt2017b}
{de Pont}, L.~R. 2017, El Observador de Estrellas Dobles, 19

\bibitem[{{Deich} \& {Orlova}(1977)}]{61cyg_residual_search4}
{Deich}, A.~N., \& {Orlova}, O.~N. 1977, \sovast, 21, 182

\bibitem[{{Dembowski}(1884)}]{D__1884}
{Dembowski}, E. 1884, {Misure micrometriche di stelle doppie e multiple: Le
  osservazioni fatte a Gallarate sopra le stelle del Catalogo di Dorpat e delle
  appendici di W. Struve} (Rome: Coi Tipi Del Salviucci)

\bibitem[{{Doberck}(1876)}]{1876AN.....88...45D}
{Doberck}, W. 1876, Astronomische Nachrichten, 88, 45,
  \dodoi{10.1002/asna.18760880303}

\bibitem[{{Doberck}(1927)}]{Dob1927}
---. 1927, {Double star observations, 1875-1927} (Germany: (n.p.).)

\bibitem[{{Dotter}(2016)}]{MIST0}
{Dotter}, A. 2016, \apjs, 222, 8, \dodoi{10.3847/0067-0049/222/1/8}

\bibitem[{{Duńer}(1876)}]{Du_1876}
{Duńer}, N.~C. 1876, {Mesures micrométriquues d'étoiles doubles, faites à
  l'Observatoire de Lund, suivies de notes sur leurs mouvements relatifs}
  (Upsala: Berling)

\bibitem[{{El-Badry} {et~al.}(2021){El-Badry}, {Rix}, \&
  {Heintz}}]{gaia_wide_binaries}
{El-Badry}, K., {Rix}, H.-W., \& {Heintz}, T.~M. 2021, \mnras, 506, 2269,
  \dodoi{10.1093/mnras/stab323}

\bibitem[{{Fekel} {et~al.}(1986){Fekel}, {Moffett}, \&
  {Henry}}]{hd24496_activity_I}
{Fekel}, F.~C., {Moffett}, T.~J., \& {Henry}, G.~W. 1986, \apjs, 60, 551,
  \dodoi{10.1086/191097}

\bibitem[{{Feng} {et~al.}(2021){Feng}, {Butler}, {Jones}, {Phillips}, {Vogt},
  {Oppenheimer}, {Holden}, {Burt}, \& {Boss}}]{feng_190360}
{Feng}, F., {Butler}, R.~P., {Jones}, H. R.~A., {et~al.} 2021, \mnras, 507,
  2856, \dodoi{10.1093/mnras/stab2225}

\bibitem[{{Fischer} {et~al.}(1999){Fischer}, {Marcy}, {Butler}, {Vogt}, \&
  {Apps}}]{1999PASP..111...50F}
{Fischer}, D.~A., {Marcy}, G.~W., {Butler}, R.~P., {Vogt}, S.~S., \& {Apps}, K.
  1999, \pasp, 111, 50, \dodoi{10.1086/316304}

\bibitem[{{Fletcher}(1931)}]{1931MNRAS..92..121F}
{Fletcher}, A. 1931, \mnras, 92, 121, \dodoi{10.1093/mnras/92.2.121}

\bibitem[{{Ford} {et~al.}(2024){Ford}, {Bender}, {Blake}, {Gupta}, {Kanodia},
  {Lin}, {Logsdon}, {Luhn}, {Mahadevan}, {Palumbo}, {Terrien}, {Wright},
  {Zhao}, {Halverson}, {Hunting}, {Robertson}, {Roy}, \&
  {Stefansson}}]{2024arXiv240813318F}
{Ford}, E.~B., {Bender}, C.~F., {Blake}, C.~H., {et~al.} 2024, arXiv e-prints,
  arXiv:2408.13318, \dodoi{10.48550/arXiv.2408.13318}

\bibitem[{{Foreman-Mackey} {et~al.}(2013){Foreman-Mackey}, {Hogg}, {Lang}, \&
  {Goodman}}]{emcee}
{Foreman-Mackey}, D., {Hogg}, D.~W., {Lang}, D., \& {Goodman}, J. 2013, \pasp,
  125, 306, \dodoi{10.1086/670067}

\bibitem[{{Franz}(1964{\natexlab{a}})}]{USN1963}
{Franz}, O.~G. 1964{\natexlab{a}}, Publications of the U.S. Naval Observatory
  Second Series, 18, 29

\bibitem[{{Franz}(1964{\natexlab{b}})}]{USN1969}
---. 1964{\natexlab{b}}, Publications of the U.S. Naval Observatory Second
  Series, 18, 29

\bibitem[{{Gaia Collaboration} {et~al.}(2016){Gaia Collaboration}, {Prusti},
  {de Bruijne}, {Brown}, {Vallenari}, {Babusiaux}, {Bailer-Jones}, {Bastian},
  {Biermann}, {Evans}, {Eyer}, {Jansen}, {Jordi}, {Klioner}, {Lammers},
  {Lindegren}, {Luri}, {Mignard}, {Milligan}, {Panem}, {Poinsignon},
  {Pourbaix}, {Randich}, {Sarri}, {Sartoretti}, {Siddiqui}, {Soubiran},
  {Valette}, {van Leeuwen}, {Walton}, {Aerts}, {Arenou}, {Cropper}, {Drimmel},
  {H{\o}g}, {Katz}, {Lattanzi}, {O'Mullane}, {Grebel}, {Holland}, {Huc},
  {Passot}, {Bramante}, {Cacciari}, {Casta{\~n}eda}, {Chaoul}, {Cheek}, {De
  Angeli}, {Fabricius}, {Guerra}, {Hern{\'a}ndez}, {Jean-Antoine-Piccolo},
  {Masana}, {Messineo}, {Mowlavi}, {Nienartowicz}, {Ord{\'o}{\~n}ez-Blanco},
  {Panuzzo}, {Portell}, {Richards}, {Riello}, {Seabroke}, {Tanga},
  {Th{\'e}venin}, {Torra}, {Els}, {Gracia-Abril}, {Comoretto},
  {Garcia-Reinaldos}, {Lock}, {Mercier}, {Altmann}, {Andrae}, {Astraatmadja},
  {Bellas-Velidis}, {Benson}, {Berthier}, {Blomme}, {Busso}, {Carry},
  {Cellino}, {Clementini}, {Cowell}, {Creevey}, {Cuypers}, {Davidson}, {De
  Ridder}, {de Torres}, {Delchambre}, {Dell'Oro}, {Ducourant}, {Fr{\'e}mat},
  {Garc{\'\i}a-Torres}, {Gosset}, {Halbwachs}, {Hambly}, {Harrison}, {Hauser},
  {Hestroffer}, {Hodgkin}, {Huckle}, {Hutton}, {Jasniewicz}, {Jordan},
  {Kontizas}, {Korn}, {Lanzafame}, {Manteiga}, {Moitinho}, {Muinonen},
  {Osinde}, {Pancino}, {Pauwels}, {Petit}, {Recio-Blanco}, {Robin}, {Sarro},
  {Siopis}, {Smith}, {Smith}, {Sozzetti}, {Thuillot}, {van Reeven}, {Viala},
  {Abbas}, {Abreu Aramburu}, {Accart}, {Aguado}, {Allan}, {Allasia},
  {Altavilla}, {{\'A}lvarez}, {Alves}, {Anderson}, {Andrei}, {Anglada Varela},
  {Antiche}, {Antoja}, {Ant{\'o}n}, {Arcay}, {Atzei}, {Ayache}, {Bach},
  {Baker}, {Balaguer-N{\'u}{\~n}ez}, {Barache}, {Barata}, {Barbier}, {Barblan},
  {Baroni}, {Barrado y Navascu{\'e}s}, {Barros}, {Barstow}, {Becciani},
  {Bellazzini}, {Bellei}, {Bello Garc{\'\i}a}, {Belokurov}, {Bendjoya},
  {Berihuete}, {Bianchi}, {Bienaym{\'e}}, {Billebaud}, {Blagorodnova},
  {Blanco-Cuaresma}, {Boch}, {Bombrun}, {Borrachero}, {Bouquillon}, {Bourda},
  {Bouy}, {Bragaglia}, {Breddels}, {Brouillet}, {Br{\"u}semeister},
  {Bucciarelli}, {Budnik}, {Burgess}, {Burgon}, {Burlacu}, {Busonero}, {Buzzi},
  {Caffau}, {Cambras}, {Campbell}, {Cancelliere}, {Cantat-Gaudin}, {Carlucci},
  {Carrasco}, {Castellani}, {Charlot}, {Charnas}, {Charvet}, {Chassat},
  {Chiavassa}, {Clotet}, {Cocozza}, {Collins}, {Collins}, {Costigan}, {Crifo},
  {Cross}, {Crosta}, {Crowley}, {Dafonte}, {Damerdji}, {Dapergolas}, {David},
  {David}, {De Cat}, {de Felice}, {de Laverny}, {De Luise}, {De March}, {de
  Martino}, {de Souza}, {Debosscher}, {del Pozo}, {Delbo}, {Delgado},
  {Delgado}, {di Marco}, {Di Matteo}, {Diakite}, {Distefano}, {Dolding}, {Dos
  Anjos}, {Drazinos}, {Dur{\'a}n}, {Dzigan}, {Ecale}, {Edvardsson}, {Enke},
  {Erdmann}, {Escolar}, {Espina}, {Evans}, {Eynard Bontemps}, {Fabre},
  {Fabrizio}, {Faigler}, {Falc{\~a}o}, {Farr{\`a}s Casas}, {Faye}, {Federici},
  {Fedorets}, {Fern{\'a}ndez-Hern{\'a}ndez}, {Fernique}, {Fienga}, {Figueras},
  {Filippi}, {Findeisen}, {Fonti}, {Fouesneau}, {Fraile}, {Fraser}, {Fuchs},
  {Furnell}, {Gai}, {Galleti}, {Galluccio}, {Garabato}, {Garc{\'\i}a-Sedano},
  {Gar{\'e}}, {Garofalo}, {Garralda}, {Gavras}, {Gerssen}, {Geyer}, {Gilmore},
  {Girona}, {Giuffrida}, {Gomes}, {Gonz{\'a}lez-Marcos},
  {Gonz{\'a}lez-N{\'u}{\~n}ez}, {Gonz{\'a}lez-Vidal}, {Granvik}, {Guerrier},
  {Guillout}, {Guiraud}, {G{\'u}rpide}, {Guti{\'e}rrez-S{\'a}nchez}, {Guy},
  {Haigron}, {Hatzidimitriou}, {Haywood}, {Heiter}, {Helmi}, {Hobbs},
  {Hofmann}, {Holl}, {Holland}, {Hunt}, {Hypki}, {Icardi}, {Irwin}, {Jevardat
  de Fombelle}, {Jofr{\'e}}, {Jonker}, {Jorissen}, {Julbe}, {Karampelas},
  {Kochoska}, {Kohley}, {Kolenberg}, {Kontizas}, {Koposov}, {Kordopatis},
  {Koubsky}, {Kowalczyk}, {Krone-Martins}, {Kudryashova}, {Kull}, {Bachchan},
  {Lacoste-Seris}, {Lanza}, {Lavigne}, {Le Poncin-Lafitte}, {Lebreton},
  {Lebzelter}, {Leccia}, {Leclerc}, {Lecoeur-Taibi}, {Lemaitre}, {Lenhardt},
  {Leroux}, {Liao}, {Licata}, {Lindstr{\o}m}, {Lister}, {Livanou}, {Lobel},
  {L{\"o}ffler}, {L{\'o}pez}, {Lopez-Lozano}, {Lorenz}, {Loureiro},
  {MacDonald}, {Magalh{\~a}es Fernandes}, {Managau}, {Mann}, {Mantelet},
  {Marchal}, {Marchant}, {Marconi}, {Marie}, {Marinoni}, {Marrese},
  {Marschalk{\'o}}, {Marshall}, {Mart{\'\i}n-Fleitas}, {Martino}, {Mary},
  {Matijevi{\v{c}}}, {Mazeh}, {McMillan}, {Messina}, {Mestre}, {Michalik},
  {Millar}, {Miranda}, {Molina}, {Molinaro}, {Molinaro}, {Moln{\'a}r},
  {Moniez}, {Montegriffo}, {Monteiro}, {Mor}, {Mora}, {Morbidelli}, {Morel},
  {Morgenthaler}, {Morley}, {Morris}, {Mulone}, {Muraveva}, {Musella},
  {Narbonne}, {Nelemans}, {Nicastro}, {Noval}, {Ord{\'e}novic},
  {Ordieres-Mer{\'e}}, {Osborne}, {Pagani}, {Pagano}, {Pailler}, {Palacin},
  {Palaversa}, {Parsons}, {Paulsen}, {Pecoraro}, {Pedrosa}, {Pentik{\"a}inen},
  {Pereira}, {Pichon}, {Piersimoni}, {Pineau}, {Plachy}, {Plum}, {Poujoulet},
  {Pr{\v{s}}a}, {Pulone}, {Ragaini}, {Rago}, {Rambaux}, {Ramos-Lerate},
  {Ranalli}, {Rauw}, {Read}, {Regibo}, {Renk}, {Reyl{\'e}}, {Ribeiro},
  {Rimoldini}, {Ripepi}, {Riva}, {Rixon}, {Roelens}, {Romero-G{\'o}mez},
  {Rowell}, {Royer}, {Rudolph}, {Ruiz-Dern}, {Sadowski}, {Sagrist{\`a}
  Sell{\'e}s}, {Sahlmann}, {Salgado}, {Salguero}, {Sarasso}, {Savietto},
  {Schnorhk}, {Schultheis}, {Sciacca}, {Segol}, {Segovia}, {Segransan},
  {Serpell}, {Shih}, {Smareglia}, {Smart}, {Smith}, {Solano}, {Solitro},
  {Sordo}, {Soria Nieto}, {Souchay}, {Spagna}, {Spoto}, {Stampa}, {Steele},
  {Steidelm{\"u}ller}, {Stephenson}, {Stoev}, {Suess}, {S{\"u}veges}, {Surdej},
  {Szabados}, {Szegedi-Elek}, {Tapiador}, {Taris}, {Tauran}, {Taylor},
  {Teixeira}, {Terrett}, {Tingley}, {Trager}, {Turon}, {Ulla}, {Utrilla},
  {Valentini}, {van Elteren}, {Van Hemelryck}, {van Leeuwen}, {Varadi},
  {Vecchiato}, {Veljanoski}, {Via}, {Vicente}, {Vogt}, {Voss}, {Votruba},
  {Voutsinas}, {Walmsley}, {Weiler}, {Weingrill}, {Werner}, {Wevers},
  {Whitehead}, {Wyrzykowski}, {Yoldas}, {{\v{Z}}erjal}, {Zucker}, {Zurbach},
  {Zwitter}, {Alecu}, {Allen}, {Allende Prieto}, {Amorim},
  {Anglada-Escud{\'e}}, {Arsenijevic}, {Azaz}, {Balm}, {Beck}, {Bernstein},
  {Bigot}, {Bijaoui}, {Blasco}, {Bonfigli}, {Bono}, {Boudreault}, {Bressan},
  {Brown}, {Brunet}, {Bunclark}, {Buonanno}, {Butkevich}, {Carret}, {Carrion},
  {Chemin}, {Ch{\'e}reau}, {Corcione}, {Darmigny}, {de Boer}, {de Teodoro}, {de
  Zeeuw}, {Delle Luche}, {Domingues}, {Dubath}, {Fodor}, {Fr{\'e}zouls},
  {Fries}, {Fustes}, {Fyfe}, {Gallardo}, {Gallegos}, {Gardiol}, {Gebran},
  {Gomboc}, {G{\'o}mez}, {Grux}, {Gueguen}, {Heyrovsky}, {Hoar}, {Iannicola},
  {Isasi Parache}, {Janotto}, {Joliet}, {Jonckheere}, {Keil}, {Kim},
  {Klagyivik}, {Klar}, {Knude}, {Kochukhov}, {Kolka}, {Kos}, {Kutka}, {Lainey},
  {LeBouquin}, {Liu}, {Loreggia}, {Makarov}, {Marseille}, {Martayan},
  {Martinez-Rubi}, {Massart}, {Meynadier}, {Mignot}, {Munari}, {Nguyen},
  {Nordlander}, {Ocvirk}, {O'Flaherty}, {Olias Sanz}, {Ortiz}, {Osorio},
  {Oszkiewicz}, {Ouzounis}, {Palmer}, {Park}, {Pasquato}, {Peltzer}, {Peralta},
  {P{\'e}turaud}, {Pieniluoma}, {Pigozzi}, {Poels}, {Prat}, {Prod'homme},
  {Raison}, {Rebordao}, {Risquez}, {Rocca-Volmerange}, {Rosen}, {Ruiz-Fuertes},
  {Russo}, {Sembay}, {Serraller Vizcaino}, {Short}, {Siebert}, {Silva},
  {Sinachopoulos}, {Slezak}, {Soffel}, {Sosnowska}, {Strai{\v{z}}ys}, {ter
  Linden}, {Terrell}, {Theil}, {Tiede}, {Troisi}, {Tsalmantza}, {Tur},
  {Vaccari}, {Vachier}, {Valles}, {Van Hamme}, {Veltz}, {Virtanen}, {Wallut},
  {Wichmann}, {Wilkinson}, {Ziaeepour}, \& {Zschocke}}]{Gaia_mission}
{Gaia Collaboration}, {Prusti}, T., {de Bruijne}, J.~H.~J., {et~al.} 2016,
  \aap, 595, A1, \dodoi{10.1051/0004-6361/201629272}

\bibitem[{{Gaia Collaboration} {et~al.}(2018){Gaia Collaboration}, {Brown},
  {Vallenari}, {Prusti}, {de Bruijne}, {Babusiaux}, {Bailer-Jones}, {Biermann},
  {Evans}, {Eyer}, {Jansen}, {Jordi}, {Klioner}, {Lammers}, {Lindegren},
  {Luri}, {Mignard}, {Panem}, {Pourbaix}, {Randich}, {Sartoretti}, {Siddiqui},
  {Soubiran}, {van Leeuwen}, {Walton}, {Arenou}, {Bastian}, {Cropper},
  {Drimmel}, {Katz}, {Lattanzi}, {Bakker}, {Cacciari}, {Casta{\~n}eda},
  {Chaoul}, {Cheek}, {De Angeli}, {Fabricius}, {Guerra}, {Holl}, {Masana},
  {Messineo}, {Mowlavi}, {Nienartowicz}, {Panuzzo}, {Portell}, {Riello},
  {Seabroke}, {Tanga}, {Th{\'e}venin}, {Gracia-Abril}, {Comoretto},
  {Garcia-Reinaldos}, {Teyssier}, {Altmann}, {Andrae}, {Audard},
  {Bellas-Velidis}, {Benson}, {Berthier}, {Blomme}, {Burgess}, {Busso},
  {Carry}, {Cellino}, {Clementini}, {Clotet}, {Creevey}, {Davidson}, {De
  Ridder}, {Delchambre}, {Dell'Oro}, {Ducourant},
  {Fern{\'a}ndez-Hern{\'a}ndez}, {Fouesneau}, {Fr{\'e}mat}, {Galluccio},
  {Garc{\'\i}a-Torres}, {Gonz{\'a}lez-N{\'u}{\~n}ez}, {Gonz{\'a}lez-Vidal},
  {Gosset}, {Guy}, {Halbwachs}, {Hambly}, {Harrison}, {Hern{\'a}ndez},
  {Hestroffer}, {Hodgkin}, {Hutton}, {Jasniewicz}, {Jean-Antoine-Piccolo},
  {Jordan}, {Korn}, {Krone-Martins}, {Lanzafame}, {Lebzelter}, {L{\"o}ffler},
  {Manteiga}, {Marrese}, {Mart{\'\i}n-Fleitas}, {Moitinho}, {Mora}, {Muinonen},
  {Osinde}, {Pancino}, {Pauwels}, {Petit}, {Recio-Blanco}, {Richards},
  {Rimoldini}, {Robin}, {Sarro}, {Siopis}, {Smith}, {Sozzetti}, {S{\"u}veges},
  {Torra}, {van Reeven}, {Abbas}, {Abreu Aramburu}, {Accart}, {Aerts},
  {Altavilla}, {{\'A}lvarez}, {Alvarez}, {Alves}, {Anderson}, {Andrei},
  {Anglada Varela}, {Antiche}, {Antoja}, {Arcay}, {Astraatmadja}, {Bach},
  {Baker}, {Balaguer-N{\'u}{\~n}ez}, {Balm}, {Barache}, {Barata}, {Barbato},
  {Barblan}, {Barklem}, {Barrado}, {Barros}, {Barstow}, {Bartholom{\'e}
  Mu{\~n}oz}, {Bassilana}, {Becciani}, {Bellazzini}, {Berihuete}, {Bertone},
  {Bianchi}, {Bienaym{\'e}}, {Blanco-Cuaresma}, {Boch}, {Boeche}, {Bombrun},
  {Borrachero}, {Bossini}, {Bouquillon}, {Bourda}, {Bragaglia}, {Bramante},
  {Breddels}, {Bressan}, {Brouillet}, {Br{\"u}semeister}, {Brugaletta},
  {Bucciarelli}, {Burlacu}, {Busonero}, {Butkevich}, {Buzzi}, {Caffau},
  {Cancelliere}, {Cannizzaro}, {Cantat-Gaudin}, {Carballo}, {Carlucci},
  {Carrasco}, {Casamiquela}, {Castellani}, {Castro-Ginard}, {Charlot},
  {Chemin}, {Chiavassa}, {Cocozza}, {Costigan}, {Cowell}, {Crifo}, {Crosta},
  {Crowley}, {Cuypers}, {Dafonte}, {Damerdji}, {Dapergolas}, {David}, {David},
  {de Laverny}, {De Luise}, {De March}, {de Martino}, {de Souza}, {de Torres},
  {Debosscher}, {del Pozo}, {Delbo}, {Delgado}, {Delgado}, {Di Matteo},
  {Diakite}, {Diener}, {Distefano}, {Dolding}, {Drazinos}, {Dur{\'a}n},
  {Edvardsson}, {Enke}, {Eriksson}, {Esquej}, {Eynard Bontemps}, {Fabre},
  {Fabrizio}, {Faigler}, {Falc{\~a}o}, {Farr{\`a}s Casas}, {Federici},
  {Fedorets}, {Fernique}, {Figueras}, {Filippi}, {Findeisen}, {Fonti},
  {Fraile}, {Fraser}, {Fr{\'e}zouls}, {Gai}, {Galleti}, {Garabato},
  {Garc{\'\i}a-Sedano}, {Garofalo}, {Garralda}, {Gavel}, {Gavras}, {Gerssen},
  {Geyer}, {Giacobbe}, {Gilmore}, {Girona}, {Giuffrida}, {Glass}, {Gomes},
  {Granvik}, {Gueguen}, {Guerrier}, {Guiraud}, {Guti{\'e}rrez-S{\'a}nchez},
  {Haigron}, {Hatzidimitriou}, {Hauser}, {Haywood}, {Heiter}, {Helmi}, {Heu},
  {Hilger}, {Hobbs}, {Hofmann}, {Holland}, {Huckle}, {Hypki}, {Icardi},
  {Jan{\ss}en}, {Jevardat de Fombelle}, {Jonker}, {Juh{\'a}sz}, {Julbe},
  {Karampelas}, {Kewley}, {Klar}, {Kochoska}, {Kohley}, {Kolenberg},
  {Kontizas}, {Kontizas}, {Koposov}, {Kordopatis}, {Kostrzewa-Rutkowska},
  {Koubsky}, {Lambert}, {Lanza}, {Lasne}, {Lavigne}, {Le Fustec}, {Le
  Poncin-Lafitte}, {Lebreton}, {Leccia}, {Leclerc}, {Lecoeur-Taibi},
  {Lenhardt}, {Leroux}, {Liao}, {Licata}, {Lindstr{\o}m}, {Lister}, {Livanou},
  {Lobel}, {L{\'o}pez}, {Managau}, {Mann}, {Mantelet}, {Marchal}, {Marchant},
  {Marconi}, {Marinoni}, {Marschalk{\'o}}, {Marshall}, {Martino}, {Marton},
  {Mary}, {Massari}, {Matijevi{\v{c}}}, {Mazeh}, {McMillan}, {Messina},
  {Michalik}, {Millar}, {Molina}, {Molinaro}, {Moln{\'a}r}, {Montegriffo},
  {Mor}, {Morbidelli}, {Morel}, {Morris}, {Mulone}, {Muraveva}, {Musella},
  {Nelemans}, {Nicastro}, {Noval}, {O'Mullane}, {Ord{\'e}novic},
  {Ord{\'o}{\~n}ez-Blanco}, {Osborne}, {Pagani}, {Pagano}, {Pailler},
  {Palacin}, {Palaversa}, {Panahi}, {Pawlak}, {Piersimoni}, {Pineau}, {Plachy},
  {Plum}, {Poggio}, {Poujoulet}, {Pr{\v{s}}a}, {Pulone}, {Racero}, {Ragaini},
  {Rambaux}, {Ramos-Lerate}, {Regibo}, {Reyl{\'e}}, {Riclet}, {Ripepi}, {Riva},
  {Rivard}, {Rixon}, {Roegiers}, {Roelens}, {Romero-G{\'o}mez}, {Rowell},
  {Royer}, {Ruiz-Dern}, {Sadowski}, {Sagrist{\`a} Sell{\'e}s}, {Sahlmann},
  {Salgado}, {Salguero}, {Sanna}, {Santana-Ros}, {Sarasso}, {Savietto},
  {Schultheis}, {Sciacca}, {Segol}, {Segovia}, {S{\'e}gransan}, {Shih},
  {Siltala}, {Silva}, {Smart}, {Smith}, {Solano}, {Solitro}, {Sordo}, {Soria
  Nieto}, {Souchay}, {Spagna}, {Spoto}, {Stampa}, {Steele},
  {Steidelm{\"u}ller}, {Stephenson}, {Stoev}, {Suess}, {Surdej}, {Szabados},
  {Szegedi-Elek}, {Tapiador}, {Taris}, {Tauran}, {Taylor}, {Teixeira},
  {Terrett}, {Teyssandier}, {Thuillot}, {Titarenko}, {Torra Clotet}, {Turon},
  {Ulla}, {Utrilla}, {Uzzi}, {Vaillant}, {Valentini}, {Valette}, {van Elteren},
  {Van Hemelryck}, {van Leeuwen}, {Vaschetto}, {Vecchiato}, {Veljanoski},
  {Viala}, {Vicente}, {Vogt}, {von Essen}, {Voss}, {Votruba}, {Voutsinas},
  {Walmsley}, {Weiler}, {Wertz}, {Wevers}, {Wyrzykowski}, {Yoldas},
  {{\v{Z}}erjal}, {Ziaeepour}, {Zorec}, {Zschocke}, {Zucker}, {Zurbach}, \&
  {Zwitter}}]{gaia_dr2}
{Gaia Collaboration}, {Brown}, A.~G.~A., {Vallenari}, A., {et~al.} 2018, \aap,
  616, A1, \dodoi{10.1051/0004-6361/201833051}

\bibitem[{{Gaia Collaboration} {et~al.}(2022){Gaia Collaboration}, {Vallenari},
  {Brown}, {Prusti}, {de Bruijne}, {Arenou}, {Babusiaux}, {Biermann},
  {Creevey}, {Ducourant}, {Evans}, {Eyer}, {Guerra}, {Hutton}, {Jordi},
  {Klioner}, {Lammers}, {Lindegren}, {Luri}, {Mignard}, {Panem}, {Pourbaix},
  {Randich}, {Sartoretti}, {Soubiran}, {Tanga}, {Walton}, {Bailer-Jones},
  {Bastian}, {Drimmel}, {Jansen}, {Katz}, {Lattanzi}, {van Leeuwen}, {Bakker},
  {Cacciari}, {Casta{\~n}eda}, {De Angeli}, {Fabricius}, {Fouesneau},
  {Fr{\'e}mat}, {Galluccio}, {Guerrier}, {Heiter}, {Masana}, {Messineo},
  {Mowlavi}, {Nicolas}, {Nienartowicz}, {Pailler}, {Panuzzo}, {Riclet}, {Roux},
  {Seabroke}, {Sordo{\o}rcit}, {Th{\'e}venin}, {Gracia-Abril}, {Portell},
  {Teyssier}, {Altmann}, {Andrae}, {Audard}, {Bellas-Velidis}, {Benson},
  {Berthier}, {Blomme}, {Burgess}, {Busonero}, {Busso}, {C{\'a}novas}, {Carry},
  {Cellino}, {Cheek}, {Clementini}, {Damerdji}, {Davidson}, {de Teodoro},
  {Nu{\~n}ez Campos}, {Delchambre}, {Dell'Oro}, {Esquej},
  {Fern{\'a}ndez-Hern{\'a}ndez}, {Fraile}, {Garabato}, {Garc{\'\i}a-Lario},
  {Gosset}, {Haigron}, {Halbwachs}, {Hambly}, {Harrison}, {Hern{\'a}ndez},
  {Hestroffer}, {Hodgkin}, {Holl}, {Jan{\ss}en}, {Jevardat de Fombelle},
  {Jordan}, {Krone-Martins}, {Lanzafame}, {L{\"o}ffler}, {Marchal}, {Marrese},
  {Moitinho}, {Muinonen}, {Osborne}, {Pancino}, {Pauwels}, {Recio-Blanco},
  {Reyl{\'e}}, {Riello}, {Rimoldini}, {Roegiers}, {Rybizki}, {Sarro}, {Siopis},
  {Smith}, {Sozzetti}, {Utrilla}, {van Leeuwen}, {Abbas}, {{\'A}brah{\'a}m},
  {Abreu Aramburu}, {Aerts}, {Aguado}, {Ajaj}, {Aldea-Montero}, {Altavilla},
  {{\'A}lvarez}, {Alves}, {Anders}, {Anderson}, {Anglada Varela}, {Antoja},
  {Baines}, {Baker}, {Balaguer-N{\'u}{\~n}ez}, {Balbinot}, {Balog}, {Barache},
  {Barbato}, {Barros}, {Barstow}, {Bartolom{\'e}}, {Bassilana}, {Bauchet},
  {Becciani}, {Bellazzini}, {Berihuete}, {Bernet}, {Bertone}, {Bianchi},
  {Binnenfeld}, {Blanco-Cuaresma}, {Blazere}, {Boch}, {Bombrun}, {Bossini},
  {Bouquillon}, {Bragaglia}, {Bramante}, {Breedt}, {Bressan}, {Brouillet},
  {Brugaletta}, {Bucciarelli}, {Burlacu}, {Butkevich}, {Buzzi}, {Caffau},
  {Cancelliere}, {Cantat-Gaudin}, {Carballo}, {Carlucci}, {Carnerero},
  {Carrasco}, {Casamiquela}, {Castellani}, {Castro-Ginard}, {Chaoul},
  {Charlot}, {Chemin}, {Chiaramida}, {Chiavassa}, {Chornay}, {Comoretto},
  {Contursi}, {Cooper}, {Cornez}, {Cowell}, {Crifo}, {Cropper}, {Crosta},
  {Crowley}, {Dafonte}, {Dapergolas}, {David}, {David}, {de Laverny}, {De
  Luise}, {De March}, {De Ridder}, {de Souza}, {de Torres}, {del Peloso}, {del
  Pozo}, {Delbo}, {Delgado}, {Delisle}, {Demouchy}, {Dharmawardena}, {Di
  Matteo}, {Diakite}, {Diener}, {Distefano}, {Dolding}, {Edvardsson}, {Enke},
  {Fabre}, {Fabrizio}, {Faigler}, {Fedorets}, {Fernique}, {Fienga}, {Figueras},
  {Fournier}, {Fouron}, {Fragkoudi}, {Gai}, {Garcia-Gutierrez},
  {Garcia-Reinaldos}, {Garc{\'\i}a-Torres}, {Garofalo}, {Gavel}, {Gavras},
  {Gerlach}, {Geyer}, {Giacobbe}, {Gilmore}, {Girona}, {Giuffrida}, {Gomel},
  {Gomez}, {Gonz{\'a}lez-N{\'u}{\~n}ez}, {Gonz{\'a}lez-Santamar{\'\i}a},
  {Gonz{\'a}lez-Vidal}, {Granvik}, {Guillout}, {Guiraud},
  {Guti{\'e}rrez-S{\'a}nchez}, {Guy}, {Hatzidimitriou}, {Hauser}, {Haywood},
  {Helmer}, {Helmi}, {Sarmiento}, {Hidalgo}, {Hilger}, {H{\l}adczuk}, {Hobbs},
  {Holland}, {Huckle}, {Jardine}, {Jasniewicz}, {Jean-Antoine Piccolo},
  {Jim{\'e}nez-Arranz}, {Jorissen}, {Juaristi Campillo}, {Julbe}, {Karbevska},
  {Kervella}, {Khanna}, {Kontizas}, {Kordopatis}, {Korn}, {K{\'o}sp{\'a}l},
  {Kostrzewa-Rutkowska}, {Kruszy{\'n}ska}, {Kun}, {Laizeau}, {Lambert},
  {Lanza}, {Lasne}, {Le Campion}, {Lebreton}, {Lebzelter}, {Leccia}, {Leclerc},
  {Lecoeur-Taibi}, {Liao}, {Licata}, {Lindstr{\o}m}, {Lister}, {Livanou},
  {Lobel}, {Lorca}, {Loup}, {Madrero Pardo}, {Magdaleno Romeo}, {Managau},
  {Mann}, {Manteiga}, {Marchant}, {Marconi}, {Marcos}, {Marcos Santos},
  {Mar{\'\i}n Pina}, {Marinoni}, {Marocco}, {Marshall}, {Polo},
  {Mart{\'\i}n-Fleitas}, {Marton}, {Mary}, {Masip}, {Massari},
  {Mastrobuono-Battisti}, {Mazeh}, {McMillan}, {Messina}, {Michalik}, {Millar},
  {Mints}, {Molina}, {Molinaro}, {Moln{\'a}r}, {Monari}, {Mongui{\'o}},
  {Montegriffo}, {Montero}, {Mor}, {Mora}, {Morbidelli}, {Morel}, {Morris},
  {Muraveva}, {Murphy}, {Musella}, {Nagy}, {Noval}, {Oca{\~n}a}, {Ogden},
  {Ordenovic}, {Osinde}, {Pagani}, {Pagano}, {Palaversa}, {Palicio},
  {Pallas-Quintela}, {Panahi}, {Payne-Wardenaar}, {Pe{\~n}alosa Esteller},
  {Penttil{\"a}}, {Pichon}, {Piersimoni}, {Pineau}, {Plachy}, {Plum}, {Poggio},
  {Pr{\v{s}}a}, {Pulone}, {Racero}, {Ragaini}, {Rainer}, {Raiteri}, {Rambaux},
  {Ramos}, {Ramos-Lerate}, {Re Fiorentin}, {Regibo}, {Richards}, {Rios Diaz},
  {Ripepi}, {Riva}, {Rix}, {Rixon}, {Robichon}, {Robin}, {Robin}, {Roelens},
  {Rogues}, {Rohrbasser}, {Romero-G{\'o}mez}, {Rowell}, {Royer}, {Ruz Mieres},
  {Rybicki}, {Sadowski}, {S{\'a}ez N{\'u}{\~n}ez}, {Sagrist{\`a} Sell{\'e}s},
  {Sahlmann}, {Salguero}, {Samaras}, {Sanchez Gimenez}, {Sanna},
  {Santove{\~n}a}, {Sarasso}, {Schultheis}, {Sciacca}, {Segol}, {Segovia},
  {S{\'e}gransan}, {Semeux}, {Shahaf}, {Siddiqui}, {Siebert}, {Siltala},
  {Silvelo}, {Slezak}, {Slezak}, {Smart}, {Snaith}, {Solano}, {Solitro},
  {Souami}, {Souchay}, {Spagna}, {Spina}, {Spoto}, {Steele},
  {Steidelm{\"u}ller}, {Stephenson}, {S{\"u}veges}, {Surdej}, {Szabados},
  {Szegedi-Elek}, {Taris}, {Taylo}, {Teixeira}, {Tolomei}, {Tonello}, {Torra},
  {Torra}, {Torralba Elipe}, {Trabucchi}, {Tsounis}, {Turon}, {Ulla}, {Unger},
  {Vaillant}, {van Dillen}, {van Reeven}, {Vanel}, {Vecchiato}, {Viala},
  {Vicente}, {Voutsinas}, {Weiler}, {Wevers}, {Wyrzykowski}, {Yoldas}, {Yvard},
  {Zhao}, {Zorec}, {Zucker}, \& {Zwitter}}]{2022arXiv220800211G}
{Gaia Collaboration}, {Vallenari}, A., {Brown}, A.~G.~A., {et~al.} 2022, arXiv
  e-prints, arXiv:2208.00211, \dodoi{10.48550/arXiv.2208.00211}

\bibitem[{{Gaia Collaboration} {et~al.}(2023){Gaia Collaboration}, {Vallenari},
  {Brown}, {Prusti}, {de Bruijne}, {Arenou}, {Babusiaux}, {Biermann},
  {Creevey}, {Ducourant}, {Evans}, {Eyer}, {Guerra}, {Hutton}, {Jordi},
  {Klioner}, {Lammers}, {Lindegren}, {Luri}, {Mignard}, {Panem}, {Pourbaix},
  {Randich}, {Sartoretti}, {Soubiran}, {Tanga}, {Walton}, {Bailer-Jones},
  {Bastian}, {Drimmel}, {Jansen}, {Katz}, {Lattanzi}, {van Leeuwen}, {Bakker},
  {Cacciari}, {Casta{\~n}eda}, {De Angeli}, {Fabricius}, {Fouesneau},
  {Fr{\'e}mat}, {Galluccio}, {Guerrier}, {Heiter}, {Masana}, {Messineo},
  {Mowlavi}, {Nicolas}, {Nienartowicz}, {Pailler}, {Panuzzo}, {Riclet}, {Roux},
  {Seabroke}, {Sordo}, {Th{\'e}venin}, {Gracia-Abril}, {Portell}, {Teyssier},
  {Altmann}, {Andrae}, {Audard}, {Bellas-Velidis}, {Benson}, {Berthier},
  {Blomme}, {Burgess}, {Busonero}, {Busso}, {C{\'a}novas}, {Carry}, {Cellino},
  {Cheek}, {Clementini}, {Damerdji}, {Davidson}, {de Teodoro}, {Nu{\~n}ez
  Campos}, {Delchambre}, {Dell'Oro}, {Esquej}, {Fern{\'a}ndez-Hern{\'a}ndez},
  {Fraile}, {Garabato}, {Garc{\'\i}a-Lario}, {Gosset}, {Haigron}, {Halbwachs},
  {Hambly}, {Harrison}, {Hern{\'a}ndez}, {Hestroffer}, {Hodgkin}, {Holl},
  {Jan{\ss}en}, {Jevardat de Fombelle}, {Jordan}, {Krone-Martins}, {Lanzafame},
  {L{\"o}ffler}, {Marchal}, {Marrese}, {Moitinho}, {Muinonen}, {Osborne},
  {Pancino}, {Pauwels}, {Recio-Blanco}, {Reyl{\'e}}, {Riello}, {Rimoldini},
  {Roegiers}, {Rybizki}, {Sarro}, {Siopis}, {Smith}, {Sozzetti}, {Utrilla},
  {van Leeuwen}, {Abbas}, {{\'A}brah{\'a}m}, {Abreu Aramburu}, {Aerts},
  {Aguado}, {Ajaj}, {Aldea-Montero}, {Altavilla}, {{\'A}lvarez}, {Alves},
  {Anders}, {Anderson}, {Anglada Varela}, {Antoja}, {Baines}, {Baker},
  {Balaguer-N{\'u}{\~n}ez}, {Balbinot}, {Balog}, {Barache}, {Barbato},
  {Barros}, {Barstow}, {Bartolom{\'e}}, {Bassilana}, {Bauchet}, {Becciani},
  {Bellazzini}, {Berihuete}, {Bernet}, {Bertone}, {Bianchi}, {Binnenfeld},
  {Blanco-Cuaresma}, {Blazere}, {Boch}, {Bombrun}, {Bossini}, {Bouquillon},
  {Bragaglia}, {Bramante}, {Breedt}, {Bressan}, {Brouillet}, {Brugaletta},
  {Bucciarelli}, {Burlacu}, {Butkevich}, {Buzzi}, {Caffau}, {Cancelliere},
  {Cantat-Gaudin}, {Carballo}, {Carlucci}, {Carnerero}, {Carrasco},
  {Casamiquela}, {Castellani}, {Castro-Ginard}, {Chaoul}, {Charlot}, {Chemin},
  {Chiaramida}, {Chiavassa}, {Chornay}, {Comoretto}, {Contursi}, {Cooper},
  {Cornez}, {Cowell}, {Crifo}, {Cropper}, {Crosta}, {Crowley}, {Dafonte},
  {Dapergolas}, {David}, {David}, {de Laverny}, {De Luise}, {De March}, {De
  Ridder}, {de Souza}, {de Torres}, {del Peloso}, {del Pozo}, {Delbo},
  {Delgado}, {Delisle}, {Demouchy}, {Dharmawardena}, {Di Matteo}, {Diakite},
  {Diener}, {Distefano}, {Dolding}, {Edvardsson}, {Enke}, {Fabre}, {Fabrizio},
  {Faigler}, {Fedorets}, {Fernique}, {Fienga}, {Figueras}, {Fournier},
  {Fouron}, {Fragkoudi}, {Gai}, {Garcia-Gutierrez}, {Garcia-Reinaldos},
  {Garc{\'\i}a-Torres}, {Garofalo}, {Gavel}, {Gavras}, {Gerlach}, {Geyer},
  {Giacobbe}, {Gilmore}, {Girona}, {Giuffrida}, {Gomel}, {Gomez},
  {Gonz{\'a}lez-N{\'u}{\~n}ez}, {Gonz{\'a}lez-Santamar{\'\i}a},
  {Gonz{\'a}lez-Vidal}, {Granvik}, {Guillout}, {Guiraud},
  {Guti{\'e}rrez-S{\'a}nchez}, {Guy}, {Hatzidimitriou}, {Hauser}, {Haywood},
  {Helmer}, {Helmi}, {Sarmiento}, {Hidalgo}, {Hilger}, {H{\l}adczuk}, {Hobbs},
  {Holland}, {Huckle}, {Jardine}, {Jasniewicz}, {Jean-Antoine Piccolo},
  {Jim{\'e}nez-Arranz}, {Jorissen}, {Juaristi Campillo}, {Julbe}, {Karbevska},
  {Kervella}, {Khanna}, {Kontizas}, {Kordopatis}, {Korn}, {K{\'o}sp{\'a}l},
  {Kostrzewa-Rutkowska}, {Kruszy{\'n}ska}, {Kun}, {Laizeau}, {Lambert},
  {Lanza}, {Lasne}, {Le Campion}, {Lebreton}, {Lebzelter}, {Leccia}, {Leclerc},
  {Lecoeur-Taibi}, {Liao}, {Licata}, {Lindstr{\o}m}, {Lister}, {Livanou},
  {Lobel}, {Lorca}, {Loup}, {Madrero Pardo}, {Magdaleno Romeo}, {Managau},
  {Mann}, {Manteiga}, {Marchant}, {Marconi}, {Marcos}, {Marcos Santos},
  {Mar{\'\i}n Pina}, {Marinoni}, {Marocco}, {Marshall}, {Martin Polo},
  {Mart{\'\i}n-Fleitas}, {Marton}, {Mary}, {Masip}, {Massari},
  {Mastrobuono-Battisti}, {Mazeh}, {McMillan}, {Messina}, {Michalik}, {Millar},
  {Mints}, {Molina}, {Molinaro}, {Moln{\'a}r}, {Monari}, {Mongui{\'o}},
  {Montegriffo}, {Montero}, {Mor}, {Mora}, {Morbidelli}, {Morel}, {Morris},
  {Muraveva}, {Murphy}, {Musella}, {Nagy}, {Noval}, {Oca{\~n}a}, {Ogden},
  {Ordenovic}, {Osinde}, {Pagani}, {Pagano}, {Palaversa}, {Palicio},
  {Pallas-Quintela}, {Panahi}, {Payne-Wardenaar}, {Pe{\~n}alosa Esteller},
  {Penttil{\"a}}, {Pichon}, {Piersimoni}, {Pineau}, {Plachy}, {Plum}, {Poggio},
  {Pr{\v{s}}a}, {Pulone}, {Racero}, {Ragaini}, {Rainer}, {Raiteri}, {Rambaux},
  {Ramos}, {Ramos-Lerate}, {Re Fiorentin}, {Regibo}, {Richards}, {Rios Diaz},
  {Ripepi}, {Riva}, {Rix}, {Rixon}, {Robichon}, {Robin}, {Robin}, {Roelens},
  {Rogues}, {Rohrbasser}, {Romero-G{\'o}mez}, {Rowell}, {Royer}, {Ruz Mieres},
  {Rybicki}, {Sadowski}, {S{\'a}ez N{\'u}{\~n}ez}, {Sagrist{\`a} Sell{\'e}s},
  {Sahlmann}, {Salguero}, {Samaras}, {Sanchez Gimenez}, {Sanna},
  {Santove{\~n}a}, {Sarasso}, {Schultheis}, {Sciacca}, {Segol}, {Segovia},
  {S{\'e}gransan}, {Semeux}, {Shahaf}, {Siddiqui}, {Siebert}, {Siltala},
  {Silvelo}, {Slezak}, {Slezak}, {Smart}, {Snaith}, {Solano}, {Solitro},
  {Souami}, {Souchay}, {Spagna}, {Spina}, {Spoto}, {Steele},
  {Steidelm{\"u}ller}, {Stephenson}, {S{\"u}veges}, {Surdej}, {Szabados},
  {Szegedi-Elek}, {Taris}, {Taylor}, {Teixeira}, {Tolomei}, {Tonello}, {Torra},
  {Torra}, {Torralba Elipe}, {Trabucchi}, {Tsounis}, {Turon}, {Ulla}, {Unger},
  {Vaillant}, {van Dillen}, {van Reeven}, {Vanel}, {Vecchiato}, {Viala},
  {Vicente}, {Voutsinas}, {Weiler}, {Wevers}, {Wyrzykowski}, {Yoldas}, {Yvard},
  {Zhao}, {Zorec}, {Zucker}, \& {Zwitter}}]{gaia_dr3}
---. 2023, \aap, 674, A1, \dodoi{10.1051/0004-6361/202243940}

\bibitem[{{Gardner} {et~al.}(2006){Gardner}, {Mather}, {Clampin}, {Doyon},
  {Greenhouse}, {Hammel}, {Hutchings}, {Jakobsen}, {Lilly}, {Long}, {Lunine},
  {McCaughrean}, {Mountain}, {Nella}, {Rieke}, {Rieke}, {Rix}, {Smith},
  {Sonneborn}, {Stiavelli}, {Stockman}, {Windhorst}, \& {Wright}}]{jwst}
{Gardner}, J.~P., {Mather}, J.~C., {Clampin}, M., {et~al.} 2006, \ssr, 123,
  485, \dodoi{10.1007/s11214-006-8315-7}

\bibitem[{Giovinazzi(2025)}]{zenodo15352084}
Giovinazzi, M. 2025, markgiovinazzi/binary\_mc: binary\_mcmc v1.0.0, 1.0.0,
  Zenodo, \dodoi{10.5281/zenodo.15352084}

\bibitem[{{Giovinazzi} \& {Blake}(2022)}]{2022AJ....164..164G}
{Giovinazzi}, M.~R., \& {Blake}, C.~H. 2022, \aj, 164, 164,
  \dodoi{10.3847/1538-3881/ac8cf710.48550/arXiv.2208.12112}

\bibitem[{{Giovinazzi} {et~al.}(2020){Giovinazzi}, {Blake}, {Eastman},
  {Wright}, {McCrady}, {Wittenmyer}, {Johnson}, {Plavchan}, {Sliski}, {Wilson},
  {Johnson}, {Horner}, {Kane}, {Houghton}, {Garc{\'\i}a-Mej{\'\i}a}, \&
  {Glaser}}]{2020AN....341..870G}
{Giovinazzi}, M.~R., {Blake}, C.~H., {Eastman}, J.~D., {et~al.} 2020,
  Astronomische Nachrichten, 341, 870, \dodoi{10.1002/asna.202013830}

\bibitem[{{Giovinazzi} {et~al.}(2024){Giovinazzi}, {Cale}, {Eastman},
  {Rodriguez}, {Blake}, {Stassun}, {Vanderburg}, {Kunimoto}, {Kraus},
  {Twicken}, {Beatty}, {Dedrick}, {Horner}, {Johnson}, {Johnson}, {McCrady},
  {Plavchan}, {Sliski}, {Wilson}, {Wittenmyer}, {Wright}, {Johnson}, {Rose}, \&
  {Cornachione}}]{KELT24_stellar_params}
{Giovinazzi}, M.~R., {Cale}, B., {Eastman}, J.~D., {et~al.} 2024, \aj, 168,
  118, \dodoi{10.3847/1538-3881/ad55ec}

\bibitem[{{Gledhill}(1901)}]{Gld1901}
{Gledhill}, J. 1901, \mnras, 61, 556, \dodoi{10.1093/mnras/61.8.556}

\bibitem[{{Gorshanov} {et~al.}(2006){Gorshanov}, {Shakht}, \&
  {Kisselev}}]{61cyg_residual_search5}
{Gorshanov}, D.~L., {Shakht}, N.~A., \& {Kisselev}, A.~A. 2006, Astrophysics,
  49, 386, \dodoi{10.1007/s10511-006-0038-7}

\bibitem[{{Gupta} {et~al.}(2024{\natexlab{a}}){Gupta}, {Millholland}, {Im},
  {Dong}, {Jackson}, {Libby-Roberts}, {Delamer}, {the NEID Team}, {the HPF
  Team}, \& {TESS Single-Tranist Planet Candidate Working Group}}]{Gupta_TIC}
{Gupta}, A., {Millholland}, S., {Im}, H., {et~al.} 2024{\natexlab{a}}, in
  AAS/Division for Extreme Solar Systems Abstracts, Vol.~56, AAS/Division for
  Extreme Solar Systems Abstracts, 615.05

\bibitem[{{Gupta} {et~al.}(2021){Gupta}, {Wright}, {Robertson}, {Halverson},
  {Luhn}, {Roy}, {Mahadevan}, {Ford}, {Bender}, {Blake}, {Hearty}, {Kanodia},
  {Logsdon}, {McElwain}, {Monson}, {Ninan}, {Schwab}, {Stef{\'a}nsson}, \&
  {Terrien}}]{NETS}
{Gupta}, A.~F., {Wright}, J.~T., {Robertson}, P., {et~al.} 2021, \aj, 161, 130,
  \dodoi{10.3847/1538-3881/abd79e}

\bibitem[{{Gupta} {et~al.}(2023){Gupta}, {Jackson}, {H{\'e}brard}, {Lin},
  {Stassun}, {Dong}, {Villanueva}, {Dragomir}, {Mahadevan}, {Wright},
  {Almenara}, {Blake}, {Boisse}, {Cort{\'e}s-Zuleta}, {Dalba}, {D{\'\i}az},
  {Ford}, {Forveille}, {Gagliano}, {Halverson}, {Heidari}, {Kanodia}, {Kiefer},
  {Latham}, {McElwain}, {Mireles}, {Moutou}, {Pepper}, {Ricker}, {Robertson},
  {Roy}, {Schlecker}, {Schwab}, {Seager}, {Shporer}, {Stef{\'a}nsson},
  {Terrien}, {Ting}, {Winn}, \& {Youngblood}}]{TOI4127}
{Gupta}, A.~F., {Jackson}, J.~M., {H{\'e}brard}, G., {et~al.} 2023, \aj, 165,
  234, \dodoi{10.3847/1538-3881/accb9b}

\bibitem[{{Gupta} {et~al.}(2024{\natexlab{b}}){Gupta}, {Luhn}, {Wright},
  {Mahadevan}, {Robertson}, {Krolikowski}, {Ford}, {Ca{\~n}as}, {Halverson},
  {Lin}, {Kanodia}, {Fitzmaurice}, {Gilbertson}, {Bender}, {Blake}, {Dong},
  {Giovinazzi}, {Logsdon}, {Monson}, {Ninan}, {Rajagopal}, {Roy}, {Schwab}, \&
  {Stef{\'a}nsson}}]{HD86728}
{Gupta}, A.~F., {Luhn}, J.~K., {Wright}, J.~T., {et~al.} 2024{\natexlab{b}},
  arXiv e-prints, arXiv:2409.12315, \dodoi{10.48550/arXiv.2409.12315}

\bibitem[{{Halbwachs} {et~al.}(2023){Halbwachs}, {Pourbaix}, {Arenou},
  {Galluccio}, {Guillout}, {Bauchet}, {Marchal}, {Sadowski}, \&
  {Teyssier}}]{GaiaDR3_astrometrc_binaries}
{Halbwachs}, J.-L., {Pourbaix}, D., {Arenou}, F., {et~al.} 2023, \aap, 674, A9,
  \dodoi{10.1051/0004-6361/202243969}

\bibitem[{{Hale}(1995)}]{HD114762_BDconf2}
{Hale}, A. 1995, \pasp, 107, 22, \dodoi{10.1086/133511}

\bibitem[{{Hall}(1892)}]{Hl_1892c}
{Hall}, A. 1892, Observations made at the U.S. Naval Observatory, 6, E.1

\bibitem[{{Halverson} {et~al.}(2016){Halverson}, {Terrien}, {Mahadevan}, {Roy},
  {Bender}, {Stef{\'a}nsson}, {Monson}, {Levi}, {Hearty}, {Blake}, {McElwain},
  {Schwab}, {Ramsey}, {Wright}, {Wang}, {Gong}, \&
  {Roberston}}]{NEID_error_budget}
{Halverson}, S., {Terrien}, R., {Mahadevan}, S., {et~al.} 2016, in Society of
  Photo-Optical Instrumentation Engineers (SPIE) Conference Series, Vol. 9908,
  Ground-based and Airborne Instrumentation for Astronomy VI, ed. C.~J.
  {Evans}, L.~{Simard}, \& H.~{Takami}, 99086P, \dodoi{10.1117/12.2232761}

\bibitem[{{Heintz}(1980)}]{1980ApJS...44..111H}
{Heintz}, W.~D. 1980, \apjs, 44, 111, \dodoi{10.1086/190686}

\bibitem[{{Heintz}(1990)}]{Hei1990b}
---. 1990, \apjs, 74, 275, \dodoi{10.1086/191499}

\bibitem[{{Hempelmann} {et~al.}(2006){Hempelmann}, {Robrade}, {Schmitt},
  {Favata}, {Baliunas}, \& {Hall}}]{2006A&A...460..261H}
{Hempelmann}, A., {Robrade}, J., {Schmitt}, J.~H.~M.~M., {et~al.} 2006, \aap,
  460, 261, \dodoi{10.1051/0004-6361:20065459}

\bibitem[{{Henry} {et~al.}(1994){Henry}, {Kirkpatrick}, \&
  {Simons}}]{1994AJ....108.1437H}
{Henry}, T.~J., {Kirkpatrick}, J.~D., \& {Simons}, D.~A. 1994, \aj, 108, 1437,
  \dodoi{10.1086/117167}

\bibitem[{{Henry} {et~al.}(2018){Henry}, {Jao}, {Winters}, {Dieterich},
  {Finch}, {Ianna}, {Riedel}, {Silverstein}, {Subasavage}, \&
  {Vrijmoet}}]{2018AJ....155..265H}
{Henry}, T.~J., {Jao}, W.-C., {Winters}, J.~G., {et~al.} 2018, \aj, 155, 265,
  \dodoi{10.3847/1538-3881/aac262}

\bibitem[{{Herschel}(1833)}]{1833MmRAS...5...13H}
{Herschel}, J.~F.~W. 1833, \memras, 5, 13

\bibitem[{{Hirsch} {et~al.}(2021){Hirsch}, {Rosenthal}, {Fulton}, {Howard},
  {Ciardi}, {Marcy}, {Nielsen}, {Petigura}, {de Rosa}, {Isaacson}, {Weiss},
  {Sinukoff}, \& {Macintosh}}]{Hirsch_AO}
{Hirsch}, L.~A., {Rosenthal}, L., {Fulton}, B.~J., {et~al.} 2021, \aj, 161,
  134, \dodoi{10.3847/1538-3881/abd639}

\bibitem[{{Holden}(1888)}]{Lick_Observatory}
{Holden}, E.~S. 1888, Sidereal Messenger, 7, 49

\bibitem[{{Hopkins}(1915)}]{61cyg_residual_search1}
{Hopkins}, M.~M. 1915, Contributions from the Rutherford Observatory of
  Columbia University New York, 29, 1

\bibitem[{{Isaacson} {et~al.}(2024){Isaacson}, {Howard}, {Fulton}, {Petigura},
  {Weiss}, {Kane}, {Carter}, {Beard}, {Giacalone}, {Van Zandt}, {Akana Murphy},
  {Dai}, {Chontos}, {Polanski}, {Rice}, {Lubin}, {Brinkman}, {Rubenzahl},
  {Blunt}, {Yee}, {MacDougall}, {Dalba}, {Tyler}, {Behmard}, {Angelo},
  {Pidhorodetska}, {Mayo}, {Holcomb}, {Turtelboom}, {Hill}, {Bouma}, {Zhang},
  {Crossfield}, \& {Saunders}}]{2024arXiv240617332I}
{Isaacson}, H., {Howard}, A.~W., {Fulton}, B., {et~al.} 2024, arXiv e-prints,
  arXiv:2406.17332, \dodoi{10.48550/arXiv.2406.17332}

\bibitem[{{Izmailov} {et~al.}(2020){Izmailov}, {Rublevsky}, \&
  {Apetyan}}]{Izm2020}
{Izmailov}, I., {Rublevsky}, A., \& {Apetyan}, A. 2020, Astronomische
  Nachrichten, 341, 762, \dodoi{10.1002/asna.202013815}

\bibitem[{{Izmailov}(2019)}]{IZM2019}
{Izmailov}, I.~S. 2019, Astronomy Letters, 45, 30,
  \dodoi{10.1134/S106377371901002X}

\bibitem[{{Izmailov} \& {Roshchina}(2016{\natexlab{a}})}]{IZM2015}
{Izmailov}, I.~S., \& {Roshchina}, E.~A. 2016{\natexlab{a}}, Astrophysical
  Bulletin, 71, 225, \dodoi{10.1134/S1990341316020097}

\bibitem[{{Izmailov} \& {Roshchina}(2016{\natexlab{b}})}]{Pulkovo2009}
---. 2016{\natexlab{b}}, Astrophysical Bulletin, 71, 225,
  \dodoi{10.1134/S1990341316020097}

\bibitem[{{Izmailov} {et~al.}(2021){Izmailov}, {Shakht}, {Polyakov},
  {Gorshanov}, \& {Pogodin}}]{2021Ap.....64..160I}
{Izmailov}, I.~S., {Shakht}, N.~A., {Polyakov}, E.~V., {Gorshanov}, D.~L., \&
  {Pogodin}, M.~A. 2021, Astrophysics, 64, 160,
  \dodoi{10.1007/s10511-021-09677-0}

\bibitem[{{Izmailov} {et~al.}(2010){Izmailov}, {Khovricheva}, {Khovrichev},
  {Kiyaeva}, {Khrutskaya}, {Romanenko}, {Grosheva}, {Maslennikov}, \&
  {Kalinichenko}}]{Izm2010}
{Izmailov}, I.~S., {Khovricheva}, M.~L., {Khovrichev}, M.~Y., {et~al.} 2010,
  Astronomy Letters, 36, 349, \dodoi{10.1134/S1063773710050051}

\bibitem[{{Jeffers}(1951{\natexlab{a}})}]{Jef1951}
{Jeffers}, H.~M. 1951{\natexlab{a}}, Lick Observatory Bulletin, 524, 51

\bibitem[{{Jeffers}(1951{\natexlab{b}})}]{DeO1957}
---. 1951{\natexlab{b}}, Lick Observatory Bulletin, 524, 51

\bibitem[{{Johns} {et~al.}(2018){Johns}, {Marti}, {Huff}, {McCann},
  {Wittenmyer}, {Horner}, \& {Wright}}]{2018ApJS..239...14J}
{Johns}, D., {Marti}, C., {Huff}, M., {et~al.} 2018, \apjs, 239, 14,
  \dodoi{10.3847/1538-4365/aae5fb}

\bibitem[{{Josties}(1978)}]{USN1978}
{Josties}, F.~J. 1978, Publications of the U.S. Naval Observatory Second
  Series, 24, 1

\bibitem[{{Josties} {et~al.}(1974){Josties}, {Dahn}, {Kallarakal}, {Miranian},
  {Douglass}, {Christy}, {Behall}, \& {Harrington}}]{USN1974}
{Josties}, F.~J., {Dahn}, C.~C., {Kallarakal}, V.~V., {et~al.} 1974,
  Publications of the U.S. Naval Observatory Second Series, 22, 7

\bibitem[{{Josties} \& {Harrington}(1984{\natexlab{a}})}]{USN1984}
{Josties}, F.~J., \& {Harrington}, R.~S. 1984{\natexlab{a}}, \apjs, 54, 103,
  \dodoi{10.1086/190920}

\bibitem[{{Josties} \& {Harrington}(1984{\natexlab{b}})}]{Vie1992}
---. 1984{\natexlab{b}}, \apjs, 54, 103, \dodoi{10.1086/190920}

\bibitem[{{Kanodia} {et~al.}(2022){Kanodia}, {Libby-Roberts}, {Ca{\~n}as},
  {Ninan}, {Mahadevan}, {Stefansson}, {Lin}, {Jones}, {Monson}, {Parker},
  {Kobulnicky}, {Swaby}, {Powers}, {Beard}, {Bender}, {Blake}, {Cochran},
  {Dong}, {Diddams}, {Fredrick}, {Gupta}, {Halverson}, {Hearty}, {Logsdon},
  {Metcalf}, {McElwain}, {Morley}, {Rajagopal}, {Ramsey}, {Robertson}, {Roy},
  {Schwab}, {Terrien}, {Wisniewski}, \& {Wright}}]{TOI3757}
{Kanodia}, S., {Libby-Roberts}, J., {Ca{\~n}as}, C.~I., {et~al.} 2022, \aj,
  164, 81, \dodoi{10.3847/1538-3881/ac7c20}

\bibitem[{{Kervella} {et~al.}(2008){Kervella}, {M{\'e}rand}, {Pichon},
  {Th{\'e}venin}, {Heiter}, {Bigot}, {ten Brummelaar}, {McAlister}, {Ridgway},
  {Turner}, {Sturmann}, {Sturmann}, {Goldfinger}, \&
  {Farrington}}]{kervella_61cyg}
{Kervella}, P., {M{\'e}rand}, A., {Pichon}, B., {et~al.} 2008, \aap, 488, 667,
  \dodoi{10.1051/0004-6361:200810080}

\bibitem[{{Kiefer}(2019)}]{HD114762_Gaiamass}
{Kiefer}, F. 2019, \aap, 632, L9, \dodoi{10.1051/0004-6361/201936942}

\bibitem[{{Knapp}(2017)}]{Kpp2017j}
{Knapp}, W.~R. 2017, The Webb Deep-Sky Society, 21

\bibitem[{{Knobel}(1877)}]{1877MmRAS..43....1K}
{Knobel}, E.~B. 1877, \memras, 43, 1

\bibitem[{{Kuiper}(1938)}]{old_mass_lum}
{Kuiper}, G.~P. 1938, \apj, 88, 472, \dodoi{10.1086/143999}

\bibitem[{{Labitzke} \& {Przbyllok}(1929)}]{Lbz1929}
{Labitzke}, P., \& {Przbyllok}, E. 1929, {Koenigsberg Obs. 45, 65}

\bibitem[{{Lagarde} {et~al.}(2012){Lagarde}, {Decressin}, {Charbonnel},
  {Eggenberger}, {Ekstr{\"o}m}, \& {Palacios}}]{starevol_I}
{Lagarde}, N., {Decressin}, T., {Charbonnel}, C., {et~al.} 2012, \aap, 543,
  A108, \dodoi{10.1051/0004-6361/201118331}

\bibitem[{{Lagarde} {et~al.}(2017){Lagarde}, {Robin}, {Reyl{\'e}}, \&
  {Nasello}}]{starevol_II}
{Lagarde}, N., {Robin}, A.~C., {Reyl{\'e}}, C., \& {Nasello}, G. 2017, \aap,
  601, A27, \dodoi{10.1051/0004-6361/201630253}

\bibitem[{{Latham} {et~al.}(1989){Latham}, {Mazeh}, {Stefanik}, {Mayor}, \&
  {Burki}}]{HD114762b}
{Latham}, D.~W., {Mazeh}, T., {Stefanik}, R.~P., {Mayor}, M., \& {Burki}, G.
  1989, \nat, 339, 38, \dodoi{10.1038/339038a0}

\bibitem[{{Li} {et~al.}(2021){Li}, {Brandt}, {Brandt}, {Dupuy}, {Michalik},
  {Jensen-Clem}, {Zeng}, {Faherty}, \& {Mitra}}]{nine_RV_exoplanets}
{Li}, Y., {Brandt}, T.~D., {Brandt}, G.~M., {et~al.} 2021, \aj, 162, 266,
  \dodoi{10.3847/1538-3881/ac27ab}

\bibitem[{{Li} {et~al.}(2023){Li}, {Brandt}, {Brandt}, {An}, {Franson},
  {Dupuy}, {Chen}, {Bowens-Rubin}, {Lewis}, {Bowler}, {Gibbs}, {Kiman},
  {Faherty}, {Currie}, {Jensen-Clem}, {Zhang}, {Contreras-Martinez},
  {Fitzgerald}, {Mazin}, \& {Millar-Blanchaer}}]{2023MNRAS.522.5622L}
---. 2023, \mnras, 522, 5622, \dodoi{10.1093/mnras/stad1315}

\bibitem[{{Lin}(2024)}]{SNEAK}
{Lin}, A. S.~J. 2024, PhD thesis, Pennsylvania State University, Department of
  Astronomy

\bibitem[{{Lindegren}(1997)}]{HIP_double2}
{Lindegren}, L. 1997, in ESA Special Publication, Vol. 402, Hipparcos - Venice
  '97, ed. R.~M. {Bonnet}, E.~{H{\o}g}, P.~L. {Bernacca}, L.~{Emiliani},
  A.~{Blaauw}, C.~{Turon}, J.~{Kovalevsky}, L.~{Lindegren}, H.~{Hassan},
  M.~{Bouffard}, B.~{Strim}, D.~{Heger}, M.~A.~C. {Perryman}, \& L.~{Woltjer},
  13--18

\bibitem[{{Lindegren} {et~al.}(1997){Lindegren}, {Mignard}, {S{\"o}derhjelm},
  {Badiali}, {Bernstein}, {Lampens}, {Pannunzio}, {Arenou}, {Bernacca},
  {Falin}, {Froeschl{\'e}}, {Kovalevsky}, {Martin}, {Perryman}, \&
  {Wielen}}]{HIP_double1}
{Lindegren}, L., {Mignard}, F., {S{\"o}derhjelm}, S., {et~al.} 1997, \aap, 323,
  L53

\bibitem[{{Lindegren} {et~al.}(2018){Lindegren}, {Hern{\'a}ndez}, {Bombrun},
  {Klioner}, {Bastian}, {Ramos-Lerate}, {de Torres}, {Steidelm{\"u}ller},
  {Stephenson}, {Hobbs}, {Lammers}, {Biermann}, {Geyer}, {Hilger}, {Michalik},
  {Stampa}, {McMillan}, {Casta{\~n}eda}, {Clotet}, {Comoretto}, {Davidson},
  {Fabricius}, {Gracia}, {Hambly}, {Hutton}, {Mora}, {Portell}, {van Leeuwen},
  {Abbas}, {Abreu}, {Altmann}, {Andrei}, {Anglada}, {Balaguer-N{\'u}{\~n}ez},
  {Barache}, {Becciani}, {Bertone}, {Bianchi}, {Bouquillon}, {Bourda},
  {Br{\"u}semeister}, {Bucciarelli}, {Busonero}, {Buzzi}, {Cancelliere},
  {Carlucci}, {Charlot}, {Cheek}, {Crosta}, {Crowley}, {de Bruijne}, {de
  Felice}, {Drimmel}, {Esquej}, {Fienga}, {Fraile}, {Gai}, {Garralda},
  {Gonz{\'a}lez-Vidal}, {Guerra}, {Hauser}, {Hofmann}, {Holl}, {Jordan},
  {Lattanzi}, {Lenhardt}, {Liao}, {Licata}, {Lister}, {L{\"o}ffler},
  {Marchant}, {Martin-Fleitas}, {Messineo}, {Mignard}, {Morbidelli}, {Poggio},
  {Riva}, {Rowell}, {Salguero}, {Sarasso}, {Sciacca}, {Siddiqui}, {Smart},
  {Spagna}, {Steele}, {Taris}, {Torra}, {van Elteren}, {van Reeven}, \&
  {Vecchiato}}]{gaia_astrometric_params}
{Lindegren}, L., {Hern{\'a}ndez}, J., {Bombrun}, A., {et~al.} 2018, \aap, 616,
  A2, \dodoi{10.1051/0004-6361/201832727}

\bibitem[{{Lomb}(1976)}]{lomb}
{Lomb}, N.~R. 1976, \apss, 39, 447, \dodoi{10.1007/BF00648343}

\bibitem[{{Mamajek} \& {Hillenbrand}(2008)}]{mamajek_activity_age_estimation}
{Mamajek}, E.~E., \& {Hillenbrand}, L.~A. 2008, \apj, 687, 1264,
  \dodoi{10.1086/591785}

\bibitem[{{Mann} {et~al.}(2015){Mann}, {Feiden}, {Gaidos}, {Boyajian}, \& {von
  Braun}}]{2015ApJ...804...64M}
{Mann}, A.~W., {Feiden}, G.~A., {Gaidos}, E., {Boyajian}, T., \& {von Braun},
  K. 2015, \apj, 804, 64, \dodoi{10.1088/0004-637X/804/1/64}

\bibitem[{{Mann} {et~al.}(2019){Mann}, {Dupuy}, {Kraus}, {Gaidos}, {Ansdell},
  {Ireland}, {Rizzuto}, {Hung}, {Dittmann}, {Factor}, {Feiden}, {Martinez},
  {Ru{\'\i}z-Rodr{\'\i}guez}, \& {Thao}}]{2019ApJ...871...63M}
{Mann}, A.~W., {Dupuy}, T., {Kraus}, A.~L., {et~al.} 2019, \apj, 871, 63,
  \dodoi{10.3847/1538-4357/aaf3bc10.48550/arXiv.1811.06938}

\bibitem[{{Mason} \& {Hartkopf}(2017)}]{WSI2017b}
{Mason}, B.~D., \& {Hartkopf}, W.~I. 2017, \aj, 154, 183,
  \dodoi{10.3847/1538-3881/aa8038}

\bibitem[{{Mason} {et~al.}(2018{\natexlab{a}}){Mason}, {Hartkopf}, {Urban}, \&
  {Josties}}]{WSI2018a}
{Mason}, B.~D., {Hartkopf}, W.~I., {Urban}, S.~E., \& {Josties}, J.~D.
  2018{\natexlab{a}}, \aj, 156, 240, \dodoi{10.3847/1538-3881/aae484}

\bibitem[{{Mason} {et~al.}(2018{\natexlab{b}}){Mason}, {Hartkopf}, {Urban}, \&
  {Josties}}]{Smr2018c}
---. 2018{\natexlab{b}}, \aj, 156, 240, \dodoi{10.3847/1538-3881/aae484}

\bibitem[{{Mason} {et~al.}(2021){Mason}, {Williams}, {Matson}, {Josties},
  {Eakens}, {Justice}, {Kilian}, \& {Warner}}]{WSI2021}
{Mason}, B.~D., {Williams}, S.~J., {Matson}, R.~A., {et~al.} 2021, \aj, 162,
  53, \dodoi{10.3847/1538-3881/abfaa2}

\bibitem[{{Mason} {et~al.}(2001){Mason}, {Wycoff}, {Hartkopf}, {Douglass}, \&
  {Worley}}]{WDS}
{Mason}, B.~D., {Wycoff}, G.~L., {Hartkopf}, W.~I., {Douglass}, G.~G., \&
  {Worley}, C.~E. 2001, \aj, 122, 3466, \dodoi{10.1086/323920}

\bibitem[{{Munoz}(2020)}]{MuR2020a}
{Munoz}, R.~S. 2020, El Observador de Estrellas Dobles, 24

\bibitem[{{Naef} {et~al.}(2001){Naef}, {Mayor}, {Pepe}, {Queloz}, {Santos},
  {Udry}, \& {Burnet}}]{HD2171707_coralie}
{Naef}, D., {Mayor}, M., {Pepe}, F., {et~al.} 2001, \aap, 375, 205,
  \dodoi{10.1051/0004-6361:20010841}

\bibitem[{{Naef} {et~al.}(2003){Naef}, {Mayor}, {Korzennik}, {Queloz}, {Udry},
  {Nisenson}, {Noyes}, {Brown}, {Beuzit}, {Perrier}, \&
  {Sivan}}]{HD190360_elodie_afoe}
{Naef}, D., {Mayor}, M., {Korzennik}, S.~G., {et~al.} 2003, \aap, 410, 1051,
  \dodoi{10.1051/0004-6361:20031341}

\bibitem[{{NASA Exoplanet Archive}(2019)}]{https://doi.org/10.26133/nea1}
{NASA Exoplanet Archive}. 2019, Confirmed Planets Table,  IPAC,
  \dodoi{10.26133/NEA1}

\bibitem[{{Nugent} \& {Iverson}(2012)}]{Nug2012}
{Nugent}, R.~L., \& {Iverson}, E.~W. 2012, Journal of Double Star Observations,
  8, 213

\bibitem[{{Perruchot} {et~al.}(2008){Perruchot}, {Kohler}, {Bouchy}, {Richaud},
  {Richaud}, {Moreaux}, {Merzougui}, {Sottile}, {Hill}, {Knispel}, {Regal},
  {Meunier}, {Ilovaisky}, {Le Coroller}, {Gillet}, {Schmitt}, {Pepe}, {Fleury},
  {Sosnowska}, {Vors}, {M{\'e}gevand}, {Blanc}, {Carol}, {Point}, {Laloge}, \&
  {Brunel}}]{SOPHIE}
{Perruchot}, S., {Kohler}, D., {Bouchy}, F., {et~al.} 2008, in Society of
  Photo-Optical Instrumentation Engineers (SPIE) Conference Series, Vol. 7014,
  Ground-based and Airborne Instrumentation for Astronomy II, ed. I.~S.
  {McLean} \& M.~M. {Casali}, 70140J, \dodoi{10.1117/12.787379}

\bibitem[{{Perryman} {et~al.}(1997){Perryman}, {Lindegren}, {Kovalevsky},
  {Hoeg}, {Bastian}, {Bernacca}, {Cr{\'e}z{\'e}}, {Donati}, {Grenon},
  {Grewing}, {van Leeuwen}, {van der Marel}, {Mignard}, {Murray}, {Le Poole},
  {Schrijver}, {Turon}, {Arenou}, {Froeschl{\'e}}, \&
  {Petersen}}]{Hipparcos_catalog}
{Perryman}, M.~A.~C., {Lindegren}, L., {Kovalevsky}, J., {et~al.} 1997, \aap,
  323, L49

\bibitem[{{Peters}(1886)}]{1886AN....113..321P}
{Peters}, C.~F.~W. 1886, Astronomische Nachrichten, 113, 321,
  \dodoi{10.1002/asna.18861132002}

\bibitem[{{Petigura} {et~al.}(2017){Petigura}, {Howard}, {Marcy}, {Johnson},
  {Isaacson}, {Cargile}, {Hebb}, {Fulton}, {Weiss}, {Morton}, {Winn}, {Rogers},
  {Sinukoff}, {Hirsch}, \& {Crossfield}}]{specmatch_syn}
{Petigura}, E.~A., {Howard}, A.~W., {Marcy}, G.~W., {et~al.} 2017, \aj, 154,
  107, \dodoi{10.3847/1538-3881/aa80de}

\bibitem[{{Pietrinferni} {et~al.}(2004){Pietrinferni}, {Cassisi}, {Salaris}, \&
  {Castelli}}]{BaSTI_I}
{Pietrinferni}, A., {Cassisi}, S., {Salaris}, M., \& {Castelli}, F. 2004, \apj,
  612, 168, \dodoi{10.1086/422498}

\bibitem[{{Pietrinferni} {et~al.}(2006){Pietrinferni}, {Cassisi}, {Salaris}, \&
  {Castelli}}]{BaSTI_II}
---. 2006, \apj, 642, 797, \dodoi{10.1086/501344}

\bibitem[{{Przybyllok}(1926)}]{Prz1926}
{Przybyllok}, E., L.~P.~. M.~P. 1926, Astronomische Nachrichten

\bibitem[{{Rabe}(1923)}]{Rab1923}
{Rabe}, W. 1923, Astronomische Nachrichten, 217, 413,
  \dodoi{10.1002/asna.19222172102}

\bibitem[{{Rabe}(1939)}]{Rab1939}
---. 1939, {Pub. Munich Obs. 2, \#1}

\bibitem[{{Rabe}(1953)}]{Rab1953}
---. 1953, {Mikrometermessungen von Doppelsternen in den Jahren 1932 bis 1946.}

\bibitem[{{Radovan} {et~al.}(2014){Radovan}, {Lanclos}, {Holden}, {Kibrick},
  {Allen}, {Deich}, {Rivera}, {Burt}, {Fulton}, {Butler}, \&
  {Vogt}}]{2014SPIE.9145E..2BR}
{Radovan}, M.~V., {Lanclos}, K., {Holden}, B.~P., {et~al.} 2014, in Society of
  Photo-Optical Instrumentation Engineers (SPIE) Conference Series, Vol. 9145,
  Ground-based and Airborne Telescopes V, ed. L.~M. {Stepp}, R.~{Gilmozzi}, \&
  H.~J. {Hall}, 91452B, \dodoi{10.1117/12.2057310}

\bibitem[{{Raghavan} {et~al.}(2010){Raghavan}, {McAlister}, {Henry}, {Latham},
  {Marcy}, {Mason}, {Gies}, {White}, \& {ten Brummelaar}}]{raghavan_2010}
{Raghavan}, D., {McAlister}, H.~A., {Henry}, T.~J., {et~al.} 2010, \apjs, 190,
  1, \dodoi{10.1088/0067-0049/190/1/1}

\bibitem[{{Ricker} {et~al.}(2015){Ricker}, {Winn}, {Vanderspek}, {Latham},
  {Bakos}, {Bean}, {Berta-Thompson}, {Brown}, {Buchhave}, {Butler}, {Butler},
  {Chaplin}, {Charbonneau}, {Christensen-Dalsgaard}, {Clampin}, {Deming},
  {Doty}, {De Lee}, {Dressing}, {Dunham}, {Endl}, {Fressin}, {Ge}, {Henning},
  {Holman}, {Howard}, {Ida}, {Jenkins}, {Jernigan}, {Johnson}, {Kaltenegger},
  {Kawai}, {Kjeldsen}, {Laughlin}, {Levine}, {Lin}, {Lissauer}, {MacQueen},
  {Marcy}, {McCullough}, {Morton}, {Narita}, {Paegert}, {Palle}, {Pepe},
  {Pepper}, {Quirrenbach}, {Rinehart}, {Sasselov}, {Sato}, {Seager},
  {Sozzetti}, {Stassun}, {Sullivan}, {Szentgyorgyi}, {Torres}, {Udry}, \&
  {Villasenor}}]{TESS}
{Ricker}, G.~R., {Winn}, J.~N., {Vanderspek}, R., {et~al.} 2015, Journal of
  Astronomical Telescopes, Instruments, and Systems, 1, 014003,
  \dodoi{10.1117/1.JATIS.1.1.014003}

\bibitem[{{Risin} {et~al.}(2022){Risin}, {Altunin}, {Wasson}, {Genet}, \&
  {Dye}}]{SHS2022b}
{Risin}, S., {Altunin}, I., {Wasson}, R., {Genet}, R., \& {Dye}, S. 2022,
  Journal of Double Star Observations, 18, 30

\bibitem[{{Rosenthal} {et~al.}(2021){Rosenthal}, {Fulton}, {Hirsch},
  {Isaacson}, {Howard}, {Dedrick}, {Sherstyuk}, {Blunt}, {Petigura}, {Knutson},
  {Behmard}, {Chontos}, {Crepp}, {Crossfield}, {Dalba}, {Fischer}, {Henry},
  {Kane}, {Kosiarek}, {Marcy}, {Rubenzahl}, {Weiss}, \&
  {Wright}}]{2021ApJS..255....8R}
{Rosenthal}, L.~J., {Fulton}, B.~J., {Hirsch}, L.~A., {et~al.} 2021, \apjs,
  255, 8, \dodoi{10.3847/1538-4365/abe23c}

\bibitem[{{Rouan} {et~al.}(2000){Rouan}, {Riaud}, {Boccaletti}, {Cl{\'e}net},
  \& {Labeyrie}}]{4QPM}
{Rouan}, D., {Riaud}, P., {Boccaletti}, A., {Cl{\'e}net}, Y., \& {Labeyrie}, A.
  2000, \pasp, 112, 1479, \dodoi{10.1086/317707}

\bibitem[{{Salama} {et~al.}(2022){Salama}, {Ziegler}, {Baranec}, {Liu}, {Law},
  {Riddle}, {Henry}, {Winters}, {Jao}, {Ou}, \& {Hermosillo
  Ruiz}}]{2022AJ....163..200S}
{Salama}, M., {Ziegler}, C., {Baranec}, C., {et~al.} 2022, \aj, 163, 200,
  \dodoi{10.3847/1538-3881/ac53fc}

\bibitem[{{Scargle}(1982)}]{scargle}
{Scargle}, J.~D. 1982, \apj, 263, 835, \dodoi{10.1086/160554}

\bibitem[{{Schiaparelli}(1909)}]{Sp_1909}
{Schiaparelli}, G.~V. 1909, {Osservazioni sulle stelle doppie. Serie seconda:
  comprendente le misure DI 636 sistemi eseguite col refrattore equatoriale
  Merz-Repsold negli anni 1886-1900} (Italy: U. Hoepli.)

\bibitem[{{Schlimmer}(2019{\natexlab{a}})}]{Smr2019}
{Schlimmer}, J. 2019{\natexlab{a}}, Journal of Double Star Observations, 15,
  536

\bibitem[{{Schlimmer}(2019{\natexlab{b}})}]{Smr2021b}
---. 2019{\natexlab{b}}, Journal of Double Star Observations, 15, 536

\bibitem[{{Schlimmer}(2019{\natexlab{c}})}]{Kpp2020f}
---. 2019{\natexlab{c}}, Journal of Double Star Observations, 15, 536

\bibitem[{{Schlimmer}(2022)}]{JDSO_link}
{Schlimmer}, J.~S. 2022, Journal of Double Star Observations, 18, 322

\bibitem[{{Schwab} {et~al.}(2016){Schwab}, {Rakich}, {Gong}, {Mahadevan},
  {Halverson}, {Roy}, {Terrien}, {Robertson}, {Hearty}, {Levi}, {Monson},
  {Wright}, {McElwain}, {Bender}, {Blake}, {St{\"u}rmer}, {Gurevich},
  {Chakraborty}, \& {Ramsey}}]{NEID_design}
{Schwab}, C., {Rakich}, A., {Gong}, Q., {et~al.} 2016, in Society of
  Photo-Optical Instrumentation Engineers (SPIE) Conference Series, Vol. 9908,
  Ground-based and Airborne Instrumentation for Astronomy VI, ed. C.~J.
  {Evans}, L.~{Simard}, \& H.~{Takami}, 99087H, \dodoi{10.1117/12.2234411}

\bibitem[{{Shakht} {et~al.}(2017){Shakht}, {Gorshanov}, \&
  {Vasilkova}}]{PkO2017b}
{Shakht}, N.~A., {Gorshanov}, D.~L., \& {Vasilkova}, O.~O. 2017, Astrophysics,
  60, 507, \dodoi{10.1007/s10511-017-9502-9}

\bibitem[{{Simpson} {et~al.}(2010){Simpson}, {Baliunas}, {Henry}, \&
  {Watson}}]{2010MNRAS.408.1666S}
{Simpson}, E.~K., {Baliunas}, S.~L., {Henry}, G.~W., \& {Watson}, C.~A. 2010,
  \mnras, 408, 1666, \dodoi{10.1111/j.1365-2966.2010.17230.x}

\bibitem[{{Sivo} {et~al.}(2022){Sivo}, {Scharw{\"a}chter}, \&
  {Sivanandam}}]{gemini_AO}
{Sivo}, G., {Scharw{\"a}chter}, J., \& {Sivanandam}, S. 2022, The NOIRLab
  Mirror, 3, 17

\bibitem[{{Skrutskie} {et~al.}(2006){Skrutskie}, {Cutri}, {Stiening},
  {Weinberg}, {Schneider}, {Carpenter}, {Beichman}, {Capps}, {Chester},
  {Elias}, {Huchra}, {Liebert}, {Lonsdale}, {Monet}, {Price}, {Seitzer},
  {Jarrett}, {Kirkpatrick}, {Gizis}, {Howard}, {Evans}, {Fowler}, {Fullmer},
  {Hurt}, {Light}, {Kopan}, {Marsh}, {McCallon}, {Tam}, {Van Dyk}, \&
  {Wheelock}}]{2mass}
{Skrutskie}, M.~F., {Cutri}, R.~M., {Stiening}, R., {et~al.} 2006, \aj, 131,
  1163, \dodoi{10.1086/498708}

\bibitem[{{Smyth}(1844)}]{Smy1844}
{Smyth}, W.~H. 1844, {A Cycle of Celestial Objects} (London: J.W. Parker)

\bibitem[{{Soubiran} {et~al.}(2024){Soubiran}, {Creevey}, {Lagarde},
  {Brouillet}, {Jofr{\'e}}, {Casamiquela}, {Heiter}, {Aguilera-G{\'o}mez},
  {Vitali}, {Worley}, \& {de Brito Silva}}]{gaia_fgk_standards}
{Soubiran}, C., {Creevey}, O.~L., {Lagarde}, N., {et~al.} 2024, \aap, 682,
  A145, \dodoi{10.1051/0004-6361/202347136}

\bibitem[{{Spergel} {et~al.}(2015){Spergel}, {Gehrels}, {Baltay}, {Bennett},
  {Breckinridge}, {Donahue}, {Dressler}, {Gaudi}, {Greene}, {Guyon}, {Hirata},
  {Kalirai}, {Kasdin}, {Macintosh}, {Moos}, {Perlmutter}, {Postman},
  {Rauscher}, {Rhodes}, {Wang}, {Weinberg}, {Benford}, {Hudson}, {Jeong},
  {Mellier}, {Traub}, {Yamada}, {Capak}, {Colbert}, {Masters}, {Penny},
  {Savransky}, {Stern}, {Zimmerman}, {Barry}, {Bartusek}, {Carpenter}, {Cheng},
  {Content}, {Dekens}, {Demers}, {Grady}, {Jackson}, {Kuan}, {Kruk}, {Melton},
  {Nemati}, {Parvin}, {Poberezhskiy}, {Peddie}, {Ruffa}, {Wallace}, {Whipple},
  {Wollack}, \& {Zhao}}]{2015arXiv150303757S}
{Spergel}, D., {Gehrels}, N., {Baltay}, C., {et~al.} 2015, arXiv e-prints,
  arXiv:1503.03757, \dodoi{10.48550/arXiv.1503.03757}

\bibitem[{{Stassun} {et~al.}(2017){Stassun}, {Collins}, \&
  {Gaudi}}]{2017AJ....153..136S}
{Stassun}, K.~G., {Collins}, K.~A., \& {Gaudi}, B.~S. 2017, \aj, 153, 136,
  \dodoi{10.3847/1538-3881/aa5df3}

\bibitem[{{Stef{\'a}nsson} {et~al.}(2025){Stef{\'a}nsson}, {Mahadevan}, {Winn},
  {Marcussen}, {Kanodia}, {Albrecht}, {Fitzmaurice}, {Mikulskyt{\.{e}}},
  {Ca{\~n}as}, {Espinoza-Retamal}, {Zwart}, {Krolikowski}, {Hotnisky},
  {Robertson}, {Alvarado-Montes}, {Bender}, {Blake}, {Callingham}, {Cochran},
  {Delamer}, {Diddams}, {Dong}, {Fernandes}, {Giovinazzi}, {Halverson},
  {Libby-Roberts}, {Logsdon}, {McElwain}, {Ninan}, {Rajagopal}, {Reji}, {Roy},
  {Schwab}, \& {Wright}}]{2025AJ....169..107S}
{Stef{\'a}nsson}, G., {Mahadevan}, S., {Winn}, J.~N., {et~al.} 2025, \aj, 169,
  107, \dodoi{10.3847/1538-3881/ada9e1}

\bibitem[{{Stolte} {et~al.}(2010){Stolte}, {Morris}, {Ghez}, {Do}, {Lu},
  {Wright}, {Ballard}, {Mills}, \& {Matthews}}]{2010ApJ...718..810S}
{Stolte}, A., {Morris}, M.~R., {Ghez}, A.~M., {et~al.} 2010, \apj, 718, 810,
  \dodoi{10.1088/0004-637X/718/2/810}

\bibitem[{{Strand}(1937)}]{1937AnLei..18B...1S}
{Strand}, K.~A. 1937, Annalen van de Sterrewacht te Leiden, 18, B1

\bibitem[{{Strand}(1943)}]{61cyg_residual_search2}
---. 1943, \pasp, 55, 29, \dodoi{10.1086/125484}

\bibitem[{{Strand}(1969)}]{1969AJ.....74..760S}
---. 1969, \aj, 74, 760, \dodoi{10.1086/110853}

\bibitem[{{Struve}(1837)}]{StF1837}
{Struve}, F.~G.~W. 1837, Astronomische Nachrichten, 14, 249,
  \dodoi{10.1002/asna.18370141609}

\bibitem[{{Struve}(1962)}]{StG1962a}
{Struve}, G. 1962, {Pub. Berlin Babelsberg Obs. 14, Pt.1}

\bibitem[{{Struve}(1911)}]{StH1911a}
{Struve}, H. 1911, Astronomische Nachrichten, 188, 5,
  \dodoi{10.1002/asna.19111880103}

\bibitem[{{Struve}(1879)}]{Stt1878}
{Struve}, O. 1879, The Observatory, 2, 347

\bibitem[{{Swift} {et~al.}(2015){Swift}, {Bottom}, {Johnson}, {Wright},
  {McCrady}, {Wittenmyer}, {Plavchan}, {Riddle}, {Muirhead}, {Herzig}, {Myles},
  {Blake}, {Eastman}, {Beatty}, {Barnes}, {Gibson}, {Lin}, {Zhao}, {Gardner},
  {Falco}, {Criswell}, {Nava}, {Robinson}, {Sliski}, {Hedrick}, {Ivarsen},
  {Hjelstrom}, {de Vera}, \& {Szentgyorgyi}}]{swift2015}
{Swift}, J.~J., {Bottom}, M., {Johnson}, J.~A., {et~al.} 2015, Journal of
  Astronomical Telescopes, Instruments, and Systems, 1, 027002,
  \dodoi{10.1117/1.JATIS.1.2.027002}

\bibitem[{{Takeda} {et~al.}(2007){Takeda}, {Ford}, {Sills}, {Rasio}, {Fischer},
  \& {Valenti}}]{2007ApJS..168..297T}
{Takeda}, G., {Ford}, E.~B., {Sills}, A., {et~al.} 2007, \apjs, 168, 297,
  \dodoi{10.1086/509763}

\bibitem[{{Tal-Or} {et~al.}(2019{\natexlab{a}}){Tal-Or}, {Trifonov}, {Zucker},
  {Mazeh}, \& {Zechmeister}}]{HIRES_RV_corrections}
{Tal-Or}, L., {Trifonov}, T., {Zucker}, S., {Mazeh}, T., \& {Zechmeister}, M.
  2019{\natexlab{a}}, \mnras, 484, L8, \dodoi{10.1093/mnrasl/sly227}

\bibitem[{{Tal-Or} {et~al.}(2019{\natexlab{b}}){Tal-Or}, {Trifonov}, {Zucker},
  {Mazeh}, \& {Zechmeister}}]{2019MNRAS.484L...8T}
---. 2019{\natexlab{b}}, \mnras, 484, L8, \dodoi{10.1093/mnrasl/sly227}

\bibitem[{{ten Brummelaar} {et~al.}(2005){ten Brummelaar}, {McAlister},
  {Ridgway}, {Bagnuolo}, {Turner}, {Sturmann}, {Sturmann}, {Berger}, {Ogden},
  {Cadman}, {Hartkopf}, {Hopper}, \& {Shure}}]{chara}
{ten Brummelaar}, T.~A., {McAlister}, H.~A., {Ridgway}, S.~T., {et~al.} 2005,
  \apj, 628, 453, \dodoi{10.1086/430729}

\bibitem[{{Tokovinin}(2014{\natexlab{a}})}]{2014AJ....147...86T}
{Tokovinin}, A. 2014{\natexlab{a}}, \aj, 147, 86,
  \dodoi{10.1088/0004-6256/147/4/86}

\bibitem[{{Tokovinin}(2014{\natexlab{b}})}]{2014AJ....147...87T}
---. 2014{\natexlab{b}}, \aj, 147, 87, \dodoi{10.1088/0004-6256/147/4/87}

\bibitem[{{Tokovinin} \& {Shatskii}(1995)}]{Tok1995}
{Tokovinin}, A.~A., \& {Shatskii}, N.~I. 1995, Astronomy Letters, 21, 464

\bibitem[{{Tull} {et~al.}(1995){Tull}, {MacQueen}, {Sneden}, \&
  {Lambert}}]{Tull}
{Tull}, R.~G., {MacQueen}, P.~J., {Sneden}, C., \& {Lambert}, D.~L. 1995,
  \pasp, 107, 251, \dodoi{10.1086/133548}

\bibitem[{{Tuomi} {et~al.}(2009){Tuomi}, {Kotiranta}, \&
  {Kaasalainen}}]{2009A&A...494..769T}
{Tuomi}, M., {Kotiranta}, S., \& {Kaasalainen}, M. 2009, \aap, 494, 769,
  \dodoi{10.1051/0004-6361:200810288}

\bibitem[{{van Biesbroeck}(1927)}]{VBs1927a}
{van Biesbroeck}, G. 1927, Publications of the Yerkes Observatory, 5, 1.vii

\bibitem[{{Vaughan} {et~al.}(1981){Vaughan}, {Baliunas}, {Middelkoop},
  {Hartmann}, {Mihalas}, {Noyes}, \& {Preston}}]{1981ApJ...250..276V}
{Vaughan}, A.~H., {Baliunas}, S.~L., {Middelkoop}, F., {et~al.} 1981, \apj,
  250, 276, \dodoi{10.1086/159372}

\bibitem[{{Velghe}(1957)}]{61cyg_residual_search3}
{Velghe}, A.~G. 1957, \aj, 62, 35, \dodoi{10.1086/107628}

\bibitem[{{Vogt}(1987)}]{Hamilton}
{Vogt}, S.~S. 1987, \pasp, 99, 1214, \dodoi{10.1086/132107}

\bibitem[{{Vogt} {et~al.}(2005){Vogt}, {Butler}, {Marcy}, {Fischer}, {Henry},
  {Laughlin}, {Wright}, \& {Johnson}}]{2005ApJ...632..638V}
{Vogt}, S.~S., {Butler}, R.~P., {Marcy}, G.~W., {et~al.} 2005, \apj, 632, 638,
  \dodoi{10.1086/432901}

\bibitem[{{Vogt} {et~al.}(1994){Vogt}, {Allen}, {Bigelow}, {Bresee}, {Brown},
  {Cantrall}, {Conrad}, {Couture}, {Delaney}, {Epps}, {Hilyard}, {Hilyard},
  {Horn}, {Jern}, {Kanto}, {Keane}, {Kibrick}, {Lewis}, {Osborne},
  {Pardeilhan}, {Pfister}, {Ricketts}, {Robinson}, {Stover}, {Tucker}, {Ward},
  \& {Wei}}]{1994SPIE.2198..362V}
{Vogt}, S.~S., {Allen}, S.~L., {Bigelow}, B.~C., {et~al.} 1994, in Society of
  Photo-Optical Instrumentation Engineers (SPIE) Conference Series, Vol. 2198,
  Instrumentation in Astronomy VIII, ed. D.~L. {Crawford} \& E.~R. {Craine},
  362, \dodoi{10.1117/12.176725}

\bibitem[{{Vollmann}(2008)}]{VLM2008}
{Vollmann}, W. 2008, Journal of Double Star Observations, 4, 74

\bibitem[{{Vousden} {et~al.}(2016){Vousden}, {Farr}, \& {Mandel}}]{ptemcee}
{Vousden}, W.~D., {Farr}, W.~M., \& {Mandel}, I. 2016, \mnras, 455, 1919,
  \dodoi{10.1093/mnras/stv2422}

\bibitem[{{Webster}(2017{\natexlab{a}})}]{Wbt2018}
{Webster}, N. 2017{\natexlab{a}}, The Webb Deep-Sky Society, 32

\bibitem[{{Webster}(2017{\natexlab{b}})}]{Wbt2017}
---. 2017{\natexlab{b}}, The Webb Deep-Sky Society, 51

\bibitem[{{Webster}(2020)}]{Wbt2020}
---. 2020, The Webb Deep-Sky Society, 31

\bibitem[{{Wilson} \& {Seabroke}(1877)}]{WS_1877}
{Wilson}, J.~M., \& {Seabroke}, G.~M. 1877, \memras, 43, 105

\bibitem[{{Wilson} {et~al.}(2019){Wilson}, {Eastman}, {Cornachione}, {Wang},
  {Johnson}, {Sliski}, {Schap}, {Morton}, {Johnson}, {McCrady}, {Wright},
  {Wittenmyer}, {Plavchan}, {Blake}, {Swift}, {Bottom}, {Baker}, {Barnes},
  {Berlind}, {Blackhurst}, {Beatty}, {Bolton}, {Cale}, {Calkins}, {Col{\'o}n},
  {de Vera}, {Esquerdo}, {Falco}, {Fortin}, {Garcia-Mejia}, {Geneser},
  {Gibson}, {Grell}, {Groner}, {Halverson}, {Hamlin}, {Henderson}, {Horner},
  {Houghton}, {Janssens}, {Jonas}, {Jones}, {Kirby}, {Lawrence}, {Luebbers},
  {Muirhead}, {Myles}, {Nava}, {Rivera-Garc{\'\i}a}, {Reed}, {Relles},
  {Riddle}, {Robinson}, {Chaput de Saintonge}, \& {Sergi}}]{wilson2019}
{Wilson}, M.~L., {Eastman}, J.~D., {Cornachione}, M.~A., {et~al.} 2019, \pasp,
  131, 115001, \dodoi{10.1088/1538-3873/ab33c5}

\bibitem[{{Wilson}(1978)}]{Mt_Wilson_I}
{Wilson}, O.~C. 1978, \apj, 226, 379, \dodoi{10.1086/156618}

\bibitem[{{Winn}(2022)}]{2022AJ....164..196W}
{Winn}, J.~N. 2022, \aj, 164, 196, \dodoi{10.3847/1538-3881/ac9126}

\bibitem[{{Wittenmyer} {et~al.}(2007){Wittenmyer}, {Endl}, \&
  {Cochran}}]{hd217107_tull}
{Wittenmyer}, R.~A., {Endl}, M., \& {Cochran}, W.~D. 2007, \apj, 654, 625,
  \dodoi{10.1086/509110}

\bibitem[{{Wittenmyer} {et~al.}(2020){Wittenmyer}, {Clark}, {Sharma}, {Stello},
  {Horner}, {Kane}, {Stevens}, {Wright}, {Spina}, {{\v{C}}otar}, {Asplund},
  {Bland-Hawthorn}, {Buder}, {Casey}, {De Silva}, {D'Orazi}, {Freeman}, {Kos},
  {Lewis}, {Lin}, {Lind}, {Martell}, {Simpson}, {Zucker}, \&
  {Zwitter}}]{2020MNRAS.496..851W}
{Wittenmyer}, R.~A., {Clark}, J.~T., {Sharma}, S., {et~al.} 2020, \mnras, 496,
  851, \dodoi{10.1093/mnras/staa1528}

\bibitem[{{Wizinowich} {et~al.}(2000){Wizinowich}, {Acton}, {Shelton},
  {Stomski}, {Gathright}, {Ho}, {Lupton}, {Tsubota}, {Lai}, {Max}, {Brase},
  {An}, {Avicola}, {Olivier}, {Gavel}, {Macintosh}, {Ghez}, \&
  {Larkin}}]{Keck_AO}
{Wizinowich}, P., {Acton}, D.~S., {Shelton}, C., {et~al.} 2000, \pasp, 112,
  315, \dodoi{10.1086/316543}

\bibitem[{{Wright} {et~al.}(2023){Wright}, {Rieke}, {Glasse}, {Ressler},
  {Garc{\'\i}a Mar{\'\i}n}, {Aguilar}, {Alberts}, {{\'A}lvarez-M{\'a}rquez},
  {Argyriou}, {Banks}, {Baudoz}, {Boccaletti}, {Bouchet}, {Bouwman}, {Brandl},
  {Breda}, {Bright}, {Cale}, {Colina}, {Cossou}, {Coulais}, {Cracraft}, {De
  Meester}, {Dicken}, {Engesser}, {Etxaluze}, {Fox}, {Friedman}, {Fu},
  {Gasman}, {G{\'a}sp{\'a}r}, {Gastaud}, {Geers}, {Glauser}, {Gordon},
  {Greene}, {Greve}, {Grundy}, {G{\"u}del}, {Guillard}, {Haderlein},
  {Hashimoto}, {Henning}, {Hines}, {Holler}, {Detre}, {Jahromi}, {James},
  {Jones}, {Justtanont}, {Kavanagh}, {Kendrew}, {Klaassen}, {Krause},
  {Labiano}, {Lagage}, {Lambros}, {Larson}, {Law}, {Lee}, {Libralato}, {Lorenzo
  Alverez}, {Meixner}, {Morrison}, {Mueller}, {Murray}, {Mycroft}, {Myers},
  {Nayak}, {Naylor}, {Nickson}, {Noriega-Crespo}, {{\"O}stlin}, {O'Sullivan},
  {Ottens}, {Patapis}, {Penanen}, {Pietraszkiewicz}, {Ray}, {Regan},
  {Roteliuk}, {Royer}, {Samara-Ratna}, {Samuelson}, {Sargent}, {Scheithauer},
  {Schneider}, {Schreiber}, {Shaughnessy}, {Sheehan}, {Shivaei}, {Sloan},
  {Tamas}, {Teague}, {Temim}, {Tikkanen}, {Tustain}, {van Dishoeck},
  {Vandenbussche}, {Weilert}, {Whitehouse}, \& {Wolff}}]{2023PASP..135d8003W}
{Wright}, G.~S., {Rieke}, G.~H., {Glasse}, A., {et~al.} 2023, \pasp, 135,
  048003, \dodoi{10.1088/1538-3873/acbe66}

\bibitem[{{Wright} {et~al.}(2004){Wright}, {Marcy}, {Butler}, \&
  {Vogt}}]{2004ApJS..152..261W}
{Wright}, J.~T., {Marcy}, G.~W., {Butler}, R.~P., \& {Vogt}, S.~S. 2004, \apjs,
  152, 261, \dodoi{10.1086/386283}

\bibitem[{{Wright} {et~al.}(2008){Wright}, {Marcy}, {Butler}, {Vogt}, {Henry},
  {Isaacson}, \& {Howard}}]{2008ApJ...683L..63W}
{Wright}, J.~T., {Marcy}, G.~W., {Butler}, R.~P., {et~al.} 2008, \apjl, 683,
  L63, \dodoi{10.1086/587461}

\bibitem[{{Xiao} {et~al.}(2023){Xiao}, {Liu}, {Teng}, {Wang}, {Brandt}, {Zhao},
  {Zhao}, {Zhai}, \& {Gao}}]{xiao_154345}
{Xiao}, G.-Y., {Liu}, Y.-J., {Teng}, H.-Y., {et~al.} 2023, Research in
  Astronomy and Astrophysics, 23, 055022, \dodoi{10.1088/1674-4527/accb7e}

\end{thebibliography}
\bibliographystyle{aasjournal}

\end{document}